\documentclass[bibyear]{aa}  
\usepackage{lscape}
\usepackage{graphicx}
\usepackage{txfonts}
\usepackage{multirow} 
\usepackage{longfigure}
\usepackage{supertabular}	
\usepackage{changepage}
\usepackage{nopageno}
\usepackage{afterpage}
\begin{document} 
\pagestyle{fancy}

   \title{The detection and characterisation of 54 massive companions with the SOPHIE spectrograph}
   \subtitle{7 new brown dwarfs and constraints on the BD desert}
   \titlerunning{7 new brown dwarfs}   
   \authorrunning{F. Kiefer et al.}

   \author{F.~Kiefer
          \inst{1}\fnmsep\thanks{\email{flavien.kiefer@iap.fr}},
          G.~H\'ebrard\inst{1,2}, 
          J.~Sahlmann\inst{3}, 
          S.\,G.~Sousa\inst{4,5}
          T.~Forveille\inst{6},
          N.~Santos\inst{4,5},
          M.~Mayor\inst{7}, 
          M.~Deleuil\inst{8},
          P.~A.~Wilson\inst{1,9},
          S.~Dalal\inst{1},
          R.\,F.~D\'{\i}az\inst{10,11},
          G.\,W.~Henry\inst{12},
          J.~Hagelberg\inst{13}, 
          M.\,J.~Hobson\inst{8},
          O.~Demangeon\inst{4},
          V.~Bourrier\inst{7},
          X.~Delfosse\inst{6},  
          L.~Arnold\inst{2},
          N.~Astudillo-Defru\inst{14},
          J.-L.~Beuzit\inst{8},
          I.~Boisse\inst{8},
          X.~Bonfils\inst{6},
          S.~Borgniet\inst{15},
          F.~Bouchy\inst{7},
          B.~Courcol,
          D.~Ehrenreich\inst{7},
          N.~Hara\inst{7},
          A.-M.~Lagrange\inst{6},
          C.~Lovis\inst{7},
          G.~Montagnier\inst{1,2},
          C.~Moutou\inst{8},
          F.~Pepe\inst{7},
          C.~Perrier\inst{6},
          J.~Rey\inst{16},
          A.~Santerne\inst{8},
          D.~S\'egransan\inst{7}, 
          S.~Udry\inst{7}, \and
          A.~Vidal-Madjar\inst{1}
          }

          \institute{\inst{1}Institut d'Astrophysique de Paris,   98 bis, boulevard Arago,  75014, Paris;
   	          \inst{2}Observatoire de Haute-Provence, CNRS, Universit\'e d'Aix-Marseille, 04870 Saint-Michel-l'Observatoire, France;
   	          \inst{3}Space Telescope Science Institute, 3700 San Martin Drive, Baltimore, MD 21218, USA;
   	          \inst{4}Instituto de Astrof{\'\i}sica e Ci\^encias do Espa\c{c}o, Universidade do Porto, CAUP, Rua das Estrelas, 4150-762 Porto, Portugal;
                   \inst{5}Departamento\,de\,F{\'\i}sica\,e\,Astronomia,\,Faculdade\,de\,Ci\^encias,\,Universidade\,do\,Porto,\,Rua\,Campo\,Alegre,\,4169-007\,Porto,\,Portugal;
                   \inst{6}Univ. Grenoble Alpes, CNRS, IPAG, 38000 Grenoble, France;
                   \inst{7}Observatoire de Gen\`eve,  Universit\'e de Gen\`eve, 51 Chemin des Maillettes, 1290 Sauverny, Switzerland;
   	          \inst{8}Aix Marseille Univ, CNRS, CNES, LAM, Marseille, France;
   	          \inst{9}Department of Physics, University of Warwick, Coventry, CV4 7AL;
   	          \inst{10}Universidad de Buenos Aires, Facultad de Ciencias Exactas y Naturales. Buenos Aires, Argentina;
                   \inst{11}CONICET - Universidad de Buenos Aires. Instituto de Astronom\'ia y F\'isica del Espacio (IAFE). Buenos Aires, Argentina;
   	          \inst{12}Center of Excellence in Information Systems, Tennessee State University, Nashville, TN 37209, USA;
   	          \inst{13}Institute for Particle Physics and Astrophysics, ETH Zurich, CH-8093 Zurich, Switzerland;
   	          \inst{14}Departamento de Matem\'atica y F\'isica Aplicadas, Universidad Cat\'olica de la Sant\'isima Concepci\'on, Alonso de Rivera 2850, Concepci\'on, Chile;
   	          \inst{15}LESIA, Observatoire de Paris, Universit\'{e} PSL, CNRS, Sorbonne Universit\'{e}, Universit\'{e} de Paris, 5 Place Jules Janssen, 92195 Meudon, France;
   	          \inst{16}Las Campanas Observatory, Carnegie Institution of Washington, Colina el Pino, Casilla 601 La Serena, Chile.
             }

   \date{Received ; accepted }

% \abstract{}{}{}{}{} 
% 5 {} token are mandatory
 
  \abstract
{ Brown-dwarfs (BD) are substellar objects with masses intermediate between planets and stars within about 13-80\,M$_\text{J}$. While isolated brown-dwarfs are most likely produced 
 by gravitational collapse in molecular clouds down to masses of a few\,M$_\text{J}$, a non-negligible fraction of low-mass companions might be formed through 
 the planet formation channel in protoplanetary disks. The upper mass limit of objects formed within disks is still observationnally unknown, the main reason being the strong 
 dearth of BD companions at orbital periods shorter than 10\,years, a.k.a. the brown-dwarf desert. }
   {We aim at determining the best statistics of secondary companions within the 10-100 M$_\text{Jup}$ range within $\sim$10\,au from the primary star, while minimising 
   observational bias. This can help determining the mass limit separating planet-formed from star-formed brown-dwarfs. Moreover, the exact shape of the BD desert
   in a mass-period space is still underdetermined, and can strongly constrain the companion-star interactions mechanisms at work in close binary systems at small mass ratio. }
   {We made an extensive use of the radial velocity (RV) surveys of FGK stars below 60\,pc distance to the Sun and in the northern hemisphere performed with the SOPHIE spectrograph at
   Observatoire de Haute-Provence. We derived the Keplerian solutions of the RV variations of 54 sources. Public astrometric data of the Hipparcos and Gaia missions 
   allowed deriving direct astrometric solution of orbital motion and constraining the mass of the companion for 
   most sources. We introduce GASTON, a new method to derive inclination combining RVs Keplerian and astrometric excess noise from Gaia DR1.}
   % results
   {We report the discovery of 12 new BD candidates. For 5 of them, additional astrometric data led to revise their mass in the 
   M-dwarf regime. Among the 7 remaining objects, 4 are confirmed BD companions, and 3 others are likely also in this mass regime. Moreover, we report 
   the detection of 42 objects in the M-dwarf mass regime 90\,M$_\text{J}$--0.52\,M$_\odot$. 
   The resulting $M\sin i$-$P$ distribution of BD candidates shows a clear drop in the detection rate below $80$-day orbital period. Above that limit, the BD 
   desert reveals rather wet, with a uniform distribution of the $M\sin i$. We derive a minimum BD-detection frequency around Solar-like stars of 2.0$\pm$0.5\%.}
   {}


   \maketitle
%
%-------------------------------------------------------------------

\section{Introduction}
According to the classical convention, brown dwarfs are substellar objects whose mass is too small to maintain hydrostatic equilibrium thanks to hydrogen-based nuclear 
reactions, while massive enough to ignite Deuterium nuclear reactions in the core, at least for few million years. Following this definition, the brown-dwarf domain is framed 
within the mass range 13--80 $M_\text{J}$. These boundaries may vary according to intrinsic stellar properties, such as metallicity (Chabrier \& Baraffe 1997, Spiegel et al. 2011). 
Defining these limits is the subject of many debates (in e.g. Saumon et al., 1996, Chabrier \& Baraffe 2000, Luhman et al. 2007, Luhman 2012, Chabrier et al. 2014) out of which it is 
proposed that the existence of nuclear reactions in its core is not the crucial parameter to define the nature of a substellar body. 

The observation of objects with masses as low as 5\,M$_\text{J}$ in young stellar clusters is a strong evidence that molecular cloud fragmentation is not 
limited in mass and can form objects in the brown dwarf and giant planet mass regime (see e.g. de Marchi, Paresce \& Portegies Zwart 2010). This is well reproduced by star 
formation simulations (Chabrier 2003, Luhman 2012, Lee \& Hennebelle 2018). Moreover, the mass distribution of widely separated binaries extends well within the 
brown dwarf domain (see e.g. Burgasser et al. 2007 and reference therein). On the other hand, a dearth of detections of brown dwarf companions with orbital periods shorter than
10\,yrs, the so-called brown dwarf desert (Halbwachs et al. 2000, Grether \& Lineweaver 2006), is followed by an increase of detection frequency at masses lower than 10\,M$_\text{J}$ 
(Marcy \& Butler 2000, Udry et al. 2002). This shows that giant planets and substellar objects that were formed like stars overlap on a few tens of Jupiter masses. 

Planet formation pathways, such as disk instability or core accretion, could in principle allow to form bodies up to 40 M$_\text{J}$ within protoplanetary 
disks (Pollack 1996, Boss 1997, Ida \& Lin 2004, Alibert et al. 2005, Mordasini et al. 2009). Knowing the extent of the tail of the distribution of giant planets within the BD domain 
could thus help constraining the planet formation models. This tail is yet undetermined because the statistics of detections of substellar companions in the 5-40 M$_\text{J}$ are still 
poor, though the observational efforts made in the recent years have led to abundant detections of brown dwarf companions with diverse instrumental methods (Sozzetti \& Desidera 2010, 
Sahlmann et al. 2011, Diaz et al. 2012, Ranc et al. 2015, Wilson et al. 2016). 

A difficulty arises due to the dearth of brown dwarf companion detected at short orbital periods, i.e. the brown dwarf desert. For some reason, the presence of substellar 
companions is forbidden at close distance of a more massive primary star. This implies that the mass-period distribution of brown dwarf companions to sun-like stars are 
deformed by possibly several perturbing effects, such as tidal interactions, magnetic braking and tidal dissipation (Guillot et al. 2014). This strongly bias the determination of 
the real mass distribution of giant planets and very low-mass stars.

It is thus necessary to constrain the minimum orbital period above which this effect becomes negligible. Mixing the results from several surveys done with diverse detection 
methods, Ma \& Ge et al. (2014) proposed a restricted BD desert enclosed within  P$<$100\,d and $30<M<60$\,M$_\text{J}$, with a mass separation between star-like and 
planet-like BD at 43\,M$_\text{J}$. More recent micro-lensing detections (Ranc et al. 2015), and the results of RV and astrometry (Wilson et al. 2016)
added to already published detections tends to confirm the framing of the desert at periods lower than 100 days. But the use of detections arising from several diversely biased 
or incomplete surveys is perilous. To our knowledge, there exists no fully complete non-biased statistical sample of detected brown-dwarfs companions. 

It would be most valuable to achieve a survey of brown-dwarf companions that is non-biased, or at least for which the selection function of the followed-up sample is well known and
allow deriving a meaningful statistics of BD population. 
Some of the most problematic issues with gathering detections from multiple surveys, apart from instrumental bias, are the diverse observers own interests. Observations are usually
stopped as soon as the followed-up target is not anymore of interest regarding the given study. Typically, on one side, sources with a companion that is not within the planetary 
mass domain, beyond about 20\,M$_\text{J}$, are not continued and not always published. On the other side, orbits and mass ratio of obvious stellar binaries are easily 
characterised and published. It follows that brown dwarfs within the BD desert and especially at periods larger than 1 year are under-sampled. 

The volume-limited FGK stars survey program for searching giant planets with the SOPHIE spectrograph installed at the Observatoire de Haute-Provence (Bouchy et al. 2009, 
H\'ebrard et al. 2016) offers a well-constrained framework for characterising the statistics of BD companions around solar-like stars. The target sample includes about 
2350 sources among all 2950 known FGK stars of the northern sky ($\delta$$>$$+00$:$00$:$00$) in the neighbourhood of the Sun below 60\,pc, and  
in the main sequence ($\pm$2 mag.), with +0.35$<$$B$-$V$$<$+1 (Dalal et al., in prep).  
To this date, around 2050 sources were observed at least 3 epochs each, with an aimed signal-to-noise ratio (SNR) per spectrum of at least 50.

The reflex-motion due to brown-dwarfs within the desert leads to RV amplitudes larger than 100\,m\,s$^{-1}$. Since the SOPHIE spectrograph is able to 
detect RV signals as low as a few\,m\,s$^{-1}$ (Courcol et al. 2015) on a time baseline of 13 years, brown dwarfs companions can easily be detected around nearby bright stars. 
Therefore, we expect to reach eventually almost 100\,\% completion of RV-detected brown-dwarf candidates with orbital period less than 10,000\,days around FGK 
stars in this volume-limited sample of stars, which visual magnitude is brighter than 11. 

In the continuation of the work of D\'iaz et al. (2012) and Wilson et al. (2016) which published several new objects in the brown dwarf desert with SOPHIE, we present here the 
latest results of this radial velocity survey on 54 solar-like sources with spectral types ranging from K5 to F5. With RV only, we report 12 BD candidates with $M\sin i$ within 
15--90\,M$_\text{J}$, among which 8 never published. 

We conservatively extend the BD-domain above 80\,M$_\text{J}$ in order to include objects in the grey zone 
80-90\,M$_\text{J}$, separating M-dwarf from brown dwarfs. We think this is justified for essentially three reasons. First, 
there is always an uncertainty (up to few M$_\text{J}$) on the $M\sin i$ derived with RV. Second, the mass limit for Hydrogen-burning is not a strict one, 
and may vary according to e.g. metallicity from 83 to 75\,M$_\text{J}$ within M/H$\sim$[-1 ; 0] (Chabrier \& Baraffe 1997). And third, extending towards low-mass M-dwarfs
allows exploring the tail of the BD-mass distribution on the stellar side. 

Although velocimetry is an efficient mean to detect companions, either stellar, sub-stellar or planetary, to stars, it also comes with a drawback. The inclination of the system 
being unknown, the derivation of orbital parameters of the star can only lead to determine the companion mass up to a factor depending on inclination. We present in this work 
exact mass derivations using astrometry with Hipparcos and Gaia. In particular, Gaia's intermediate data being yet unpublished, we developed the GASTON method to 
make use of Gaia released data to constrain the inclination of the systems studied here.

In Section~\ref{sec:motiv} we present the target selection. In Section~\ref{sec:obs} we review the observations performed and the targets observed. In Section~\ref{sec:spectro}, 
the spectroscopic analysis of the SOPHIE's observations is discussed, including the result of Keplerian fitting to the radial velocity variations. In Sections~\ref{sec:astrometry} and
~\ref{sec:gaia_astro} we study the astrometric measurements made with Hipparcos and Gaia. In Section~\ref{sec:details} we review the 7 discovered brown dwarfs. And finally, 
in Section~\ref{sec:discussion}, we discuss the implication of the presented results on the brown dwarf desert localisation. We conclude in Section~\ref{sec:conclusion}.

\section{Target selection}
\label{sec:motiv}

The goal of the programme in which this study takes place, is to complete a meaningful unbiased statistic of companions detected within and about the brown-dwarf mass regime, 
and up to 10\,yrs period. Extracting brown dwarf candidates out of a sample of stars which selection function is well controlled, gives us the opportunity to constrain the location of the BD desert in 
terms of period and mass. 

In the framework of the volume-limited FGK stars survey program for searching giant planets with the SOPHIE spectrograph (Bouchy et al. 2009, H\'ebrard et al. 2016), observers 
have collected RVs for many massive objects, including companions with $M\sin i$$>$$15$\,M$_\text{J}$, on a time span larger than 10 years. 
This could allow the determination of the orbit of BD companions with period as large as 10 years. 

In order to gather the largest possible number of brown-dwarfs in the BD desert, and to be able to compare the brown-dwarf population to the low-mass star population, we were especially 
focused on sources with companion masses in the broad $M\sin i$ range of 20--150\,M$_\text{J}$. This range includes the whole BD regime, from the upper end of the 
giant planets domain, but also extends up to the late M-dwarfs domain.  
We thus continued the radial velocity monitoring of the sources that present any clue of a companion within 20-150\,M$_\text{J}$ with SOPHIE, along with the giant planet candidates
below 20\,M$_\text{J}$. Interested only in massive companions producing RV signals with large amplitudes, we aimed at a signal-to-noise ratio (SNR) per spectrum of at least 
30.

Table~\ref{tab:targets} summarises the basic informations on the 54 targets covered by the present study. It includes only sources for which we gathered more than 6 RV data 
and for which a meaningful Keplerian solution of the RV variations, or a lower mass limit beyond $150$\,M$_\text{J}$, could be derived. We excluded SB2 sources from this 
publication.

%--------------------------------------------------------------------
\section{Observations}
\label{sec:obs}
The observations were performed with the SOPHIE spectrograph, fiber-fed from the Cassegrain focus of the 1.93-m telescope at the Haute-Provence Observatory (OHP, France). 
It is installed in a temperature-stabilised environment and the dispersive elements are kept at constant pressure in order to provide high-precision radial velocities 
(Perruchot et al. 2008). The 39 spectral orders of SOPHIE cover the visible range between 3872\,\AA~and 6943\,\AA. The spectra were collected in high-resolution mode, 
which leads to a resolving power of $\sim$75,000 at 550 nm.
During exposition of the spectrograph to the stellar photons in the science fiber, the instrument is also exposed to the background sky in a second fiber allowing subtraction of 
scattered light contamination in the science spectrum. The exposure time was varied to reach a signal-to-noise ratio (S/N) of at least 30 per resolving element under varying 
weather conditions.

Radial velocities are derived by the standard data reduction pipeline (Bouchy et al. 2009), including spectrum extraction, telluric lines removal, sky spectrum removal, CTI correction, 
CCF computation, and barycentric earth radial velocity correction. In the reduction software, the CCFs are fitted by Gaussians to calculate the sources radial velocities (Baranne et al. 1996; Pepe et al. 2002). 
Moreover, the bisector spans (BIS) and full-width-at-half-maximum (FWHM) of each CCF are computed following Queloz et al. (2001). We did not correct the seasonal RV zero point variation, from standard 
stars variation (see e.g. Courcol et al. 2015) that are being tiny ($\sim$m/s) compared to the expected velocity variations amplitude ($\sim$km/s).

In June 2011, the SOPHIE spectrograph hexagonal fibers were installed, greatly improving the precision of the RV (Perruchot et al. 2011, Bouchy et al. 2013). Additionally a shift up to 
about 50\,m\,s$^{-1}$ in the measured velocities was observed on standard stars (Bouchy et al. 2013). We therefore separated the data about June 2011 (JD 2455731.5). Before that date, the data will be referred to as 
SOPHIE, and after that date, they will be referred to as SOPHIE+. Moreover a systematic noise of 5\,m\,s$^{-1}$ was quadratically added to the measured RV uncertainty of the SOPHIE 
data before June 2011 (Hebrard et al. 2016). 

Additional non-SOPHIE data were found in the literature, with occasionally, already published orbits. These are summarised in Table~\ref{tab:pub}. We make use of these additional 
data to maximise the precision on the derived companion mass and period, and present relevant refinements of the already published companions parameters. Some of the public data
were found in the SB9 catalogue\footnote{http://sb9.astro.ulb.ac.be} (Pourbaix et al. 2004).

Data points with less than half of the median S/N and large uncertainty on the radial velocity measurements were treated as outliers and discarded. The number of points 
given in Table~\ref{tab:targets} takes this into account, with an average of 18 SOPHIE spectra per star. Adding the other published data, the average number of RV points per source 
rises up to 27, with a minimum of 8 RV points per star and a maximum of 103. The RV coverage of all the stars spans between 475 days and 47 years, with a median at 8 years.

\section{Spectroscopic analysis}
\label{sec:spectro}
\subsection{Stellar parameters}
The stellar parameters, effective temperature, surface gravity, microturbulence and metallicity, were derived using the spectroscopic analysis methods described in Santos et al. (2013), 
and references therein; see also Sousa et al. (2018) for more recent updates. The method makes use of the equivalent widths of a list of Fe I and Fe II lines in the SOPHIE spectra, which number is given in Table~\ref{tab:spec_type}, and 
assuming local thermodynamical equilibrium (LTE). The software used for the parameter derivation is the 2014 version of the \verb+MOOG+ software (Sneden 1973) with one-dimensional Kurucz model atmospheres.
All derived parameters are given in Table~\ref{tab:spec_type}. We estimated the stellar mass and radius of the primary star using the Torres et al.~(2010) empirical relation. 
The $\log(g)$ were corrected in order to be calibrated on $\log(g)$ derived using asterosismology, following equation (4) in Mortier et al. (2014):
\begin{equation}
\log(g)_\text{sismo} = \log(g)_\text{spectro} -3.89 \pm 0.23 \times 10^{-4} T_\text{eff} + 2.10 \pm 0.14
\end{equation}

The host stars presented in this paper are of type K5 to F8 (from the SIMBAD catalog) on the main sequence with metallicities [Fe/H] ranging from $-0.3$ to $+0.3$\,dex. HD24505, HD109157 and HD204613 
that were reported as (sub)giants in Simbad are rather located in the dwarf regime according to the present derivation. In particular, the spectral type of HD204613 is reported in 
Simbad with the spectral type of a giant CH-star, G1IIIa:CH1.5 according to the analysis of photographic spectrogram done by Keenan \& McNeil (1989). Interestingly, the 
photometry and colorimetry of this star tends to be more compatible with a dwarf (Ginestet et al. 2000). In agreement with the most recent published analysis of spectra of this source done 
with MOOG by Karinkuzhi \& Goswami (2015), the present derivation leads to a G1V-IV, with an effective temperature of 5870 K, a surface gravity of 4.1 and metallicity of 
$-0.3$\,dex. 

We appended to the table the average activity indicator $\log R'_{HK}$ calculated using all spectra of each target, with the SOPHIE reduction software (Boisse et al. 2010). The uncertainties are 
estimated from the standard deviation of the mean, and an error of 0.1 dex was quadratically added to account for typical uncertainty of $\log R'_{HK}$ in SOPHIE spectra (Boisse et al. 2010).
Our targets show medium stellar activity level in general, with 17 targets having $\log R'_{HK}$$<$$-4.75$, a classical limit for separating active from weakly active stars (Santos et al. 2000). 

Among the 39 more active sources, 9 can be considered as highly active with $\log R'_{HK}$$>$$-4.5$. Nevertheless, the amplitude of the derived RVs are all larger than a 
few hundred m\,s$^{-1}$, while activity is expected to influence RV measurements only at the scale of a few tens of m\,s$^{-1}$ (Campbell et al. 1991, Saar \& Donahue 1997, 
Saar et al. 1998, Santos et al. 2000, Boisse et al. 2010). All the detections presented in this paper are securely those of true companions but the magnetic jitter of the most 
active stars will add scatter and imply larger uncertainty to the measurement of orbital parameters for the hosted companions.

\subsection{Search for activity and binarity indicators in the CCF}

We calculated the full width at half maximum (FWHM), the bissector span (BIS) and the signal-to-noise ratio (SNR) variations for all sources. This allows us to check whether the radial velocity 
variations are polluted by the light of the secondary. This typically happens if the mass ratio $q$ is greater than 0.6 (Halbwachs et al. 2014, Santerne et al. 2015). Any spectrum for which the FWHM or 
the bissector span showed anomalously large variation uncorrelated with SNR variations, was systematically verified for a secondary peak. In the sample presented here, we selected 
only targets for which no spectra showed obvious secondary peaks.

To verify the absence of weaker secondary peak pollution, we used the 2 indicators being the variations of the FWHM and of the bissector span (Santerne et al. 2014, Santerne et al. 2015). 
For each indicator, we performed a $\chi^2$-test of the "no-variation" null hypothesis, and calculated the Pearson correlation coefficient $R$  of FWHM or BIS with RVs. SOPHIE and SOPHIE+ 
datasets were considered separately, since the instrument update could have introduced changes in the reduction and the quality of the spectra (D\'iaz et al. 2016). Among all datasets, 
8 with less than 4 spectra were not analysed regarding these diagnostics, since any variations would hardly be meaningful. 

Initially the errorbars of the FWHM and the bissector were calculated using the correspondence with RV errors $\sigma_\text{FWHM}$$\sim $$(2-4)\times\sigma_\text{RV}$ and $\sigma_\text{BIS}$$\sim $$2\times\sigma_\text{BIS}$  
proposed in Santerne et al. (2014) and Santerne et al. (2015).  These multiplications factors can be refined here, 
comparing the scatter of the FWHM or BIS to the median RV uncertainty for every sources. The median factors found lead to
\begin{align}
& \sigma_\text{FWHM} \sim 5.8 \times \sigma_\text{RV} \nonumber \\
& \sigma_\text{BIS} \sim 2.1 \times \sigma_\text{RV} 
\label{eq:factors}
\end{align}

\begin{figure}
\centering
\includegraphics[width=89mm]{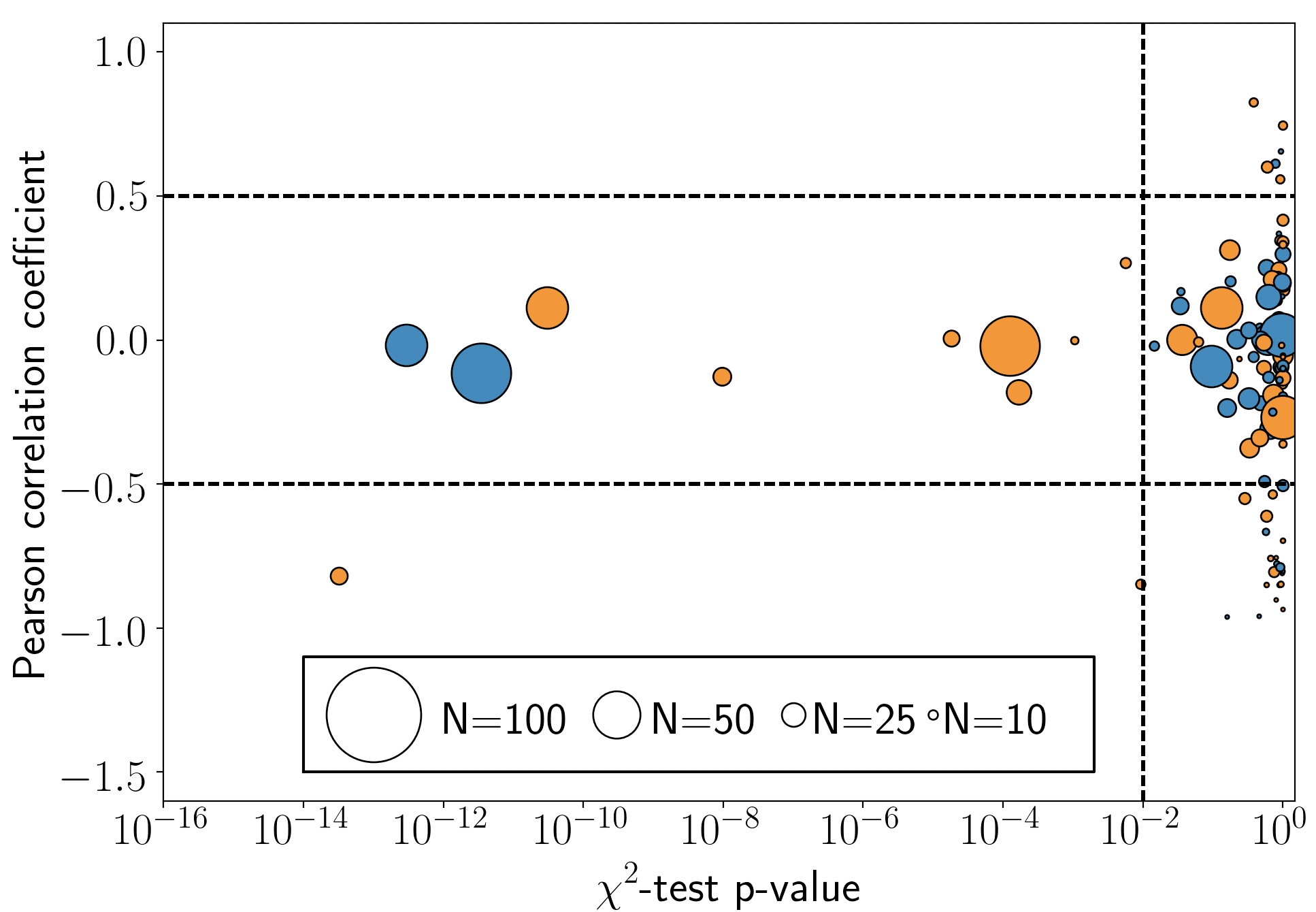}
\caption{\label{fig:SB2_variations} The p-value of the $\chi^2$-test vs the Pearson correlation coefficient, $R$, for both FWHM (in orange) and bissector span (in blue) indicators. 
The radii of the symbols are linearly scaled with the number of points considered. The dashed lines show the limits of significance for $R$ ($\pm$0.5) and for the p-value at 3$\sigma$ ($<$0.01).}
\end{figure}

Thus for the BIS we confirm the result of Santerne et al. (2015). For the FWHM, we found that the multiplication factor rather stands higher. These corrected factors were used 
eventually to calculate the $\chi^2$ test and the Pearson correlation coefficient that are summarised in Table~\ref{tab:SB2_variations} and presented in Figure~\ref{fig:SB2_variations}.

Only 10 systems show an FWHM or BIS dispersion that is significant with p-values lower than the 3$\sigma$ limit. But a single source shows also a strong correlation 
coefficient of FWHM with RV variations, $R(\text{FWHM, RV})$=-0.82. This system is HD77712, a K-type star with a medium activity level at $\log R'_{HK}$$\sim $$-4.7$. On 
the other hand, it presents no significant variations of the bissector. This is similar to the case of a triple system studied in Section 2.9 of Santerne et al. (2015) with the pollution of 
the CCF from a weak secondary peak always present at fixed radial velocity. A possible explanation is therefore that HD77712 is a triple system with a long period binary which 
secondary is polluting the CCF and a shorter period binary with a dark companion. The RV amplitude and $M\sin i$ we report for this system are likely underestimated. 

\begin{figure}
\centering
\includegraphics[width=89mm]{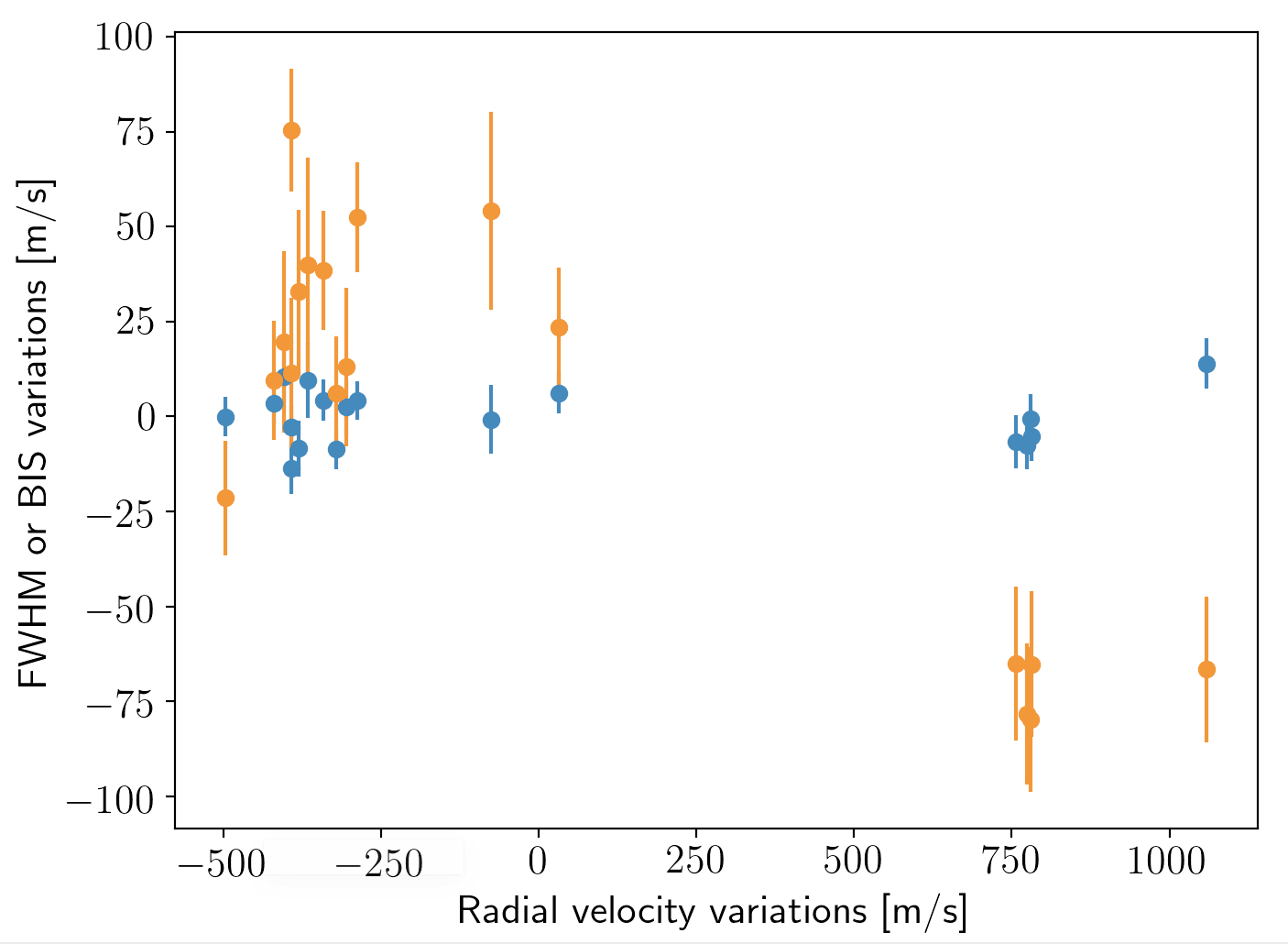}
\caption{\label{fig:HD77712} FWHM (orange) and bissector span (blue) variations of the cross-correlation function of HD77712 spectra. The error bars are calculated from the 
RV uncertainties, multiplied by the factors given in Equation~\ref{eq:factors}.}
\end{figure}

\subsection{Keplerian orbits fitting}
The \verb+yorbit+ software (Segransan et al. 2011) was used to calculate the solution by genetic algorithm refining of initial parameters for a Levenberg-Marquardt optimisation. This leads
to priors for an MCMC estimation of errorbars following D\'iaz et al. (2014,2016). MCMC was applied on 1,000 iterations. The varied parameters are the period $P$, the RV amplitude 
$K$, the eccentricity $e$, the angle of periastron $\omega$, the periastron passage time $T_0$, and the offsets $\gamma_\text{S}$ and $\gamma_\text{S,+}$ for SOPHIE and SOPHIE+ 
datasets respectively. Specific additional offsets are used for supplementary datasets as indicated in Table~\ref{tab:pub}. When the RV errorbars of the previously published 
data are not given, they are uniformly fixed to the unbiased standard deviation of the residuals as soon as a good orbital solution is found. Following Anderson et al. (2012), 
eccentricities compatible with zero at 2$\sigma$ were subsequently fixed to zero and the solution recalculated with these new constraints. In this case, $T_p$ indicates the 
epoch of the transit - if the system were to be edge-on.

The final parameters values given in the results section are the median of the MCMC distribution, and the symmetric error bars calculated by the standard deviation of the MCMC 
distribution. The errorbars defined by the confidence interval (CI) at 68.3\% around the best-value are barely asymmetric, while the difference between median 
and best-value is not significant. We found that the standard deviation gives more conservative uncertainties than the CI at 68.3\%. We thus uses the standard deviation as 
errorbar in order to keep on the conservative side, especially for cases with inaccurate derivation of the orbital parameters. For incompletely covered orbits the MCMC distributions 
have non-Gaussian tails: the interval for a confidence interval equivalent to 3-$\sigma$ is almost certainly not just 3 times broader than that for 1-$\sigma$.

For 7 stars, HD5470, HD7747, HD153376, HD193554, HD207992, HD212735 and BD+212816, SOPHIE and SOPHIE+  data are not sufficiently numerous separately to derive a meaningful 
solution with $\Delta\gamma$=$\gamma_\text{S,+}-\gamma_\text{S}$ on the order of $50$\,m\,s$^{-1}$ at most. In those cases, we fixed $\Delta\gamma$=0 to derive the solution. 

All the results of the Keplerian fits are summarised in Tables~\ref{tab:orbits},~\ref{tab:orbits_SB1} and~\ref{tab:trip}. We find 51 binary systems and 3 triple systems. These 
divide subsequently in 2 categories of companions, BD candidates and M-dwarfs. We have 11 binary BD candidates in the mass range 15-90\,M$_\text{J}$ (Tables~\ref{tab:orbits}), 
and 40 binary companions in the M-dwarf regime (Table~\ref{tab:orbits_SB1}). These are presented in more details in Section~\ref{sec:binary} below. The results for the 3 triple 
systems are presented in Table~\ref{tab:trip} and in Section~\ref{sec:multi}. They include 1 BD candidate and 2 M-dwarfs, and in both cases, a drift which requires a companion 
mass above 5\,M$_\text{J}$. For the targets with additional public data from the literature, residuals $O-C$ and RV center-of-mass offset $\gamma$ of each additional 
dataset are given in Table~\ref{tab:res}. Finally for the objects in the 15-90\,M$_\text{J}$ mass range, Keplerian solutions and residuals are plotted in Fig.~\ref{fig:solutions_BD}. 
For the objects beyond 90\,M$_\text{J}$ the solutions are shown individually in Fig.~\ref{fig:solutions_dM}. 

In general, the fits are accurate with precision on the orbital elements better than 7\% in 90\% of the cases, and a median precision of at most 1\%. A few cases show however a 
highly inaccurate derivation of orbital elements that is due mainly to an incomplete covering of the full orbital phase. For HD85533, although the uncertainty on the period is 
$\sim$100\%, the given value is a lower-limit, and the companion should be at least as massive as 450\,M$_\text{J}$. On the other hand, the eccentricity is surprisingly accurate
with an error of only 20\%. This results from a better coverage of an inflexion in the RV curve that leads to a good fit only for eccentricities larger than 0.44. This stands also for 
HD13014, as well as HD40647, HD60846 and HD146735, for which the period, RV amplitude and companion mass, already in the M-dwarf domain, are likely underestimated, 
while the eccentricity is conversely better constrained. 

Finally, the O-C residuals lies below 10\,m\,s$^{-1}$ except for a few active sources that show much larger dispersion of residuals close to 40\,m\,s$^{-1}$. We discuss the distribution
of the residuals in more details, especially comparing the SOPHIE and SOPHIE+ datasets below in Section~\ref{sec:compS+S-}, and confronting to activity indices for the observed
sources in Section~\ref{sec:OC_activ}.

\subsection{Comparing SOPHIE and SOPHIE+}
\label{sec:compS+S-}

\begin{figure}
\includegraphics[width=89mm]{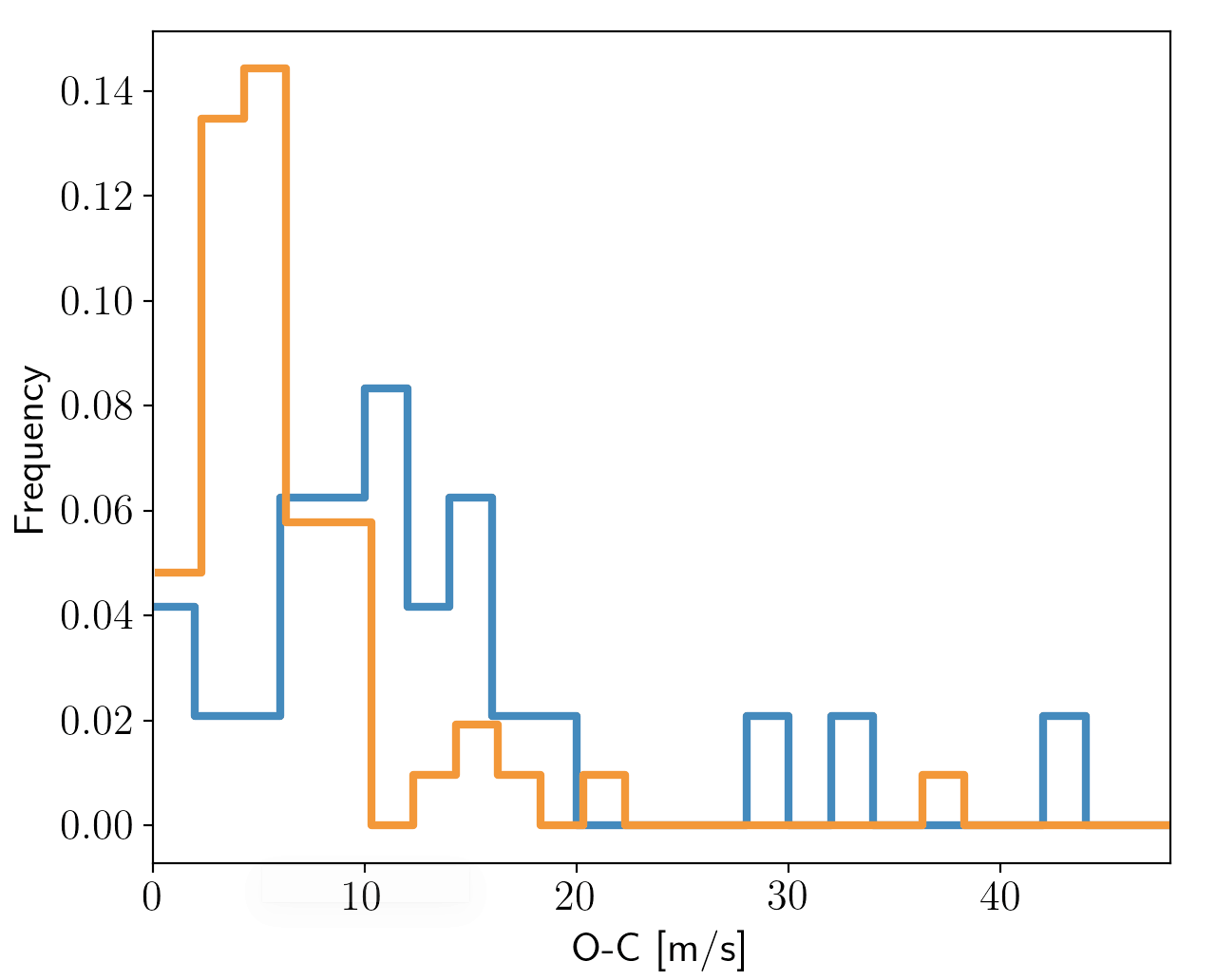}
\includegraphics[width=89mm]{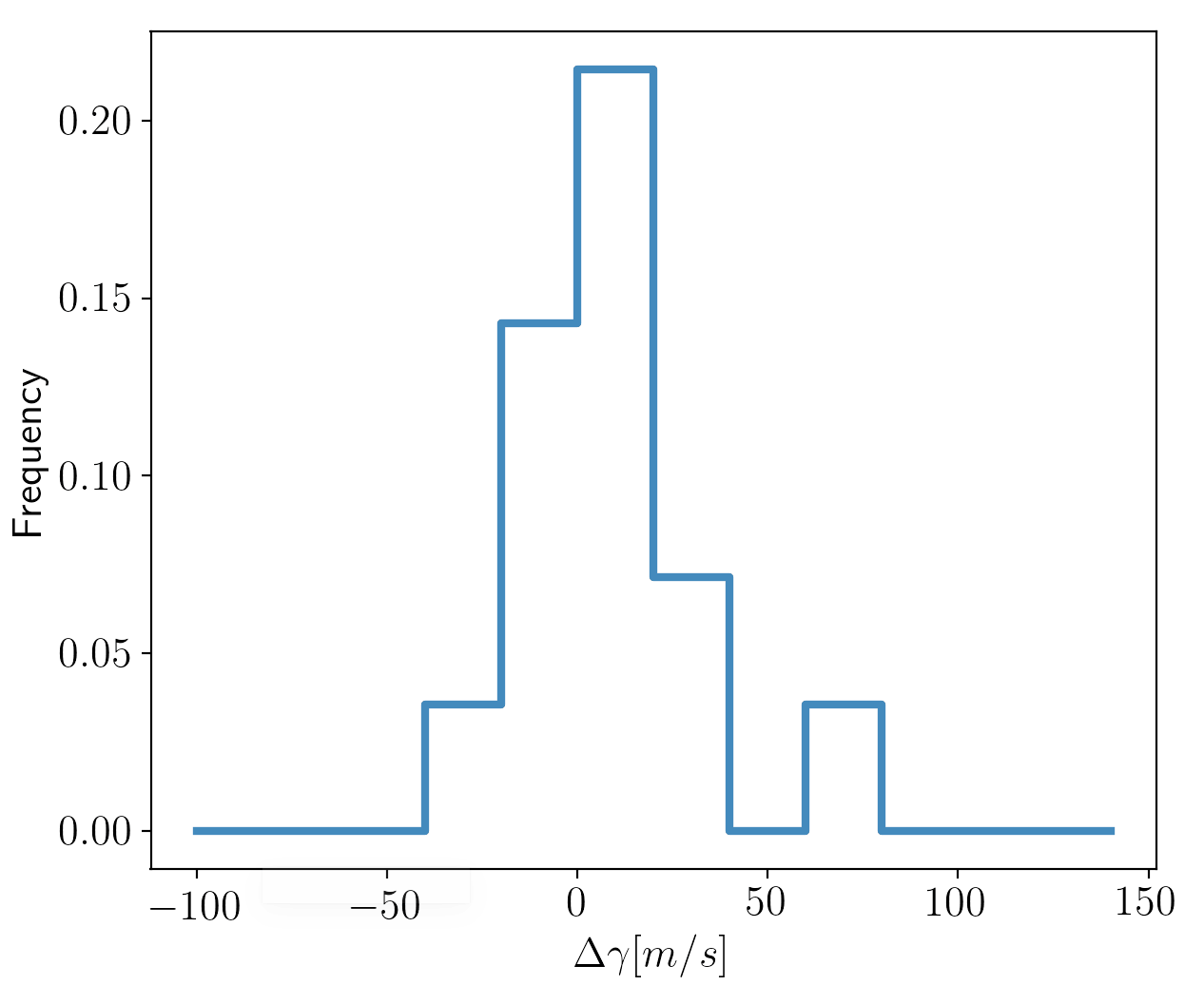}
\caption{\label{fig:O-C}Upper panel: O-C distribution of SOPHIE (blue) and SOPHIE+ (orange) datasets. Lower panel: RV offset distribution between SOPHIE and SOPHIE+ datasets.}
\end{figure}

Analysing the O-C residuals of the Keplerian fits allows us to verify the accuracy of SOPHIE data and in particular comparing the quality of the measurements before and after the 
instrument upgrade in June 2011. The standard deviation of the residuals can give the actual precision of the RV
measurements, because the targets in the sample are of similar spectral type, with similar CCF shape and FWHM below $\sim$10\,km\,s$^{-1}$, as shown in Table~\ref{tab:SB2_variations}. 
Moreover, with 23 targets observed with both SOPHIE and SOPHIE+ instruments, we can characterise the typical RV offsets between the 2 datasets. 
Summarising the data gathered in Tables~\ref{tab:orbits} and~\ref{tab:orbits_SB1}, we show the distribution of O-C values in SOPHIE and SOPHIE+ configurations separately, 
and the RV offset distribution in Figure~\ref{fig:O-C}. 

The distribution of RV offsets between SOPHIE and SOPHIE+ is centred on
\begin{equation}
\Delta \gamma \sim 11 \pm 27 \,\text{m\,s}^{-1}
\end{equation}

This is compatible with the results found by Bouchy et al. (2013) that bounds the RV shift due to the upgrade to 0-50\,m\,s$^{-1}$. To obtain this distribution, we assumed that the
offset should not exceed 100\,m\,s$^{-1}$, in which case the offset should be better explained by a slow drift, due to third companion. This led to consider a few systems as rather 
multiple than binary, as shown in Section~\ref{sec:multi} above.

The core of the distribution of O-C residuals standard deviation leads to the following estimation of RV accuracy for SOPHIE and SOPHIE+ configurations:
\begin{align}
&\sigma_\text{SOPHIE} \sim 11\,\text{m\,s}^{-1} \nonumber \\
&\sigma_\text{SOPHIE+} \sim 5 \,\text{m\,s}^{-1}
\end{align}

These values are in line with the results obtained by H\'ebrard et al. (2016) where a median RV accuracy of about 7\,m\,s$^{-1}$ is derived for observations before June 2011,
and of about 3.5\,m\,s$^{-1}$ for observations taken after the instrument update. The values derived here are higher, which might be explained by the activity index greater
than -4.65 for about half of our sample, the absence of instrumental drift ($\sim$m\,s$^{-1}$) corrections, as derived in Courcol et al. (2015), in the present study. 
Moreover, our observed sample also includes spectra with SNRs down to 30, while the H\'ebrard et al. (2016) sample only includes spectra with SNR$>$50.

Comparing the O-C individually for every targets confirms that the general tendency is of a reduction in the RV dispersion after the upgrade of the instrument towards a value
close to 5\,m\,s$^{-1}$ . If not the case, it should be explained in terms of supplementary signal in the RV, 
due to either activity jitter, or planetary signal. Among our sample, only HD23965, HD40647, and HD161479, admit large O-C 
dispersion $>$14\,m\,s$^{-1}$ in both datasets. This is however most likely explained by their significant activity index $\log R'_{HK}$$>$$-4.5$ (Table~\ref{tab:spec_type}). We can
thus exclude that more planetary signals are hidden in any residuals beyond an amplitude of about 8\,m\,s$^{-1}$.

\subsection{Residuals dispersion and magnetic activity}
\label{sec:OC_activ}

\begin{figure}
\includegraphics[width=89mm]{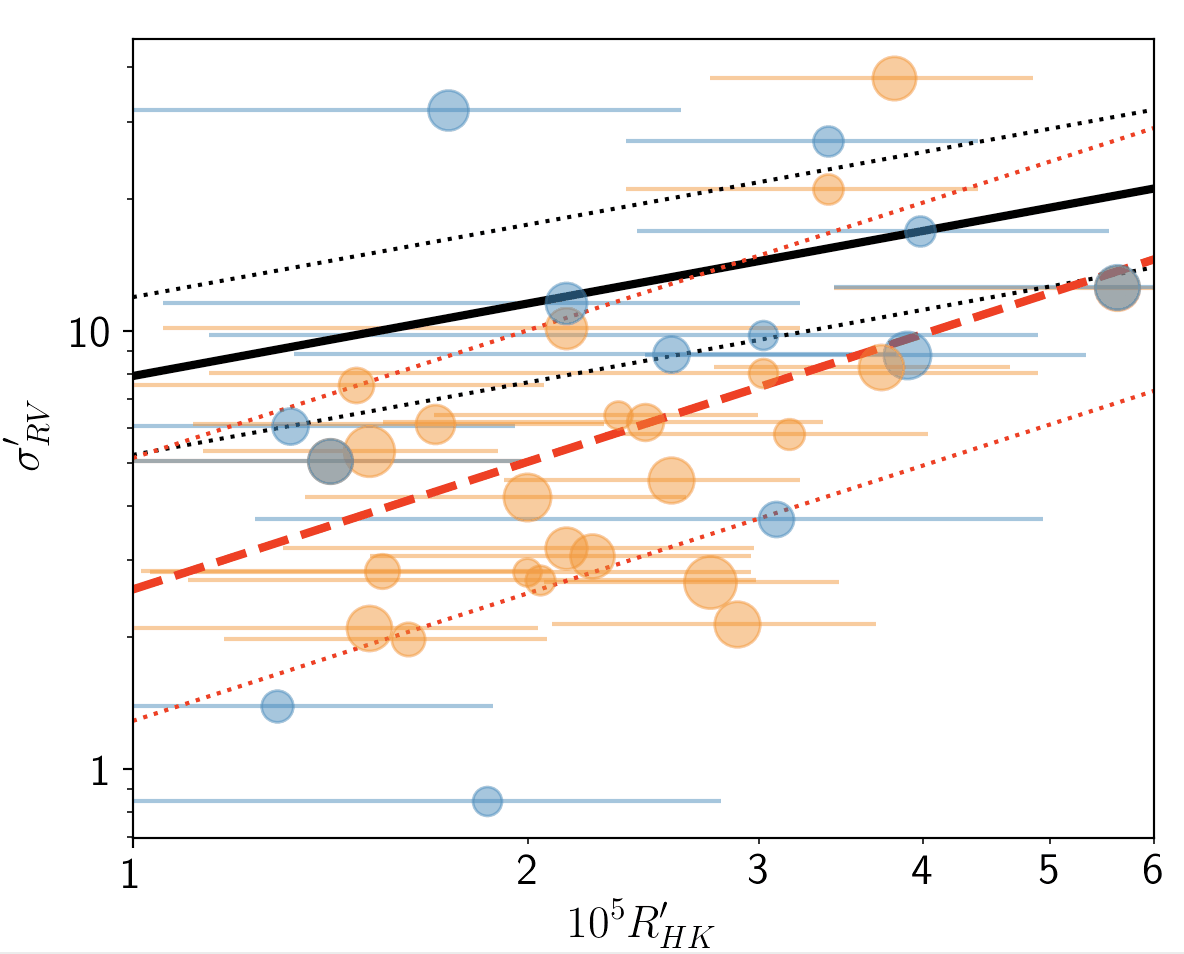}
\caption{\label{fig:sigma_RHK} $\sigma'_{v}$ (in m\,s$^{-1}$) plotted vs $10^5 R'_{HK}$ for SOPHIE (in blue) and SOPHIE+ (in orange) datasets. The symbol size is 
proportional to $B-V$ with values between 0.5 and 1.2. The black lines represent the relation derived by Santos et al. (2000) for G-type stars 
$\sigma'_v=7.8\,(10^5 R'_{HK})^{0.55}$ with a fit uncertainty of 0.18\,dex. The red lines represent the relation derived from the datasets of this work, 
$\sigma'_v=2.6\,(10^5 R'_{HK})^{1.0}$ with a fit uncertainty of 0.3\,dex.}
\end{figure}

The residuals of the Keplerian fit, $\sigma_v$, can be compared to the activity index $\log R'_{HK}$ to verify if it can explains the amplitude of the residuals, in general and in 
specific cases where it is exceptionally large. Previous study of the correlation between RV dispersion and magnetic activity was done  in Saar et al. (1998) and in 
Santos et al. (2000). We followed here the same procedure in order to make a point comparison. We first exclude datasets with less than 7\,pts ; then we quadratically subtract 
the mean internal RV error of all SOPHIE exposures $<\sigma_i>$ from the dispersion of the RV residuals for every targets, $\sigma'_v = \sqrt{\sigma_v^2 - <\sigma_i>^2}$. 
This should let only variations from the instrument itself and magnetic activity. Sources for which $\sigma_v$ is smaller than $<\sigma_i>$ were excluded from this analysis. 
Figure~\ref{fig:sigma_RHK} plots $\log R'_{HK}$ and $\sigma'_v$ as derived from our sample and compares to the relation obtain in Santos et al. (2000).

We observe that the dispersion of the residuals correlates well with the magnetic activity, with only few outlying points. But we see a discrepancy between our values and 
the relation derived for G-type stars in the CORALIE sample by Santos et al. (2000), where they find that $\sigma'_{v,G}=7.8\times(10^5 R'_{HK})^{0.55}$. In our case, the 
slope is stronger, with a linear fit of the log-log relation leading rather to
\begin{equation}
\sigma'_v=2.6\times(10^5 R'_{HK})^{1.0}
\end{equation} 

The uncertainty of the fit is $\sigma_\text{fit}$=0.3\,dex. The slope is closer to the relation obtained by Saar et al. (1998), $\sigma'_v\propto R_{HK}^{'1.1}$. After the exclusion
of the F and K type stars of our sample, keeping 0.6$<$$B$-$V$$<$0.8, there remains 18 G-type stars. It leads to a similar relation 
$\sigma'_{v,G}$$=$$3.6\times(10^5 R'_{HK})^{0.9}$ but a larger fit uncertainty of 0.4\,dex. 

The most significant outlier in Fig.~\ref{fig:sigma_RHK} at $\log R'_{HK}$$\sim $$4.75$ and $\sigma'_v$$\sim $$30$\,m\,s$^{-1}$ is HD207992. We collected 11 RV points 
in the SOPHIE configuration, but only 2 with SOPHIE+ for this source. The RV curve in Fig.~\ref{fig:solutions_dM} shows indeed variability in the residuals, which could be due to a 
supplementary signal for this relatively low activity star. In Table~\ref{tab:SB2_variations} we do not see any significant BIS nor FWHM variations. We conclude that this signal 
could be a tentative evidence of a third object in the system of HD207992.  

One other case is HD161479 with $\sigma_{v,S+}$=36\,m\,s$^{-1}$ and $\sigma_{v,S-}$=42\,m\,s$^{-1}$. This residuals dispersion is large, but might be compatible with 
magnetic activity since $\log R'_{HK}$=-4.42 for this K0 star. Moreover, according to Table~\ref{tab:SB2_variations} the bissector and FWHM variations are relatively significant. 
We measure a p-value of 0.001 for the no-var model of the FWHM in the SOPHIE+ dataset, and a strong correlation of  -0.95 for the bissector in the SOPHIE dataset, although 
based on only 4 points. We conclude that the supplementary RV variability of HD161479 is most likely due to magnetic activity.

\subsection{Results of the Keplerian fit}

In total, we characterised 54 massive companions in 54 different systems. We report the Keplerian orbit and $M\sin i$ measurements of 
12 brown dwarf candidates in the extended range 15-90\,M$_\text{J}$. One among the 12 is part of a triple system, HD71827, which discovery is reported here. We also 
characterised the orbit of 42 stellar companions with a mass in the M-dwarf regime 90\,M$_\text{J}$--0.52\,M$_\odot$. 
Two brown-dwarf candidates lie in the grey zone between the classical upper-limit of brown-dwarfs and the lower-limit of M-dwarf, 80--90\,M$_\text{J}$.

Moreover, we recall that the constraint on the mass obtained from velocimetry is only a lower limit because of the uncertainty on the inclination of 
the systems implying an unknown value of $\sin i$. We will see in Sections~\ref{sec:astrometry} and~\ref{sec:gaia_astro} that thanks to Hipparcos and Gaia astrometry we are 
able to add constraints on the inclination and thus the true mass for 46 of the candidates presented here. 

\begin{figure}
\centering
\includegraphics[width=89mm]{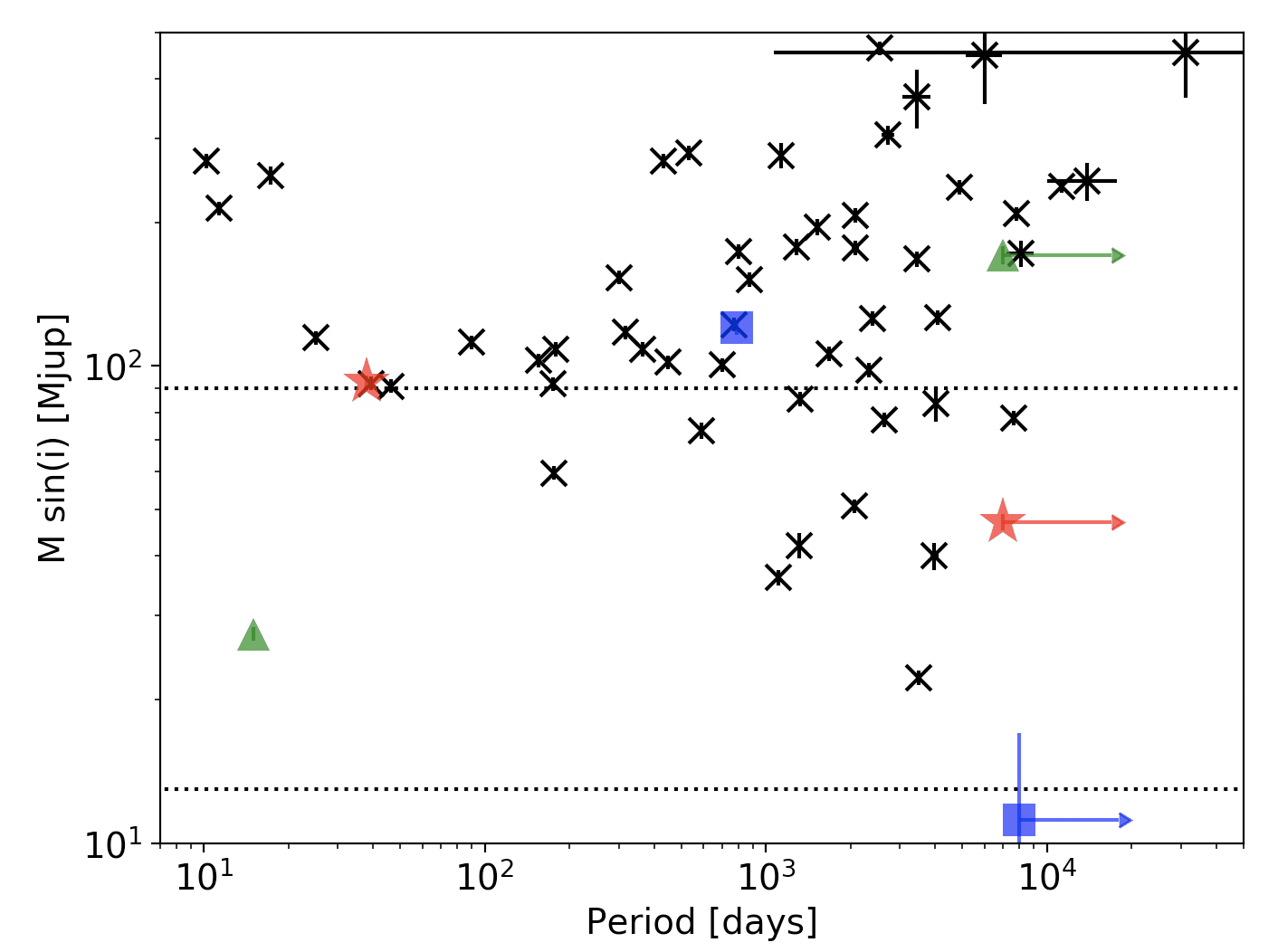}
\caption{\label{fig:P_msini} Period vs $M\sin i$ for the systems studied in this paper. The crosses show the new results of this paper. The 13 and 90\,M$_\text{J}$  limits
 are drawn as dotted lines. The plain white symbols represent stars with 2 companions. HD71827 b and c are represented as green triangles, HD212735 b and c as red stars, and 
 BD+212816 b and c as blue squares. Most errorbars are smaller than the symbols.}
\end{figure}

\begin{figure}
\centering
\includegraphics[width=89mm]{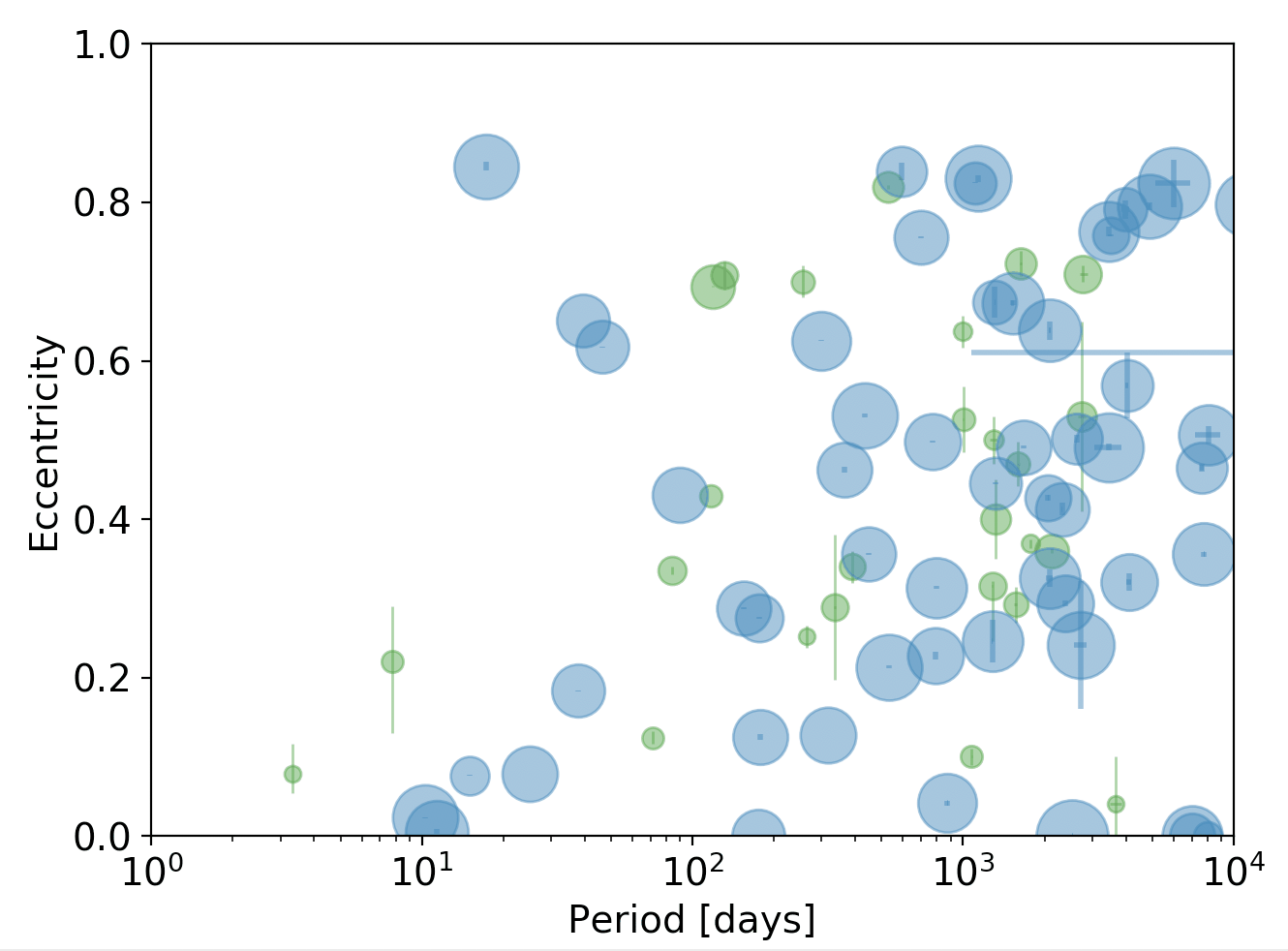}
\caption{\label{fig:P_ecc} Period vs eccentricity for the systems studied in this paper. The blue circles show the new results of this paper, while the green circles represent the 
Exoplanet.eu database with primaries verifying that $\delta$>0$^\circ$, +0.35$<$B-V$<$+1, $d$$<$$60$\,pc. The symbol size is proportional to the logarithm of $M\sin i$.}
\end{figure}

The $M\sin i$-period diagram summarising the results is shown in Fig.~\ref{fig:P_msini}. The period of the derived orbits are large in general, with only 9 companions below 100-days 
period. Among the latters, one is member of a triple system and has an $M\sin i$ within the BD regime. Eccentricities are large as well, with only 7 orbits with $e$$<$$0.1$. The eccentricities 
are dispersed around 0.42$\pm$0.27. Fig.~\ref{fig:P_ecc} shows the period-eccentricity distribution of our results, and compare it to the massive planets collected in the
Exoplanet.eu database with $M$$>$$4$\,M$_\text{J}$. We selected systems exclusively compatible with the constraints of our survey 
($\delta$>0$^\circ$, +0.35$<$B-V$<$+1, $d$$<$$60$\,pc, $\pm$2 mag from MS). The period--eccentricity distribution of the brown dwarfs reported in this work agrees 
with that of giant exoplanets. The eccentricities of giant exoplanets are fully compatible in average with the eccentricities of brown dwarfs with $e_\text{GP}$\,\!$\sim$0.42$\pm$0.22. 
This is in line with the conclusions of Sozzetti \& Desidera (2010) finding strong similarities in terms of eccentricity distributions between massive planets and BD. 

One candidate stands apart at small period and large eccentricity, BD+362641, for which the orbit is actually not well constrained because of the 
small number of points (N$_\text{RV}$=9). However, the large radial velocity variation observed of $\sim$40\,km\,s$^{-1}$ places it in the M-dwarf regime with a mass most 
likely larger than 200\,M$_\text{J}$. 

\subsubsection{Binary companions in the BD and M-dwarf regime}
\label{sec:binary}

Among the 12 detected BD candidates, 11 are components of a binary system. They have orbital periods shorter than 30 yr, or semi-major axis smaller than 10\,au. Eight 
of these brown dwarf candidates are brand new discoveries, among which we report 6 of them with $M\sin i$ strictly below 80\,M$_\text{J}$. This is a significant increase of 
the number of known BD candidates. We notice that the orbital period of all these companions is larger than 100\,days, even though massive companions with a minimum mass 
close to but larger than 90\,M$_\text{J}$ with an orbital period as low as 40\,days are also reported. Interestingly, adding BD detections around solar-like stars in the solar neighbourhood that 
are reported in previous papers tend to confirm this distribution. We discuss all the consequent improvements these new detections bring on the statistics of objects in the 
BD regime in Section~\ref{sec:discussion}. 

Four objects, HD28635, HD210631, HD211681, and HD217850 were already published as brown dwarf candidates. Improvements on their orbital parameters and $M\sin i$ are
summarised here:

\paragraph{\textbf{HD28635.}} Also known as "vB 88", it was reported hosting a BD companion with an approximate spectroscopic 
mass of 70$\,M_J$ using Keck/HIRES data (Paulson et al. 2004). Adding 13 SOPHIE and 3 Elodie data, we find that the RVs are compatible with a BD-mass companion at a 
period of 2636.8$\pm$2.2\,days with $M_2\sin i$$\sim $$77.1$$\pm $$2.7$\,M$_\text{J}$ and $a_2$$\sim $$4.014$$\pm $$0.068$\,au. 

\paragraph{\textbf{HD210631.}} Latham et al. (2002) reported a 82$\pm$6\,M$_\text{J}$ companion in this system. Adding SOPHIE data, we confirm this result, finding compatible 
minimum mass of 83.4$\pm$6.9\,M$_\text{J}$, with a period of 4030$\pm$40\,days at a separation of 4.976$\pm$0.085\,au. 

\paragraph{\textbf{HD211681.}} The companion of this sub-giant G5 was reported as a low-mass star with a minimum mass in the range 72-100$\,M_J$ by Patel et al. (2007) 
using Keck/HIRES data. Adding 30 SOPHIE and 12 ELODIE measurements, we are able to narrow down the $M\sin i$ range of the companion to 77.8$\pm$2.6\,M$_\text{J}$ with 
a period of 7612$\pm$131\,days at a semi-major axis of 8.28$\pm$0.16\,au. 

\paragraph{\textbf{HD217850.}} The radial velocities variations of this G8-type star were reported to be compatible with an 11\,M$_\text{J}$ companion in Butler et al. (2017) using 
an incomplete coverage of the orbit with Keck/HIRES data. Adding 41 SOPHIE data we find the lowest mass BD of our sample with an orbital period of 3508.2$\pm$2.6\,days, 
an $M_2\sin i$$\sim $$22.27$$\pm $$0.77$\,M$_\text{J}$ and $a_2$$\sim $$4.672$$\pm $$0.079$\,au. This is the candidate BD with lowest $M\sin i$ in our sample. \\

Finally, among the 42 massive companions in the M-dwarf regime, 40 form a binary system with their host star. For 24 of them, to our knowledge, this is the first publication 
of an RV orbital solution. For the 6 systems for which an RV orbit was already published, the last 2 columns of Table~\ref{tab:pub} summarises the improvement on the 
$M\sin i$ for these stellar companions.

\subsubsection{Triple systems}
\label{sec:multi}

We found evidence for a secondary drift signal in the RV data of 3 stars, BD+212816, HD71827, and HD212735. The result of fitting a single Keplerian and 
a drift for each system are summarised in Table~\ref{tab:trip}. In order to derive a minimum estimation of the mass of the second companion, we fitted the drift signal with a 
Keplerian with the shortest period possible compatible with a drift. 

For the 3 sources, the RV offset between SOPHIE or ELODIE, and SOPHIE+ datasets is significantly larger than 100\,m\,s$^{-1}$. It should be on the order of 10$\pm$30\,m\,s$^{-1}$ 
between SOPHIE and SOPHIE+ measurements (see Section~\ref{sec:compS+S-} below) and on the order of 50-100\,m\,s$^{-1}$ between ELODIE and SOPHIE+ (Boisse et al. 2013). 
This is the sign of a real drift due to a third companion in the system. We had to fix the RV offset to $\gamma_{S,+}-\gamma_{S}$=0\,km\,s$^{-1}$ in 
order to derive a Keplerian solution with a supplementary linear drift. Since the Keplerian of the drift signals cannot be constrained, we only report them here and do not 
include them in any other analysis in the rest of the paper.

\paragraph{\textbf{HD71827.}} It is a triple system composed of an F8-primary surrounded by one BD and a low-mass star. The 26\,M$_\text{J}$-BD stands at a short period 
of 15\,days. This is the only BD in our present sample with a period shorter than 100 days, and it is interesting to note that it is also part of a triple system with possible dynamical 
interaction. There are clear evidences in SOPHIE+ data of a second signal with a large period and confirm the presence of a cubic drift. The shortest period orbit found 
compatible with the drift leads to a minimum mass $\sim$163$\pm$7\,M$_\text{J}$ for a companion on a 20\,yrs orbit. 

\paragraph{\textbf{HD212735.}} Apart from an obvious 38-days period signal, the RVs of this system display a significant linear drift during the 10 years of data. Fitting a second 
long-period Keplerian to the drift signal leads to a minimum estimation of the period and the mass beyond 20\,yrs and 47\,M$_\text{J}$ for the tertiary. The outer companion is thus 
possibly a brown dwarf, but most likely an M-dwarf with a much larger period. 

\paragraph{\textbf{BD+212816}} The secondary companion of this K0-type star is an M-dwarf, but a supplementary long-period signal might be present as a drift. However, this 
linear drift is compatible at 2$\sigma$ with a constant. It should be considered as a possible, yet unconfirmed, triple system. The mass of the outer companion could be 
as low as 5\,M$_\text{J}$, but is likely much higher. \\

For all these triple systems future Gaia data releases or direct imaging could help probing for the third companion. In every cases, the semi-major axis of the outer orbit is larger 
than 7\,au, with a parallax on the order of $20$\,mas. Thus it could be seen with adaptive optics that can probe down to about 100\,mas in the neighbourhood of stars.

\section{Hipparcos astrometry}
\label{sec:astrometry}

In complement to the RV orbital derivation, the Hipparcos astrometry can allow to constrain the inclination of the systems as was performed in e.g. 
Sahlmann et al. (2011), D\'iaz et al. (2013), and Wilson et al. (2016). 

For all 54 systems of our sample, the new Hipparcos reduction catalog (van Leeuwen 2007) provides informations on the type of fitting solution ('5' for standard, 'X' for stochastic, 
and 'G' for accelerated solutions; see e.g. Perryman et al. (1997) or Lindegren et al. (1997), number of field-of-view transit, measurement time span, and abscissa measurement errors. 
A summary of these informations is presented in Table~\ref{tab:HIPparams}.  
 
After a preliminary analysis of all systems, we found 16 of them for which there are indications of significant orbital motions in the Intermediate Astrometric Data 
(IAD), plus 1 system, HD193554, already solved in the Hipparcos double star catalog (ESA 1997), and 18 systems for which it could be possible to derive an upper-limit on the 
astrometric motion due to the massive companion. Outliers in the IAD had to be removed because  they can substantially alter the outcome of the astrometric analysis.
The result of the Keplerian fit of the HD193554 astrometric motion analysis done by the Hipparcos team is given in Table~\ref{tab:HIP_double}. It compares well with our RV 
derivation. The true mass estimation for the companion is beyond the 90\,M$_\text{J}$ limit.

\begin{table}  \centering
\caption{\label{tab:HIP_double} The Hipparcos double star catalog orbital solution for HD193554, including estimation of the inclination $I_c$. This allows us to derive the 
true mass of the companions out of RV results here recalled for comparison.}
\begin{tabular}{@{}lc@{}}
 Parameters		& Values \\
\hline
$P_\text{HIP}$ [day]	&	832$\pm$50\\
$T_\text{HIP}$ [JD]	&	48574$\pm$51\\
$e_\text{HIP}$	&	0.33$\pm$0.14 \\
$\omega_\text{HIP}$ [$^\circ$]	& 147$\pm$39	\\
$I_c$ [$^\circ$]	&36$\pm$11	\\
$\Omega$ [$^\circ$]	&	98$\pm$21 \\
$a_\text{ph}$ [mas]	& 15.0$\pm$1.5	\\
\hline \\
$P_\text{RV}$ [day]	& 708.60$\pm$0.29 \\
$T_\text{RV}$ [JD]	& 56564.9$\pm$1.6 \\
$e_\text{RV}$		& 0.3093$\pm$0.0022	\\
$\omega_\text{RV}$ [$^\circ$]	& 140.12$\pm$0.64	\\
$f(m)$ [$10^{-6}$\,M$_\odot$]	&	4870$\pm$77	 \\
$M_2\sin i$ [M$_\text{J}$] 	& 181.6$\pm$5.8 \\
$a_1\sin i$   [mas]  & 7.78$\pm$0.86 \\
\hline \\
$M_2$	[M$_\text{J}$] & 309$\pm$82\\
\hline
\end{tabular}
\end{table}

For the 16 systems with a significant orbital solution, we fitted the astrometric measurements with a seven-parameter model, in which the free parameters are the inclination 
$i$, the longitude of the ascending node $\Omega$, the parallax $\varpi$, and offsets to the coordinates ($\Delta \alpha^{\star}$, $\Delta \delta$) and proper motions 
($\Delta \mu_{\alpha^\star}$, $\Delta \mu_{\delta}$). The other orbital parameters are fixed according to the radial velocity results given in Tables \ref{tab:orbits} and 
\ref{tab:orbits_SB1}. A two-dimensional grid in $i$ and $\Omega$ was searched for its global $\chi^2$-minimum. The statistical significance of the derived astrometric orbit was 
determined with a permutation test employing 1000 pseudo-orbits (Sahlmann et al. 2011). 

For all 16 sources except two, we detect the astrometric orbit with a significance $>$$2$$\sigma$. Those are listed in Table \ref{tab:hip_masses_16} with their orbital solution. 
Table \ref{tab:hip_ppm_16} lists updated parallaxes, proper motion, coordinates offsets, inclination and ascending node of the orbits. The updated parallaxes are compared to 
the DR2 parallaxes given in Table~\ref{tab:targets}. Moreover, the updated proper motions are compared to the Tycho-Gaia Astrometric Solution (TGAS) proper motion that should be closer to the actual proper 
motion of systems since based on a 24-years baseline of astrometric data. Finally, Figures \ref{fig:orbits_3sigma}, and \ref{fig:orbits_2sigma} show the significant orbits. 

In general, the updated Hipparcos-2 parallax are not compatible with the Gaia DR2 parallax at the 1-$\sigma$ level. Even for 2 systems, HD133621 and HD155228, the discrepancy
is larger than 3-$\sigma$. This shows that accounting for the orbital motion can lead to strong corrections of the published parallax on the order of $\sim$10\%. Besides, the 
comparison of the Hipparcos-2 proper motion corrections, after fitting the orbital motion, with the Gaia DR1 proper motion shows a global nice agreement, validating the 
solutions and corrections proposed in Tables~\ref{tab:hip_masses_16} and~\ref{tab:hip_ppm_16}. Indeed, we expect the proper motion derived in the TGAS sample of the DR1 
to be closer to the true linear proper motion of the system, since for those sources the astrometric solution takes into account Hipparcos-2 and Gaia measurements along a 
24-years baseline. 

In only one case, HD87899, there is no strong agreement of the proper motion corrections. With a long period of 4.2 years, the phase coverage of the Hipparcos-2 
measurements is only partial, and the proper motion corrections are quite uncertain with a given error of 2.8\,mas/yr. Thus, the derived orbital parameters for HD87899 should be 
considered only conservatively within their 3-$\sigma$ errorbars. If the Gaia DR1 proper motion is correct, the semi-major axis as derived with the Hipparcos-2 data should be 
rather around 10-15\,mas and the mass closer to 0.2\,M$_\odot$, which is more in line with what will be derived using Gaia data only in Section~\ref{sec:gaia_astro}.

For the two stars HD110376 and HD155228, the F-test of the orbital model and the permutation test yield significantly discrepant results. The F-test indicates orbit detection 
whereas the permutation test is inconclusive. Usually, this is caused by strong fit-parameter correlations that skew the 
average semi-major axis estimation and therefore the result of the permutation test. For HD110376 and HD155228, however, this is not the case and the exact reason for the 
failure of the permutation test is unclear. Because the orbit sizes are relatively large, the F-test null probabilities are very small (2.2\,10$^{-12}$ and 3.2\,10$^{-7}$, respectively), 
and the acceptable $i$-$\Omega$ parameter space is well constrained, we present orbit solutions for these two sources as well. As can be seen in 
Table~\ref{tab:hip_masses_16} the significance is lower than 1$\sigma$ for these 2 systems. Note also that the parallax change caused by fitting the orbit model is large 
(almost 3 mas) for HD110376.

For HD225239  and HD62923, the derived secondary mass is larger than the primary mass, which could be caused by light contribution by the secondary, shifting the 
position of the photocenter out of the primary star center. Indeed, our model assumes that the companion is dark (Sahlmann et al, 2011). We are developing supplementary methods to treat these cases and will report results in an 
upcoming publication. Here, we note that the orbit detection in both cases is significant but that the semi-major axis $a$ refers to the photocentric orbit and that the derived secondary 
masses are incorrect. Other possibilities could be that the companions of these stars are actually massive dead stars such as white dwarf, neutron stars or black holes ; they 
could also be couples of low mass stars. This can be consistent with the $\log (g)$ of these two primaries (Table~\ref{tab:spec_type}) that are small ($\sim$4.1-4.2) compared to the expected 
value of surface gravity for G2-3 dwarf stars ($\sim$4.4-4.5). Therefore these 2 primaries might rather be more evolved sub-giants. Precise astrometry with Gaia and imaging
can allow determining the exact mass and nature of these companions.

Two brown dwarf candidates, BD+210055 and HD210631, have their mass re-evaluated above 90\,M$_\text{J}$. For BD+210055, as guaranteed by the good coverage 
of the orbit ($N_\text{orb}$=0.9), the fit by the astrometric model is excellent, with a significance close to 100\%. It leads to a real mass that is significantly larger than the 
$M\sin i$$\sim $$85$\,M$_\text{J}$ derived thanks to RV only, with $M_2$ between 140 and 290\,M$_\text{J}$, well within the M-dwarf regime.  For HD210631, the orbital
coverage of 0.3 is not ideal, owing to the long 11-yrs orbital period. Still, the fit of the astrometric motion could catch some significant acceleration in Hipparcos data points and 
led to a 2$\sigma$-detection. It shows that the $M \sin i$$\sim $$83$\,M$_\text{J}$  of HD210631's companion derived by RV was strongly underestimated compared to its real mass, here 
constrained to lie between 140\,M$_\text{J}$ and 1.5\,M$_\odot$. We emphasise that the upper mass-range (about $>$0.6\,M$_\odot$) neglects that in such domain the 
secondary might contribute light and even produce a secondary peak in the CCF that we actually do not detect in our observations. 

Finally, we derived upper limits on the astrometric semi-major axis of the primary and the mass of the companion for 18 sources. Provided that at least about 80\% of the orbit is covered the 
upper-limit of an undetected semi-major axis can be deduced from the value of the median measurement precision $\sigma_\Lambda$.  The formula is the one used in 
Sahlmann et al. (2011), but moreover assumes the most unfavourable case of an edge-on orbit which projection on the plane of the sky only presents its minor-axis:
\begin{equation}
a_\text{prim} \lesssim \frac{\sigma_\Lambda}{\sqrt{1-e^2}}
\end{equation}

The value of the upper limits on semi-major axis of the primary and the corresponding companion mass are added to Table~\ref{tab:hip_masses_16}. For the triple
systems HD71827, HD212735 and BD+212816, only the inner companion $b$ orbit was considered, since the outer companion is not  constrained by RVs. 
Unfortunately, because of the loose constraints on the mass of the orbiting companions, we cannot exclude that all these systems are stellar binaries in the M-dwarfs regime. 
We will see in Section~\ref{sec:gaia_astro}, that using Gaia's published astrometric data allows tightening up the constraints on the mass for several of these systems, including 
systems for which $N_\text{orb}$$<$$0.8$.

\section{Gaia astrometry}
\label{sec:gaia_astro}

To overcome the large uncertainties on the true masses obtained using Hipparcos, we also cross match our sample with the Gaia catalog. 
We found 51 of our 54 targets in the Gaia DR1 catalog (Gaia collaboration et al. 2016). Among these 51 systems, we could measure the companion mass for 33 of them, and 
derive upper-limits for 6 others. The companion mass of the 12 that remain out if the 51 systems could not be constrained further with Gaia data. Their mass was nonetheless 
already bounded from below, thanks to RV, well within the M-dwarf domain.  

The DR1 of Gaia does not provide the individual positional measurements for all our sample stars. However, two binarity indicators are published in the released catalogues, 
the astrometric excess noise $\epsilon$ and the TGAS discrepancy factor $\Delta Q$ (Lindegren et al. 2012, Michalik et al. 2014, Rey et al. 2015).

The astrometric excess noise $\epsilon$ is a measure of scatter around the 5-parameters astrometric solution as resolved by the Gaia Reduction Software. Given the knowledge of the 
RV orbital parameters and the dates of Gaia data collection for the DR1, it can be used to derive an estimation of the inclination of the system, and thus of the true mass of the companion. 
To do so, we applied an MCMC method, presented in the following Section~\ref{sec:method}, that is able to output possible inclinations for a given astrometric excess noise and 
fixed orbital parameters. 

As published in the DR1, the dimensionless quantity $\Delta Q$ calculates the difference between the proper motion published in the Hipparcos-2 catalog and the proper motion 
derived in the TGAS sample by Gaia (Lindegren et al. 2016). We note that this differs from the original $\Delta Q$ definition as given in Michalik et al. (2014). The proper motion 
derived in the TGAS is based on a 24-year baseline of astrometric measurements, the 4-years monitoring of Hipparcos-2 and the 14-months monitoring of Gaia for the DR1. 
Measuring a significant long-term astrometric displacement, it can be used as a binarity diagnosis (Michalik et al. 2014, Lindegren et al. 2016). Comparing the value of $\Delta Q$ 
for every sources in our sample with the typical value obtained for any source in the DR1 allows us to determine if a system is a likely astrometric binary. This analysis is performed in Section~\ref{sec:deltaq} 

The DR1 archive provides both quantities, while only $\epsilon$ can be found in the DR2 (Gaia collaboration et al. 2018). Moreover, the excess noise values from the DR1, although 
based on a shorter timeline of astrometric measurements (25 July 2014 -- 16 September 2015, or 416\,days) are more reliable than in DR2 because of the so-called 
"DOF-bug" that directly affected the measurement of the dispersion of the final astrometric solution (Lindegren et al. 2018). For these reasons, we have only used the DR1 
results for $\epsilon$ and $\Delta Q$, as extracted from Gaia's DR1 archives\footnote{http://gea.esac.esa.int/archive/}. They are presented in Table~\ref{tab:gaia}. 

\subsection{GASTON: Gaia Astrometric Noise Simulation To derive Orbit incliNation}
\label{sec:method}

As such, it is not possible to directly interpret $\epsilon$ as a measure of the semi-major axis of an astrometric orbit, because it highly depends on the inclination of the 
orbit of the system that is seen projected on the plane of the sky. But since the fit of the RVs leads to precise orbital parameters, the inclination is also the only remaining free 
parameter that could have an impact on the value of $\epsilon$. 

We introduce here the new GASTON method based on Gaia data simulation to derive the inclination of the system from the measure of astrometric excess noise. The 
photocenter semi-major axis and secondary masses derived using this method are given in Table~\ref{tab:ratio_excess_noise}.

\subsubsection{Basic principle}

The principle of this method is to simulate Gaia photocenter measurements along the derived RV orbits presented in Section~\ref{sec:discussion}. 
Measurements epochs and Gaia along-scan (AL) axis orientations are randomised along the RV orbit, bounded by the DR1 data collection epochs. Real measurement epochs and 
AL axis orientations available on-line\footnote{https://gaia.esac.esa.int/gost/} compare well with random values. Considering random epochs and AL-axis orientation is thus sufficient 
for applying the method we present here that makes use of the excess noise, a quantity that cannot be considered as accurate. 

Different inclinations can be tested, each leading to a simulated astrometric excess noise $\epsilon_s$. We then constrain the different possible inclinations by comparing the 
whole set of $\epsilon_s$ with its actual measurement in the DR1, $\epsilon_\text{DR1}$.  As a result of applying this method on our targets sample, we found that the 
astrometric excess noise follows a one-to-one correspondence with inclination, owing to the increase of the photocenter semi-major axis with decreasing inclination. 

A few effects introduce scatter into this relation. First of all, the DR1 excess noise may incorporate bad spacecraft attitude modelling, which means that the value of 
$\epsilon$ does not account only for binary motion (Lindegren et al. 2012). The amplitude of the bad attitude modelling within $\epsilon$ could be estimated from its median value in the full sample
of objects observed with Gaia (Lindegren et al. 2016), $\epsilon_\text{med}$=0.5\,mas. Any value of excess noise below that value cannot be trusted to be genuinely 
astrophysical, although it could be considered an upper-limit. Conversely any value of $\epsilon$ above that level likely contains true binary astrometric motion. To take this effect 
into account, we added a bad-attitude-modelling noise of 0.5\,mas to Gaia's measurements in the simulation. 

Second of all, the astrometric motion of sources whose orbit has a period close to 1\,yr could be modelled by an excess parallax if the orientation of the system coincides with that of 
parallax motion. Moreover, slow orbital motion with period on the order of Gaia-Hipparcos baseline ($\sim$25\,yrs) can be absorbed into an erroneous proper motion. These effects cannot 
be properly taken into account in our simulations, since we have no prior knowledge of the orientation $\Omega$ for any of our targets. Thus the simulated $\epsilon_s$ could be 
overestimated compared to $\epsilon_\text{DR1}$ for the sources with period close to 1\,yr or larger than $\sim$20 years. This will tend to underestimate the inclination and overestimate the exact mass of 
the companion.

We will first describe the method to calculate the orbital model and the simulated along-scan Gaia measurements, and then how to derive a simulated excess noise for a given inclination.
Once this relation established, we will be able to derive an interval of inclinations compatible with a given value of $\epsilon_\text{DR1}$.

\subsubsection{Modelling of the along-scan data}

In a fixed non-accelerating reference frame, any source has a position vector ${\bf u_\star}$$=$$(x_\star,y_\star)$ in the plane of the sky. 
We will assume that proper motion, annual parallax and attitude of the spacecraft have been properly modelled and subtracted with only the orbital motion of the star remaining. 
Using the RV orbital model derived in this paper, it is fairly easy to obtain the projection of 
the star's orbit with a given inclination on the plane of the sky, and to derive $u_\star$ with respect to Keplerian parameters and inclination $I_c$:
\begin{equation}
{\bf u_\star}(I_c) = \left( R_x(I_c) \cdot R_z(\omega) \cdot {\bf k}(t|P,a,e,T_p) \right)\cdot \left({\bf u_x}.{\bf u_x} + {\bf u_y}.{\bf u_y} \right) 
\end{equation} 

where ${\bf u_x}$ is an arbitrary direction in the plane of the sky ; the direction
${\bf u_z}$ is orthogonal to the plane of the sky and oriented toward the observer ; the remaining direction ${\bf u_y}$ is directly oriented with respect to ${\bf u_x}$ and ${\bf u_z}$
composing the triad $({\bf u_x, u_y, u_z})$. The position vector ${\bf k}(t|P,a,e,T_p)$ at epoch $t$ is that of a Keplerian orbit which periastron is oriented along the ${\bf u_x}$ direction 
before applying the rotation matrices $R_z(\omega)$ and $R_x(I_c)$.

A number of locations are randomly selected along the above orbit by drawing random epochs between the bounds of Gaia DR1 data collection 
($t_\text{min}$=$2456863.0$, $t_\text{max}$=$2457282.0$). The number of these locations is given by the number of "matched observations" in the DR1 ; it is the total 
number of field-of-view $N_\text{FoV}$ CCD transits of a given star captured by Gaia.This value is given in Table~\ref{tab:gaia}. At each epoch, there are between 1 and 9 
measurements of along-scan (AL) angle $\eta$ per FoV transit (Lindegren et al. 2016). For simplicity, we assume a uniform number of CCD
transits per epoch, given by $N_\text{rec}$$=$$\text{round}(N_\text{tot}/N_\text{obs})$. Therefore, for each of these locations, we simulate $N_\text{rec}$ measurements of 
$\eta$ along the AL direction. 

The AL axis ${\bf u_\text{AL}}(\theta)$ is defined independently at each location with a random orientation $\theta$. 
The measurements are picked randomly along this axis accounting for the uncertainty on the spacecraft direction at any epoch (0.5\,mas; see the preceding section), and the 
uncertainty on AL measurements during the transit of the target on the CCD ($\sim$0.4\,mas; see Lindegren et al. 2018). Thus for 
any FoV transit observation indexed $i$, at an epoch $t_i$, and $N_i$ CCD measurements indexed $j$, the simulated read-out AL angles are:
\begin{equation}\label{eq:simu_mes}
\eta^{(i)}_{j}(I_c) = {\bf u_\star}^{(i)}(I_c)\cdot {\bf u_\text{AL}}(\theta_i) + \xi_{\text{inst}, i} + \xi_{\text{AL}, j}
\end{equation}  

where ${\bf u_\star}$ is projected on the AL direction, and $\xi_{\text{inst}, i}$ and $\xi_{\text{AL}, j}$ are the instrumental and AL errors introduced above. 

When dealing with an unresolved binary star, Gaia actually measures the position of the photocenter on the plane of the sky. The photocenter motion has the same orbital parameters 
than the primary, except for the semi-major axis. At a fixed system inclination, we use the RV orbital solution, and a mass-luminosity 
empirical model for both binary components, in order to calculate the photocenter semi-major axis, as described in the Appendix~\ref{sec:photocenter}. 
Since no secondary peaks was seen in any CCF for all targets, we always assumed that $M_{V,2}-M_{V,1}$ had to be greater than $2.5$, and thus the luminosity fraction in the optical range
is $<$10\%. 

\begin{figure*}
\includegraphics[width=\linewidth]{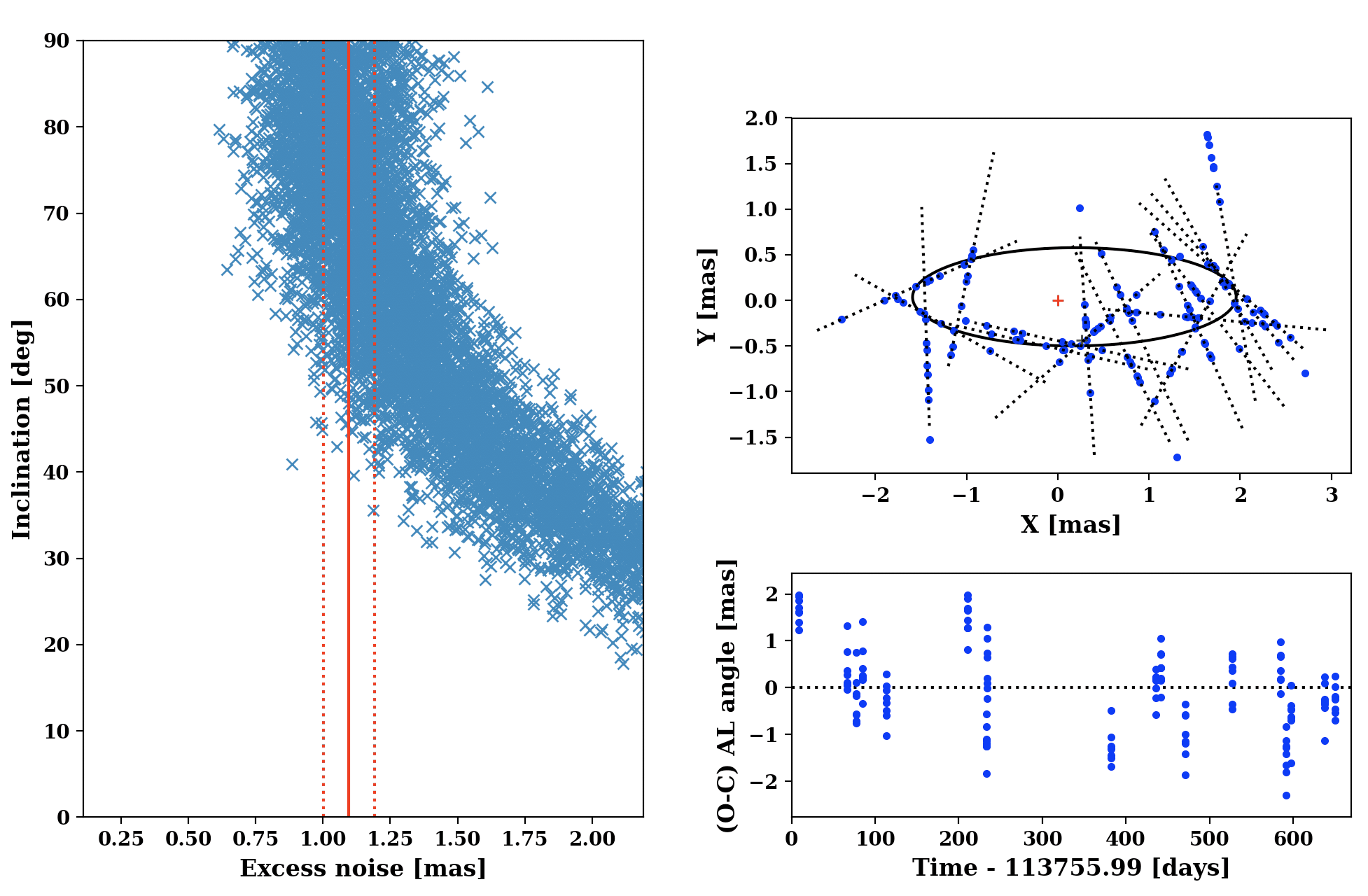}
\caption{\label{fig:gaia_simu_example_1} Left: Inclination vs simulated excess noise for BD+192536. The red lines mark the value of $\sqrt{\epsilon_\text{DR1}^2 + 0.5^2}$ for this 
star $\pm$10\%. Top-right: An example of simulation of Gaia measurements (blue points) for the peculiar case of BD+192536 that fits the value of $\epsilon_\text{DR1}$=1.02\,mas. 
The true proper motion and true parallax are assumed subtracted. The purple line is the residual proper motion (moving centroid) fitted to the simulated measurements and the 
red cross mark the true center of gravity of the system. Bottom-right: Residuals with respect to the moving centroid.}
\end{figure*}

\begin{figure*}
\includegraphics[width=\linewidth]{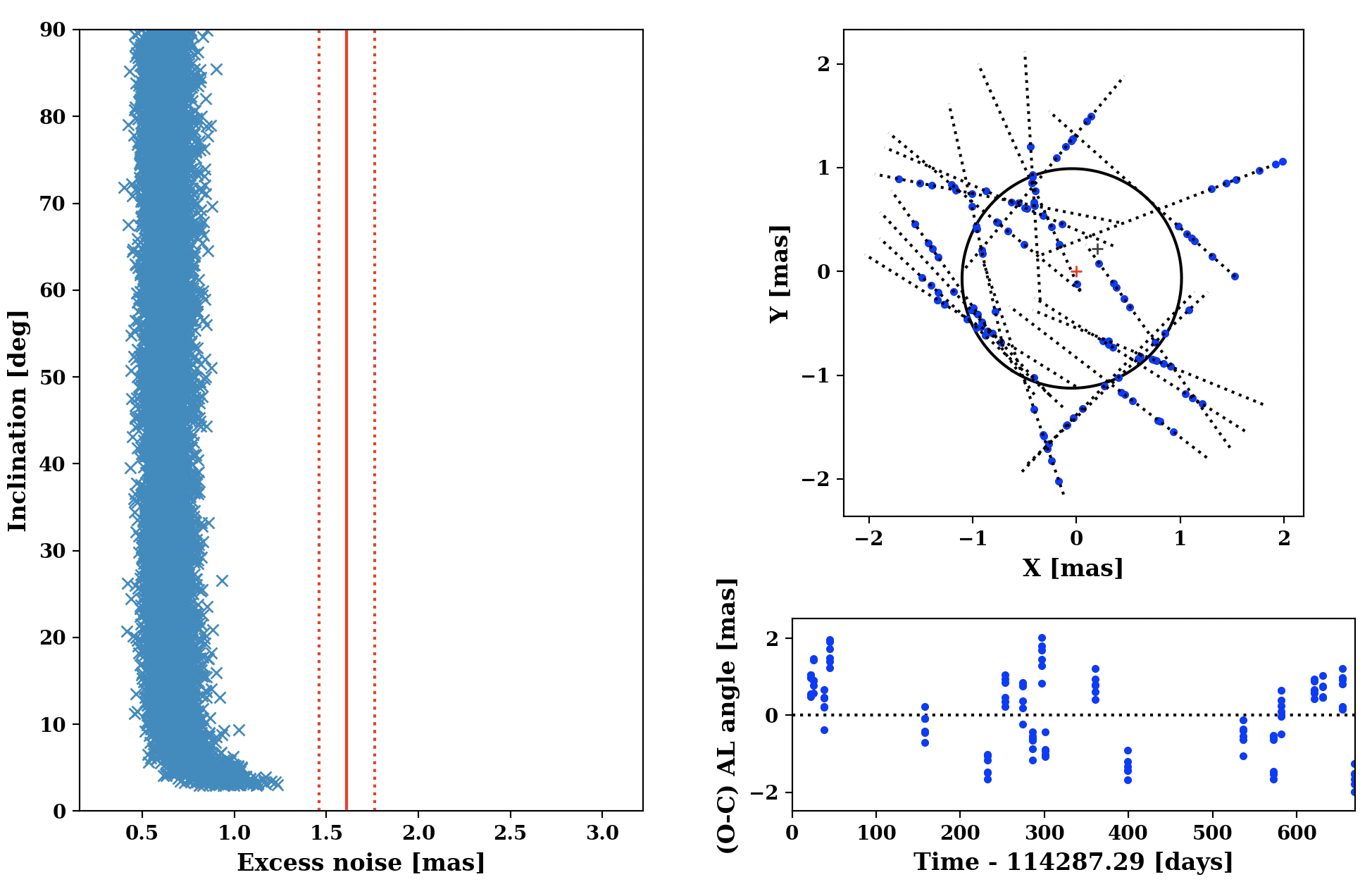}
\caption{\label{fig:gaia_simu_example_2} Same as Figure~\ref{fig:gaia_simu_example_1} for HD71827. Here the value of $\epsilon_\text{DR1}$ is larger than any of the simulations 
provided that the secondary companion is dark (i.e. not visible as a secondary peak in the CCF). Most likely the assumption that the inner companion is responsible for the 
astrometric motion of HD71827 is wrong (see text for explanation).}
\end{figure*}

Using the photocenter semi-major axis, along with all other orbital parameters from the RV Keplerian solution, in Equation~\ref{eq:simu_mes} leads to the final simulated Gaia's measurements. 
Examples are given in Figs.~\ref{fig:gaia_simu_example_1} and~\ref{fig:gaia_simu_example_2}. Out of these simulated data, we are now able to calculate an astrometric excess 
noise.

\subsubsection{Simulated excess noise} 
\label{sec:simexcess}
The astrometric excess noise is obtained by estimating the $\chi^2$ of its $\eta_\text{AL}$-residuals around the 5-parameters solution derived by Gaia's reduction software 
(Lindegren et al. 2012). In the simulations, we did not account for the true proper motion and the parallax, assuming they have been already modelled out. It results in 
only 2 remaining parameters to model out of our simulations, \textit{i.e.} the $(x,y)$-position of the photocenter on the plane of the sky.

The "average" target position published in the DR1 is given by the centroid of $\eta_\text{AL}$ measurements ${\bf u_c} = (x_{c}, y_{c})$. 
Assuming that most systematic positional errors have been accounted for, with only remaining the uncertainties $\xi_{\text{inst}, i}$ and $\xi_{\text{AL}, j}$ introduced above, 
the centroid position is found by minimising the squared sum of residuals $R_\ell$ with $\ell$ a given observation
\begin{align}\label{eq:residuals}
\sum_{\text{obs}\,\ell} R_\ell^2 &= \sum_{i=1}^{N_{FoV}}\sum_{j=1}^{N_i} \left( {\bf u_c} \cdot{\bf u_\text{AL}}(\theta_i) - \eta^{(i)}_j\right)^2 \nonumber \\
&= \sum_{i=1}^{N_{FoV}}\sum_{j=1}^{N_i} \left( x_c \cos \theta_i + y_c \sin\theta_i - \eta^{(i)}_j\right)^2
\end{align}

This leads to a simple system of four linear equations, which can be inverted, solving for $(x_{c},y_{c})$.
Once ${\bf u_c}$ is derived, this expression leads also to the $\chi^2$ of the residuals. In Lindegren et al. (2012) the expression of the 
$\chi^2$ and of the excess noise with respect to residuals and the AL uncertainty is given by
\begin{equation}
\chi^2 = \sum_{\text{obs}\,\ell} w_\ell \frac{R_\ell^2}{\sigma_{\text{AL}, \ell}^2 + \epsilon^2}
\end{equation}

Here we will assume that the down-weighting factors $w_\ell$$=$$1$ since we are only interested in the good AL measurements (with $w_\ell$$\sim $$1$). The $\chi^2$ 
should follow a $\chi^2$ distribution with a mean value equal to the number of degrees of freedom, i.e. the total number of points minus the number of parameters of the 
astrometric model derived by Gaia, thus $N_\text{DOF}$=$N_\text{tot}$$-$$5$. Therefore, at a given inclination $I_c$, and assuming a uniform value of $\sigma_\text{AL}$ 
along all observations, we should solve
\begin{equation}\label{eq:epsilon_chi2}
\sigma_{\text{AL}}^2 + \epsilon^2 = \frac{\chi^2 (I_c)}{N_\text{tot}-5}
\end{equation}

The above equation can be solved for $I_c$ by performing the simulations at various inclinations and comparing the right-hand side of Eq.~\ref{eq:epsilon_chi2} to the value of
$\epsilon^2+\sigma_\text{AL}^2$ that is measured by Gaia in the DR1. We sampled the inclination on a grid of 10,000 values uniformly distributed between 0 and $\pi/2$. 
Each time, the full set of inclinations compatible with $\epsilon$ ($\pm $10\,\%) leads to a range of possible values of the semi-major axis of the photocenter and the companion mass. 

The bounds $\gamma_{\pm}$ of a given parameter $\gamma$ compatible with $\epsilon$ are obtained by solving the following Bayesian equation for diverse posterior probabilities $p$
\begin{equation}
P(\gamma>\gamma_{\pm} | \epsilon_\text{DR1}) = \frac{P(\gamma>\gamma_\pm) P(\epsilon_\text{DR1} | \gamma>\gamma_\pm)}{P(\epsilon)} = p
\end{equation}

with e.g. $p$=0.68 leading to the 1-$\sigma$ bound and $p$=0.5 the median.
To solve this equation, we assumed that $\epsilon_\text{DR1}$ is conservatively known at $\pm$10\% $\sim \epsilon_\text{DR1}/\sqrt{N_\text{tot}}$. The prior 
$P(\gamma$$>$$\gamma_\pm)$ is calculated by assuming that the unknown inclination is uniformly distributed between 90$^\circ$ and the inclination at which the secondary is too 
massive for not being observed in the spectra, i.e. verifying $M_{V,2}-M_{V,1}$$=$$2.5$. In this case 

\begin{equation}
P(\gamma>\gamma_\pm) = P(I_c<I_{c,\pm}) = \frac{I_{c,\pm} - I_{c,\text{min}}}{90-I_{c,\text{min}}}
\end{equation}

The likelihood $P(\epsilon | \gamma$$>$$\gamma_\pm)$ sums all simulations compatible with $\epsilon$($\pm$10\%) divided by the total number of simulations such that 
$\gamma$$>$$\gamma_\pm$. Finally, the marginal probability $P(\epsilon)$ is the sum of all simulations compatible with $\epsilon$($\pm$10\%) divided by the total number of simulations.

\subsubsection{A few caveats}
\label{sec:caveats}

\begin{itemize}
\item To begin with, as was mentioned in the preceding section, we did not account for the true proper motion and the parallax in the simulations. In reality, with an additional 
accelerated motion unaccounted for in the 5-parameters model, the Gaia's reduction software could have derived an excess of proper motion and an excess of parallax. The 
excess parallax modelled is maximal when the orbital motion is aligned with the parallax direction and the orbital period is close to 365\,days. Unfortunately, the 
orientation of the orbits of our targets compared to the parallax direction are generally unknown, so this cannot be taken into account properly. This could lead to 
underestimate the photocenter semi-major axis and thus the mass of companions with $P$$\sim $$365$\,days. This concerns 5 systems with orbital period within 25\% of 
365\,days. On the other hand, the issue of excess proper motion is only relevant for those sources that are member of the secondary dataset of the DR1, i.e. not members of the TGAS sample. 
Indeed, for those, the time baseline of the astrometric measurement is not 24 years but rather $<$416 days, i.e. less than the duration of the DR1 campaign. But in these cases, 
the proper motion can be fitted out easily since it is purely linear. This is done by slightly modifying the system of equations (\ref{eq:residuals}) with a moving centroid 
${\bf u_c}(t) = (x_{c} + \mu_x t, y_{c} + \mu_y t)$ and inverting the system solving for 4 parameters rather than 2. We incorporated this correction for 8 sources in the 
secondary dataset: BD+132550, BD+210055, BD+680971, HD147847, HD155228, HD207992, HD225239, HD24505, and HD62923. \\

\item For triple systems, the Keplerian solution of the outer, long-period, companion being unknown, we could not simulate its effect on the motion of the photocenter.  
We could only simulate the astrometric excess noise derived by Gaia by assuming that the reflex motion of the primary star is mainly explained by the innermost better 
constrained companion. This could be a wrong assumption (see below), but still allows deriving a strict upper-limit on the mass of the inner companion. \\

\item Finally, we must warn that the excess noise is intrinsically sensitive to outliers of which there are probably quite a few in the DR1 state of the Gaia processing. Added to the 
issues of attitude modelling errors, the astrometric excess noise measured by Gaia might be in a few cases overestimated. On the other hand, the level of bad calibration in DR1 
could also be underestimated, for the targets with the fewest effective epochs and in cases in which the companion is stellar in nature and contributes light. Nevertheless, we 
found in general that the value given for the excess noise in the DR1 is relevant and agrees with the $\chi^2$ published in the DR2 accounting for a longer baseline of 
astrometric monitoring (see following section).
\end{itemize}

\subsubsection{A comparison with Gaia DR2}
\label{sec:comp_DR2}

We added in Table~\ref{tab:gaia} the DR2 normalised unit weight errors, $\text{RUWE}$$=$$\sqrt{\chi^2/2}$, as defined in Lindegren et al. (2018) and accounting for an
average renormalization factor $1$/$\sqrt{2}$ due to the DOF-bug on bright (G$<$11) targets. Like $\epsilon$, the RUWE is a measurement of the astrometric scatter around the 
5-parameters solution. As the DR2 is based on 670 days of astrometric measurements, we would expect the astrometric noise to be of better quality than in the DR1, or at 
least generally agree with the excess noise measurements in the DR1. Unfortunately, we cannot use the values 
of RUWE for individual sources to directly apply the GASTON method, since neither the individual renormalization factor nor the typical excess attitude noise 
are known.

\begin{figure}
\includegraphics[width=84mm]{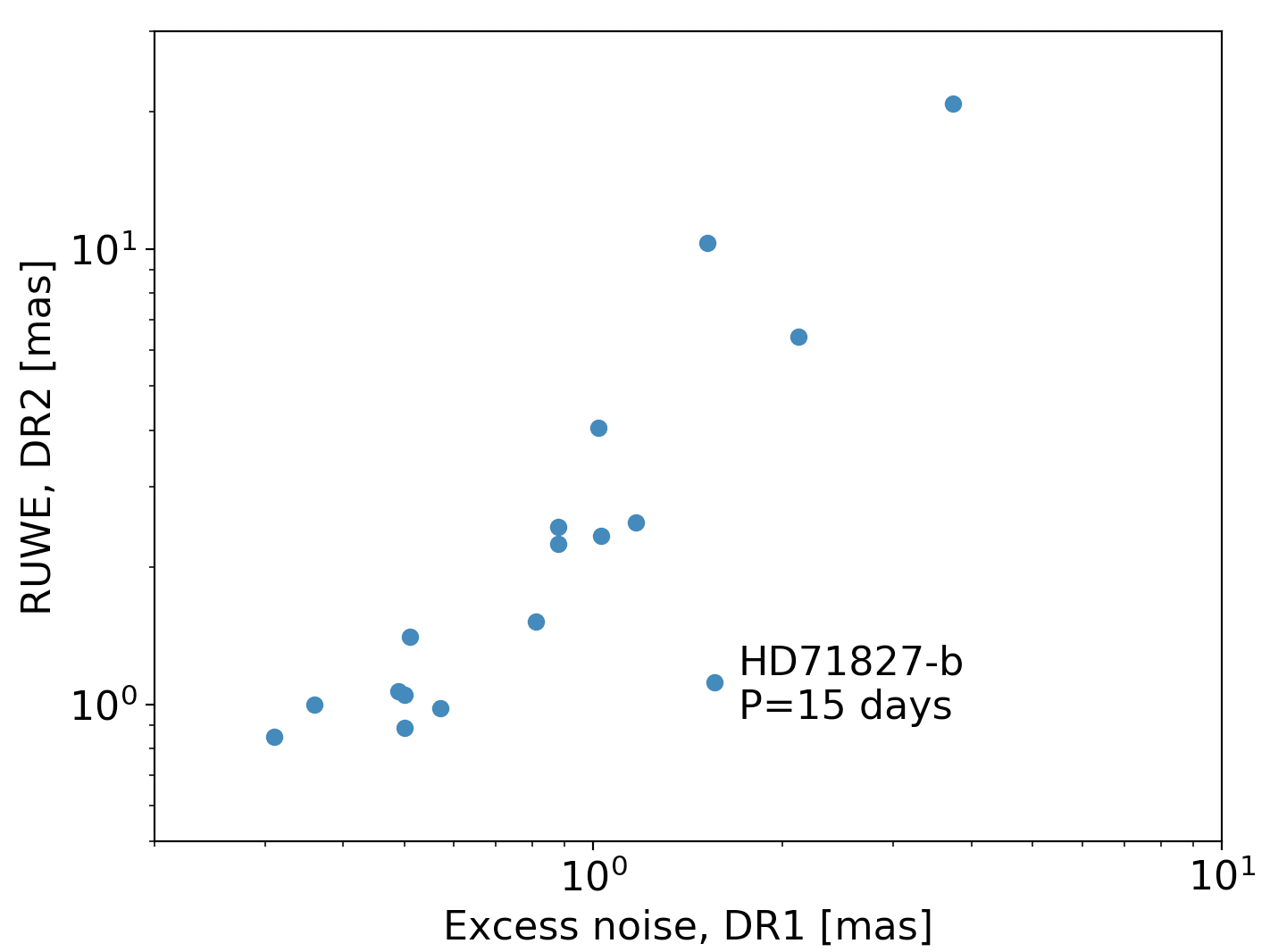}
\caption{\label{fig:compareGaiaDR2} A comparison of Gaia DR1 excess noise with Gaia DR2 renormalized unit weight error (see text for explanation). Only systems with short orbital period
($<$670\,days) and which are part of the TGAS sample are compared. For longer period systems or for those of which Hipparcos positioning was not taken into account, the orbital motion 
could have been modelled out of the DR2 or the DR1 data by fitting the proper motion. The outlying HD71827 case in discussed in the text.}
\end{figure}

A global comparison of the DR1 excess noise to the DR2 RUWE for the sources of our sample reveals a nice positive correlation, as shown in 
Figure~\ref{fig:compareGaiaDR2}. This strengthen the reliability of $\epsilon_\text{DR1}$ as a measurement of the astrometric scatter.

Here, we focused only on short period systems with $P$$<$$670$\,days and on sources of the DR1 that are members of the TGAS sample. 
For long period orbits the astrometric displacement as seen by Gaia in the DR2, with a 670-days long baseline, could have been modelled by "instantaneous" proper motion, 
thus biasing the measurement of astrometric scatter. This stands also for sources of the DR1 that are not members of the TGAS sample, since for those only Gaia 
measurements were used and the time baseline is 416 days. On the contrary, for sources in the TGAS sample, the time baseline is 24 years, and for short period systems the 
DR2 could not have confused orbital motion with proper motion.  

In this figure, the outlier HD71827 is a triple system, with an inner companion at a period of 15 days and an outer companion with a period larger than 20 years. The large 
discrepancy between the DR2 RUWE and the DR1 excess noise shows that Gaia rather caught the motion of the star due to the outer companion rather than the inner one. In 
the DR2, this long-term motion could have been confused with proper motion and thus modelled out, while in the DR1, since HD71827 is a member of the TGAS sample, the 
whole motion participates to the scatter accounted in the excess noise. Despite this issue, we will keep on assuming that the excess noise of triple systems is due to the inner 
companion in order to derive an upper-limit on its mass.

\subsubsection{Results} 

In Table~\ref{tab:ratio_excess_noise}, we give the 1-$\sigma$ bounds of semi-major axis of the photocenter and companion mass, as defined in the preceding section. We only 
consider systems for which a well-constrained Keplerian was derived from the RV data, i.e. for which 
the uncertainty on the orbital parameters does not exceed 10\%. This led to reject 5 more stars among the 51 considered in this analysis, HD13014, HD146735, HD40647, HD60846 
and HD85533, all of which admit a companion well within the M-dwarf regime. We will thus only consider now the remaining 46 targets with well-constrained RV orbit. 
Moreover, for 13 out of them, $\epsilon$ is less than the typical "normal" $\epsilon$ value for a single star in Gaia's DR1, i.e. $0.5$\,mas. In these cases, the derived masses 
should only be considered as upper-limits. 

For 43 out of the 46 systems, we found that the marginal probability of $\epsilon_\text{DR1}$$\pm $$10\%$ being 
produced by simulations with any inclinations is larger than 0.001. This is a positive sign that the GASTON method produce sensible values of the astrometric excess noise 
compatible with real Gaia measurements, in more than 90\% of the cases. Conversely for 3 systems, HD62923, HD71827 and HD156728, the value of $\epsilon_\text{DR1}$ 
was difficult to produce, and needed strong fine tuning of the simulated data. The excess noise is either wrongly estimated by Gaia, either one of our assumptions is 
incorrect, either the astrometry is polluted by one of the companion. 

Indeed, HD71827 is a triple system for which the Gaia's DR1-DR2 comparison in Section~\ref{sec:caveats} suggested that the astrometry 
recorded by Gaia was rather due to the outer companion. Our assumption that only the inner companion participates to the excess noise was therefore certainly wrong. HD62923 was 
shown to be massive in Section~\ref{sec:astrometry} with a companion that could not be assumed as dark. Finally, the HD156728's phase curve is not fully covered by radial 
velocity measurements with a derived orbital period of $\sim$4100\,days. Our solution is possibly inexact, or the value of $\epsilon_\text{DR1}$=0.48\,mas is underestimated. 

The case of the triple system HD71827 is worth commenting further. For this source, we could not produce many $\epsilon_s$ as large as what measured by Gaia for 
this system $\epsilon_\text{DR1}$=1.56\,mas. The largest excess noise simulated is obtained for an inclination of 3.3$^\circ$ leading to $M_2$=0.6\,M$_\odot$. Beyond this 
limit, the magnitude difference between the companion and the primary must be less than 2.5, implying a secondary peak present in the CCF of HD71827 spectra, which is not 
the case. This confirms that the excess noise measured in the DR1 is more likely due to the outer companion rather than the inner one. The inner companion mass 
$M_2$=0.6\,M$_\odot$ has thus to be interpreted as an upper-limit only, and the inclination of the system is likely (much) larger than 3.3$^\circ$.

\begin{figure}
\includegraphics[width=84mm]{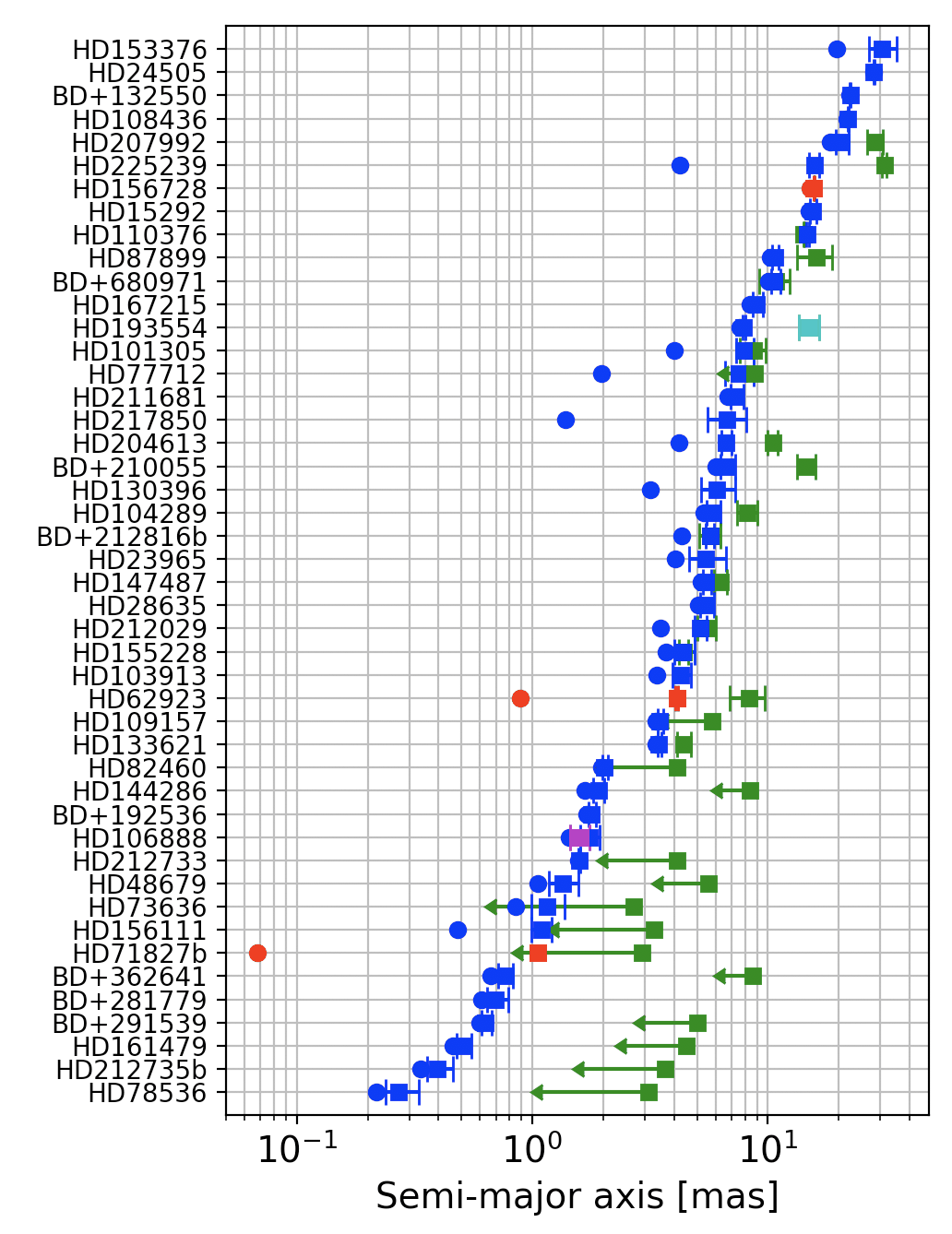}
\caption{\label{fig:compareGaiaHIP} Semi-major axis of the photocenter as measured with Gaia from the GASTON method (blue squares). The circles represent the 
semi-major axis derived if an inclination of 90$^\circ$ (edge-on system) is assumed. 
Green points are Hipparcos measurements or upper-limits, as explained in Section~\ref{sec:astrometry}. The red squares correspond to a measured Gaia excess noise with a marginal
probability smaller than 0.001. The cyan point is the result found in the Hipparcos double and multiple systems catalog. The magenta square is the 
speckle measurement of Tokovinin et al. (2014) for HD106888.}
\end{figure}

Figure~\ref{fig:compareGaiaHIP} summarises the results. It includes a comparison between the semi-major axis obtained with Gaia to those obtained by using the Hipparcos 
Data (Section~\ref{sec:astrometry}). We also found ground-based speckle interferometry for  HD106888 that led to a separation of 32$\pm$3\,mas and a magnitude difference 
of $\Delta M_V$=1  between the 2 components of this system (Tokovinin et al. 2014). If we assume that the 2 components were at apoastron, this is equivalent 
to a semi-major axis of the photocenter of about 1.6$\pm$0.15\,mas. 

\begin{figure}
\includegraphics[width=84mm]{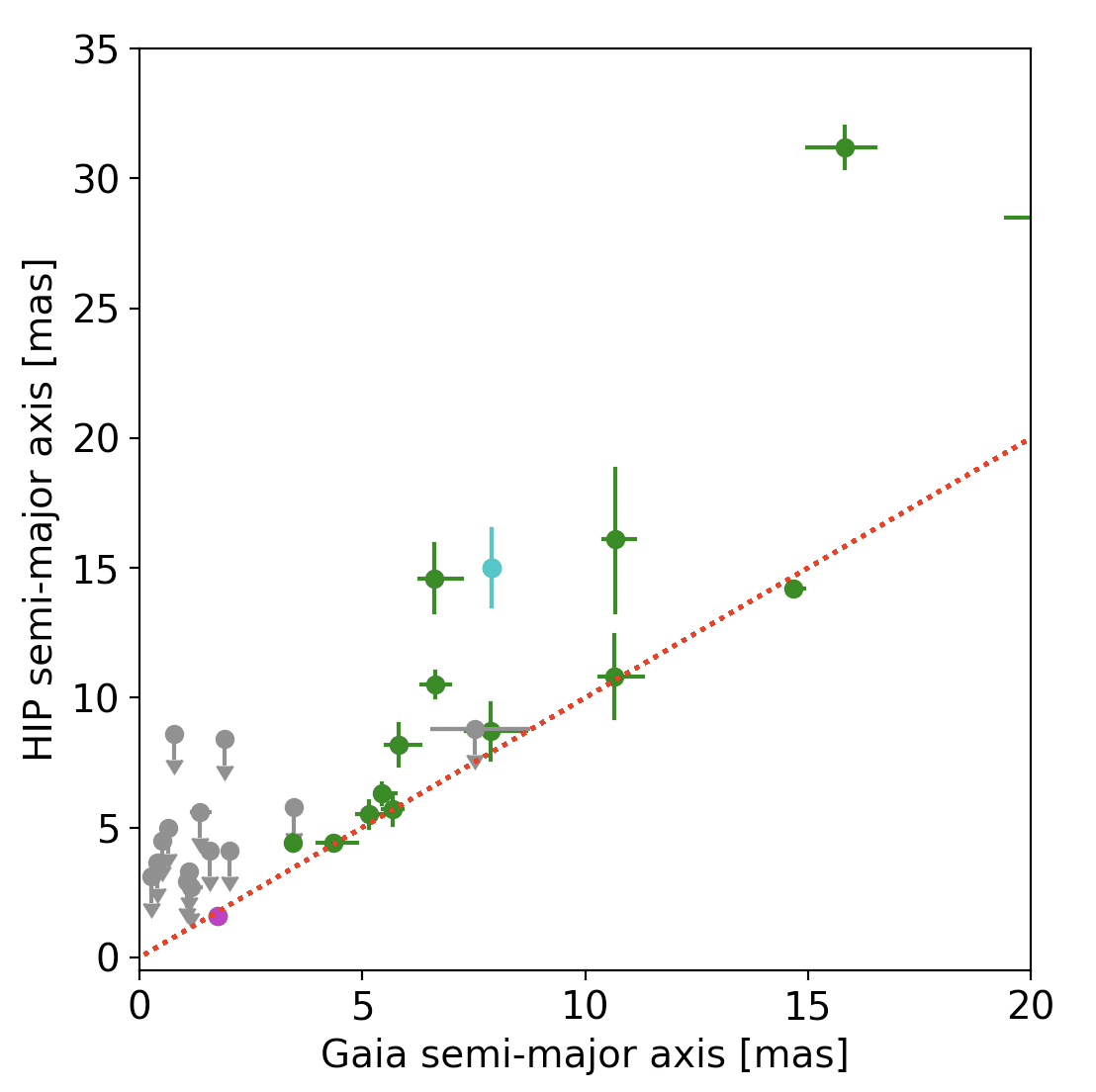}
\caption{\label{fig:GaiaHIP_emph} Direct comparison of Gaia and Hipparcos derivation of the semi-major axis of the photocenter when both are available. The colour code is the same
as in Fig.~\ref{fig:compareGaiaHIP}. Gray points indicate upper-limits derived with Hipparcos. The red dotted line represents the equality $a_\text{HIP}$=$a_\text{Gaia}$. The 
outlying point with $a_\text{HIP}$=32\,mas is HD225239. This case is discussed in Section~\ref{sec:astrometry}.}
\end{figure}

The comparison with Hipparcos and interferometry is quite satisfying. In all the cases where a significant non edge-on inclination is measured and a corresponding Hipparcos 
solution is derived, the revised semi-major axis of the photocenter $a_\text{ph}$ tends to always be much closer to the Hipparcos result than the semi-major axis derived with RV results only. This is 
emphasised in Fig.~\ref{fig:GaiaHIP_emph}. Most importantly, the GASTON method always leads to a value of $a_\text{ph}$ that is 
larger than the Hipparcos and interferometric measurements. Therefore it looks relatively safe to consider the results of this method as an improved 
measurement of the inclination and of the true mass of the companion compared to RV fitting only. 

We can take a particular look at the BD candidate BD+210055\,b that was well-constrained using Hipparcos astrometry to be an actual M-dwarf, with 
$M_2$=140-290\,M$_\text{J}$ at 3$\sigma$. We found here that the large value of $\epsilon_\text{DR1}$=1.3\,mas measured for this system led with GASTON to derive a 
companion mass of 96-110\,M$_\text{J}$ at 1$\sigma$. Although not exactly compatible with the Hipparcos result, it is remarkable that we reach to the same conclusion 
concerning the real stellar nature of this object. 

We found 12 systems that were not already constrained with Hipparcos and for which GASTON led to an inclination significantly different than 90$^\circ$ at 3--$\sigma$. These are 
BD+362641, HD23965, HD48679, HD73636, HD77712, HD103913, HD106888, HD130396, HD144286, HD153376, HD156111, and HD217850. Five of them are particularly 
interesting to us since their companion was determined to be in the BD mass regime thanks to radial velocities: HD23965, HD48679, HD77712, HD130396 and HD217850.

For all five, the simulations could produce many $\epsilon_s$ values compatible with the DR1 excess noise with $P(\epsilon_\text{DR1}\pm10\%)$$>$$1\%$.
The measurements of their inclination thanks to the GASTON method is thus pretty robust. We can safely state that the companions of HD77712, HD130396 and HD217850 
are M-dwarfs with masses well above 80\,M$_\text{J}$. We also recall that the $M\sin i$ of HD77712\,b was underestimated in Section~\ref{sec:binary} due to the deformation of the CCFs 
by a hidden component. HD77712\,b is thus well within the M-dwarf domain.  On the other hand, for HD48679 and HD23965, while the inclination is significantly different 
than 90$^\circ$, the companion mass does not exceed 90\,M$_\text{J}$. These are likely to be brown dwarfs.

Finally, we measured that the mass of 7 companions, among the 46 considered here, are compatible with the BD regime at 1$\sigma$. They are BD+291539 b, HD23965 b, 
HD28635 b, HD48679 b, HD71827\,b, HD82460 b, and HD211681 b. We discuss them in more details in Section~\ref{sec:details}.

Although it is not free of possible systematics, we conclude that the GASTON method is able to derive reliable estimations of systems inclination without the use of the 
definitive Gaia intermediate data. It proves to be a useful method allowing the characterisation of binaries mass and discarding massive companions with short periods in 
exoplanet RV survey. We are now applying that method to other catalogues of RV-detected binary stars and exoplanets in order to remove the inclination degeneracy on their 
$M\sin i$ measurements, and thus constrain their true masses. This should show that some bodies now considered as exoplanets actually are face-on binaries.  

For the largest period orbits, if virtually nothing can be said using this method, the discrepancy between Hipparcos and Gaia's DR1
proper motions, the $\Delta Q$ factor, will be more relevant to these cases, with a time baseline larger than 25\,yrs. We explore this option in the following section.

\subsection{The TGAS discrepancy factor $\Delta Q$}
\label{sec:deltaq}

While it was pointed out that $\Delta Q$, as produced in DR1, does not take into account the perspective acceleration (Michalik et al. 2014), the stars 
in our sample are too distant and the proper motions too small for perspective acceleration to be significant. In principle, a value of $\Delta Q$, 
typically larger than 90\% of Gaia primary sample, i.e. $\Delta Q$$>$$10$ (Lindegren et al. 2016), could be considered as significant, and we should conclude that a non-zero 
acceleration is being detected, advocating for binarity of the system. 

\begin{figure}
\includegraphics[width=89mm]{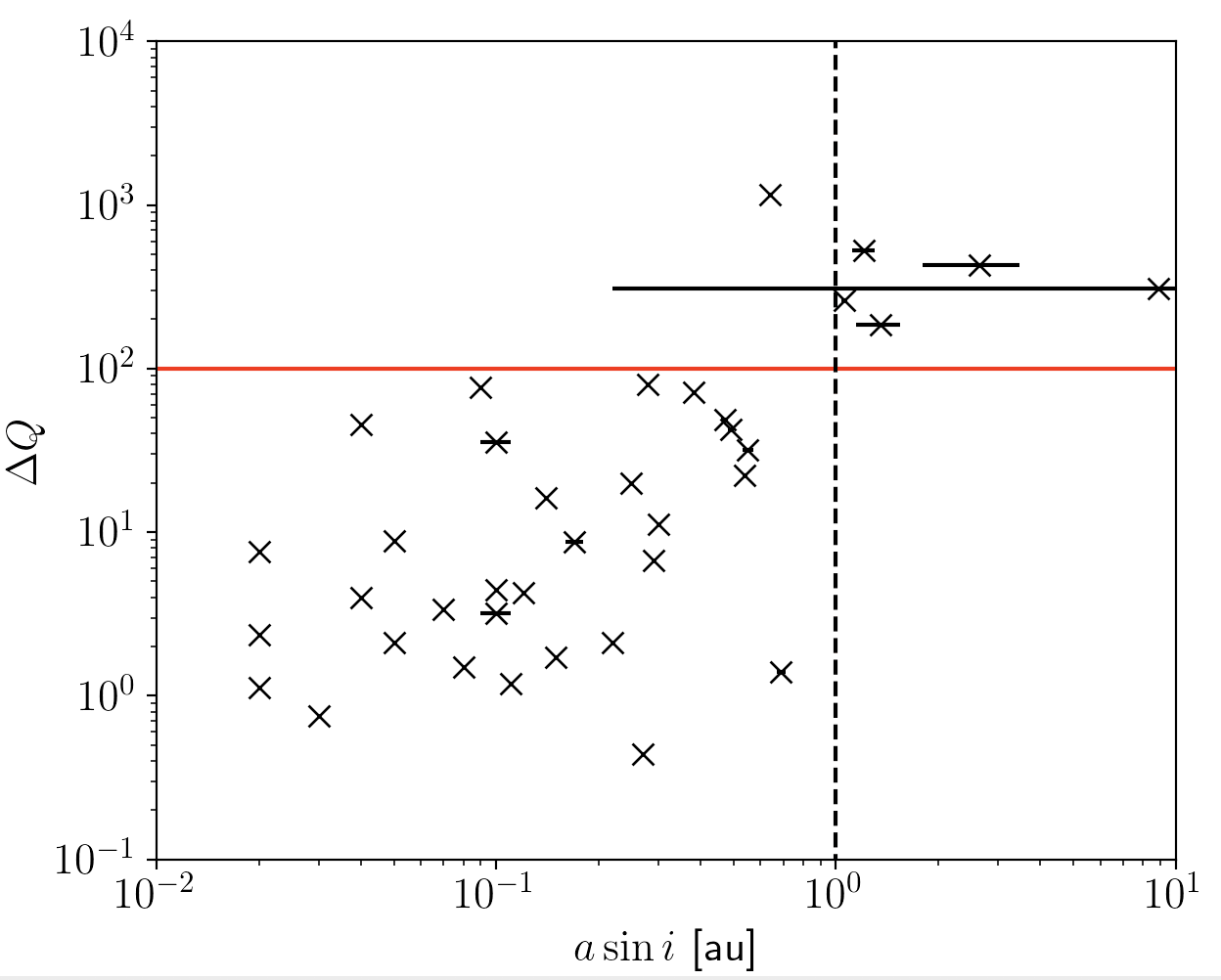}
\caption{\label{fig:DeltaQ_asini} The TGAS discrepancy factor $\Delta Q$ vs the $a_1 \sin i$ (in AU), the minimum semi-major axis of the primary star orbit derived from RV. 
The solid red line indicates the $\Delta Q$=100 limit, and the dotted black line represents the $1$\,au limit. Some error bars are smaller than the size of the symbols.}
\end{figure}

As showed in Table~\ref{tab:gaia}, 19 targets have a value of $\Delta Q$ larger than 10, while 7 have $\Delta Q$$>$$100$ and 1 of them has $\Delta Q$$>$$1000$. 
The value of $\Delta Q$ must be related to the amplitude of the orbital motion, and thus should present a correlation with the semi-major axis of the primary star $a_1$. 
Indeed, $\Delta Q$ should be more sensible to large orbital period ($P$$>$$4$\,yr; the Hipparcos baseline) that lead to larger differences between the proper motions 
measured on a 24-yr baseline and those measured on a 4-yr baseline ; while larger companion mass also increase the astrometric acceleration.
Because of the degeneracy on the inclination of the systems in RV solutions, only the minimum estimation $a_1\sin i$ is known. Figure~\ref{fig:DeltaQ_asini} 
display the relation between $\Delta Q$ and $a_1\sin i$ for the present sample, only considering binaries and excluding triples. We find that values of $\Delta Q$ larger than 
100 are exclusively found for primaries with a semi-major axis of at least $\sim$0.7\,au. Moreover, values of $a_1\sin i$ greater than 1\,au systematically lead to 
$\Delta Q$$>$$100$. On the other hand, below 1\,au the values of $\Delta Q$ are scattered uniformly between 0 and 100. We conclude that only values of $\Delta Q$$>$$100$ 
should be trusted as a positive detection of binarity, with $a_1$$>$$1$\,au. 

The system with $\Delta Q$$>$$1000$ is HD156728. Its period is larger than 10\,yrs, and the companion mass stands in the stellar domain above 126\,M$_\text{J}$. The primary  
semi-major axis derived from the radial velocity solution is greater than 0.73\,au. The detection of a large value of $\Delta Q$ advocates for an underestimation of $a_1$ 
and of the mass of the companion in this system that is most likely seen nearly face-on. The non-detection in $\epsilon_\text{DR1}$, lower than 0.5\,mas, could be compatible with this result
since it allows the mass to be as large as 250\,M$_\text{J}$, as shown in Table~\ref{tab:ratio_excess_noise}. 

The other 6 systems with $\Delta Q$$>$$100$ are HD108436, HD13014, HD153376, HD60846, HD71827, and HD85533.  Their long-period companions stand beyond 100\,M$_\text{J}$ and have 
all $a_1\sin i$$>$$1$\,au. No Hipparcos astrometric solution could be derived for any of these sources owing to the short span of the Hipparcos measurements, smaller than 0.4 
orbital periods for all of them. Interestingly, the astrometric excess noise of these 6 systems is significantly greater than 0.5\,mas. However, for 4 them, the RV orbit is not 
well constrained, having a large period unknown at more than 10\% uncertainty. Moreover, the orbital phase is in most cases not fully spanned by the RV measurements. Only 
HD108436 and HD153376 are well fitted and the application of the GASTON method leads to re-evaluate the mass at a larger value, and to semi-major axis of the photocenter larger 
than 20\,mas. This is on the order of magnitude of the displacement $\Delta Q$ could be able to detect since the precision of Hipparcos is $\sim$10\,mas. 

The case of the triple system HD71827 is interesting, since a clear motion is detected by Gaia, with $\epsilon_\text{DR1}$=1.6\,mas, while $\Delta Q$=358. This suggests that the motion of the 
star under the influence of its companions is detected by both indicators. While it remains possible that they do not detect the motion due to the same companion, we found in the 
preceding section that the DR2 $\chi^2$ and DR1 excess noise agree if $\epsilon_\text{DR1}$ measures the astrometric motion of the long-period 
outer companion. Undoubtedly, the same motion was measured by $\Delta Q$. With a photometric semi-major axis on the order of 1\,mas, the motion due to the inner companion 
is clearly out of the detection zone of $\Delta Q$.

We conclude that $\Delta Q$ is a useful binarity indicator, provided that $\Delta Q$$>$$100$, leading to the detection of a primary star motion with a semi-major axis greater 
than 1\,au. It has however a limited usage since it does not allow for deriving the exact mass of the companion.

\section{Detailing the 7 brown-dwarf companions}
\label{sec:details}

Among the initial sample of 12 BD candidates derived by RV in Section~\ref{sec:spectro}, we excluded 5 of them by using astrometric data of Hipparcos and Gaia.  
HD210631\,b and BD+210055\,b were found to be M-dwarfs using the Sahlmann et al. (2011) method on Hipparcos data in Section~\ref{sec:astrometry}. Moreover, the mass of 
HD130396\,b, HD217850\,b, and HD77712\,b could be constrained beyond 90\,M$_\text{J}$ thanks to the GASTON method applied on Gaia's DR1 astrometric excess noise in 
Section~\ref{sec:gaia_astro}. 

Most importantly, we derived in Section~\ref{sec:gaia_astro} that the mass of the 7 remaining companions could be constrained below 90\,M$_\text{J}$. All these
7 companions are thus likely brown-dwarfs. We list some details on their detection below.

\paragraph{\textbf{BD+291539.}} We observed for this G-type star radial velocity variations compatible with a 60-M$_\text{J}$ BD companion on a 176\,days orbit at a semi-major
axis of 0.6\,au. The fit of the 17 RV points is of good quality with a residual dispersion of 4.7\,m\,s$^{-1}$.  Probably owing to the short period, the values of $\epsilon$ and
$\Delta Q$ are too small to indicate any significant astrometric motion in Gaia data. Unsurprisingly it was not detected either by Hipparcos. This companion is likely a 
brown-dwarf with a maximum mass about 69\,M$_\text{J}$.

\paragraph{\textbf{HD211681.}} From the orbital parameters and minimum mass of the secondary companion in this system, $M_2\sin i$=77.8$\pm$2.6\,M$_\text{J}$ with 
$P$=7612$\pm$131\,days and $a_2$=8.28$\pm$0.16\,au, we deduce an inferior limit of 7 mas for the semi-major axis of the primary's astrometric orbit. 
Neither Hipparcos nor Gaia detect any significant motion. Moreover, the comparison between Hipparcos and Gaia astrometry is barely significant with $\Delta Q$=42, which 
is not surprising considering the Hipparcos precision of about 10\,mas. We conclude that given the metallicity of HD211681 (Fe/H$\sim$0.36), its companion is likely an object 
probing a mass regime between star and brown-dwarf, around 80\,M$_\text{J}$.

\paragraph{\textbf{HD23965.}} Using 84 RV measurements obtained with SOPHIE, we derived for this active F-type star a Keplerian compatible with a 40-M$_\text{J}$
BD candidate on an 11-yrs orbit at 5\,au from the star. The large dispersion of the residuals $\sim$30\,m\,s$^{-1}$ is compatible with the strong activity that is measured for this source, with 
$\log R'_{HK}$=-4.47. The RV jitter tends to magnify the uncertainties of the derived parameters, but they remain known with a precision better than 10\%.  However, since the 
full orbital phase has not been covered yet, the period is still not constrained above 3974\,days. The Gaia DR1 astrometry, measuring $\epsilon$=0.6\,mas and 
an insignificant $\Delta Q$, is suggestive of a system close to edge-on. Applying GASTON on this system, assuming the 3974-days period, lead to an inclination of 76$\pm$5$^\circ$ 
and a companion mass of 42$\pm$1\,M$_\text{J}$. HD23965\,b is thus a strong brown dwarf candidate.

\paragraph{\textbf{HD28635.}} Paulson et al. (2004) argued that the mass of the companion could be significantly higher than what found with RVs ($\sim$77\,M$_\text{J}$). They proposed
 $0.86$$\pm$$0.31$$M_\odot$. Evidence for small inclination was drawn according to the $v\sin i$$\sim $$1$\,km\,s$^{-1}$ of the primary compared to the estimation of its 
true rotation velocity. However, the astrometric data presented here, with $\epsilon$=0.51\,mas and 
$\Delta Q$=20, do not tend to confirm this result. The value of $\epsilon$, given the RV solution derived, rather lead to an inclination of 66-80$^\circ$ and a companion mass of 
82-88\,M$_\text{J}$ at 1$\sigma$. With a Fe/H$\sim$0.16, this companion is located slightly above the BD-M\,dwarf limit. Nevertheless, the phase coverage with Gaia is only 
partial along the 7-years orbit. Excess proper motion could tend to lower the value of $\epsilon$ that was measured for the DR1. 

\paragraph{\textbf{HD48679.}} According to the 26 RV measurements obtained with SOPHIE, the companion of this G0 star is a brown dwarf candidate with an $M_2\sin i$$\sim $$36$\,M$_\text{J}$, 
an orbital period of 1111\,days and a semi-major axis of 2.1\,au. The Keplerian fit is of good quality with a small residual dispersion of 4.6\,m\,s$^{-1}$.
For this star, the astrometric excess noise measured by Gaia of 0.8\,mas leads to derive an inclination between 41 and 65$^\circ$, a true companion mass 
of 41-57\,M$_\text{J}$, and $a_\text{ph}$$\sim $$1.3$\,mas. Moreover, the small extent of the astrometric motion of the photocenter is compatible with a non-detection from 
comparing Gaia and Hipparcos, with $\Delta Q$=3. Thus, HD48679 b is a likely brown dwarf companion.  

\paragraph{\textbf{HD71827.}} The inclination of this triple system and the true mass of the inner companion ($M\sin i$=26\,M$_\text{J}$ and $P$=$15$\,days) could not be 
constrained by astrometry. Indeed, it was shown in Section~\ref{sec:comp_DR2} that the astrometric scatter derived in Gaia DR2 and DR1 strongly disagree. Moreover, it was 
also difficult to model with GASTON an excess noise as large as measured by Gaia in the DR1 for this system up to inclinations that could not be compatible with a dark companion. 
This shows that the astrometric scatters measured by both the Gaia DR1 and DR2 for this system cannot be explained by the inner companion, but rather by a long period object 
such as the outer M-dwarf companion of HD71827 ($P$$>$$20$\,yrs). It follows that the real mass of HD71827-b is not constrained and could still be compatible with a mass in the
brown-dwarf domain, although it could also still be more massive. 

\paragraph{\textbf{HD82460.}} Using 17 RV measurements obtained with SOPHIE, we report for this early G-type star the detection of a BD candidate with $M\sin i$$\sim $$73$-M$_\text{J}$
on a 590-days orbit, at 1.4\,au from the primary. The Keplerian fit is of fairly good quality with medium residual dispersion of 7\,m\,s$^{-1}$ and orbital elements known with errors 
smaller than 5\%. Hipparcos intermediate data does not lead to astrometric motion detection, and only allow deriving a upper-limit on the mass about 270\,M$_\text{J}$.  
On the other han, Gaia measures $\epsilon$=0.51\,mas, which is also barely significant. The Gaia-Hipparcos discrepancy factor is not much informative with $\Delta Q$$<$$10$. 
This cannot be surprising since the orbital period is short compared to the Gaia-Hipparcos baseline of 25\,yrs. Applying the GASTON method on the value of $\epsilon$ leads 
to a mass of the companion next to the classical Hydrogen-burning limit at $\sim $80\,M$_\text{J}$. This is most likely an upper-limit on the mass. Therefore HD82460\,b 
should be a brown-dwarf.

\section{Discussion}
\label{sec:discussion}

The present results allow us to complete the statistics of brown dwarfs companions candidates around solar-like stars. 
In the following, we consider the $M\sin i$ of BD candidates rather than the exact mass, because considering only companions for which the true mass is derived would 
introduce bias, favouring the inclusion of transiting edge-on systems and systems closer to the Sun which astrometric motion is easier to measure. Moreover, many 
companions detected as exoplanets in RV survey might actually have a true mass in the BD-regime and the derived statistics would miss them.
Already published and new companions in the northern sky were compiled in Wilson et al. (2016), Sozzetti \& Desidera (2010) and the SB9 catalog (Pourbaix et al. 2004). 
We included all companions of these publications, even truly stellar, for which the $M\sin i$ is within 13.5-90\,M$_\text{J}$.  

\begin{figure}
\includegraphics[width=89mm]{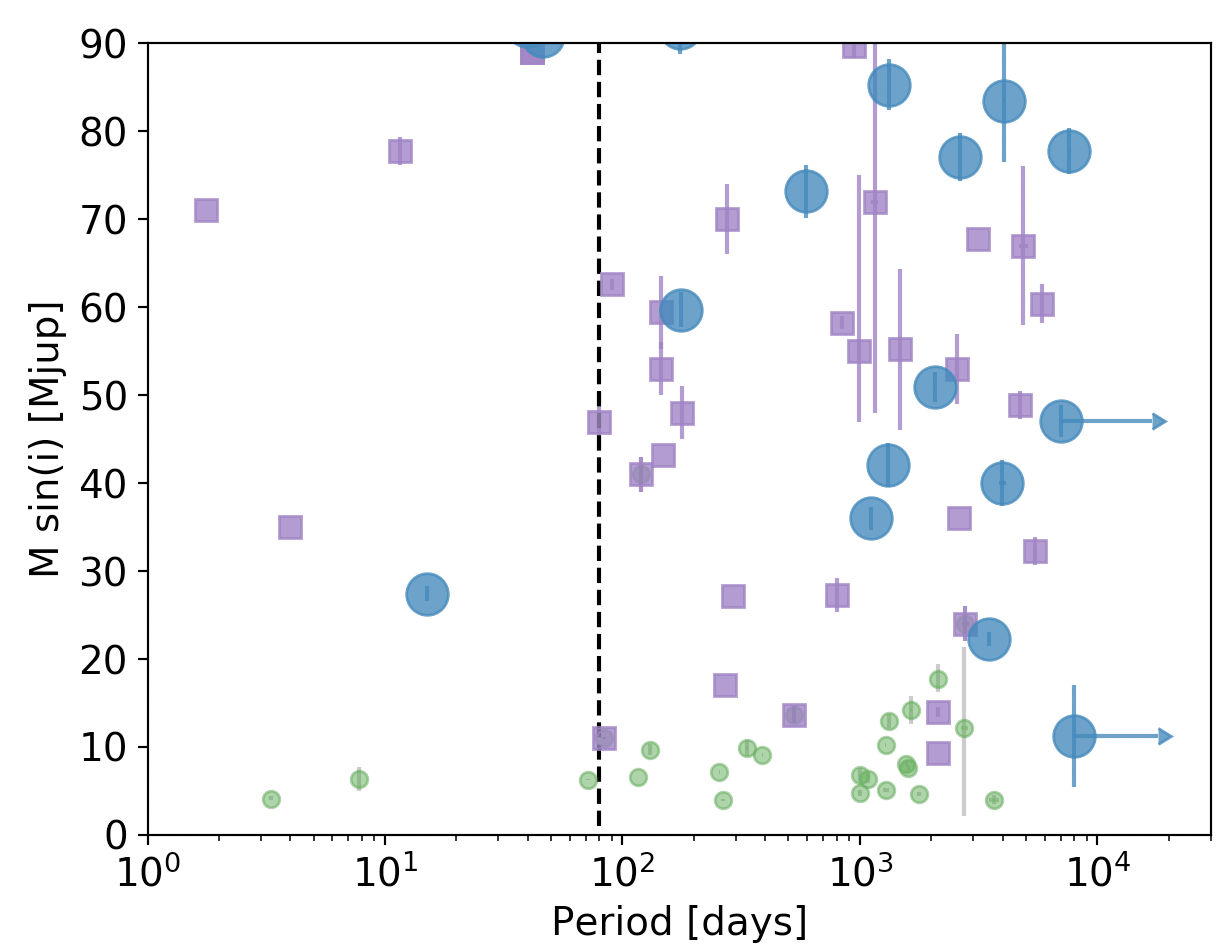}
\caption{\label{fig:BD_MP}$M\sin i$--$P$ diagram of all companions reported in this study (blue circles) compared to RV-detected BDs of Table A.1 in Wilson et al. (2016) (purple squares) and 
giant exoplanets ($M\sin i$$>$$1$\,M$_\text{J}$) from the Exoplanet.eu database (green circles). We selected only objects with $\delta$$>$$0^\circ$ and that satisfy the 
constraint of +0.35$<$$B-V$$<$+1, and $d$$<$$60$\,pc.}
\end{figure}

Our initial sample consists in a selection of 2350 targets which have $\delta$$>$$0^\circ$, $d$$<$$60$\,pc, +0.35$<$$B$-$V$$<$+1, and located at less than 
$\pm$2 mag from the main sequence (Dalal et al., in prep). These criterions were also used on the additional previously published data. Table~\ref{tab:sup_targets} gives a 
summary of all additional systems used here. Adding those to the new detections in this work, we obtain the $M\sin i$-period diagram plotted in Figure~\ref{fig:BD_MP}. This 
diagram shows a clear dearth of detection below $\sim$80-days period (0.4\,au semi-major axis), for the whole 15-90\,M$_\text{J}$ mass regime. On the contrary, the 
detected companions are more uniformly distributed in mass above this limit. 

\subsection{Brown dwarf frequency and a desert below 80\,days period}

The work presented here add a consequent number of new BD companions candidates, with orbital period shorter than 10,000 days and $M\sin i$ within 
13.5-90\,M$_\text{J}$. The full sample of main-sequence FGK stars within 60\,pc in the northern sky gathers $\sim$2950\,stars in the new Hipparcos catalog (van Leeuwen et 
al. 2007). Apart from the 12 new BD candidates reported in this paper, we collected 32 other BD candidates in the literature that are companions to main-sequence FGK stars 
at less than 60\,pc from the Sun in the Northern sky. This leads to a minimum of 44 BD candidates among the 2950\,systems identified by Hipparcos.

The monitoring program for the search of Giant planets with Sophie already gathered more than 3 RV points per star for 2050 of them. About 300 
sources still have less than 3 RV points and are still uncharacterised, but this number decreases yearly. More observing time being devoted to sources with more than 3 RV points 
presenting interesting variations, less observing time is available to complete the monitoring of a random set of stars.

Inspecting the RV variations of the 2050 systems with at least 3 RV points, apart from those published in this paper, we found that there could remain as much 
as $\sim$30 more BD candidates with a period up to 10,000 days to characterise. The time span of RV measurements in this sample of 2050 systems ranges from 2 to 
4200\,days for 99\% of them, with a median at 850\,days. Among these 30 yet unconstrained companions, about 15 have a long unconstrained-period orbit, which RV 
measurements probe a drift-like variation on a baseline of 300 to 5000\,days. Their period and $M\sin i$ could be considerably larger than these time spans. We thus estimate 
that between 0 and at least 30 BD companions, within the 2950 main-sequence and nearby FGK systems of the northern hemisphere, remain to be discovered in addition to 
the 44 brown-dwarfs gathered here.

This corresponds to a lower limit on the detection frequency of BD candidates within $M\sin i$=13.5-90\,M$_\text{J}$ and with orbital period less than 10,000 days 
of 2.0$\pm$0.5\%. This value remains compatible with the upper-limit of this frequency obtained by Guenther et al. (2005) with $f_\text{BD}$$<$$2\%$ for BD companions 
in the Hyades cluster with a semi-major axis (sma) $<$8\,AU, but larger than the estimation of Sahlmann et al. (2011), $f_\text{BD}$=1.3\% for candidates with sma$<$10\,AU.

Being obtained from RV and $M\sin i$ only, this frequency is overestimated due to the uncertainty on the system's inclination. Sahlmann et al. (2011) proposed a correction to this number 
by only considering companions that were not found to be real M-dwarf using astrometry, which  decreases this determination to $f_\text{BD,corr}$$<$$0.6$\%. This compared well to the rate 
estimation $<$0.5\% of Marcy \& Butler (2000) and the one derived by Santerne et al. (2016) of 0.29$\pm$0.17\% for transiting brown dwarfs with orbital periods smaller than 
400\,days observed with Kepler (Borucki et al. 2010). Applying the same procedure as in Sahlmann et al. (2011), we find that 35 companions over the 44 considered here 
are compatible with the BD domain. This leads to a revised lower-limit on the BD frequency $f_\text{BD,corr}$$>$$1.7$$\pm $$0.5$\%, which remains higher than 
expected. Still, this number is most likely overestimated since 25 of the 32 additional systems in Table~\ref{tab:sup_targets} were not constrained by astrometry yet. We are 
now applying the GASTON method systematically on all systems with BD candidates to constrain their mass. 

Below 80-days period, we find that the detection frequency drops significantly, with only 6 BD candidates found in this region. Within the 2050 systems mentioned above, 
we found only two possibly missing BDs with period less than 80\,days. This leads to a lower-limit $f_\text{BD, low}$$\sim $$0.24$$\pm $$0.04\%$, a factor of 8 lower than above 80\,days. 
This is in 
line with the findings of Ma \& Ge (2014) that brown dwarfs companions tend to avoid a mass-period region bounded by the limits 30-60\,M$_\text{J}$ and $P$$<$$100$\,days. 
Guillot et al. (2014) showed that this dearth of detections below 100-days period, especially for G-dwarfs, can be explained by the dynamical interactions between the stars 
and close-in companions, including tidal interactions, stellar evolution, magnetic braking of stars, and tidal dissipation by gravity waves. 

\subsection{The companion mass distribution beyond 80-days period}

\begin{figure}
\includegraphics[width=89mm]{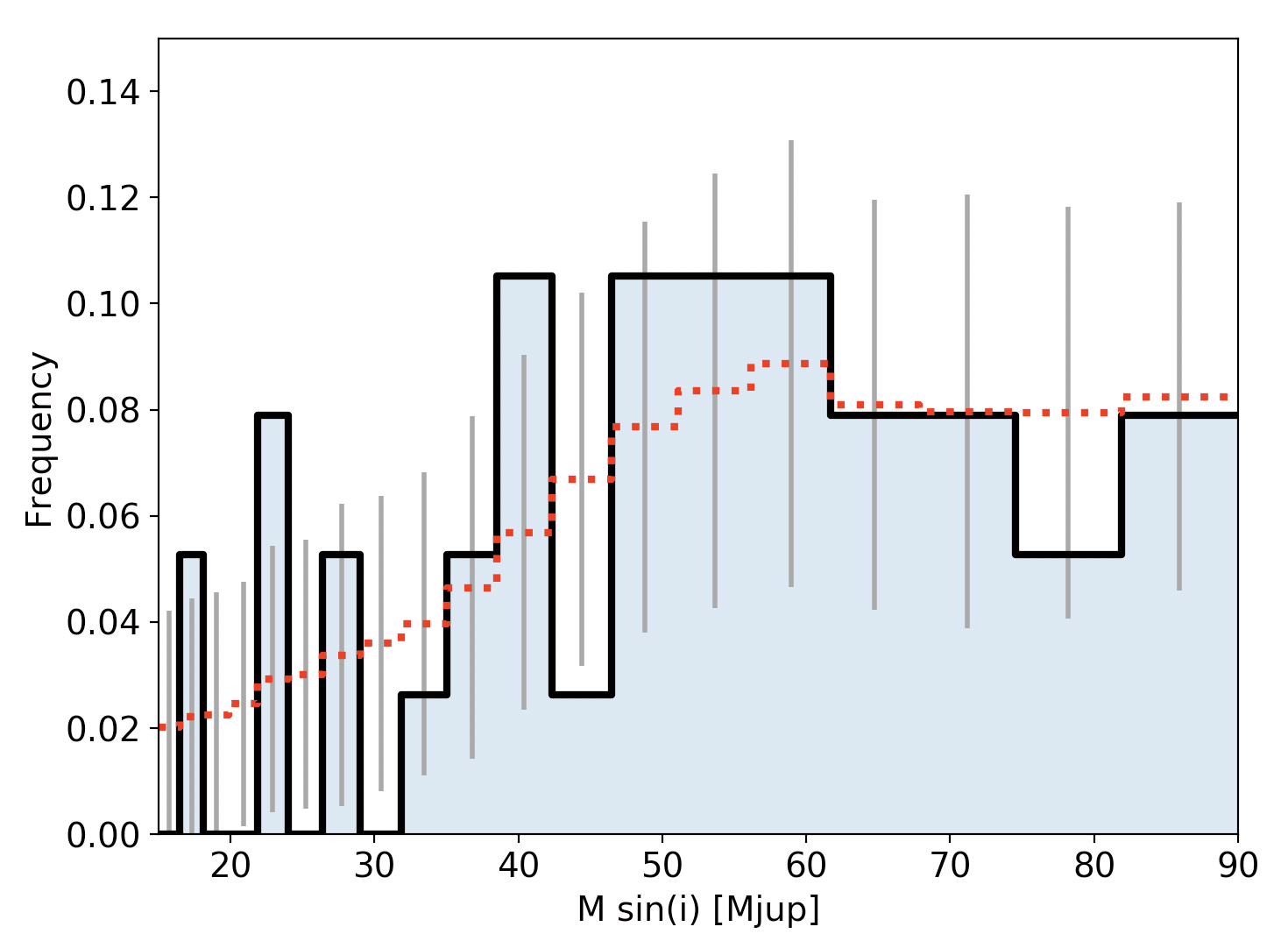}
\caption{\label{fig:histo}Histogram of the detections with $M\sin i$ between 15 and 80\,M$_\text{J}$ and $P$$>$$80$\,days. The bins are logarithmically spaced. It is compared, in red, 
 to the best-fitting distribution obtained by simulating the effect of the inclination on a random population of objects uniformly distributed in mass from 15\,M$_\text{J}$ to 
0.52\,M$_\odot$.}
\end{figure}

Studying in more details the $M\sin i$-histogram of detections beyond 80-days period, Fig.~\ref{fig:histo} shows that the $M\sin i$ distribution is possibly not exactly uniform, 
with a decrease in detection rate below $\sim$50\,M$_\text{J}$. This could be the sign of a lower bound in the true companion mass distribution that was suggested by the 
work of Halbwachs et al. (2000). We thus tried to reproduce this $M\sin i$ distribution out of simulated companions which masses are uniformly distributed from some lower-bound 
up until 0.52\,M$_\odot$. We introduced several different lower bounds on the mass from 15 to 100\,M$_\text{J}$, with a uniform density 
function above that limit and zero below. A random inclination is assigned to all these simulated objects, from which we can deduce $M\sin i$. We compare these to the 
38 detections between 15 and 90\,M$_\text{J}$ and with $P$$>$$80$\,days. We thus randomly selected 38 simulated values of $M\sin i$ and perform a two-samples 
Kolmogorov-Smirnov test between the simulated sample and the observed one. 

\begin{figure}
\includegraphics[width=89mm]{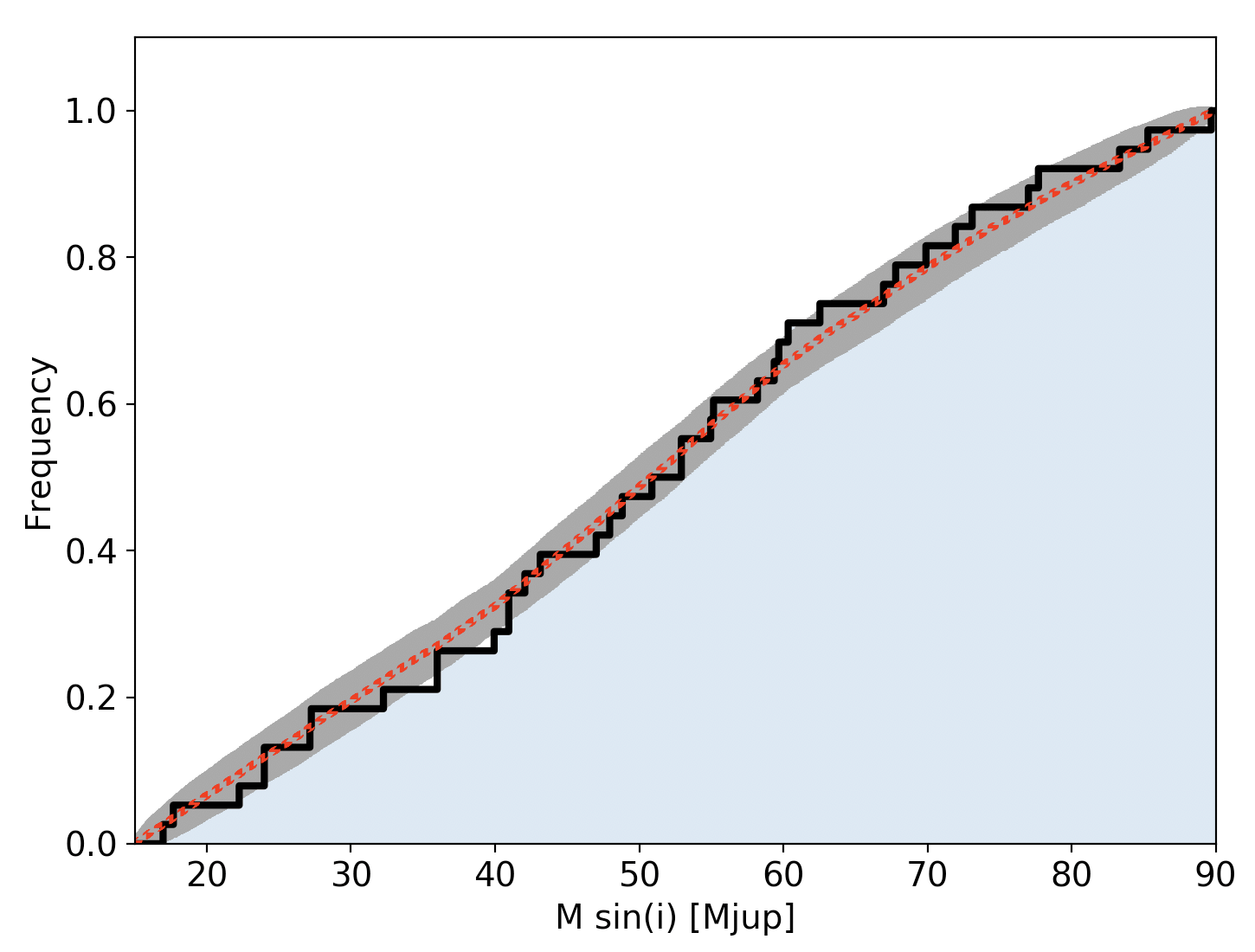}
\caption{\label{fig:cumul} Cumulative distribution of detections with $M\sin i$ between 15 and 80\,M$_\text{J}$ and $P$$>$$80$\,days. The red curve corresponds to the 
best-fitting distribution plotted in Fig.~\ref{fig:histo}.}
\end{figure}

This simulation is performed 1000 times. We then counted the number of simulated samples that are incompatible with the 
observed data, assuming the null-hypothesis rejection for a p-value$<$0.05. Two samples drawn from the same distribution should reject the null-hypothesis in average 
5\% of the time. If a mass distribution leads to a good modelling of the detection statistics, then about 95\% of the simulated samples out of this mass distribution should 
accept the null-hypothesis. We obtain that any uniform mass-distribution with a cut-off between 15 and 100\,M$_\text{J}$ leads to compatibility with observations for more 
than 95\% of the simulations. The $M\sin i$ histogram and cumulative plot of detections are plotted in Fig.~\ref{fig:histo} and~\ref{fig:cumul} and compared to the best of the 
simulations with the 15-M$_\text{J}$ lower bound model. 

Conversely to the intuition based on the $M\sin i$ distribution, the actual mass distribution of brown dwarf beyond 0.4\,au might thus be uniform all the way down to 
15\,M$_\text{J}$, as is found at wider separation (Reid et al. 2002). We do not find evidence of a lower mass limit in the brown-dwarf companions population, but
cannot exclude the existence of such bound in the mass distribution (Halbwachs et al. 2000, Sahlmann et al. 2011). 

Moreover, a second distribution coming from lower masses, 
i.e. the massive planets, does not appear necessary to explain the actual $M\sin i$ distribution beyond 80\,days. This suggests that the mass distribution of massive planets 
does not spread much within the BD domain. The tail of the distribution probably stops around 20\,M$_\text{J}$ but not much beyond. Independent direct imaging surveys 
of long-period brown-dwarfs (Brandt et al. 2014) reached similar conclusions, with low-mass BD, even those below the deuterium burning limit, more likely arising, as more 
massive objects, from gravitational collapse in disk or fragmenting cloud.


\section{Summary and conclusions}
\label{sec:conclusion}
We reported here the detection of 54 companions to FGK stars in the neighbourhood of the Sun using radial velocities observations of the SOPHIE spectrograph. 
Among them, 12 were detected as brown dwarfs candidates according to their projected mass $M\sin i$, and 42 as M-dwarfs. 

Using Hipparcos and Gaia, we could reconsider the values of the mass derived for several of the companions, as summarised in Table~\ref{tab:mass_summary}. 
We introduced a new method, GASTON, to derive inclination by combining the astrometric excess noise published in the Gaia DR1 and the Keplerian orbit derived from
RV variations. This allowed us to reconsider the mass of the 12 BD candidates derived thanks to SOPHIE's radial velocities. We found that 5 of them actually stand in the M-dwarf 
regime, and confirmed that the 7 remaining companions are likely brown-dwarfs. BD+291539\,b, HD23965\,b, HD48679\,b, and HD82460\,b are strongly constrained below 
90\,M$_\text{J}$. While HD28635\,b, HD211681\,b and HD71827\,b are possible brown dwarfs, although still remains candidates. Moreover, we obtained a stellar 
mass for the companion of HD210631 that was previously published as a BD candidate by Latham et al. (2002). 

Our 12 BD detected with RV, added to those reviewed and discovered in Wilson et al. (2016), in the SB9 (Pourbaix et al. 2004) and in Sozzetti \& Desidera (2010), place a strong 
lower limit on the orbital period of brown dwarfs at 80\,days. The short period region below 80\,days appears 4 times less populated than at a larger period. On the other hand, 
above 80\,days the $M\sin i$ detection density is well reproduced by a flat distribution of mass that goes down to as low as 15\,M$_\text{J}$. Moreover, a long extension of the 
massive planet distribution beyond 20\,M$_\text{J}$ does not appear necessary to reproduce the detected $M\sin i$ of all companions in the BD regime. 

These conclusions should however not be understood as definitive. This statistics is not completely free of biases yet, since the monitoring of the volume limited sample is not 
completed. However, we estimated that only few short-period radial-velocity BD could have been missed in our sample, and a conservative number of up to 30 
companions could still be missing in the 15-90\,M$_\text{J}$ regime at orbital period shorter than 10,000\,days. 

As was demonstrated in Sahlmann et al. (2011), accounting for the exact mass rather than minimum mass can strongly change the picture. Moreover, we showed that only being based on 
$M\sin i$ cannot lead to a firm conclusion on the presence or absence of a lower bound on mass for brown dwarfs. Only a systematic search for constraints on both mass and 
inclination for every system with a candidate BD will lead to the ultimate unbiased mass distribution of brown-dwarfs. 

In this work, we showed that such goal will be fulfilled thanks to the combination of Gaia and high-resolution spectroscopy. By 2022, it is likely that acceleration solutions and 
actual orbital solutions for many of the detected companions presented in this work (and many new brown dwarf and stellar companions) will become available in 
the Gaia DR3. In the meanwhile, the GASTON method presented here will allow using the already published data release of Gaia to constraint the inclination and mass of many systems and 
companions, including planets, brown dwarfs and binary stars.

\begin{acknowledgements}
We are thankful to the anonymous referee for his constructive comments that help to significantly improve the quality of this paper. We thank all the staff of Haute-Provence 
Observatory for their support at the 1.93 m telescope and on SOPHIE. This work is based on observations made with SOPHIE in the context
of the programme "Recherche de Planètes Extrasolaire" (PI: I. Boisse, F. Bouchy) and the Programme National de Physique Stellaire (PNPS), in particular the programme "Exploring 
the Brown Dwarf Desert around FGK stars in the Solar neighborhood" (PI: F. Kiefer ; IDs: 2017B\_PNPS008, 2018A\_PNPS008, 2018B\_PNPS013). F.K. acknowledges support by 
a fellowship grant from the Centre National d'Etude Spatiale (CNES). This work was also supported by the PNPS of CNRS/INSU co-funded by CEA and CNES. A.V-M and 
F.K thank the CNES for financial support. P.A.W. acknowledge support from the European Research Council under the European Unions Horizon 2020 research and innovation 
programme under grant agreement No. 694513. V. B. and J. H. acknowledge support by the Swiss National Science Foundation (SNSF) in the frame of the National Centre for 
Competence in Research PlanetS. V. B. has received funding from the European Research Council (ERC) under the European Union’s Horizon 2020 research and innovation 
programme (project Four Aces; grant agreement No 724427). This work was supported by FCT - Funda\c c\,\!$\tilde{\text{a}}$\,\!o para a Ci\^encia e a Tecnologia through national funds and by FEDER 
through COMPETE2020 - Programa Operacional Competitividade e Internacionaliza\c c\,\!$\tilde{\text{a}}$\,\!o by these grants: UID/FIS/04434/2013, UID/FIS/04434/2019 \& POCI-01-0145-FEDER-007672; PTDC/FIS-AST
/28953/2017 \& POCI-01-0145-FEDER-028953 and PTDC/FIS-AST/32113/2017 \& POCI-01-0145-FEDER-032113. N. A-D. acknowledges the support of FONDECYT project 3180063. 
This work presents results from the European Space Agency (ESA) space mission Gaia. Gaia data are being processed by the Gaia Data Processing and Analysis Consortium (DPAC). Funding for the DPAC is provided by national 
institutions, in particular the institutions participating in the Gaia MultiLateral Agreement (MLA). The Gaia mission website is https://www.cosmos.esa.int/gaia. The Gaia archive 
website is https://archives.esac.esa.int/gaia.
\end{acknowledgements}


\begin{appendix}

\section{Photocenter semi-major axis} 
\label{sec:photocenter}

To fix the value of the photocenter semi-major axis, we use the following formula (van de Kamp 1975):
\begin{equation}
a_\text{phot} = (B - \beta)\,a_\text{tot}  \qquad \text{with } a_\text{tot} = P^{2/3}\,M_\star^{1/3}\left(1+q\right)^{1/3}\,\varpi
\end{equation}

The period $P$ is expressed in years ; the masses are given in $M_\odot$ ; the parallax $\varpi$ unit is mas. $\beta$ is the luminosity fraction $\beta$=$L_2/(L_1+L_2)$ and $B$ the 
mass fraction $B$=$q/(1+q)$. The mass ratio $q$ can be found by solving the equation of the mass function (see e.g. Halbwachs et al. 2014)
\begin{equation}
q^3 = \left(\frac{M_\text{RV}/\sin i}{M_\star}\right)^3 \, \left(1+q\right)^{2}
\end{equation} 

with $M_\text{RV}$=$M_\star^{2/3} K (P/2\pi G)^{1/3}$. The solution of this equation, fixing $q_0= M_\text{RV}/\sin(i)\,M_\star$, is
\begin{align}
q = & \frac{q_0^3}{3} +\frac{\sqrt[3]{2}}{3}  \frac{q_0^2\left(6+q_0^3\right)}{\left(27 + 18 \, q_0^3 + 2\,q_0^6 + 3 \sqrt{3} \sqrt{27+4\,q_0^3}\right)^{1/3}} \nonumber \\  
& + \frac{q_0}{3 \sqrt[3]{2}}  \left(27 + 18 \, q_0^3 + 2\,q_0^6 + 3 \sqrt{3} \sqrt{27+4\,q_0^3}\right)^{1/3}
\end{align}

Knowing the primary mass $M_1$ from Table~\ref{tab:spec_type} and deducing $M_2$ from $q$, the luminosity fraction at optical wavelength is derived by
\begin{equation}
\beta=\frac{1}{1+10^{(M_{V,2}-M_{V,1})/2.5}}
\end{equation}

Then, the visual luminosity fraction can be calculated thanks to the empirical relation existing between absolute visual magnitude and star mass (Kroupa et al. 1993) which can be 
approached by
\begin{equation}
M_V  = 5.69\,\left(\frac{M_{1,2}}{M_\odot}\right)^2 - 17.54\,\left(\frac{M_{1,2}}{M_\odot}\right) + 16.43
\end{equation}

Since no secondary peaks was seen in any CCF for all targets, we assumed in the simulations that $M_{V,2}-M_{V,1}$ had to be greater than $2.5$, and thus $\beta$$<$$0.1$. 
This led us to discard certain values of $I_c$ implying too large values of the luminosity fraction.

\onecolumn
\section{Tables}
\pagestyle{plain}

\vspace*{1cm}
\begin{table*}[htp]\tiny\centering
\caption{\label{tab:targets}Table of the 54 observed targets. Coordinates, magnitudes, color and spectral types are taken from SIMBAD. Parallaxes were obtained from the Gaia DR2 
(http://gea.esac.esa.int/archive/).}
\begin{tabular}{lcccccccc}
Name & RA         & DEC       & $V$ & $B-V$ & $\pi$   & Sp. Type      &  Time span & $N_\text{meas}$                \\
           &(J2000) & (J2000) &    &         &  (mas) &                      &    (days)       &  Total (SOPHIE/SOPHIE+)  \\
\hline
HD225239  	& 00:04:53.7604 & +34:39:35.259 & 6.11 	& 0.63 	& 28.28$\pm$0.23  	& G2V    	& 1324.26    	&	45 (0/45)	\\
BD+210055 	& 00:30:31.0151 & +22:46:08.282 & 9.24 	& 0.94 	& 26.375$\pm$0.097  & K5    	& 2303.79    	&	20 (0/20)	\\
HD5470 		& 00:56:40.2179 & +17:57:35.427	& 8.33	& 0.64	& 14.853$\pm$0.097	& G0    	& 5812.53    	&	24 (1/3)	\\
HD13014   	& 02:08:26.0644 & +43:11:27.653 & 7.59 	& 0.62 	& 15.075$\pm$0.076  & F5     	& 1354.27    	&	16 (0/16)	\\
HD15292  	& 02:33:40.2378 & +77:40:02.217 & 7.66 	& 0.69 	& 27.063$\pm$0.028  & G5     	& 761.78    	&	12 (0/12)	\\
HD18450  	& 02:58:52.4290 & +26:46:26.675 & 8.22 	& 0.90 	& 34.022$\pm$0.066 	& K2V     	& 2263.87    	&	11 (0/11)	\\
HD23965  	& 03:50:03.4617 & +22:35:29.896 & 7.27 	& 0.54 	& 23.183$\pm$0.054  & F7     	& 3779.79    	&	84 (19/65)	\\
HD24505		& 03:54:59.8438 & +28:11:17.160 & 8.03	& 0.70 	& 13.65$\pm$0.10 	& G5III 	& 5135.40    	&	29 (0/8)	\\
HD28635		& 04:31:29.3459 & +13:54:12.510 & 7.75 	& 0.55 	& 20.384$\pm$0.084 	& F9V   	& 7703.28    	&	29 (0/13)	\\
HD40647	 	& 06:06:05.7202 & +69:28:34.069 & 8.26	& 0.80	& 31.86$\pm$0.16	& G5V   	& 4462.55    	&	34 (18/7)	\\
HD48679  	& 06:58:18.0025 & +80:55:42.319 & 8.85 	& 0.75 	& 14.964$\pm$0.078  & G0     	& 1288.64    	&	26 (0/26)	\\
BD+291539 	& 07:30:22.3337 & +29:22:50.355 & 9.34 	& 0.79 	& 16.016$\pm$0.040 	& G5     	& 1317.21    	&	17 (0/17)	\\
HD60846  	& 07:38:17.9491 & +42:27:35.051 & 7.92 	& 0.62 	& 13.736$\pm$0.067  & F8    	& 2194.99    	&	10 (0/10)	\\
HD62923  	& 07:48:31.8080 & +47:45:53.769 & 8.03 	& 0.75 	& 17.22$\pm$0.74 	& G5    	& 1837.98    	&	18 (3/0)	\\
HD71827   	& 08:37:14.4992 & +77:02:48.470 & 7.29 	& 0.53 	& 22.703$\pm$0.024 	& F8     	& 4837.82    	&	46 (0/44)	\\
HD73636  	& 08:41:14.5774 & +47:28:50.118 & 7.57 	& 0.57 	& 17.50$\pm$0.12  	& G0     	& 4167.73    	&	22 (20/2)	\\
HD77712   	& 09:04:15.0684 & +03:01:34.932 & 8.93 	& 0.85 	& 19.57$\pm$0.12  	& K1/2(V) 	& 1508.85    	&	19 (0/19)	\\
HD78536  	& 09:08:53.9250 & +03:57:33.093 & 8.30 	& 0.65 	& 12.257$\pm$0.054  & G3V    	& 4068.99    	&	8 (3/5)	\\
HD82460  	& 09:33:28.7365 & +46:13:43.274 & 8.37 	& 0.66 	& 19.826$\pm$0.091  & G0    	& 1819.03    	&	17 (0/17)	\\
BD+281779 	& 09:36:50.0041 & +27:58:22.406 & 9.17 	& 0.82 	& 23.415$\pm$0.053 	& G5V   	& 2419.36    	&	15 (4/11)	\\
HD85533  	& 09:55:46.9216 & +70:02:28.078 & 8.46 	& 0.68 	& 18.765$\pm$0.037  & G5    	& 3141.28    	&	11 (5/6)	\\
HD87899  	& 10:09:14.2011 & +46:17:02.355 & 8.88 	& 0.65 	& 19.15$\pm$0.14  	& G5     	& 1168.96    	&	20 (20/0)	\\
HD101305  	& 11:39:28.4419 & +02:50:47.630 & 8.33 	& 0.54 	& 14.187$\pm$0.072 	& F6V    	& 1524.87    	&	19 (0/19)	\\
HD103913 	& 11:58:04.3185 & +25:08:16.182 & 8.28 	& 0.52 	& 11.371$\pm$0.063  & F8    	& 4089.76    	&	21 (13/8)	\\
HD104289 	& 12:00:41.2765 & +59:21:11.190 & 8.07 	& 0.52 	& 14.182$\pm$0.043  & F8     	& 5136.81    	&	16 (0/9)	\\
BD+192536 	& 12:10:04.4512 & +18:58:36.151 & 10.08 & 1.22 	& 21.86$\pm$0.15  	& K5     	& 4105.73    	&	11 (4/7)	\\
HD106888 	& 12:17:36.1848 & +14:26:34.187 & 8.18 	& 0.54 	& 14.89$\pm$0.13 	& F8       	& 4304.19    	&	20 (14/4)	\\
HD108436 	& 12:26:53.7835 & +69:43:46.205 & 8.46 	& 0.63 	& 18.392$\pm$0.082  & G0       	& 2043.26    	&	8 (0/8)	\\
HD109157 	& 12:32:27.4354 & +28:05:04.636 & 9.16 	& 0.82 	& 22.84$\pm$0.10  	& G7IV     	& 1935.93    	&	11 (0/11)	\\
BD+132550 	& 12:34:52.7685 & +12:27:33.399 & 8.94	& 0.75	& 15.66$\pm$0.12 	& G5      	& 1281.74    	&	10 (0/10)	\\
HD110376 	& 12:41:37.0772	& +19:51:04.687	& 8.99	& 0.95	& 30.95$\pm$0.11	& K3V     	& 17201.33    	&	103 (0/8)	\\
HD130396 	& 14:47:31.8899 & +19:03:00.114 & 7.45 	& 0.50 	& 22.127$\pm$0.076  & F8V     	& 2155.90    	&	30 (0/30)	\\
HD133621 	& 15:00:26.9508 & +71:45:55.645 & 6.66 	& 0.61 	& 28.73$\pm$0.26  	& G0      	& 10222.62    	&	38 (0/12)	\\
BD+362641 	& 15:45:00.2842 & +35:57:40.797 & 10.11 & 1.15 	& 18.499$\pm$0.032  & K4/5V  	& 979.22    	&	9 (9/0)	\\
BD+212816 	& 15:45:30.0338 & +21:10:43.015 & 9.21 	& 0.85 	& 19.49$\pm$0.21  	& K0      	& 4003.96    	&	17 (12/5)	\\
HD144286 	& 16:03:55.0314 & +31:02:34.943 & 9.42 	& 0.75 	& 14.912$\pm$0.080  & K0      	& 2317.76    	&	15 (0/15)	\\
HD146735	& 16:14:44.6766 & +57:01:34.537 & 8.38	& 0.60	& 11.912$\pm$0.032	& G0    	& 2295.82    	&	12 (0/12)	\\
HD147487 	& 16:21:15.5930 & +27:22:32.054 & 8.56 	& 0.57 	& 14.93$\pm$0.31 	& G0V    	& 1087.97    	&	11 (0/11)	\\
HD153376 	& 16:58:37.9267 & +15:27:15.681 & 6.90 	& 0.63 	& 18.526$\pm$0.027  & F8V    	& 4059.90    	&	20 (18/2)	\\
HD155228 	& 17:09:27.2020 & +22:05:30.558 & 7.04 	& 0.49 	& 14.75$\pm$0.18 	& F6V    	& 928.59    	&	13 (0/13)	\\
HD156111  	& 17:14:57.1366 & +19:40:57.353 & 7.22 	& 0.81 	& 22.576$\pm$0.054  & G8V    	& 475.75    	&	21 (0/21)	\\
HD156728 	& 17:16:39.1391 & +50:36:23.329 & 8.03 	& 0.64 	& 23.63$\pm$0.12 	& G5     	& 2264.84    	&	12 (0/12)	\\
HD161479 	& 17:45:02.9275 & +19:17:25.650 & 8.11 	& 0.78 	& 20.353$\pm$0.039  & K0     	& 5080.98    	&	19 (4/8)	\\
BD+680971	& 18:00:36.1034	& +68:33:24.238	& 9.75	& 0.81	& 16.405$\pm$0.068	& K2    	& 2476.15    	&	16 (0/16)	\\
HD167215 	& 18:12:59.4025 & +28:15:27.357 & 8.10 	& 0.52 	& 12.201$\pm$0.032  & F8    	& 14262.80    	&	54 (14/5)	\\
HD193554 	& 20:20:03.6455 & +23:38:17.172 & 8.26 	& 0.63 	& 26.60$\pm$0.30  	& G5    	& 5752.59    	&	62 (6/4)	\\
HD204613 	& 21:27:42.9669 & +57:19:18.864 & 8.22 	& 0.65 	& 15.91$\pm$0.29  	& G1IIIa:CH1.5	& 14876.74 	&	43 (11/5)	\\
HD207992 	& 21:52:19.6628 & +39:48:06.213 & 8.28 	& 0.72 	& 26.069$\pm$0.066  & G5    	& 5218.37    	&	19 (12/2)	\\
HD211681 	& 22:06:49.0856	& +85:24:33.746	& 8.09	& 0.74	& 13.822$\pm$0.030	& G5    	& 5942.22    	&	56 (23/7)	\\
HD210631 	& 22:11:39.3642	& +06:11:36.405	& 8.51	& 0.60	& 14.04$\pm$0.15	& G0     	& 12929.42    	&	93 (0/8)	\\
HD212029 	& 22:20:23.8494	& +46:25:05.719 & 8.51 	& 0.50 	& 16.12$\pm$0.14 	& G0     	& 1796.08   	&	21 (0/21)	\\
HD212733	& 22:25:55.0226	& +35:21:53.473	& 8.30	& 0.91	& 32.956$\pm$0.061	& K2    	& 5306.19    	&	38 (2/22)	\\
HD212735 	& 22:26:21.5485 & +10:45:27.260 & 8.25 	& 0.75 	& 18.174$\pm$0.072  & G5     	& 3416.71    	&	11 (6/5)	\\
HD217850 	& 23:02:36.6571 & +58:52:33.315 & 8.50 	& 0.80 	& 15.17$\pm$0.20 	& G8V     	& 4480.33    	&	64 (9/32)	\\
\hline
\end{tabular}
\end{table*}

\begin{table*}[hbt]
\caption{\label{tab:pub}Published data and public non-SOPHIE data for 19 systems. In Butler et al. (2017), some data were published but no orbit and companion mass were explicitly derived.
For the published orbits, we appended the $M\sin i$ derived hereafter for comparison.}
\centering
\begin{tabular}{l@{~~}c@{~~}c@{~~}c@{~~}c@{~~}|c}
Name 		& Instrument 			&	Reference					& $N_\text{meas}$ 	&	$M\sin i\,_\text{pub} $   &  $M\sin i \,_\text{here}$\\ 
				&	 							&									&								&	(M$_\text{J}$)		&	(M$_\text{J}$) \\
\hline
HD104289	&	Elodie				&	---							&	7		&	      &\\      
HD106888	&	Elodie				&	---							&	2		&	      & \\ 
HD110376	& 	Coravel/RVS\tablefootmark{$a$}			&	Griffin et al. (2006)		&	95	&	145$\pm$2\tablefootmark{$c$}	 & 177.7$\pm$7.1\\
HD133621 	&	CfA\tablefootmark{$a$}					& 	Latham et al. (2002) 		& 	26	&	92$\pm$7	 & 101.8$\pm$3.5 \\
HD161479	&	HIRES				&	Butler et al. (2017)		&	3  		&       & \\%268.7$\pm$9.4\\
			&	Elodie				&	---							&	4		&	      & \\
HD167215	&	SOPHIE				&	D\'iaz et al. (2012)			&	14		&	74$-$121 & 167.5$\pm$6.3\\
			& 	Coravel\tablefootmark{$a$}				&	Halbwachs et al. (2012)		&	22		& 	141$\pm$20 & 	\\
HD193554 	& 	Coravel\tablefootmark{$a$}			&	Griffin et al. (2013)		&	52		&	171$\pm$4	& 173.6$\pm$6.0 \\
HD204613 	&	RVS\tablefootmark{$a$,$b$}				& 	McCLure et al. (1997)		&	27		&	147$\pm$7	& 151.7$\pm$5.5 \\
HD207992	&	HIRES				&	Butler et al. (2017)		&	5  		&       & \\
HD210631	&	CfA\tablefootmark{$a$}					&	Latham et al. (2002) 		& 	85	&	82$\pm$6	& 83.4$\pm$6.9	\\
HD211681	& 	HIRES				&	Patel et al. (2007)			&	9		&	72$-$102 & 77.8$\pm$2.6 \\
			&	Elodie				&	---							&	12		&	     	&  \\
HD212733	&	HIRES				&	Butler et al. (2017)		&	6		&	      & \\%112.1$\pm$3.6 \\
			&	Elodie				&	---							&	8		&	      \\
HD217850 	&	HIRES				& 	Butler et al. (2017)		&	23  	&	11  & 22.16$\pm$0.73 \\ 
HD24505		&	HIRES				&	Butler et al. (2017)		&	18  	&	     &	\\
			&	Elodie				&	---							&	1		&	      	&\\
HD28635		&	HIRES\tablefootmark{$a$}				&	Paulson et al. (2004)		&	13		&	70 & 77.1$\pm$2.7\\
			& 	Elodie				&	---							&	3		&	      &\\
HD40647		&	HIRES				&	Butler et al. (2017)		&	20  	&	     	& \\
HD5470		& 	HIRES				&	Patel et al. (2007)			&	10		&	163$-$465 & 208.5$\pm$7.0\\
			&	HIRES				&	Butler et al. (2017)		&	20  	&	     	&	\\
HD62923		&	Elodie				&	---							& 15		&	      & \\
HD71827		& 	Elodie				&	---							&	2		&	      &\\
\hline 
\end{tabular} 
\tablefoot{\\
\tablefoottext{$a$}{Data archived on the SB9 catalog (http://sb9.astro.ulb.ac.be).} \\
\tablefoottext{$b$}{Radial-velocity spectrometer of the Domininon Astrophysical Observatory in Victoria.} \\
\tablefoottext{$c$}{Using a K3-star primary mass of 0.75\,$M_\odot$.}}
\end{table*}

\begin{table*}\tiny\centering
\caption{\label{tab:spec_type}Table of stellar parameters for the 54 stars in Table~\ref{tab:targets}. See explanations in Section~\ref{sec:spectro}. } 
\begin{tabular}{l@{\quad}c@{\quad}c@{\quad}c@{\quad}c@{\quad}c@{\quad}c@{\quad}c@{\quad}c@{\quad}c}
Star	& $T_\text{eff}$	& $\log g$ 	& $v_\text{turb}$	& [Fe/H]		& N$_\text{lines}$(Fe I)& N$_\text{lines}$(Fe II)& $M_\text{Torres}$	& $\log R'_{HK}$  & $v\sin i$ \\
		&		($^\circ$K)	&	(s.i.)				&	(km\,s$^{-1}$)	&	(dex)		&						&						&	(M$_\odot$)		&	(dex) 			& (km/s)\\
\hline
BD+132550	&	5529$\pm$38	& 4.12$\pm$0.20	&	 0.83$\pm$0.06	&	 0.21$\pm$0.03	&	244	&	31	&	 1.05$\pm$0.08 	& 	-4.59$\pm$0.11 	& 3.4$\pm$1.0 \\
BD+192536	&	4609$\pm$182& 4.52$\pm$0.52	&	 0.29$\pm$1.16	&	 0.08$\pm$0.10	&	108	&	13	&	 0.75$\pm$0.18 	& 	-4.52$\pm$0.10 	& 2.1$\pm$1.0 \\
BD+210055	&	4833$\pm$73	& 4.38$\pm$0.24	&	 0.15$\pm$1.30	&	-0.22$\pm$0.09	&	110	&	13	&	 0.77$\pm$0.05 	& 	-4.56$\pm$0.11 	& 2.1$\pm$1.0 \\
BD+212816	&	5263$\pm$36	& 4.56$\pm$0.21	&	 0.59$\pm$0.10	&	-0.08$\pm$0.02	&	223	&	33	&	 0.81$\pm$0.03 	& 	-4.41$\pm$0.16 	& 2.5$\pm$1.0 \\
BD+281779	&	5136$\pm$42	& 4.48$\pm$0.20	&	 0.34$\pm$0.15	&	-0.33$\pm$0.03	&	108	&	13	&	 0.77$\pm$0.03 	& 	-4.54$\pm$0.12 	& 3.1$\pm$1.0 \\
BD+291539	&	5445$\pm$38	& 4.41$\pm$0.20	&	 0.69$\pm$0.07	&	 0.14$\pm$0.03	&	242	&	33	&	 0.91$\pm$0.03 	& 	-4.65$\pm$0.14 	& 3.3$\pm$1.0 \\
BD+362641	&	4753$\pm$190& 4.44$\pm$0.47	&	 0.06$\pm$3.33	&	-0.09$\pm$0.04	&	 72	&	 5	&	 0.79$\pm$0.17 	& 	-4.86$\pm$0.16 	& 0 \\
BD+680971	&	5286$\pm$31	& 4.41$\pm$0.20	&	 0.74$\pm$0.06	&	-0.05$\pm$0.02	&	236	&	33	&	 0.85$\pm$0.06 	& 	-4.43$\pm$0.11 	& 3.0$\pm$1.0 \\
HD101305	&	6040$\pm$27	& 4.12$\pm$0.20	&	 0.92$\pm$0.04	&	-0.28$\pm$0.02	&	229	&	32	&	 0.99$\pm$0.03 	& 	-4.54$\pm$0.18 	& 1.7$\pm$1.0 \\
HD103913	&	5964$\pm$27	& 3.93$\pm$0.20	&	 1.09$\pm$0.04	&	-0.10$\pm$0.02	&	236	&	35	&	 1.12$\pm$0.04 	& 	-4.52$\pm$0.27 	& 3.5$\pm$1.0 \\
HD104289	&	6231$\pm$39	& 4.03$\pm$0.21	&	 1.20$\pm$0.05	&	 0.07$\pm$0.03	&	218	&	29	&	 1.18$\pm$0.04 	& 	-4.62$\pm$0.23 	& 4.4$\pm$1.0 \\
HD106888	&	6249$\pm$49	& 4.29$\pm$0.21	&	 1.17$\pm$0.07	&	 0.14$\pm$0.04	&	221	&	28	&	 1.12$\pm$0.04 	& 	-4.40$\pm$0.17 	& 5.5$\pm$1.0 \\
HD108436	&	5651$\pm$23	& 4.23$\pm$0.19	&	 0.56$\pm$0.05	&	-0.38$\pm$0.02	&	241	&	32	&	 0.88$\pm$0.06 	& 	-4.67$\pm$0.13 	& 2.3$\pm$1.0 \\
HD109157	&	5184$\pm$34	& 4.45$\pm$0.20	&	 0.61$\pm$0.08	&	-0.10$\pm$0.02	&	237	&	33	&	 0.81$\pm$0.06 	& 	-4.59$\pm$0.11 	& 3.2$\pm$1.0 \\
HD110376	&	4826$\pm$59	& 4.37$\pm$0.23	&	 0.30$\pm$0.31	&	-0.23$\pm$0.03	&	115	&	13	&	 0.77$\pm$0.07 	& 	-4.68$\pm$0.10 	& 1.9$\pm$1.0 \\
HD13014 	&	6075$\pm$34	& 3.95$\pm$0.20	&	 1.22$\pm$0.04	&	 0.20$\pm$0.03	&	242	&	32	&	 1.25$\pm$0.09 	& 	-4.76$\pm$0.13 	& 6.1$\pm$1.0 \\
HD130396	&	6349$\pm$26	& 4.18$\pm$0.21	&	 1.16$\pm$0.04	&	-0.03$\pm$0.02	&	232	&	30	&	 1.11$\pm$0.03 	& 	-4.63$\pm$0.12 	& 3.2$\pm$1.0 \\
HD133621	&	5711$\pm$23	& 4.06$\pm$0.19	&	 0.89$\pm$0.03	&	-0.43$\pm$0.02	&	234	&	33	&	 0.93$\pm$0.03 	& 	-4.85$\pm$0.13 	& 2.4$\pm$1.0 \\
HD144286	&	5353$\pm$41	& 4.40$\pm$0.20	&	 0.54$\pm$0.09	&	-0.02$\pm$0.03	&	238	&	33	&	 0.87$\pm$0.03 	& 	-4.56$\pm$0.12 	& 2.2$\pm$1.0 \\
HD146735	&	5974$\pm$19	& 3.98$\pm$0.20	&	 1.10$\pm$0.02	&	 0.11$\pm$0.01	&	248	&	32	&	 1.17$\pm$0.08 	& 	-4.79$\pm$0.12 	& 3.2$\pm$1.0 \\
HD147487	&	5865$\pm$26	& 4.19$\pm$0.20	&	 0.77$\pm$0.05	&	-0.21$\pm$0.02	&	226	&	27	&	 0.96$\pm$0.07 	& 	-4.62$\pm$0.15 	& 2.6$\pm$1.0 \\
HD15292 	&	5679$\pm$25	& 4.27$\pm$0.20	&	 0.79$\pm$0.04	&	 0.02$\pm$0.02	&	240	&	34	&	 0.96$\pm$0.03 	& 	-4.85$\pm$0.13 	& 3.0$\pm$1.0 \\
HD153376	&	5944$\pm$26	& 3.81$\pm$0.20	&	 1.18$\pm$0.03	&	 0.14$\pm$0.02	&	238	&	26	&	 1.31$\pm$0.09 	& 	-4.51$\pm$0.26 	& 4.7$\pm$1.0 \\
HD155228	&	6272$\pm$31	& 3.81$\pm$0.21	&	 1.47$\pm$0.04	&	-0.13$\pm$0.02	&	213	&	30	&	 1.23$\pm$0.08 	& 	-4.90$\pm$0.19 	& 5.6$\pm$1.0 \\
HD156111	&	5208$\pm$21	& 4.01$\pm$0.19	&	 0.65$\pm$0.04	&	-0.35$\pm$0.02	&	253	&	32	&	 0.90$\pm$0.04 	& 	-4.82$\pm$0.15 	& 2.9$\pm$1.0 \\
HD156728	&	5777$\pm$21	& 4.35$\pm$0.20	&	 0.74$\pm$0.04	&	-0.14$\pm$0.02	&	240	&	32	&	 0.92$\pm$0.03 	& 	-4.64$\pm$0.15 	& 2.6$\pm$1.0 \\
HD161479	&	5642$\pm$30	& 4.16$\pm$0.20	&	 1.06$\pm$0.04	&	 0.25$\pm$0.02	&	237	&	30	&	 1.06$\pm$0.08 	& 	-4.42$\pm$0.12 	& 4.4$\pm$1.0 \\
HD167215	&	6201$\pm$39	& 3.99$\pm$0.20	&	 1.31$\pm$0.05	&	-0.29$\pm$0.03	&	203	&	30	&	 1.06$\pm$0.07 	& 	-4.73$\pm$0.22 	& 4.9$\pm$1.0 \\
HD18450 	&	5016$\pm$40	& 4.39$\pm$0.21	&	 0.25$\pm$0.18	&	-0.19$\pm$0.02	&	115	&	13	&	 0.79$\pm$0.04 	& 	-4.68$\pm$0.13 	& 1.9$\pm$1.0 \\
HD193554	&	5841$\pm$22	& 4.25$\pm$0.20	&	 0.91$\pm$0.03	&	-0.12$\pm$0.02	&	247	&	31	&	 0.96$\pm$0.07 	& 	-4.40$\pm$0.13 	& 2.7$\pm$1.0 \\
HD204613	&	5868$\pm$28	& 4.10$\pm$0.20	&	 0.98$\pm$0.05	&	-0.27$\pm$0.02	&	239	&	30	&	 0.97$\pm$0.03 	& 	-4.88$\pm$0.21 	& 2.6$\pm$1.0 \\
HD207992	&	5426$\pm$21	& 4.34$\pm$0.20	&	 0.61$\pm$0.04	&	-0.27$\pm$0.02	&	245	&	33	&	 0.84$\pm$0.06 	& 	-4.76$\pm$0.22 	& 2.5$\pm$1.0 \\
HD210631	&	5725$\pm$27	& 4.09$\pm$0.20	&	 0.74$\pm$0.05	&	-0.33$\pm$0.02	&	230	&	32	&	 0.94$\pm$0.07 	& 	-4.66$\pm$0.14 	& 2.6$\pm$1.0 \\
HD211681	&	5793$\pm$30	& 4.00$\pm$0.20	&	 1.05$\pm$0.04	&	 0.36$\pm$0.02	&	247	&	33	&	 1.23$\pm$0.09 	& 	-4.67$\pm$0.22 	& 4.0$\pm$1.0 \\
HD212029	&	5927$\pm$49	& 4.14$\pm$0.20	&	 1.08$\pm$0.10	&	-0.92$\pm$0.03	&	169	&	30	&	 0.83$\pm$0.06 	& 	-4.70$\pm$0.21 	& 1.5$\pm$1.0 \\
HD212733	&	5046$\pm$49	& 4.41$\pm$0.23	&	 0.42$\pm$0.15	&	 0.04$\pm$0.03	&	112	&	14	&	 0.83$\pm$0.08 	& 	-4.82$\pm$0.11 	& 1.9$\pm$1.0 \\
HD212735	&	5693$\pm$26	& 4.17$\pm$0.20	&	 0.87$\pm$0.04	&	 0.28$\pm$0.02	&	244	&	35	&	 1.08$\pm$0.08 	& 	-4.65$\pm$0.15 	& 3.4$\pm$1.0 \\
HD217850	&	5605$\pm$30	& 4.13$\pm$0.20	&	 0.89$\pm$0.04	&	 0.28$\pm$0.02	&	239	&	31	&	 1.08$\pm$0.04 	& 	-4.85$\pm$0.19 	& 3.6$\pm$1.0 \\
HD225239	&	5705$\pm$18	& 3.93$\pm$0.19	&	 1.11$\pm$0.03	&	-0.41$\pm$0.01	&	238	&	31	&	 0.99$\pm$0.03 	& 	-4.83$\pm$0.17 	& 2.6$\pm$1.0 \\
HD23965 	&	6423$\pm$52	& 4.34$\pm$0.21	&	 1.44$\pm$0.07	&	 0.01$\pm$0.04	&	203	&	29	&	 1.12$\pm$0.03 	& 	-4.47$\pm$0.13 	& 8.7$\pm$1.0 \\
HD24505 	&	5709$\pm$21	& 3.94$\pm$0.20	&	 0.99$\pm$0.03	&	 0.07$\pm$0.02	&	247	&	34	&	 1.13$\pm$0.08 	& 	-4.77$\pm$0.15 	& 3.4$\pm$1.0 \\
HD28635 	&	6238$\pm$22	& 4.14$\pm$0.20	&	 1.12$\pm$0.03	&	 0.18$\pm$0.02	&	246	&	33	&	 1.17$\pm$0.08 	& 	-4.50$\pm$0.12 	& 3.6$\pm$1.0 \\
HD40647 	&	5297$\pm$26	& 4.50$\pm$0.20	&	 0.80$\pm$0.06	&	-0.18$\pm$0.02	&	242	&	32	&	 0.81$\pm$0.06 	& 	-4.25$\pm$0.17 	& 3.7$\pm$1.0 \\
HD48679 	&	5621$\pm$25	& 4.21$\pm$0.20	&	 0.79$\pm$0.04	&	 0.21$\pm$0.02	&	233	&	34	&	 1.03$\pm$0.03 	& 	-4.67$\pm$0.17 	& 3.4$\pm$1.0 \\
HD5470  	&	6047$\pm$29	& 4.12$\pm$0.20	&	 0.98$\pm$0.04	&	 0.31$\pm$0.02	&	248	&	28	&	 1.19$\pm$0.08 	& 	-4.62$\pm$0.19 	& 3.1$\pm$1.0 \\
HD60846 	&	5964$\pm$23	& 3.92$\pm$0.20	&	 1.09$\pm$0.03	&	-0.09$\pm$0.02	&	249	&	34	&	 1.13$\pm$0.08 	& 	-4.81$\pm$0.15 	& 4.1$\pm$1.0 \\
HD62923 	&	5678$\pm$36	& 4.19$\pm$0.20	&	 0.90$\pm$0.05	&	 0.27$\pm$0.03	&	245	&	34	&	 1.06$\pm$0.08 	& 	-4.75$\pm$0.14 	& 3.7$\pm$1.0 \\
HD71827 	&	6147$\pm$25	& 4.11$\pm$0.20	&	 1.10$\pm$0.04	&	-0.11$\pm$0.02	&	235	&	32	&	 1.06$\pm$0.07 	& 	-4.69$\pm$0.20 	& 4.2$\pm$1.0 \\
HD73636 	&	6123$\pm$31	& 4.05$\pm$0.20	&	 1.18$\pm$0.04	&	 0.25$\pm$0.02	&	236	&	31	&	 1.22$\pm$0.04 	& 	-4.89$\pm$0.20 	& 5.3$\pm$1.0 \\
HD77712 	&	5309$\pm$44	& 4.37$\pm$0.20	&	 0.63$\pm$0.09	&	 0.18$\pm$0.03	&	237	&	33	&	 0.91$\pm$0.04 	& 	-4.70$\pm$0.14 	& 3.4$\pm$1.0 \\
HD78536 	&	5896$\pm$34	& 3.95$\pm$0.20	&	 1.06$\pm$0.04	&	 0.18$\pm$0.03	&	250	&	33	&	 1.21$\pm$0.09 	& 	-4.28$\pm$0.22 	& 5.0$\pm$1.0 \\
HD82460 	&	5757$\pm$19	& 4.29$\pm$0.20	&	 0.78$\pm$0.03	&	-0.06$\pm$0.01	&	242	&	31	&	 0.95$\pm$0.03 	& 	-4.61$\pm$0.16 	& 3.4$\pm$1.0 \\
HD85533 	&	5631$\pm$17	& 4.25$\pm$0.19	&	 0.76$\pm$0.03	&	-0.00$\pm$0.01	&	247	&	34	&	 0.95$\pm$0.07 	& 	-4.75$\pm$0.16 	& 3.1$\pm$1.0 \\
HD87899 	&	5581$\pm$23	& 4.38$\pm$0.19	&	 0.61$\pm$0.05	&	-0.30$\pm$0.02	&	243	&	32	&	 0.85$\pm$0.06 	& 	-4.59$\pm$0.21 	& 2.7$\pm$1.0 \\
\hline
\end{tabular}
\end{table*}

\begin{table*}\tiny\addtolength{\tabcolsep}{-2.5pt}\centering
\caption{\label{tab:SB2_variations} The summary of the FWHM and bissector span analysis for the 54 sources in our sample. SOPHIE and SOPHIE+ datasets are 
analysed separately.}
\begin{tabular}{llcccccccccccc}
Name	&  Dataset	& $N_\text{pts}$	& $\left<\sigma_\text{RV}\right>$	& $\left<\text{FWHM}\right>$	& std(FWHM)	& $\left<\text{BIS}\right>$	& std(BIS)	& $\chi^2_\text{FHWM}$	& $\log p_\text{FWHM}$	& $R_p$(FWHM,RV)	& $\chi^2_\text{BIS}$	& $\log p_\text{BIS}$	& $R_p$(BIS,RV)	\\
\hline
BD+132550	& SOPHIE+	& 11	& 0.0045	& 7.746 	& 0.076 	& -0.038	& 0.011 	& 24.8	& -2.25 	&  0.26 	& 14.2	& -0.79	&  0.15	\\
BD+192536	& SOPHIE	& 4 	& 0.0066	& 7.226 	& 0.046 	&  0.032	& 0.006 	& 1.48	& -0.16 	& -0.26 	& 0.31	& -0.02	&  0.02	\\
        	& SOPHIE+	& 7 	& 0.0060	& 7.331 	& 0.057 	&  0.028	& 0.015 	& 2.61	& -0.06 	&  0.25 	& 4.83	& -0.24	& -0.66	\\
BD+210055	& SOPHIE+	& 19	& 0.0033	& 6.879 	& 0.051 	&  0.010	& 0.009 	& 73.9	& -8.01 	& -0.12 	& 23.8	& -0.79	& -0.23	\\
BD+212816	& SOPHIE	& 12	& 0.0059	& 7.117 	& 0.047 	& -0.022	& 0.008 	& 13.1	& -0.54 	& -0.55 	& 3.36	&  0.00	& -0.50	\\
        	& SOPHIE+	& 5 	& 0.0063	& 7.093 	& 0.034 	& -0.025	& 0.010 	& 1.64	& -0.09 	& -0.21 	& 1.31	& -0.06	&  0.36	\\
BD+281779	& SOPHIE	& 4 	& 0.0078	& 7.477 	& 0.006 	& -0.028	& 0.006 	& 0.03	&  0.00 	& -0.93 	& 0.26	& -0.01	& -0.80	\\
        	& SOPHIE+	& 11	& 0.0062	& 7.465 	& 0.030 	& -0.021	& 0.019 	& 3.52	& -0.01 	& -0.14 	& 10.6	& -0.41	& -0.06	\\
BD+291539	& SOPHIE+	& 17	& 0.0035	& 7.568 	& 0.038 	& -0.030	& 0.006 	& 50.6	& -4.73 	&  0.01 	& 14.3	& -0.24	&  0.25	\\
BD+362641	& SOPHIE	& 9 	& 0.0063	& 7.219 	& 0.035 	&  0.028	& 0.016 	& 5.50	& -0.15 	& -0.28 	& 7.72	& -0.33	&  0.05	\\
BD+680971	& SOPHIE+	& 16	& 0.0047	& 7.602 	& 0.025 	& -0.017	& 0.009 	& 9.09	& -0.05 	&  0.24 	& 9.15	& -0.06	&  0.07	\\
HD101305	& SOPHIE+	& 21	& 0.0059	& 7.748 	& 0.051 	&  0.011	& 0.019 	& 25.7	& -0.75 	&  0.31 	& 17.1	& -0.19	& -0.30	\\
HD103913	& SOPHIE	& 13	& 0.0074	& 8.524 	& 0.032 	&  0.022	& 0.014 	& 4.33	& -0.01 	& -0.07 	& 5.43	& -0.02	& -0.02	\\
        	& SOPHIE+	& 9 	& 0.0051	& 8.527 	& 0.035 	&  0.019	& 0.014 	& 5.48	& -0.15 	& -0.53 	& 4.10	& -0.07	&  0.22	\\
HD104289	& SOPHIE+	& 9 	& 0.0086	& 9.637 	& 0.029 	&  0.015	& 0.016 	& 3.24	& -0.03 	&  0.55 	& 4.74	& -0.10	&  0.61	\\
HD106888	& SOPHIE	& 14	& 0.0082	& 10.46 	& 0.055 	&  0.025	& 0.019 	& 12.6	& -0.32 	&  0.00 	& 9.33	& -0.12	&  0.04	\\
        	& SOPHIE+	& 4 	& 0.0083	& 10.42 	& 0.049 	&  0.022	& 0.000 	& 1.78	& -0.20 	&  0.02 	& 0.00	&  0.00	& -0.80	\\
HD108436	& SOPHIE+	& 8 	& 0.0043	& 7.236 	& 0.016 	& -0.024	& 0.006 	& 0.56	&  0.00 	& -0.15 	& 2.83	& -0.04	&  0.02	\\
HD109157	& SOPHIE+	& 10	& 0.0030	& 6.989 	& 0.034 	& -0.029	& 0.006 	& 21.9	& -2.04 	& -0.84 	& 5.35	& -0.09	&  0.02	\\
HD110376	& SOPHIE+	& 8 	& 0.0042	& 6.820 	& 0.016 	&  0.017	& 0.008 	& 1.28	&  0.00 	& -0.36 	& 3.33	& -0.06	& -0.01	\\
HD13014 	& SOPHIE+	& 16	& 0.0060	& 10.71 	& 0.034 	& -0.040	& 0.027 	& 9.52	& -0.07 	& -0.00 	& 7.93	& -0.03	& -0.09	\\
HD130396	& SOPHIE+	& 30	& 0.0058	& 8.853 	& 0.029 	&  0.024	& 0.011 	& 21.2	& -0.07 	&  0.03 	& 25.4	& -0.18	&  0.00	\\
HD133621	& SOPHIE+	& 12	& 0.0037	& 7.406 	& 0.013 	& -0.009	& 0.008 	& 4.36	& -0.01 	&  0.20 	& 10.6	& -0.32	&  0.00	\\
HD144286	& SOPHIE+	& 15	& 0.0040	& 6.933 	& 0.026 	& -0.034	& 0.010 	& 12.9	& -0.27 	& -0.09 	& 13.6	& -0.32	& -0.21	\\
HD146735	& SOPHIE+	& 12	& 0.0037	& 7.872 	& 0.022 	&  0.001	& 0.008 	& 9.29	& -0.22 	&  0.60 	& 9.79	& -0.26	& -0.48	\\
HD147487	& SOPHIE+	& 11	& 0.0049	& 7.695 	& 0.020 	& -0.019	& 0.016 	& 2.86	&  0.00 	& -0.04 	& 13.8	& -0.74	&  0.20	\\
HD15292 	& SOPHIE+	& 11	& 0.0030	& 7.530 	& 0.022 	& -0.030	& 0.005 	& 4.85	& -0.04 	&  0.34 	& 6.12	& -0.09	&  0.13	\\
HD153376	& SOPHIE	& 18	& 0.0058	& 9.163 	& 0.029 	&  0.001	& 0.011 	& 8.92	& -0.02 	& -0.28 	& 9.90	& -0.04	& -0.20	\\
        	& SOPHIE+	& 2 	& 0.0046	& 9.149 	& 0.002 	&  0.014	& 0.002 	& --- 	& ---   	& ---   	& --- 	&  --- 	&  --- 	\\
HD155228	& SOPHIE+	& 12	& 0.0070	& 10.62 	& 0.036 	&  0.035	& 0.010 	& 8.36	& -0.16 	& -0.21 	& 5.21	& -0.03	& -0.10	\\
HD156111	& SOPHIE+	& 20	& 0.0030	& 6.894 	& 0.020 	& -0.026	& 0.008 	& 21.1	& -0.48 	& -0.37 	& 23.5	& -0.66	&  0.00	\\
HD156728	& SOPHIE+	& 12	& 0.0039	& 7.320 	& 0.018 	& -0.021	& 0.008 	& 3.41	& -0.01 	&  0.41 	& 9.52	& -0.24	&  0.04	\\
HD161479	& SOPHIE	& 4 	& 0.0062	& 8.945 	& 0.026 	& -0.007	& 0.023 	& 1.03	& -0.10 	& -0.75 	& 2.61	& -0.34	& -0.95	\\
        	& SOPHIE+	& 8 	& 0.0069	& 8.917 	& 0.076 	& -0.005	& 0.019 	& 24.2	& -2.98 	& -0.00 	& 15.0	& -1.45	&  0.16	\\
HD167215	& SOPHIE	& 14	& 0.0089	& 9.716 	& 0.027 	&  0.030	& 0.016 	& 2.68	&  0.00 	&  0.17 	& 8.03	& -0.07	& -0.23	\\
        	& SOPHIE+	& 5 	& 0.0082	& 9.671 	& 0.009 	&  0.024	& 0.011 	& 0.08	&  0.00 	& -0.69 	& 0.52	& -0.01	& -0.20	\\
HD18450 	& SOPHIE+	& 11	& 0.0032	& 6.731 	& 0.028 	&  0.000	& 0.010 	& 9.60	& -0.32 	& -0.01 	& 14.8	& -0.86	&  0.16	\\
HD193554	& SOPHIE	& 6 	& 0.0064	& 7.590 	& 0.033 	& -0.001	& 0.009 	& 3.22	& -0.17 	& -0.75 	& 2.22	& -0.08	& -0.77	\\
        	& SOPHIE+	& 4 	& 0.0068	& 7.569 	& 0.027 	& -0.018	& 0.021 	& 1.06	& -0.10 	& -0.90 	& 5.20	& -0.80	& -0.96	\\
HD204613	& SOPHIE	& 11	& 0.0060	& 7.524 	& 0.010 	& -0.004	& 0.007 	& 0.63	&  0.00 	& -0.06 	& 2.78	&  0.00	&  0.01	\\
        	& SOPHIE+	& 5 	& 0.0039	& 7.505 	& 0.014 	&  0.000	& 0.008 	& 1.14	& -0.05 	& -0.84 	& 0.57	& -0.01	&  0.15	\\
HD207992	& SOPHIE	& 11	& 0.0059	& 7.128 	& 0.030 	& -0.035	& 0.007 	& 6.79	& -0.12 	& -0.80 	& 2.90	& -0.01	&  0.04	\\
        	& SOPHIE+	& 2 	& 0.0026	& 7.116 	& 0.006 	& -0.042	& 0.005 	& --- 	&  ---   	&  ---   	& --- 	&  --- 	&  ---	\\
HD210631	& SOPHIE+	& 9 	& 0.0056	& 7.707 	& 0.051 	& -0.037	& 0.012 	& 8.62	& -0.42 	&  0.82 	& 3.37	& -0.04	& -0.78	\\
HD211681	& SOPHIE	& 16	& 0.0059	& 8.261 	& 0.017 	& -0.013	& 0.006 	& 2.88	& -0.00 	& -0.13 	& 3.03	&  0.00	&  0.29	\\
        	& SOPHIE+	& 7 	& 0.0034	& 8.258 	& 0.022 	& -0.012	& 0.005 	& 4.11	& -0.18 	&  0.16 	& 2.71	& -0.07	&  0.06	\\
HD212029	& SOPHIE+	& 21	& 0.0079	& 7.242 	& 0.031 	&  0.002	& 0.020 	& 6.24	&  0.00 	& -0.05 	& 13.6	& -0.07	&  0.02	\\
HD212733	& SOPHIE	& 2 	& 0.0059	& 6.848 	& 0.021 	& -0.000	& 0.001 	& --- 	&  ---   	&  ---    	& --- 	&  --- 	&  ---	\\
        	& SOPHIE+	& 22	& 0.0027	& 6.794 	& 0.030 	& -0.003	& 0.006 	& 16.7	& -0.13 	& -0.18 	& 23.3	& -0.49	& -0.20	\\
HD212735	& SOPHIE	& 6 	& 0.0058	& 7.705 	& 0.020 	& -0.029	& 0.003 	& 1.38	& -0.03 	& -0.84 	& 0.03	&  0.00	&  0.01	\\
        	& SOPHIE+	& 5 	& 0.0035	& 7.665 	& 0.032 	& -0.035	& 0.004 	& 5.50	& -0.62 	& -0.07 	& 1.24	& -0.06	& -0.32	\\
HD217850	& SOPHIE	& 9 	& 0.0057	& 7.851 	& 0.013 	& -0.030	& 0.006 	& 0.32	&  0.00 	&  0.74 	& 0.95	&  0.00	& -0.19	\\
        	& SOPHIE+	& 32	& 0.0030 	& 7.860 	& 0.028 	& -0.030	& 0.007 	& 46.5	& -1.44 	&  0.00 	& 28.3	& -0.21	&  0.00	\\
HD225239	& SOPHIE+	& 44	& 0.0031	& 7.501 	& 0.024 	&  0.001	& 0.010 	& 133.	& -10.5		&  0.11 	& 146.	& -12.5	& -0.02	\\
HD23965  	& SOPHIE	& 18	& 0.0085	& 13.85 	& 0.074 	&  0.017	& 0.025 	& 22.4	& -0.77 	& -0.13 	& 29.1	& -1.47	&  0.11	\\
         	& SOPHIE+	& 63	& 0.0105	& 13.86 	& 0.074 	&  0.008	& 0.036 	& 111.	& -3.90 	& -0.02 	& 171.	& -11.4	& -0.11	\\
HD24505 	& SOPHIE+	& 8 	& 0.0034	& 7.743 	& 0.011 	& -0.011	& 0.005 	& 0.91	&  0.00 	&  0.04 	& 2.85	& -0.04	&  0.18	\\
HD28635 	& SOPHIE+	& 12	& 0.0046	& 8.874 	& 0.032 	&  0.024	& 0.010 	& 9.52	& -0.24 	& -0.61 	& 9.02	& -0.20	& -0.12	\\
HD40647 	& SOPHIE	& 18	& 0.0063	& 7.887 	& 0.038 	& -0.005	& 0.009 	& 13.5	& -0.15 	&  0.21 	& 7.63	& -0.01	&  0.19	\\
        	& SOPHIE+	& 7 	& 0.0040 	& 7.918 	& 0.033 	& -0.007	& 0.010 	& 2.19	& -0.04 	& -0.03 	& 2.37	& -0.05	& -0.13	\\
HD48679 	& SOPHIE+	& 26	& 0.0033	& 7.679 	& 0.037 	& -0.031	& 0.010 	& 58.5	& -3.77 	& -0.18 	& 22.2	& -0.20	&  0.15	\\
HD5470  	& SOPHIE	& 1 	& 0.0077	& 7.868 	& --- 	& -0.014	& --- 	& --- 	&  ---   	&  ---    	& --- 	&  --- 	&  --- 	\\
         	& SOPHIE+	& 3 	& 0.0048	& 7.910 	& 0.015 	& -0.012	& 0.003 	& --- 	&  ---   	&  ---    	& --- 	&  --- 	&  --- 	\\
HD60846 	& SOPHIE+	& 10	& 0.0052	& 8.889 	& 0.040 	&  0.005	& 0.021 	& 16.2	& -1.21 	&  0.01 	& 20.6	& -1.83	& -0.02	\\
HD62923 	& SOPHIE	& 3 	& 0.0060 	& 7.947 	& 0.034 	& -0.015	& 0.023 	& --- 	&  ---   	&  ----    	& --- 	&  --- 	&  --- 	\\
HD71827 	& SOPHIE+	& 44	& 0.0056	& 9.367 	& 0.036 	&  0.014	& 0.012 	& 53.4	& -0.88 	&  0.11 	& 55.6	& -1.02	& -0.09	\\
HD73636 	& SOPHIE	& 12	& 0.0069	& 10.02 	& 0.064 	&  0.003	& 0.007 	& 1.68	&  0.00 	&  0.34 	& 2.70	& -0.00	& -0.09	\\
        	& SOPHIE+	& 2 	& 0.0044	& 10.00 	& 0.026 	&  0.004	& 0.002 	& --- 	&  ---   	&  ---    	& --- 	&  --- 	&  ---	\\
HD77712  	& SOPHIE+	& 18	& 0.0033	& 7.126 	& 0.048 	& -0.036	& 0.007 	& 102.	& -13.4 	& -0.81 	& 16.6	& -0.32	& -0.00	\\
HD78536 	& SOPHIE+	& 5 	& 0.0090 	& 9.599 	& 0.091 	&  0.008	& 0.021 	& 2.86	& -0.23 	& -0.84 	& 0.90	& -0.03	&  0.65	\\
HD82460 	& SOPHIE+	& 17	& 0.0039	& 7.861 	& 0.028 	& -0.014	& 0.013 	& 14.8	& -0.27 	& -0.01 	& 18.0	& -0.49	&  0.04	\\
HD85533 	& SOPHIE	& 5 	& 0.0059	& 7.543 	& 0.013 	& -0.031	& 0.006 	& 0.16	& -0.00 	&  0.05 	& 0.07	&  0.00	& -0.05	\\
        	& SOPHIE+	& 6 	& 0.0034	& 7.560 	& 0.012 	& -0.022	& 0.011 	& 1.20	& -0.02 	& -0.02 	& 0.42	&  0.00	& -0.10	\\
HD87899 	& SOPHIE	& 18	& 0.0065	& 7.446 	& 0.066 	& -0.023	& 0.010 	& 16.8	& -0.33 	& -0.33 	& 7.38	& -0.01	&  0.20	\\
\hline
\end{tabular}
\end{table*}

\onecolumn
\begin{landscape}\thispagestyle{empty}
\begin{table*}\centering
 \begin{adjustwidth}{-2cm}{}
\caption{\label{tab:orbits}Table of orbital fits for the 11 single companions with $M\sin i$ within 20-90 $M_\text{Jup}$, ordered 
by values of $M_2\sin i$. The error on stellar mass was neglected to calculate the uncertainty on $M_2\sin i$ and $a$. SOPHIE measurements before and after the instrument upgrade in June 
2011 are referred to respectively as S- and S+. Residual O-C and RV $\gamma$ of supplementary data from different instruments used to calculate these orbits are given in Table~\ref{tab:res}.  
$T_p$ is the time of passage at periastron. For objects with $e$=0, $T_p$ is the time of primary transit.} 
\begin{tiny}
\begin{tabular}{@{}lcccccccccccc@{}}
Name	&	Period         	&	K               &	$e$             	&	$\omega$            	&	$T_p$ - 2.4\,10$^6 	$	& $\gamma_{S+}$           	&	$\gamma_{S}$    	&	$\sigma_\text{O-C, S+}$	&	$\sigma_\text{O-C, S}$	& $f(m)$ &	$M_2\sin i$				& $a_2$     \\
        &	(day)           &	(m\,s$^{-1}$)	&	                  	&	($^\circ$)             	&	(day)                   & (km\,s$^{-1}$)    	&	(km\,s$^{-1}$)  	  	&	(m\,s$^{-1}$)	&	(m\,s$^{-1}$)		& ($10^{-6}$\,M$_\odot$) 	&	(M$_\text{Jup}$)		& (AU)\\
\hline
\\
\multicolumn{12}{c}{\textit{$M\sin i \in 20-90\,$M$_\text{J}$ ; brown dwarfs candidates}} \\ \\
HD217850	&	3508.1$\pm$2.7	&	433.4$\pm$2.7	&	0.7584$\pm$0.0016	&	165.68$\pm$0.26	&	57552.0$\pm$1.1	&	7.3610$\pm$0.0034	&	7.3465$\pm$0.0058	&	5.87	&	7.61	&	8.20$\pm$0.13 &	22.16$\pm$0.73	&	4.656$\pm$0.075\\
HD48679 	&	1111.61$\pm$0.30	&	1224.2$\pm$4.3	&	0.82473$\pm$0.00046	&	155.71$\pm$0.16	&	58167.08$\pm$0.29	&	19.3806$\pm$0.0053	&	                &	4.60	&       	&	38.22$\pm$0.39 &36.0$\pm$1.3	&	2.145$\pm$0.037\\
HD23965	&	3974$\pm$121	&	781$\pm$52	&	0.792$\pm$0.012	&	-37.9$\pm$2.1	&	58474$\pm$124	&	7.434$\pm$0.065	&	7.446$\pm$0.062	&	21.7	&	28.4	&	45.1$\pm$7.7 & 40.2$\pm$2.6	&	5.16$\pm$0.14\\
HD77712 	&	1311.7$\pm$2.1  	&	1129$\pm$84 	&	0.674$\pm$0.020   	&	51.1$\pm$1.7	&	57950.4$\pm$3.8 	&	5.168$\pm$0.091 	&                	&	5.33	&       	&	79$\pm$12 & 42.1$\pm$2.5	&	2.306$\pm$0.039\\
HD130396	&	2060.6$\pm$7.3  	&	837.3$\pm$3.9	&	0.4275$\pm$0.0037	&	163.02$\pm$0.73	&	57385.7$\pm$3.6  	&	8.1361$\pm$0.0054	&	               	&	8.67	&       &	92.6$\pm$1.2	&	50.9$\pm$1.7	&	3.328$\pm$0.055\\
BD+291539	&	175.8700$\pm$0.0098	&	2401.0$\pm$2.2	&	0.2749$\pm$0.0012	&	-49.60$\pm$0.17	&	57358.325$\pm$0.068	&	21.4215$\pm$0.0032	&	               	&	4.65	&       	& 224.18$\pm$0.58	&	59.7$\pm$2.0	&	0.607$\pm$0.010\\
HD82460 	&	590.90$\pm$0.24    	&	3366$\pm$182	&	0.839$\pm$0.011 	&	-81.3$\pm$4.2	&	57438.0$\pm$3.0 	&	10.95$\pm$0.29  	&	               	&	7.34	&       	& 376$\pm$32	& 73.2$\pm$3.0	&	1.387$\pm$0.023\\
HD28635	&	2636.8$\pm$2.2	&	1181.6$\pm$8.2	&	0.5018$\pm$0.0049	&	142.31$\pm$0.96	&	57405.8$\pm$3.4	&	40.257$\pm$0.017	&	&	7.42	&	& 291.8$\pm$7.1&	77.1$\pm$2.7	&	4.014$\pm$0.068\\
HD211681	&	7612$\pm$131	&	789.4$\pm$3.6	&	0.4650$\pm$0.0053	&	-51.95$\pm$0.70	&	60751$\pm$136	&	-40.9817$\pm$0.0082	&-40.9731$\pm$0.0061	&	3.79	&	13.0	&	269.1$\pm$6.0 	& 77.8$\pm$2.6	& 8.28$\pm$0.16\\
HD210631	&	4030$\pm$40	&	1349$\pm$115	&	0.569$\pm$0.042	&	-100.2$\pm$4.6	&	61353$\pm$48	&	-12.59$\pm$0.12	&	&	4.62	&	&	575$\pm$135	& 83.4$\pm$6.9	&	4.976$\pm$0.085\\
BD+210055	&	1322.63$\pm$0.65	&	2101.6$\pm$2.5	&	0.4457$\pm$0.0012	&	-65.52$\pm$0.18	&	56921.49$\pm$0.50	&	22.7199$\pm$0.0030	&	               	&	4.25	&       	&	912.4$\pm$3.0	& 85.3$\pm$2.9	&	2.235$\pm$0.037\\
\hline
\end{tabular}
\end{tiny}
\end{adjustwidth}
\end{table*}
\end{landscape}
\begin{landscape}
\thispagestyle{empty}
\begin{table*}\centering
 \begin{adjustwidth}{-2cm}{}
\caption{\label{tab:orbits_SB1}Table of orbital fits for the 40 single stellar companions in the M-dwarf regime with $M\sin i$ larger than 90\,M$_\text{J}$ and lower than 0.52\,M$_\odot$, ordered 
by values of $M_2\sin i$. The explanations are identical to Table~\ref{tab:orbits}.} 
\begin{tiny}
\begin{tabular}{@{}lcccccccccccc@{}}
Name	&	Period         	&	K               &	$e$             	&	$\omega$            	&	$T_p$ - 2.4\,10$^6 	$	& $\gamma_{S+}$           	&	$\gamma_{S}$    	&	$\sigma_\text{O-C, S+}$	&	$\sigma_\text{O-C, S}$	&	$f(m)$	& $M_2\sin i$				& $a_2$     \\
        &	(day)           &	(m\,s$^{-1}$)	&	                  	&	($^\circ$)             	&	(day)                   & (km\,s$^{-1}$)    	&	(km\,s$^{-1}$)  	  	&	(m\,s$^{-1}$)	&	(m\,s$^{-1}$)	&	($10^{-6}$\,M$_\odot$) &	(M$_\text{Jup}$)		& (AU)\\
\hline
\\
\multicolumn{12}{c}{\textit{$90\,$M$_\text{J}<M\sin i<0.52\,$M$_\odot$ ; M-dwarfs}} \\
\\
BD+281779	&	46.31706$\pm$0.00022	&	7789.2$\pm$8.0	&	0.61773$\pm$0.00058	&	-68.975$\pm$0.069	&	56469.8643$\pm$0.0041	&	-18.018$\pm$0.010	&	-18.0317$\pm$0.0095	&	3.69	&	8.09	&	1103.0$\pm$2.4	& 90.8$\pm$2.9	&	0.2396$\pm$0.0037\\
HD62923  	&	175.221$\pm$0.013	&	3200.1$\pm$5.9	&	\tablefootmark{$\dagger$}0	&	---	&	58237.6304$\pm$0.2650	&	& 8.3432$\pm$0.0049	&	& 1.07		&	594.9$\pm$3.3	& 91.8$\pm$3.0	&	0.642$\pm$0.010\\
HD156111	&	39.43957$\pm$0.00060	&	7754.8$\pm$1.6	&	0.65053$\pm$0.00013	&	65.798$\pm$0.027	&	57520.1329$\pm$0.0024	&	-48.6374$\pm$0.0025	&		&	3.66	&			&	834.83$\pm$0.59	& 92.0$\pm$3.1	&	0.2259$\pm$0.0037\\
HD103913	&	2322$\pm$10	&	1534$\pm$11	&	0.4129$\pm$0.0075	&	-175.16$\pm$0.54	&	56892.6$\pm$4.8	&	7.101$\pm$0.018	& 7.092$\pm$0.016	&	6.65	&	12.3	&	656$\pm$18	& 98.1$\pm$3.4	&	3.659$\pm$0.061\\
HD225239	&	701.49$\pm$0.55	&	3532.9$\pm$6.4	&	0.7561$\pm$0.0014	&	-34.07$\pm$0.15	&	56996.16$\pm$0.20	&	4.817$\pm$0.010	&	&	8.16	&	&	898.2$\pm$4.3	& 100.5$\pm$3.2	&	1.588$\pm$0.025\\
HD133621	&	448.60$\pm$0.17	&	3039.8$\pm$3.0	&	0.35583$\pm$0.00090	&	63.39$\pm$0.27	&	56846.11$\pm$0.46	&	-49.1224$\pm$0.0037		&	&	2.99	&	&	1065.6$\pm$3.2	& 101.8$\pm$3.5	&	1.156$\pm$0.019\\
HD73636	&	155.284$\pm$0.025	&	3550.2$\pm$6.6	&	0.2874$\pm$0.0011	&	81.85$\pm$0.35	&	55201.78$\pm$0.23	&	-16.515$\pm$0.054	&	-16.5011$\pm$0.0080	&	14.1	&	7.04	&	632.6$\pm$3.6	& 102.7$\pm$3.4	&	0.620$\pm$0.010\\
HD101305	&	1677.4$\pm$2.6	&	2099.9$\pm$4.0	&	0.4911$\pm$0.0020	&	-118.60$\pm$0.19	&	58309.0$\pm$2.2	&	12.5904$\pm$0.0047	&	&	5.11	&	&	1063.3$\pm$6.9	& 106.1$\pm$3.6	&	2.843$\pm$0.046\\
HD106888	&	365.606$\pm$0.060	&	3211$\pm$43	&	0.4629$\pm$0.0038	&	-81.9$\pm$1.1	&	55572.69$\pm$0.73	&	-5.518$\pm$0.053	&-5.546$\pm$0.047	&	8.33	&	18.8	&	872$\pm$34	& 108.3$\pm$3.8	&	1.071$\pm$0.017\\
BD+192536	&	178.543$\pm$0.025	&	4779$\pm$50	&	0.1254$\pm$0.0037	&	-144.9$\pm$1.5	&	56892.99$\pm$0.88	&	-3.302$\pm$0.052	&	-3.314$\pm$0.041	&	3.26	&	2.35	&	1972$\pm$59	& 108.5$\pm$3.7	&	0.5888$\pm$0.0093\\
HD212733	&	89.85646$\pm$0.00015	&	6379.5$\pm$1.3	&	0.43071$\pm$0.00017	&	-105.221$\pm$0.033	&	55060.6477$\pm$0.0058	&	11.6822$\pm$0.0022	&	11.6802$\pm$0.0088	&	5.61	&	10.5	&	1776.78$\pm$0.89	& 112.0$\pm$3.8	&	0.3841$\pm$0.0063\\
HD18450	&	25.037329$\pm$0.000015	&	9351.19$\pm$0.98	&	0.07884$\pm$0.00014	&	27.109$\pm$0.087	&	56747.6986$\pm$0.0060	&	37.8288$\pm$0.0018	&	&	1.17	&	&	2101.50$\pm$0.70	& 114.6$\pm$3.8	&	0.1616$\pm$0.0026\\
HD144286	&	316.7767$\pm$0.0067	&	3871.5$\pm$5.4	&	0.12729$\pm$0.00062	&	-40.48$\pm$0.46	&	56625.65$\pm$0.34	&	-53.8401$\pm$0.0064	&	&	2.82	& &	1858.5$\pm$7.6	& 117.6$\pm$3.9	&	0.905$\pm$0.015\\
HD212029	&	771.02$\pm$0.28	&	3533.8$\pm$6.6	&	0.4978$\pm$0.0011	&	2.88$\pm$0.22	&	57504.95$\pm$0.46	&	-107.529$\pm$0.011	&	&	8.39	&	&	2300$\pm$12	& 122.1$\pm$4.1	&	1.616$\pm$0.026\\
HD104289	&	2389.0$\pm$1.5	&	1790$\pm$11	&	0.2935$\pm$0.0037	&	80.1$\pm$2.1	&	57862.9$\pm$8.1	&	-19.4205$\pm$0.0097	&	&	5.26	&	&	1241$\pm$26 	&	125.8$\pm$4.5	&	3.818$\pm$0.064\\
HD156728	&	4097$\pm$79	&	1792$\pm$12	&	0.321$\pm$0.011	&	103.32$\pm$0.72	&	57093.9$\pm$6.5	&	-0.8480$\pm$0.0093	&	&	3.01	&	&	2074$\pm$27	&	126.5$\pm$4.5	&	5.08$\pm$0.11\\
HD204613	&	876.84$\pm$0.41	&	3297$\pm$44	&	0.0413$\pm$0.0030	&	-54$\pm$12	&	56524$\pm$30	&	-90.816$\pm$0.043	&	-90.779$\pm$0.061	&	2.31	&	8.54	&	3248$\pm$130	&	151.7$\pm$5.5	&	1.858$\pm$0.030\\
HD109157	&	300.212$\pm$0.025	&	6860.5$\pm$3.7	&	0.62565$\pm$0.00062	&	137.84$\pm$0.11	&	56777.730$\pm$0.047	&	6.0473$\pm$0.0073	&	&	5.47	&	&	4767.9$\pm$8.2	&	153.1$\pm$5.0	&	0.864$\pm$0.013\\
HD167215	&	3460.5$\pm$3.2	&	3354$\pm$97	&	0.7635$\pm$0.0062	&	163.72$\pm$0.26	&	55146.05$\pm$0.69	&	-42.82$\pm$0.13	&	-42.82$\pm$0.13	&	3.72	&	8.94	&	3645$\pm$198	  &	167.5$\pm$6.3	&	4.783$\pm$0.075\\
HD40647	&	8078$\pm$872	&	2318$\pm$50	&	0.506$\pm$0.011	&	30.4$\pm$3.5	&	57284$\pm$30	&	-13.612$\pm$0.072	&	-13.629$\pm$0.073		&	13.0	&	14.1	&	6693$\pm$1152	&	172$\pm$11	&	7.82$\pm$0.58\\
\tablefootmark{*}HD193554	&	797.69$\pm$0.18	&	4115.2$\pm$9.1	&	0.3136$\pm$0.0017	&	140.25$\pm$0.65	&	57365.6$\pm$1.2	&	7.5276$\pm$0.0098	&	fixed	&	6.37	&	17.2 &	4870$\pm$77	&	173.6$\pm$6.0	&	1.751$\pm$0.029\\
HD15292	&	2087$\pm$58	&	3052$\pm$20	&	0.325$\pm$0.011	&	24.07$\pm$0.28	&	59802$\pm$57	&	-40.173$\pm$0.036	&	&	0.9	&	&	5202$\pm$184	&	176.5$\pm$6.2	&	3.325$\pm$0.081\\
HD110376	&	1282.2$\pm$3.3	&	4088$\pm$119	&	0.246$\pm$0.027	&	-68.4$\pm$2.9	&	57424$\pm$32	&	-10.62$\pm$0.21	&	&	3.56	&	&	8264$\pm$581	&	177.7$\pm$7.1	&	2.262$\pm$0.036\\
HD87899	&	1527$\pm$16	&	5199$\pm$99	&	0.6734$\pm$0.0028	&	-178$\pm$30	&	55211.8$\pm$2.7	&	&	33.76$\pm$0.12	&	&	11.0	&	8986$\pm$523	&	195.3$\pm$7.8	&	2.627$\pm$0.048\\
\tablefootmark{*}HD207992	&	2090$\pm$12	&	4792$\pm$33	&	0.638$\pm$0.012	&	-62.8$\pm$1.5	&	55894$\pm$12	&	-55.261$\pm$0.051	& fixed	&	0.19	& 32.5	&	10860$\pm$236	&	206.8$\pm$7.3	&	3.238$\pm$0.054\\
\tablefootmark{*}HD5470	&	7788$\pm$50	&	2041.3$\pm$3.4	&	0.3557$\pm$0.0031	&	-124.62$\pm$0.50	&	53155.6$\pm$5.9	&	-3.874$\pm$0.013	& fixed	&	8.17	&	14.7 &	5601$\pm$27	&	208.5$\pm$6.9	&	8.57$\pm$0.14\\
HD78536	&	11.345932$\pm$0.000041	&	17029$\pm$12	&	\tablefootmark{$\dagger$}0	&	---	&	58018.4661$\pm$0.0042	&	-11.507$\pm$0.041	&	-11.567$\pm$0.044	&	7.85	&	0.88	&	5805$\pm$12	&	213.7$\pm$6.9	&	0.1109$\pm$0.0017\\
\tablefootmark{*}HD153376	&	4878$\pm$37	&	3907$\pm$59	&	0.7952$\pm$0.0053	&	113.54$\pm$0.65	&	59193$\pm$38	&	-45.629$\pm$0.085	& fixed	&	2.15	& 6.89	&	6732$\pm$80	&	236.8$\pm$8.1	&	6.49$\pm$0.11\\
HD24505	&	11268$\pm$99	&	3294.9$\pm$3.5	&	0.7974$\pm$0.0014	&	157.952$\pm$0.066	&	56995.36$\pm$0.56	&	-13.5624$\pm$0.0070	&	&	7.03		&	&	9169$\pm$362 	&	238.2$\pm$7.8	&	10.89$\pm$0.18\\
HD146735	&	13932$\pm$3864	&	2161$\pm$85	&	0.507$\pm$0.068	&	84.1$\pm$7.4	&	70233$\pm$3911	&	-16.845$\pm$0.089	&	&	4.20	&	&	9364$\pm$2296	&	244$\pm$22	&	12.7$\pm	$2.4\\
BD+362641	&	17.312171$\pm$0.000071	&	43330$\pm$1702	&	0.8456$\pm$0.0053	&	15.24$\pm$0.32	&	54646.2032$\pm$0.0051	&	-1.1$\pm$2.2	&	&	1.59	&	&	21934$\pm$1555	&	251$\pm$11	&	0.1322$\pm$0.0021\\
HD161479	&	10.2417625$\pm$0.0000035	&	24175$\pm$48	&	0.02316$\pm$0.00062	&	69.3$\pm$3.5	&	54102.9$\pm$5.0	&	-17.095$\pm$0.056	&	-17.079$\pm$0.062	&	36.0	&	42.6	&	14982$\pm$90	 &	268.7$\pm$9.1	&	0.1012$\pm$0.0016\\
HD155228	&	432.976$\pm$0.022	&	7429$\pm$70	&	0.5309$\pm$0.0025	&	-149.51$\pm$0.42	&	57639.10$\pm$0.88	&	16.40$\pm$0.11	&	&	3.34	&	&	11177$\pm$364	&	268.9$\pm$9.5	&	1.278$\pm$0.020\\
BD+680971	&	1134.141$\pm$0.057	&	10749$\pm$678	&	0.8303$\pm$0.0044	&	-39.2$\pm$1.9	&	56979.3$\pm$1.0	&	-8.26$\pm$0.79	&	&	9.54	&	&	25268$\pm$3986	&	276$\pm$17	&	2.207$\pm$0.036\\
HD147487	&	533.228$\pm$0.070	&	7370$\pm$47	&	0.2134$\pm$0.0022	&	-176.59$\pm$0.51	&	57462.6$\pm$1.0	&	-63.967$\pm$0.052	&	&	2.21	&	&	20624$\pm$368	&	279.7$\pm$9.3	&	1.378$\pm$0.021\\
HD108436	&	2720$\pm$142	&	4970$\pm$217	&	0.241$\pm$0.081	&	-30.3$\pm$3.2	&	57996$\pm$54	&	-68.27$\pm$0.40	&	&	3.78	&	&	31121$\pm$3534	&	305$\pm$14	&	4.02$\pm$0.16\\
HD13014	&	3451$\pm$390	&	4881$\pm$482	&	0.4911$\pm$0.0047	&	137.5$\pm$4.2	&	57268$\pm$28	&	0.72$\pm$0.51	&	&	5.01	&	&	27555$\pm$12207	&	366$\pm$52	&	5.22$\pm$0.46\\
HD60846	&	6027$\pm$891	&	8397$\pm$2360	&	0.824$\pm$0.030	&	-4.59$\pm$0.45	&	62514$\pm$889	&	50.0$\pm$3.0	&	&	5.92	&	&	62968$\pm$41381	&	448$\pm$95	&	7.43$\pm$0.77\\
HD85533	&	31214$\pm$30144	&	3907$\pm$355	&	0.61$\pm$0.13	&	-57$\pm$17	&	58008$\pm$98	&	-26.17$\pm$0.57	&	-26.21$\pm$0.57	&	5.19	&	10.9	&	91085$\pm$54942	&	453$\pm$89	&	22$\pm$13\\
BD+132550	&	2537$\pm$17	&	6679.1$\pm$5.6	&	\tablefootmark{$\dagger$}0	&	---	&	60080$\pm$16	&	-10.872$\pm$0.016	&	&	8.30	&	&	78330$\pm$544	&	463$\pm$16	&	4.159$\pm$0.065\\
\hline
\end{tabular}
\end{tiny}
\tablefoot{ 
\tablefoottext{*}{We fixed the $\gamma$ of the S+ and S- dataset at a common value.} \\
\tablefoottext{$\dagger$}{With an eccentricity of 0, the time of periastron passage is ill-defined. In this case, $T_p$ indicates the time of primary transit.}}
\end{adjustwidth}
\end{table*}
\end{landscape}

\begin{table*}\centering
\caption{\label{tab:res}Table of RV offsets and residues $O$$-$$C$ for the supplementary datasets, from ELODIE, HIRES or available in the SB9 catalog.} 
\begin{tabular}{l@{\qquad}c@{~~}c@{\qquad}c@{~~}c@{\qquad}c@{~~}c}
Name    	&	\multicolumn{2}{c@{~~}}{Elodie}								&	\multicolumn{2}{c@{~~}}{HIRES}									&	\multicolumn{2}{c}{SB9}	\\	
			& $\gamma_{\text{ELO}}$   	&	$\sigma_\text{O-C, ELO}$		&	$\gamma_{\text{HIR}}$		&	$\sigma_\text{O-C, HIR}$		& $\gamma_{\text{SB9}}$      	&	$\sigma_\text{O-C, SB9}$			\\
          	&	(km\,s$^{-1}$)      				&	(m\,s$^{-1}$)				&  (km\,s$^{-1}$)      			 	&	(m\,s$^{-1}$)				&	(km\,s$^{-1}$)      			&	(m\,s$^{-1}$)			\\
\hline
HD104289	&	-19.443$\pm$0.012	&	14.6					&												&														&										& \\
HD106888	&	-5.561$\pm$0.068	&	8.43					&												&														&										&\\
HD110376		&									&							&												&														&	-9.27$\pm$0.12			&	776	\\	
HD133621		&									&							&												&														&		-49.30$\pm$0.12		&	556	 \\
HD161479		&	-17.494$\pm$0.072	&	21.1					&	3.810$\pm$0.070					&	5.13												&										&	\\
HD167215		&									&							&	-0.17$\pm$0.12					&			6.97										&	-42.88$\pm$0.19			&	313 \\		
HD193554	&									&							&												&														&	8.946$\pm$0.067		&	354	\\
HD204613	&									&							&												&														&	-91.00$\pm$0.14			&	575	\\
HD207992	&									&							&	2.205$\pm$0.047					&	12.1												&										&	\\
HD210631		&									&							&												&														&	-12.62$\pm$0.13			&	511	\\
HD211681		&	-41.0443$\pm$0.0095	&	21.3				&	0.2246$\pm$0.0046			&	4.64												&										&	\\
HD212733	&	11.5117$\pm$0.0070	&	11.9					&	1.5260$\pm$0.0016				&	1.45												&										&	\\
HD217850	&	                                	&	                       	& -0.0641$\pm$0.0034          	&	3.31	    	 										&										&\\
HD24505		&	-13.630$\pm$0.024	&	\tablefootmark{*}0.00					&	-0.7875$\pm$0.0058			&	3.27												&										&	\\
HD28635		&	40.137$\pm$0.045	&	14.1					&												&														&	-0.406$\pm$0.013		&	32.3 \\	
HD40647		&									&							&	0.973$\pm$0.072					&	10.7												&										&		\\	
HD5470		&									&							&	2.1493$\pm$0.0054				&	6.00												&										&		\\
HD62923		&	8.2672$\pm$0.0072		&	13.9 					&												&														&										& \\
HD71827		&	fixed	 to $\gamma_{S,+}$		&	0.10					&												&														&										& \\
\hline
\end{tabular}
\tablefoot{ 
\tablefoottext{*}{Only one ELODIE RV measurement.}}
\end{table*}

\begin{table*}\centering  
\caption{\label{tab:trip}Fitted Keplerian and drift parameters for triple systems. The minimum possible period for the third object is given with the corresponding $m\sin i$ and 
semi-major axis derived when fixing the period at this value. In both cases, the orbit of component 1 was derived assuming the drift. In all cases, we fixed the SOPHIE (or Elodie) 
and SOPHIE+ relative RV offset to 0\,m\,s$^{-1}$.} 
\begin{tabular}{@{}l@{~~}c@{~~~}c@{~~~}c@{~~~}c@{}}
Parameters								&	HD71827					& HD212735	&	BD+212816\\
			\hline
$P_{b}$ [day] 						& 15.052366$\pm$0.000051	& 37.940879$\pm$0.000092 	& 788.91$\pm$0.10\\
$K_{b}$ [m\,s$^{-1}$] 				& 2173.9$\pm$1.5					& 5406.4$\pm$1.8 	& 3137$\pm$11 \\
$e_{b}$ 								& 0.07645$\pm$0.00068	& 0.18327$\pm$0.00037		& 0.2275$\pm$0.0052\\
$\omega_{b}$ [$^\circ$] 				& 57.02$\pm$0.42	& 101.80$\pm$0.22		& 70.98$\pm$0.74\\
$T_{0,b}$-$2.4\,10^6$  [day] 			& 57424.287$\pm$0.017	& 57585.784$\pm$0.020		& 56852.8$\pm$1.1\\ 
\hline	
Linear drift [m\,s$^{-1}$\,yr$^{-1}$]	&	-438.984$\pm$0.027				& -90.12$\pm$0.11	& 15$\pm$14\\
Quad drift [m\,s$^{-1}$\,yr$^{-2}$]	&	-40.80921$\pm$0.00033				&  \\
Cubic drift [m\,s$^{-1}$\,yr$^{-2}$]	&	-1.5666752$\pm$0.0000031				&  \\
$P_{c}$ [day] 				    			& $>$7000						&  $>$7000	& $>$8000\\
\hline
$\gamma_{S+}$ [km\,s$^{-1}$] 			& -36.9407$\pm$0.0020			& -24.8571$\pm$0.0028	& -32.141$\pm$0.017\\
$\sigma_\text{O-C, S+}$ [m\,s$^{-1}$] 	& 6.22						& 1.50	& 5.13\\
$\gamma_{S}$ [km\,s$^{-1}$] 			& 			& fixed	& fixed\\
$\sigma_\text{O-C, S}$ [m\,s$^{-1}$] 	& 	& 5.75 	& 10.6\\
\hline
$f(M_{b})$ [$10^{-6}$\,M$_\odot$]	        	& 15.881$\pm$0.033	&	590.18$\pm$0.59		  &	2329$\pm$32	\\
$M_{b}\sin i$ [M$_\text{J}$] 				& 27.43$\pm$0.89			& 92.5$\pm$3.0	& 120.4$\pm$4.2\\
$a_{b}$ [AU] 							& 0.1228$\pm$0.0020			& 0.2328$\pm$0.0037	& 1.627$\pm$0.027\\
$f(M_{c})$ [$10^{-6}$\,M$_\odot$]	   	& 3805$\pm$351		&	77.9$\pm$3.7		  &	1.9$\pm$5.4	\\
$M_{c}\sin i$ [M$_\text{J}$] 			& $>$163				& $>$45	& $>$5\\
$a_{c}i$ [AU] 					& $>$7.5		& $>$7.3 	& $>$7.2\\
\hline
\end{tabular}
\end{table*}

\begin{table}\caption{Parameters of the Hipparcos astrometric observations for the 54 stars in our sample. The last column indicates the significance of the solution that was 
possible to derive.  For the triple systems HD71827, HD212735 and BD+212816, only the inner companion $b$ orbit was considered, since the outer companion is not  constrained by 
RVs. The binary system HD193554 is already solved in the Hipparcos double and multiple catalog. Upper-limits can be calculated only if at least close to the full orbital period 
is covered by the Hipparcos data.}
\label{tab:HIPparams} 
\centering  
\tiny
\begin{tabular}{r r r r r r c} 	
\hline\hline % 
Name & HIP & $S_\mathrm{n}$& $N_\mathrm{orb}$ & $\sigma_{\Lambda}$  & $N_\mathrm{Hip}$ & Orbit detection \\  
       &     &  &           & (mas)                 &           &   \\  
\hline 
 BD+132550 & 061398 & 5 & 0.7 & 4.7 & 86 &  \\ 
 BD+192536 & 059310 & 5 & 6.4 & 7.5 & 116 & upper-limit \\ 
 BD+210055 & 002397 & G & 0.9 & 4.2 & 95 & 3$\sigma$ \\ 
 BD+212816(b/c) & 077179 & 5 & 1.2/0.1 & 6.6 & 143 & 2$\sigma$ on $b$ \\ 
 BD+281779 & 047176 & 5 & 21.6 & 4.6 & 80 & upper-limit  \\ 
 BD+291539 & 036480 & 5 & 6.0 & 5.0 & 57 & upper-limit  \\ 
 BD+362641 & 077141 & 5 & 26.5 & 8.6 & 137 & upper-limit  \\ 
 BD+680971 & 088188 & 5 & 1.0 & 5.5 & 132 & 3$\sigma$ \\ 
 HD101305 & 056859 & 5 & 0.5 & 4.4 & 76 & 2$\sigma$ \\ 
 HD103913 & 058364 & 5 & 0.5 & 3.8 & 112 &  \\ 
 HD104289 & 058572 & 5 & 0.5 & 5.3 & 156 & 2$\sigma$ \\ 
 HD106888 & 059933 & 5 & 3.1 & 3.7 & 111 & upper-limit  \\ 
 HD108436 & 060739 & 5 & 0.4 & 4.6 & 127 &  \\ 
 HD109157 & 061198 & 5 & 3.7 & 5.8 & 104 & upper-limit  \\ 
 HD110376 & 061939 & X & 0.9 & 6.3 & 97 & $<$1$\sigma$ \\ 
 HD13014 & 009974 & 5 & 0.3 & 2.9 & 95 &  \\ 
 HD130396 & 072336 & 5 & 0.6 & 3.1 & 111 &  \\ 
 HD133621 & 073440 & X & 2.6 & 3.4 & 116 & 3$\sigma$ \\ 
 HD144286 & 078700 & 5 & 3.7 & 8.4 & 173 & upper-limit  \\ 
 HD146735 & 079613 & 5 & 0.1 & 3.9 & 107 &  \\ 
 HD147487 & 080118 & 5 & 2.1 & 5.8 & 202 & 2$\sigma$\\ 
 HD15292 & 011906 & G & 0.5 & 3.5 & 157 &  \\ 
 HD153376 & 083083 & 5 & 0.2 & 2.8 & 121 &  \\ 
 HD155228 & 083942 & 5 & 2.1 & 2.9 & 148 & $<$1$\sigma$\\ 
 HD156111 & 084372 & 5 & 26.1 & 3.3 & 125 & upper-limit  \\ 
 HD156728 & 084520 & 5 & 0.3 & 3.9 & 118 &  \\ 
 HD161479 & 086882 & 5 & 112.4 & 4.5 & 125 & upper-limit  \\ 
 HD167215 & 089270 & G & 0.3 & 5.3 & 144 &  \\ 
 HD18450 & 013891 & 5 & 36.4 & 4.0 & 73 & upper-limit  \\ 
 HD193554 & 100259& D & 1.4 & 3.6 & 160 & HIP double catalog  \\
 HD204613 & 105969 & X & 1.3 & 8.3 & 123 & 3$\sigma$ \\ 
 HD207992 & 107958 & G & 0.6 & 5.3 & 189 &  3$\sigma$ \\ 
 HD210631 & 109563 & 5 & 0.3 & 3.9 & 94 & 2$\sigma$\\ 
 HD211681 & 109169 & 5 & 0.2 & 4.1 & 130 &  \\ 
 HD212029 & 110291 & 5 & 1.5 & 5.2 & 158 & 3$\sigma$ \\ 
 HD212733 & 110716 & 5 & 12.8 & 4.1 & 119 & upper-limit  \\ 
 HD212735(b/c) & 110761 & 5 & 29/0.1 & 3.6 & 85 & upper-limit on $b$ \\ 
 HD217850 & 113789 & 5 & 0.3 & 4.3 & 111 &  \\ 
HD225239 & 000394 & X & 1.7 & 14.4 & 98 &3$\sigma$\\ 
 HD23965 & 017928 & 5 & 0.2 & 2.2 & 68 &  \\ 
 HD24505 & 018320 & 5 & 0.1 & 2.8 & 83 &  \\ 
 HD28635 & 021112 & 5 & 0.3 & 2.7 & 74 &  \\ 
  HD40647 & 028902 & 5 & 0.1 & 5.3 & 254 &  \\ 
 HD48679 & 033548 & 5 & 1.0 & 5.6 & 143 & upper-limit  \\ 
 HD5470 & 004423 & 5 & 0.1 & 3.5 & 65 &  \\ 
 HD60846 & 037172 & 5 & 0.1 & 3.0 & 53 &  \\ 
 HD62923 & 038104 & 5 & 4.4 & 3.5 & 48 & 2$\sigma$ \\ 
 HD71827(b/c) & 042279 & 5 & 72.8/0.1 & 2.9 & 160 & upper-limit on $b$ \\ 
 HD73636 & 042627 & 5 & 6.9 & 2.7 & 58 & upper-limit \\ 
 HD77712 & 044518 & 5 & 0.8 & 6.5 & 81 & upper-limit \\ 
 HD78536 & 044906 & 5 & 88.0 & 3.1 & 87 & upper-limit  \\ 
 HD82460 & 046903 & 5 & 2.0 & 4.1 & 137 & upper-limit  \\ 
 HD85533 & 048691 & 5 & 0.0 & 6.0 & 169 &  \\ 
 HD87899 & 049738 & G & 0.7 & 4.3 & 76 &  3$\sigma$ \\ 
  \hline
\end{tabular} 
\end{table}

\renewcommand{\arraystretch}{1.2}
\begin{table*}
\caption{Solution parameters determined for the 16 significant detections and 18 lower-limits. They are ordered by target name.}
\label{tab:hip_masses_16} 
\centering  
\tiny
\begin{tabular}{lccccccccccc} 	
\hline\hline % 
Source &   $M_2 \sin i$ & $a_1 \sin i$ & $a_1$ & $M_2$  & $M_2$ (3-$\sigma$)& \tablefootmark{$\dagger$}$a_{\mathrm{rel}}$ &$O-C_5$ &$O-C_7$ & $\chi^2_{7,red}$ & Null prob. & Significance \\  
          	&	(M$_\text{J}$) &(mas)     & (mas) & ($M_\sun$) & ($M_\sun$) &(mas) & (mas) & (mas) & & (\%)  & (\%) \\  
\hline 
\multicolumn{12}{c}{$20\,$M$_\text{J}<M\sin i<90\,$M$_\text{J}$} \\
\\
BD+210055 & 85.3	&   6.03 & $ 14.6^{+ 1.3}_{-1.3}$ & $ 0.21^{+ 0.02}_{-0.02}$ & $(0.14,0.29)$ & 69.0 & 7.66 & 5.19& 1.28& 7.4e-14 &100.0 \vspace{1mm} \\ 
 BD+291539 & 59.7	&	0.60	&	$<$5.00	&	$<$0.50	&	&	$<$14.54\\ 
 HD210631 &   83.4	&   5.77 & $ 26.3^{+ 7.5}_{-7.4}$ & $ 0.53^{+ 0.20}_{-0.20}$ & $(0.14,1.51)$ & 74.8 & 3.93 & 3.58& 0.73& 1.9e-02 &96.5 \vspace{1mm} \\ 
 HD48679 & 36.0	&	1.06	&	$<$5.60	&	$<$0.32	&	&	$<$37.32\\ 
HD71827	&	27.4 	& 0.07	& $<$2.91 	&	$<$1.07	&	&	$<$5.78\\
 HD77712	&	42.1	&	1.97	&	$<$8.80	&	$<$0.18	&	&	$<$53.34\\
 HD82460 &  73.2	&	1.97	&	$<$4.10	&	$<$0.27	&	&	$<$30.91\\ 
\hline 
\multicolumn{12}{c}{$90\,$M$_\text{J}<M\sin i<0.52\,$M$_\odot$}\\
\\
BD+212816 &  120.4	& 4.32 & $ 5.7^{+ 0.6}_{-0.6}$ & $ 0.16^{+ 0.02}_{-0.02}$ & $(0.12,0.23)$ & 34.2 & 7.66 & 7.16& 1.05& 6.6e-04 &96.7 \vspace{1mm} \\ 
 BD+192536 & 108.50 &	1.70	&	$<$7.50	&	$<$0.46	&	&	$<$19.81 \\
 BD+281779 & 90.8	&	0.61	&	$<$4.60	&	$<$0.83	&	&	$<$10.02\\ 
 BD+362641 & 251.00	&	0.68	&	$<$8.60	&	$<$5.67	&	&	$<$10.84 \\ 
BD+680971 &  276.0	&  10.24 & $ 10.8^{+ 1.6}_{-1.6}$ & $ 0.38^{+ 0.08}_{-0.08}$ & $(0.20,0.68)$ & 35.3 & 6.53 & 5.71& 0.86& 3.3e-04 &99.9 \vspace{1mm} \\ 
HD101305 & 106.1 	&   4.00 & $ 8.7^{+ 1.1}_{-1.1}$ & $ 0.27^{+ 0.04}_{-0.04}$ & $(0.17,0.42)$ & 40.6 & 5.01 & 5.00& 0.84& 8.4e-06 &96.5 \vspace{1mm} \\ 
HD104289 &  125.8	&   5.33 & $ 8.2^{+ 0.8}_{-0.8}$ & $ 0.19^{+ 0.02}_{-0.02}$ & $(0.14,0.26)$ & 59.4 & 5.14 & 4.83& 0.81& 4.3e-06 &99.4 \vspace{1mm} \\ 
 HD106888 & 108.30	&	1.42	&	$<$3.70	&	$<$0.30	&	&	$<$19.12 \\ 
 HD109157 &  153.10	&	3.37	&	$<$5.80	&	$<$0.32	&	&	$<$24.50 \\ 
HD110376 &  177.7	& 14.45 & $ 14.2^{+ 0.8}_{-0.8}$ & $ 0.22^{+ 0.02}_{-0.02}$ & $(0.17,0.29)$ & 64.1 & 7.38 & 5.43& 0.80& 2.2e-12 &38.5 \vspace{1mm} \\ 
HD133621 & 101.8 	&  3.36 & $ 4.4^{+ 0.3}_{-0.3}$ & $ 0.13^{+ 0.01}_{-0.01}$ & $(0.11,0.17)$ & 35.3 & 3.68 & 2.87& 0.62& 1.5e-11 &100.0 \vspace{1mm} \\ 
 HD144286 & 117.60	&	1.67	&	$<$8.40	&	$<$0.57	&	&	$<$21.31\\ 
 HD147487 &  279.7	&   5.27 & $ 6.3^{+ 0.4}_{-0.4}$ & $ 0.34^{+ 0.02}_{-0.02}$ & $(0.29,0.40)$ & 24.4 & 6.99 & 6.62& 1.00& 1.8e-10 &95.6 \vspace{1mm} \\ 
HD155228 &  268.9	& 3.69 & $ 4.4^{+ 0.2}_{-0.2}$ & $ 0.30^{+ 0.01}_{-0.01}$ & $(0.26,0.37)$ & 22.0 & 3.58 & 3.33& 1.18& 3.2e-07 &53.7 \vspace{1mm} \\ 
 HD156111 & 92.0	&	0.48	&	$<$3.30	&	$<$0.79	&	&	$<$8.24 \\ 
 HD161479 &  268.70	&	0.46	&	$<$4.50	&	$<$2.49	&	&	$<$6.42\\ 
 HD18450 &  114.60	&	0.73	&	$<$4.00	&	$<$0.60	&	&	$<$9.27\\ 
HD204613 & 151.7  	&   4.22 & $ 10.5^{+ 0.5}_{-0.5}$ & $ 0.51^{+ 0.04}_{-0.04}$ & $(0.40,0.67)$ & 30.3 & 8.83 & 4.77& 0.29& 8.6e-34 &100.0 \vspace{1mm} \\ 
HD207992 &  206.8	&  18.48 & $ 28.5^{+ 2.2}_{-2.2}$ & $ 0.37^{+ 0.04}_{-0.04}$ & $(0.28,0.51)$ & 92.9 & 6.75 & 5.90& 1.21& 9.4e-11 &100.0 \vspace{1mm} \\ 
HD212029 &   122.1	& 3.50 & $ 5.5^{+ 0.5}_{-0.5}$ & $ 0.21^{+ 0.03}_{-0.03}$ & $(0.14,0.31)$ & 27.3 & 6.86 & 5.44& 0.93& 3.0e-12 &100.0 \vspace{1mm} \\  
 HD212733 & 112.10	&	1.57	&	$<$4.10	&	$<$0.31	&	&	$<$6.25  \\ 
 HD212735 & 92.5	&	0.34	&	$<$3.66	&	$<$0.96	&	&	$<$7.78	\\
HD225239 \tablefootmark{a} &  100.5	&   4.22 & $ 31.2^{+ 0.8}_{-0.8}$ & $ 1.13^{+ 0.08}_{-0.08}$ & $(0.94,1.41)$ & 58.6 & 13.73 & 2.48& 0.03& 3.2e-67 &100.0 \vspace{1mm} \\ 
 HD62923 \tablefootmark{a}  & 91.8	&   0.91 & $ 8.3^{+ 1.4}_{-1.4}$ & $ 1.27^{+ 0.33}_{-0.33}$ & $(0.49,2.66)$ & 15.4 & 5.04 & 4.39& 1.68& 1.3e-04 &99.8 \vspace{1mm} \\ 
HD73636 &  102.7	&	0.85	&	$<$2.70	&	$<$0.28	&	&	$<$14.78 \\ 
 HD78536 & 213.70	&	0.22	&	$<$3.10	&	$<$2.91	&	&	$<$4.39\\ 
HD87899 &  195.3	&   10.33 & $ 16.1^{+ 2.7}_{-2.8}$ & $ 0.35^{+ 0.08}_{-0.08}$ & $(0.20,0.65)$ & 55.2 & 6.13 & 5.28& 1.47& 4.3e-04 &99.9 \vspace{1mm} \\ 
\hline 
\end{tabular} 
\tablefoot{\\
\tablefoottext{a} {Doubtful secondary mass estimation, because the Sahlmann et al. (2011) model assumes the companion to be dark.}  \\
\tablefoottext{$\dagger$}{$a_{\mathrm{rel}}$ is the relative semi-major axis, $a_{\mathrm{rel}}$=$a_1$+$a_2$.}}
\end{table*}

\begin{table*}
\caption{Updated parallax and proper motion values and the astrometric orbit parameters ($i$, $\Omega$) for 16 sources with significant orbit detections.}
\label{tab:hip_ppm_16} 
\centering  
\tiny
\begin{tabular}{@{}l@{~~}r@{~~}r@{~~}r@{~~}r@{~~}r@{~~}r@{~~}r@{~~}r@{~~}r@{~~}r@{~~}r@{~~}r@{~~}r@{~~}r@{}} 	
\hline\hline % 
Object  & \tablefootmark{$a$}$\Delta \alpha^{\star}$ & \tablefootmark{$a$}$\Delta \delta$ & \tablefootmark{$b$}$\varpi$ &\tablefootmark{$b$}$\Delta \varpi^\text{HIP2}$ &\tablefootmark{$b$}$\Delta \varpi^\text{DR2}$  & \tablefootmark{$c$}$\Delta \mu_{\alpha^\star}^\text{HIP2}$ & \tablefootmark{$c$}$\Delta \mu_{\delta}^\text{HIP2}$ & \tablefootmark{$d$}$\Delta \mu_{\alpha^\star}^\text{DR1}$ & \tablefootmark{$d$}$\Delta \mu_{\delta}^\text{DR1}$ & \tablefootmark{$e$}$i$ & \tablefootmark{$f$}$\Omega$  \\  
        &  (mas)                     & (mas)             &  (mas)    &  (mas)           & (mas)	 &  (mas $\mathrm{yr}^{-1}$)       & (mas $\mathrm{yr}^{-1}$) &  (mas $\mathrm{yr}^{-1}$)       & (mas $\mathrm{yr}^{-1}$) & (deg)	      & (deg)	    \\  

\hline
BD+210055 	& $1.2^{+ 1.7}_{-1.7}$ 		& $-11.2^{+ 1.3}_{-1.3}$  	& $29.46^{+ 1.13}_{-1.13}$ & $-0.04$ 	& $3.09$ 	& $9.9^{+ 1.4}_{-1.4}$ 		& $0.7^{+ 1.5}_{-1.5}$ 		& $            $	& $            $	& $ 152.6^{+ 2.6}_{-3.1}$ 	& $ 111.2^{+ 6.5}_{-6.5}$   	\vspace{1mm} \\ 
BD+212816 	& $-0.4^{+ 0.8}_{-0.8}$		& $1.8^{+ 1.1}_{-1.1}$  	& $20.60^{+ 1.27}_{-1.28}$ & $0.73$ 	& $1.11$ 	& $1.9^{+ 1.2}_{-1.2}$ 		& $-1.5^{+ 1.3}_{-1.3}$ 	& $0.50\pm1.07 $	& $-0.40\pm1.24$	& $ 54.3^{+ 8.9}_{-8.0}$ 	& $ 24.5^{+ 19.4}_{-19.9}$  	\vspace{1mm} \\ 
BD+680971 	& $7.5^{+ 2.6}_{-2.6}$ 		& $-11.9^{+ 2.5}_{-2.5}$  	& $15.50^{+ 0.83}_{-0.84}$ & $-0.66$ 	& $-0.91$ 	& $-2.5^{+ 1.7}_{-1.7}$ 	& $-6.8^{+ 1.5}_{-1.5}$ 	& $			   $	& $            $	& $ 47.4^{+ 16.3}_{-11.0}$ 	& $ 174.0^{+ 16.2}_{-8.9}$  	\vspace{1mm} \\ 
 HD101305 	& $0.3^{+ 0.8}_{-0.8}$ 		& $2.8^{+ 0.9}_{-0.9}$  	& $13.58^{+ 0.97}_{-0.96}$ & $0.88$ 	& $0.61$ 	& $-5.0^{+ 2.1}_{-2.1}$ 	& $9.9^{+ 1.9}_{-1.9}$ 		& $-4.54\pm1.77$	& $9.57\pm1.24$		& $ 153.9^{+ 3.3}_{-4.2}$ 	& $ 159.1^{+ 12.5}_{-12.5}$  	\vspace{1mm} \\ 
 HD104289 	& $-1.4^{+ 0.5}_{-0.5}$		& $0.4^{+ 0.5}_{-0.5}$  	& $15.29^{+ 0.66}_{-0.66}$ & $0.38$ 	& $1.11$ 	& $2.0^{+ 1.3}_{-1.3}$ 		& $-5.3^{+ 0.7}_{-0.7}$ 	& $0.20\pm0.38 $	& $-6.91\pm0.83$	& $ 135.0^{+ 5.0}_{-6.1}$ 	& $ 172.3^{+ 10.3}_{-10.3}$  	\vspace{1mm} \\ 
 HD110376 	& $0.9^{+ 1.4}_{-1.4}$ 		& $-1.9^{+ 1.0}_{-1.0}$  	& $27.88^{+ 1.21}_{-1.20}$ & $-2.79$ 	& $-3.07$ 	& $-3.4^{+ 2.3}_{-2.3}$ 	& $8.3^{+ 1.1}_{-1.1}$ 		& $0.18\pm1.56$		& $8.02\pm1.18$		& $ 111.7^{+ 8.4}_{-9.5}$ 	& $ 125.8^{+ 44.8}_{14.3}$  	\vspace{1mm} \\ 
 HD133621 	& $-0.7^{+ 0.4}_{-0.4}$ 	& $1.0^{+ 0.4}_{-0.4}$  	& $30.19^{+ 0.42}_{-0.42}$ & $0.85$ 	& $1.46$ 	& $1.3^{+ 0.4}_{-0.4}$ 		& $0.0^{+ 0.4}_{-0.4}$ 		& $0.25\pm0.24$		& $-0.07\pm0.76$	& $ 54.5^{+ 6.1}_{-5.3}$ 	& $ 281.5^{+ 6.2}_{-6.2}$  		\vspace{1mm} \\ 
HD147487 	& $-2.4^{+ 0.5}_{-0.5}$ 	& $-2.2^{+ 0.8}_{-0.8}$  	& $17.37^{+ 0.87}_{-0.87}$ & $1.24$ 	& $2.44$ 	& $0.5^{+ 0.5}_{-0.5}$ 		& $-1.1^{+ 0.8}_{-0.8}$ 	& $            $	& $            $	& $ 103.3^{+ 6.2}_{-6.5}$ 	& $ 63.4^{+ 11.9}_{-5.4}$  		\vspace{1mm} \\ 
HD155228 	& $-3.0^{+ 0.6}_{-0.6}$ 	& $-3.0^{+ 0.7}_{-0.7}$  	& $17.05^{+ 0.73}_{-0.74}$ & $0.08$ 	& $2.30$ 	& $0.2^{+ 0.6}_{-0.6}$ 		& $-0.3^{+ 0.6}_{-0.6}$ 	& $            $	& $            $	& $ 81.7^{+ 8.4}_{-8.0}$ 	& $ 38.3^{+ 9.5}_{-9.5}$  		\vspace{1mm} \\ 
 HD204613 	& $-0.2^{+ 0.6}_{-0.6}$ 	& $1.5^{+ 0.7}_{-0.7}$  	& $14.83^{+ 0.72}_{-0.72}$ & $1.00$ 	& $-1.08$ 	& $-0.9^{+ 0.6}_{-0.6}$ 	& $0.4^{+ 0.6}_{-0.6}$ 		& $-0.06\pm0.84$	& $-1.25\pm1.52$	& $ 157.8^{+ 1.4}_{-1.6}$ 	& $ 286.7^{+ 40.1}_{-75.2}$  	\vspace{1mm} \\ 
 HD207992 	& $27.4^{+ 4.4}_{-4.3}$ 	& $8.5^{+ 5.1}_{-5.1}$  	& $27.21^{+ 1.03}_{-1.03}$ & $1.02$ 	& $1.14$ 	& $10.0^{+ 2.4}_{-2.4}$ 	& $-10.9^{+ 1.9}_{-1.9}$ 	& $            $	& $            $	& $ 42.8^{+ 4.3}_{-3.7}$ 	& $ 142.8^{+ 10.9}_{-10.9}$  	\vspace{1mm} \\ 
 HD210631 	& $18.5^{+ 9.1}_{-8.9}$ 	& $-21.8^{+ 11.1}_{-11.2}$ 	& $13.40^{+ 0.86}_{-0.86}$ & $0.69$ 	& $-0.64$ 	& $-11.0^{+ 3.1}_{-3.1}$ 	& $0.1^{+ 4.3}_{-4.3}$ 		& $            $	& $            $	& $ 11.7^{+ 5.5}_{-2.9}$ 	& $ 278.0^{+ 242.9}_{-86.5}$  	\vspace{1mm} \\ 
HD212029 	& $-3.1^{+ 0.7}_{-0.7}$ 	& $-3.1^{+ 0.7}_{-0.7}$  	& $16.37^{+ 0.72}_{-0.72}$ & $-0.71$ 	& $0.25$ 	& $0.8^{+ 0.7}_{-0.7}$ 		& $0.9^{+ 0.6}_{-0.6}$ 		& $0.48\pm0.74$		& $0.86\pm1.01$		& $ 139.7^{+ 4.8}_{-5.9}$ 	& $ 231.2^{+ 7.4}_{-7.4}$  		\vspace{1mm} \\ 
HD225239 	& $36.0^{+ 1.4}_{-1.4}$ 	& $1.2^{+ 3.4}_{-3.4}$  	& $29.48^{+ 0.83}_{-0.83}$ & $3.96$ 	& $1.20$ 	& $0.3^{+ 0.6}_{-0.6}$ 		& $-4.5^{+ 0.4}_{-0.4}$ 	& $            $	& $            $	& $ 8.1^{+ 0.4}_{-0.3}$ 	& $ 131.4^{+ 5.1}_{-5.0}$  		\vspace{1mm} \\ 
 HD62923 	& $-4.9^{+ 1.7}_{-1.7}$ 	& $0.7^{+ 1.2}_{-1.2}$  	& $19.08^{+ 1.24}_{-1.24}$ & $0.35$ 	& $1.86$ 	& $2.6^{+ 1.9}_{-1.9}$ 		& $4.8^{+ 1.5}_{-1.5}$ 		& $            $	& $            $	& $ 6.9^{+ 1.4}_{-1.0}$ 	& $ 67.4^{+ 86.0}_{-208.9}$  	\vspace{1mm} \\ 
 HD87899 	& $-17.4^{+ 4.1}_{-4.1}$ 	& $-9.1^{+ 2.7}_{-2.7}$  	& $19.97^{+ 1.27}_{-1.27}$ & $-0.84$ 	& $0.82$ 	& $4.0^{+ 2.8}_{-2.8}$ 		& $-5.0^{+ 2.8}_{-2.8}$ 	& $-1.08\pm0.79$ 	& $3.58\pm0.81$		& $ 137.2^{+ 8.2}_{-11.6}$ 	& $ 66.1^{+ 24.5}_{-29.2}$  	\vspace{1mm} \\   
 \hline	
\end{tabular} 
\tablefoot{\\
\tablefoottext{$a$}{$\Delta \alpha^{\star}$ and $\Delta \delta$, the corrections on the equatorial coordinates of the star in the tangent plane of the sky with respect to Hipparcos-2 catalog.} \\
\tablefoottext{$b$}{$\varpi$ and $\Delta \varpi$, the new parallax and the corresponding corrections with respect to Hipparcos-2 catalog and Gaia DR2 catalog.}\\
\tablefoottext{$c$}{$\Delta \mu_{\alpha^\star}^\text{HIP2}$ and $\Delta \mu_{\delta}^\text{HIP2}$, the corrections on the proper-motion  with respect to Hipparcos-2 catalog.} \\
\tablefoottext{$d$}{$\Delta \mu_{\alpha^\star}^\text{DR1}$ and $\Delta \mu_{\delta}^\text{DR1}$, the differences with the published DR1 proper-motions for sources in the TGAS sample.} \\
\tablefoottext{$e$}{$i$ the system's inclination.} \\
\tablefoottext{$f$}{$\Omega$ the angle of the ascending node. }
}
\end{table*}

\begin{table}\tiny\centering
\caption{\label{tab:gaia}Gaia DR1 excess noise $\epsilon$ and $\Delta Q$ measurement of acceleration as defined in Lindegren et al. (2012) and Lindegren et al. (2016). $D_{\epsilon}$
measures the significance of the $\epsilon$ value, and should be larger than $2$ (Lindegren et al. 2012). $N_\text{good}$ is the number of reliable measurements taken into account in 
Gaia's astrometric solution and $N_\text{FoV}$ is the number of field-of-view transits on the CCD detector. $N_\text{orb}$ counts the number of orbits covered by the 416-days long 
DR1 time span.}
\begin{tabular}{@{}l|@{~~}c@{~~}c@{~~}c@{~~}c@{~~}c@{~~}c@{~~}|c@{~~}c@{~~}c@{}}
	         &   \multicolumn{6}{c}{Gaia DR1}	&  \multicolumn{3}{|c}{Gaia DR2} \\
Star     	&	N$_\text{orb}$	&	N$_\text{good}$	&	N$_\text{FoV}$	&	$\epsilon$	&	$D_{\epsilon}$	& \tablefootmark{$a$}$\Delta Q$	& $\chi^2$	& $DoF$	& \tablefootmark{$b$}RUWE \\
 		&				&					&				&   (mas)			&				&  			& 	            	& 	 		&		     \\ 
\hline
BD+132550   		& 0.16   	& 62	&	8	& 0.45	& 139.45	& 			& 2987		& 176	& 2.96	\\
BD+192536   		& 2.33   	& 149	&	18	& 1.02	& 1079.36	& 1.50		& 6727		& 210	& 4.05	\\
BD+210055   		& 0.31   	& 55	&	10	& 1.29	& 1723.79	& 			& 772		& 243	& 1.27	\\
BD+212816(b/c) 		& 0.53/0.05	& 143	&	23	& 2.27	& 7887.53	& 3.96		& 6467		& 225	& 3.83	\\
BD+281779    		& 8.98   	& 69	&	8	& 0.50	& 219.34	& 0.75		& 391		& 183	& 1.05	\\
BD+291539    		& 2.37   	& 85	&	10	& 0.31	& 137.50	& 4.00		& 173		& 125	& 0.85	\\
BD+362641    		& 24.0 		& 47	&	10	& 0.88	& 193.29	& 45.28		& 1138		& 290	& 1.41	\\
BD+680971    		& 0.37   	& 26	&	7	& 0.37	& 177.18	&			& 1457		& 157	& 2.19	\\
HD101305    		& 0.25   	& 103	&	13	& 1.72	& 1826.20	& 80.30		& 552		& 160	& 1.33	\\
HD103913    		& 0.18   	& 96	&	13	& 1.21	& 377.21	& 11.16		& 474		& 212	& 1.07	\\
HD104289    		& 0.17   	& 77	&	10	& 0.91	& 184.45	& 71.39		& 987		& 282	& 1.33	\\
HD106888    		& 1.14   	& 81	&	10	& 1.03	& 226.59	& 4.44		& 2693		& 249	& 2.35	\\
HD108436    		& 0.15   	& 57	&	13	& 0.93	& 116.66	& 530.95	& 3628		& 243	& 2.76	\\
HD109157    		& 1.39   	& 86	&	11	& 0.88	& 641.02	& 1.72		& 3888		& 389	& 2.25	\\
HD110376    		& 0.32   	& 109	&	13	& 1.76	& 2926.61	& 48.94		& 1592		& 275	& 1.72	\\
HD13014     		& 0.12   	& 34	&	5	& 1.08	& 83.74 	& 185.49	& 128		& 80	& 0.92	\\
HD130396    		& 0.20   	& 44	&	7	& 1.80	& 1535.46	& 16.23		& 429		& 96	& 1.54	\\
HD133621    		& 0.93   	& 88	&	16	& 1.52	& 2018.64	& 4.24		& 49104		& 237	& 10.29	\\
HD144286    		& 1.31   	& 130	&	20	& 1.17	& 1958.84	& 1.18		& 3995		& 321	& 2.51	\\
HD146735    		& 0.03   	& 53	&	8	& 0.22	& 26.03		&			& 434		& 184	& 1.10	\\
HD147487    		& 0.78   	& 53	&	8	& 0.35	& 73.56		&			& 116061	& 345	& 13.06	\\
HD15292     		& 0.20   	& 70	&	11	& 1.48	& 793.05	& 31.95		& 712		& 248	& 1.21	\\
HD153376    		& 0.09   	& 52	&	7	& 1.23	& 538.55	& 260.70	& 638		& 292	& 1.05	\\
HD155228    		& 0.96   	& 19	&	8	& 2.12	& 354.70	&			& 41101		& 502	& 6.43	\\
HD156111    		& 10.6 		& 264	&	34	& 0.81	& 502.28	& 7.65		& 1851		& 405	& 1.52	\\
HD156728    		& 0.10  	& 44	&	8	& 0.48	& 48.08 	& 1161.77	& 5320		& 205	& 3.65	\\
HD161479    		& 40.6 		& 131	&	17	& 0.49	& 179.57	& 2.35		& 949		& 416	& 1.07	\\
HD167215    		& 0.12  	& 113	&	18	& 0.49	& 88.37 	& 1.39		& 591		& 291	& 1.02	\\
HD18450     	 	&			& 		& 		& 		&			& 			& 387		& 134	& 1.22	\\
HD193554    		& 0.52  	& 208	&	26	& 1.78	& 2981.09	& 6.70		& 38320		& 398	& 6.98	\\
HD204613    		& 0.47  	& 131	&	18	& 2.77	& 2797.25	& 0.44		& 7871		& 186	& 4.66	\\
HD207992    		& 0.20  	& 157	&	23	& 0.33	& 107.23	&			& 2188		& 326	& 1.85	\\
HD210631    	 	&			& 		& 		& 		&			& 			& 313		& 89	& 1.36	\\
HD211681    		& 0.05  	& 98	&	15	& 0.37	& 58.97 	& 42.41		& 383		& 214	& 0.96	\\
HD212029    		& 0.54  	& 125	&	19	& 1.68	& 1965.54	& 2.10		& 7543		& 284	& 3.68	\\
HD212733    		& 4.63  	& 204	&	27	& 0.50	& 278.37	& 2.12		& 361		& 233	& 0.89	\\
HD212735(b/c)		& 11/0.06	& 44	&	6	& 0.36	& 23.17		& 41.67		& 151		& 80	& 1.00	\\
HD217850    		& 0.12   	& 45	&	8	& 1.16	& 176.95	& 76.94		& 13163		& 256	& 5.12	\\
HD225239    		& 0.59   	& 145	&	23	& 7.96	& 49785.70	&			& 1802		& 170	& 2.34	\\
HD23965     		& 0.10   	& 153	&	23	& 0.60	& 154.17	& 8.73		& 662		& 311	& 1.04	\\
HD24505     		& 0.04   	& 133	&	19	& 0.70	& 351.22	&			& 886		& 240	& 1.37	\\
HD28635     		& 0.16   	& 54	&	6	& 0.51	& 56.57 	& 19.99		& 1858		& 316	& 1.73	\\
HD40647     		& 0.05   	& 282	&	34	& 0.53	& 246.49	&			& 12272		& 428	& 3.81	\\
HD48679     		& 0.37   	& 36	&	7	& 0.74	& 209.05	& 3.35		& 2515		& 238	& 2.32	\\
HD5470      	 	&			& 		& 		& 		&			&			& 243		& 115	& 1.05	\\
HD60846     		& 0.07  	& 110	&	14	& 1.80	& 1025.11	& 432.11	& 761		& 252	& 1.24	\\
HD62923     		& 2.37  	& 69	&	8	& 3.73	& 8476.44	&			& 132977	& 158	& 20.85	\\
HD71827(b/c)	 	& 28/0.06	& 123	&	20	& 1.56	& 2526.80	& 	358.00	& 738		& 300	& 1.12	\\
HD73636     		& 2.68  	& 44	&	6	& 0.88	& 61.70 	& 8.89		& 2795		& 238	& 2.45	\\
HD77712     		& 0.32  	& 62	&	8	& 1.65	& 3008.49	& 35.66		& 863		& 133	& 1.84	\\
HD78536     		& 36.7		& 72	&	10	& 0.57	& 111.04	& 1.12		& 230		& 125	& 0.98	\\
HD82460     		& 0.70 		& 137	&	18	& 0.51	& 212.64	& 3.20		& 1185		& 304	& 1.41	\\
HD85533     		& 0.01 		& 131	&	19	& 3.35	& 6009.15	& 309.85	& 579		& 273	& 1.04	\\
HD87899     		& 0.27 		& 168	&	20	& 2.38	& 7862.44	& 22.31		& 5491		& 274	& 3.19	\\
\hline	
\end{tabular}
\tablefoot{\\
\tablefoottext{$a$}{The TGAS discrepancy factor $\Delta Q$ is only given for sources that are member of the TGAS dataset in the DR1. Sources for which $\Delta Q$ is not
given are members of the secondary dataset, for which only 14 months of Gaia data are accounted for.}\\
\tablefoottext{$b$}{RUWE is the renormalized unit weight error of the DR2 Gaia measurements as defined in Section~\ref{sec:caveats}.}
}
\end{table}

\renewcommand\arraystretch{1.2}
\begin{table*}\centering\tiny
\caption{\label{tab:ratio_excess_noise} Photocenter semi-major axis $a_\text{ph}$, inclination $I_c$ and companion mass $M_2$, as derived from Gaia DR1 astrometric excess 
noise using the GASTON method introduced in Section~\ref{sec:method}. For periods larger than 200\,days, these values could be underestimated, as explained in the text. 
Error bars give 1-$\sigma$ confidence intervals, but a value of $\epsilon<0.5$\,mas leads to an upper-limit on mass and semi-major axis and a lower limit on inclination 
(see text). Only the sources for which the uncertainty on the period is better than 10\% are shown. For triple systems, we calculate the inclination for companion $b$ orbit, 
assuming that only the inner short-period companion has a measurable effect on Gaia's astrometry. We marked with an $*$ the BD candidates which true mass is estimated 
to be above 90\,M$_\text{J}$ at 1$\sigma$. The definition of the marginal probability is explained in Section~\ref{sec:method}. The minimum inclination $I_{c,\text{min}}$ measures the 
minimum inclination a system can have while the secondary companion remains dark.}
\begin{tabular}{@{}l@{\quad}c@{\quad}c@{\quad}c@{\quad}|c@{\quad}c@{\quad}|c@{\quad}c@{\quad}c@{}}
Name & Period & $M_2\,\sin i$ 	 & $a_1\sin i$ 	&     \tablefootmark{$a$}$I_{c,\text{min}}$	&  \tablefootmark{$b$}$P(\epsilon_\text{DR1}$$\pm $$10$\%)	 	& $a_\text{ph}$~~($1\sigma$) 	&  $I_c~~(1\sigma)$   	& $M_2~~(1\sigma)$   \\
  	 & (days) &  (M$_\text{J}$)  & (mas)        &	      ($^\circ$)						&  								&    (mas)          &    ($^\circ$)			& (M$_\text{J}$)	 \\
\hline
\multicolumn{9}{c}{$20\,$M$_\text{J}<M\sin i<90\,$M$_\text{J}$} \\ 
\\
$*$\tablefootmark{$c$}
BD+210055		& 1322.63	& 85.30		& 6.03	& 13		& 0.0979 	& $6.56_{-0.29}^{+0.67}$	& $(56,73)$		& $(96,110)$	\\
BD+291539		& 175.87	& 59.70		& 0.60	& 7.4		& 0.0275	& $<$0.67					& $>$64			& $<$70			\\
$*$HD130396		& 2060.60	& 50.90		& 3.17	& 5.3		& 0.0752	& $6.13_{-0.90}^{+1.19}$	& $(26,37)$		& $(88,130)$	\\
$*$HD211681		& 7612.00	& 77.80		& 6.76	& 7.4		& 0.0611	& $<$7.97	& $>$58			& $<$100		\\
$*$HD217850		& 3508.20	& 22.27		& 1.38	& 2.4		& 0.0311	& $6.70_{-0.94}^{+1.30}$	& $(10,14)$		& $(100,140)$	\\
HD23965 		& 3975.00	& 40.00		& 4.02	& 4.1		& 0.2549	& $5.42_{-0.82}^{+1.17}$	& $(38,61)$		& $(47,68)$		\\
$*$HD28635 		& 2636.80	& 77.10		& 5.05	& 7.6		& 0.0307	& $5.40_{-0.23}^{+0.47}$	& $(59,78)$		& $(83,94)$		\\
HD48679 		& 1111.61	& 36.00		& 1.06	& 4.0		& 0.1549	& $1.34_{-0.16}^{+0.22}$	& $(43,63)$		& $(41,55)$		\\
$*$\tablefootmark{$d$}	
HD71827			& 15.05		& 27.43		& 0.07	& 2.9		& 0.0000	& 1.00						& 	$3.3$		& $640$      	\\
$*$HD77712 		& 1311.70	& 42.10		& 1.97	& 5.2		& 0.0353	& $7.39_{-0.96}^{+1.07}$	& $(13,18)$		& $(150,210)$	\\
HD82460 		& 590.90	& 73.20		& 1.97	& 8.7		& 0.0507	& $2.02_{-0.03}^{+0.06}$	& $(72,82)$		& $(78,81)$		\\
\hline	
\multicolumn{9}{c}{$90\,$M$_\text{J}<M\sin i<0.52\,$M$_\odot$} \\
\\
\tablefootmark{$c$}	
BD+132550	& 2537.00		& 463.00	& 24.39	& 61	& 0.1078 & $<$22.41	& $>$70	& $<$680	\\
BD+192536	& 178.54		& 108.50	& 1.70	& 17	& 0.2189& $1.79_{-0.05}^{+0.08}$	& $(66,78)$	& $(120,130)$		\\
BD+212816	& 788.91		& 120.40	& 4.32	& 17	& 0.1013& $5.68_{-0.20}^{+0.23}$	& $(46,51)$	& $(170,190)$		\\
BD+281779	& 46.32			& 90.80		& 0.61	& 14	& 0.1953& $<$0.78					& $>$50		& $<$130			\\
BD+362641	& 17.31			& 251.00	& 0.68	& 39	& 0.0538& $0.78_{-0.06}^{+0.05}$	& $(49,65)$	& $(350,450)$		\\
\tablefootmark{$c$}	
BD+680971	& 1134.14		& 276		& 10.24	& 40	& 0.0213& $<$11.35	& $>$60	& $<$410		\\
HD101305	& 1677.40		& 106.10	& 4.00	& 12	& 0.0686& $7.95_{-0.63}^{+0.75}$	& $(27,33)$	& $(220,270)$		\\
HD103913	& 2322.00		& 98.10		& 3.39	& 10	& 0.1545& $4.30_{-0.37}^{+0.41}$	& $(46,60)$	& $(120,150)$		\\
HD104289	& 2389.00		& 125.80	& 5.33	& 12	& 0.1814& $5.82_{-0.30}^{+0.52}$	& $(57,75)$	& $(140,160)$		\\
HD106888	& 365.61		& 108.30	& 1.42	& 11	& 0.1680& $1.75_{-0.15}^{+0.17}$	& $(48,63)$	& $(130,160)$		\\
HD108436	& 2720.00		& 305.00	& 22.18	& 43	& 0.0429 & $21.88_{-0.12}^{+0.17}$	& $(79,85)$	& $(385,395)$		\\
HD109157	& 300.21		& 153.10	& 3.37	& 22	& 0.0485& $3.47_{-0.07}^{+0.11}$	& $(69,81)$	& $(175,190)$		\\
HD110376	& 1282.20		& 177.70	& 14.45	& 27	& 0.0974& $14.69_{-0.17}^{+0.22}$	& $(74,80)$	& $(210,220)$		\\
HD133621	& 448.60		& 101.80	& 3.36	& 12	& 0.1278& $3.44_{-0.05}^{+0.10}$	& $(71,82)$	& $(110,120)$		\\
HD144286	& 316.77		& 117.60	& 1.67	& 15	& 0.2080& $1.91_{-0.10}^{+0.11}$	& $(56,67)$	& $(140,160)$		\\
\tablefootmark{$c$}	
HD147487	& 533.23		& 279.70	& 5.27	& 35	& 0.0056& $<$5.71	& $>$66	& $<$380			\\
HD15292 	& 2087.00		& 176.50	& 14.98	& 21	& 0.0127& $15.47_{-0.36}^{+0.59}$	& $(69,81)$	& $(200,220)$			\\
HD153376	& 4878.00		& 236.80	& 19.67	& 22	& 0.1063& $30.70_{-3.54}^{+4.40}$	& $(33,46)$	& $(385,540)$		\\
\tablefootmark{$c$}	
HD155228	& 432.98		& 268.90	& 3.69	& 26	& 0.1613& $4.34_{-0.35}^{+0.50}$	& $(49,67)$	& $(340,430)$		\\
HD156111	& 39.44			& 92.00		& 0.48	& 12	& 0.1069& $1.10_{-0.11}^{+0.11}$	& $(23,29)$	& $(220,280)$		\\
HD156728	& 4097.00		& 126.50	& 15.10	& 16	& 0.0004& $<$18.28					& $>$60		& $<$170				\\
HD161479	& 10.24			& 268.70	& 0.46	& 30	& 0.2743& $<$0.55					& $>$56		& $<$395			\\
HD167215	& 3460.10		& 167.40	& 8.42	& 18	& 0.2142& $<$9.47	& $>$62	& $<$215			\\
HD193554	& 797.69		& 173.60	& 7.62	& 21	& 0.1386& $7.89_{-0.12}^{+0.14}$	& $(71,78)$	& $(200,210)$			\\
HD204613	& 876.84		& 151.70	& 4.22	& 18	& 0.0901& $6.65_{-0.32}^{+0.37}$	& $(37,41)$	& $(265,305)$		\\
\tablefootmark{$c$}	
HD207992	& 2090.00		& 206.80	& 18.48	& 29	& 0.0012& $<$22.00	& $>$56	& $<$304			\\
HD212029	& 771.02		& 122.10	& 3.50	& 17	& 0.1002& $5.16_{-0.27}^{+0.31}$	& $(39,45)$	& $(195,225)$		\\
HD212733	& 89.86			& 112.10	& 1.57	& 15	& 0.0011& $<$1.62	& $>$75	& $<$130			\\
HD212735	& 37.94			& 92.50		& 0.34	& 9.8	& 0.1180& $<$0.46					& $>$47		& $<$140			\\
\tablefootmark{$c$}	
HD225239	& 701.49		& 100.50	& 4.22	& 12	& 0.0452& $15.90_{-0.81}^{+0.64}$	& $(14,16)$	& $(480,575)$		\\
\tablefootmark{$c$}	
HD24505 	& 11268.00		& 238.20	& 28.09	& 25	& 0.0365& $28.15_{-0.06}^{+0.13}$	& $(83,87)$	& $(274,276)$		\\
\tablefootmark{$c,e$}	
HD62923 	& 175.22		& 91.80		& 0.89	& 9.9	& 0.0003& $4.11_{-0.06}^{+0.03}$	& $(10,12)$	& $(595,700)$		\\
HD73636 	& 155.28		& 102.70	& 0.85	& 9.8	& 0.1373& $1.16_{-0.16}^{+0.22}$	& $(38,58)$	& $(130,180)$		\\
HD78536 	& 11.35			& 213.70	& 0.22	& 21	& 0.2610& $0.27_{-0.03}^{+0.06}$	& $(41,67)$	& $(265,390)$		\\
HD87899 	& 1527.00		& 195.30	& 10.33	& 27	& 0.1724& $10.67_{-0.26}^{+0.46}$	& $(67,80)$	& $(230,250)$		\\
\hline	
\end{tabular}
\tablefoot{\\
\tablefoottext{$a$}{The likelihood of $\epsilon$$\pm $$10\%$ given that $I_c>I_{c,-}$ (or $I_c<I_{c,+}$) can be calculated 
by using formula (12) and (13), replacing $p$ by 0.683, $p(\epsilon)$ by the marginal probability, and $I_{c,\text{min}}$ by the value given in this column.}\\
\tablefoottext{$b$}{The marginal probability as defined in Section~\ref{sec:simexcess}.}\\
\tablefoottext{$c$}{These targets were not part of the TGAS sample. Therefore, $\epsilon_\text{DR1}$ measures the scatter around the 
5-parameters solution only accounting for the 14 months of Gaia measurements. In these cases, the proper motion is fitted and modelled out of the simulations.}\\
\tablefoottext{$d$}{For HD71827, $\epsilon$ is larger than any of the simulations that assume a dark companion, i.e. $M_{V,2}-M_{V,1}$$>$$2.5$. 
The excess noise observed by Gaia is likely due to the large 20-yrs period outer companion (see text explanation in Section~\ref{sec:caveats}).} \\
\tablefoottext{$e$}{For HD62923, $\epsilon$ is at the very limit of simulations that assume a dark companion, i.e. $M_{V,2}-M_{V,1}$$>$$2.5$. In this system, the secondary likely pollutes the astrometry 
as emphasised in Section~\ref{sec:astrometry}.}
}
\end{table*}

\renewcommand\arraystretch{1.}
\begin{table*}\centering\tiny
\caption{\label{tab:mass_summary} Summary table of the 54 companion masses measured in this paper, with re-evaluation thanks to Hipparcos and Gaia astrometry. The relative errors 
on the period and the $M\sin i$ are lower than 8\% except for 6 companions for which it is larger than 10\%. In these cases, the errorbars are given for the $M\sin i$ as well as 
for the period. The last column indicates with a 'yes' the systems for which a significant binary motion is detected with $\Delta Q$$>$$100$ (see Section~\ref{sec:deltaq}).}
\begin{tabular}{@{}l@{\quad}c@{\quad}c@{\quad}c@{\quad}c@{\quad}c@{}}
 		  	&	\multicolumn{2}{c}{Radial velocities}	& +Hipparcos 	& +Gaia & Hipparcos+Gaia  \\
Name & Period & $M_2\,\sin i$ & $M_2$ $(1\sigma)$   & $M_2 $ $(1\sigma)$ & $\Delta Q>100$ ?	\\
 		  & (days) &  (M$_\text{J}$)        & (M$_\text{J}$)	 & (M$_\text{J}$)			\\
\hline\hline
\multicolumn{6}{c}{$20\,$M$_\text{J}<M\sin i<90\,$M$_\text{J}$} \\ 
\hline
\multicolumn{6}{c}{Brown dwarfs} \\
\\
BD+291539 b &   175.87		&   59.7	& $<$500			&   $<$70 			&	\\
HD23965  b	&	3975		&	40		&					&	$(47,68)$		&	\\
HD48679  b	&	1111.61		&	36		& $<$320			&	$(41,55)$		&	\\
HD82460  b	&	590.9		&	73.2	& $<$270 			&	$(78,81)$ 			&	\\
\hline
\multicolumn{6}{c}{Possible brown dwarfs} \\
\\
HD211681  b	&   7612		&   77.8	&					&   $<$100 		&	\\
HD28635  b	&	2636.8		&	77.1	& 					&	$(83,94)$			&	\\
HD71827	 b	&	15.05		&	26.3 	& $<$1070			&	\tablefootmark{$b$}$<$640 &	\\
\hline
\multicolumn{6}{c}{M-dwarfs} \\
\\
BD+210055 b	&	1322.63		&	85.3	& $(190,230)$		&	$(96,110)$		&	\\
HD130396 b	&	2060.6		&	50.9	&					&   $(88,130)$		&	\\
HD210631  b	&   4030		&   83.4	& $(330,730)$ 		&             		&	\\
HD217850 b	&	3508.2		&	22.3	& 					&	$(100,140)$		&	\\
\tablefootmark{$c$}
HD77712  b	&	1311.7		&	42.1	& $<$180 			&	$(150,210)$		&	\\
\hline\hline
\multicolumn{6}{c}{$90\,$M$_\text{J}<M\sin i<0.52\,$M$_\odot$} \\
\hline \\	
BD+132550 b	&	2537			&	463			&				&	$<$680		&	\\
BD+192536 b	&	178.54			&	108.5		& $<$460 		&	$(120,130)$		&	\\
BD+212816 b	&	788.91			&	120.4		& $(140,180)$	&	$(170,190)$		&	\\
BD+281779 b	&	46.32			&	90.8		& $<$830		&	$<$130			&	\\
BD+362641 b	&	17.31			&	251			& $<$5670		&	$(350,450)$ 	&	\\
BD+680971 b	&   1134.14			&	276			& $(300,460)$	&   $(350,410)$		&	\\
HD101305 b 	&	1677.4			&	106.1		& $(230,310)$	&	$(220,270)$		&	\\
HD103913 b 	&	2322			&	98.1		&				&	$(120,150)$		&	\\
HD104289 b	&	2389			&	125.8		& $(120,210)$	&	$(140,160)$		&	\\
HD106888 b	&	365.61			&	108.3		& $<$300		&	$(130,160)$		&	\\
HD108436 b	&	2720			&	305			&				&	$(385,395)$		&	yes \\
HD109157 b 	&	300.21			&	153.1		&	$<$320		&	$(175,190)$ 	&	\\
HD110376 b 	&	1282.2			&	177.7		& $(200,240)$	&	$(210,220)$		&	\\
HD13014 b   &	3450$\pm$390	&	370$\pm$50	&				&					&	yes \\
HD133621 b  &	448.6			&	101.8		& $(120,140)$ 	&	$(110,120)$		&	\\
HD144286 b	&	316.77			&	117.6		& $<$570 		&	$(140,160)$		&	\\
HD146735 b	&   13932$\pm$3864	&   244			&				& 					&	\\
HD147487 b	&   533.23			&   279.7		& $(320,360)$	&	$<$380		&	\\
HD15292  b  &	2087			&	176.5		&				&	$(200,220)$		&	\\
HD153376 b	&	4878			&	236.8		&				&	$(385,540)$		&	yes \\
HD155228 b	&	432.98			&	268.9		& 	$(290,310)$	&	$(340,430)$		&	\\
HD156111 b  &	39.44			&	92			& $<$790		&	$(220,280)$		&	\\
HD156728 b	&   4097 			&	126.5		&				&	$<$170 			&	yes \\
HD161479 b	&   10.24 			&   268.7		& $<$2490		&	$<$395 			&	\\
HD167215 b 	&   3460.1 			&   167.4		&			 	&   $(190,215)$		&	\\
HD18450	 b	&   25.04			&   114.6		& $<$600		& 					&	\\
HD193554 b	&	798.6			&	173.6		&	$(227,391)$	&	$(200,210)$		&	\\
HD204613 b	&	876.84			&	151.7		& $(470,550)$	&	$(265,305)$		&	\\
HD207992 b	&	2090			&	206.8		& $(330,410)$ 	&   $(265,390)$		&	\\
HD212029 b	&	771.02			&	122.1		& $(180,240)$	&	$(195,225)$		&	\\
HD212733 b	&	89.86			&	112.1		& $<$310		&	$(125,130)$		&	\\
HD212735 b	&   37.94			&   92.5		& $<$960 		&   $<$140			&	\\
HD225239 b	&	701.49			&	100.5		& \tablefootmark{$a$}$(1050,1210)$ &	$(480,575)$		&	\\
HD24505  b	&	11268			&	238.2		&				&	$(274,276)$		&	\\
HD40647  b	&   8080$\pm$870	&	172$\pm$11	&				&					&	\\
HD5470	 b	&   7788			&   208.5		&				& 					&	\\
HD60846	 b	&   6030$\pm$890	&   450$\pm$100	&				& 					&	yes \\
HD62923  b	&	175.22			&	91.8		& \tablefootmark{$a$}$(630, 710)$ 	&	$(595,700)$  	&	\\
HD73636  b	&	155.29			&	102.7		& $<$280		&	$(130,180)$		&	\\
HD78536  b	&	11.35			&	213.7		& $<$2910 		&	$(265,390)$		&	\\
HD85533	 b	&   31214$\pm$30144	&   450$\pm$100	&				& 					&	yes \\
HD87899  b	&	1527			&	195.3		& $(270,430)$ 	&	$(230,250)$		&	\\
\hline	&	
\end{tabular}
\tablefoot{\\
\tablefoottext{$a$}{Most likely overestimated, since it was assumed that the secondary is dark.} \\
\tablefoottext{$b$}{Assuming that only the inner companion of the triple system HD71827 is responsible for the 
astrometric motion of the primary star observed with Gaia is most likely wrong, as explained in Section~\ref{sec:method}. 
The astrometric noise is more likely to be explained by the outer companion. Therefore, the mass of HD71827-b is not constrained.} \\
\tablefoottext{$c$}{The orbital parameters of the primary star, except the period, are certainly wrong, due to pollution by 
a secondary component in the CCF of the SOPHIE spectra obtained for this target. The mass is therefore most likely underestimated here.}
}
\end{table*}

\begin{table*}
\tiny\centering
\caption{\label{tab:sup_targets} 32 supplementary targets, with Keplerian orbits and companion mass.}
\begin{tabular}{l@{~~~}c@{~~~}c@{~~~}c@{~~~}c@{~~~}c@{~~~}c@{~~~}c@{~~~}c@{~~~}c}
Name 	& RA\tablefootmark{$a$}  & DEC\tablefootmark{$a$}  & $\pi$\tablefootmark{$b$} & B-V\tablefootmark{$b$} & $M_\star$\tablefootmark{$c$} & Period & e & $M\sin i$ & Reference \\
		&	&	& (mas)	& 	& (M$_\odot$)	& (days) &	& (M$_\text{J}$) \\
\hline
BD+244697 	& 23:01:39.322	& +25:47:16.54	& 20.51$\pm$1.33	& 1.005$\pm$0.036	& 0.721$\pm$0.026	& 145.081$\pm$0.016   	& 0.50048$\pm$0.00043	& 53$\pm$3   		& 1 \\
BD+482155 	& 13:50:07.269	& +47:49:15.95	&  9.87$\pm$1.40	& 0.599$\pm$0.037	& 1.068$\pm$0.039	& 90.270$\pm$0.019		& 0.4375$\pm$0.0040		& 62.6$\pm$0.6		& 1 \\
HD110833 	& 12:44:14.545	& +51:45:33.37	& 67.20$\pm$0.66	& 0.936$\pm$0.014	& 0.771$\pm$0.011	& 271.17	& 0.784	& 17	& 10 \\
HD11443 	& 01:53:04.908 	& +29:34:43.79 	& 51.50$\pm$0.23 	& 0.488$\pm$0.009 	& 1.189$\pm$0.010	& 1.77              	& 0.07           		& 71              	& 2, 3 \\
HD114762 	& 13:12:19.743 	& +17:31:01.64 	& 25.87$\pm$0.76 	& 0.525$\pm$0.013 	& 1.147$\pm$0.014	& 83.9152$\pm$0.0028 	& 0.33$\pm$0.15 		& 10.99$\pm$0.09	& 1, 2, 4 \\
HD118742 	& 13:38:01.953 	& +39:10:41.10 	& 21.74$\pm$0.80 	& 0.698$\pm$0.005 	& 0.970$\pm$0.005	& 11.5896$\pm$0.0005 	& 0.084$\pm$0.019 		& 77.8$\pm$1.6  	& 2, 5  \\
HD122562 	& 14:02:21.163	& +20:52:52.74	& 18.60$\pm$0.72	& 0.962$\pm$0.010	& 0.752$\pm$0.007	& 2777$^{+100}_{-80}$	& 0.71$\pm$0.01			& 24$\pm$2			& 1 \\
HD127506	& 14:30:44.975	& +35:27:13.43	& 44.01$\pm$0.93	& 1.031$\pm$0.014	& 0.703$\pm$0.010	& 2599	& 0.716	& 36	& 10 \\
HD132032 	& 14:56:43.930	& +13:08:57.14	& 17.86$\pm$0.97	& 0.636$\pm$0.015	& 1.030$\pm$0.015	& 274.33$\pm$0.24		& 0.0844$\pm$0.0024		& 70$\pm$4			& 1 \\
HD13507 	& 02:12:54.990	& +40:40:06.22	& 37.25$\pm$0.55	& 0.672$\pm$0.007	& 0.995$\pm$0.007	& 4880$^{+210}_{-190}$	& 0.20$\pm$0.04			& 67$\pm$9			& 1 \\
HD137510 	& 15:25:53.270	& +19:28:50.55	& 24.24$\pm$0.51	& 0.618$\pm$0.012	& 1.048$\pm$0.012	& 801.30$\pm$0.45		& 0.3985$\pm$0.0073		& 27.3$\pm$1.9		& 1	\\
HD140913 	& 15:45:07.449	& +28:28:11.74	& 22.27$\pm$0.82	& 0.612$\pm$0.007	& 1.054$\pm$0.007	& 147.968	& 0.54	& 43.2	& 10	\\
HD14348 	& 02:19:52.925	& +31:20:14.92	& 17.68$\pm$0.45	& 0.596$\pm$0.015	& 1.071$\pm$0.016	& 4740$\pm$6			& 0.455$\pm$0.004		& 48.9$\pm$1.6		& 1	\\
HD14651 	& 02:22:00.854	& +04:44:48.33	& 24.65$\pm$0.94	& 0.720$\pm$0.015	& 0.950$\pm$0.014	& 79.4179$\pm$0.0021	& 0.475$\pm$0.001		& 47.0$\pm$3.4		& 1 \\
HD160508 	& 17:39:12.696	& +26:45:27.15	& 10.83$\pm$0.79	& 0.543$\pm$0.013	& 1.127$\pm$0.014	& 178.90$\pm$0.0074		& 0.5967$\pm$0.0009		& 48$\pm$3			& 1 \\
HD169822	& 18:26:10.089	& +08:46:39.28	& 34.61$\pm$1.39	& 0.699$\pm$0.005	& 0.969$\pm$0.005	& 293.1	& 0.48	& 27.2	& 10 \\
HD174457 	& 18:50:02.059 	& +15:18:41.44 	& 18.79$\pm$0.78 	& 0.621$\pm$0.015 	& 1.045$\pm$0.015	& 840.80$\pm$0.05   	& 0.23$\pm$0.01 		& 58.22$\pm$0.75	& 1, 2, 6 \\
HD209262 	& 22:01:54.121	& +04:46:13.62	& 20.12$\pm$0.79	& 0.687$\pm$0.017	& 0.981$\pm$0.016	& 5430$^{+140}_{-100}$	& 0.35$\pm$0.01			& 32.3$\pm$1.6		& 1 \\
HD221115 	& 23:29:09.297 	& +12:45:37.99 	& 18.65$\pm$0.78 	& 0.94$\pm$0.00 	& 0.768$\pm$0.000	& 941.03$\pm$0.12   	& 0.517$\pm$0.012 		& 89.7$\pm$1.4  	& 2, 7  \\
HD22468 	& 03:36:47.289 	& +00:35:15.93 	& 32.59$\pm$0.64 	& 0.885$\pm$0.007 	& 0.810$\pm$0.005	& 1152$\pm$44       	& 0.40$\pm$0.22   		& 72$\pm$24      	& 2, 8 \\
HD22781 	& 03:40:49.524	& +31:49:34.65	& 30.51$\pm$1.11	& 0.845$\pm$0.023	& 0.842$\pm$0.019	& 528.07$\pm$0.14		& 0.8191$\pm$0.0023		& 13.65$\pm$0.97	& 1 \\
HD28291 	& 04:28:37.215 	& +19:44:26.47 	& 21.15$\pm$0.77 	& 0.741$\pm$0.014 	& 0.931$\pm$0.013	& 41.66             	& 0.66           		& 89             	& 2, 9  \\
HD283668 	& 04:27:52.909	& +24:26:41.88	& 23.66$\pm$1.97	& 0.894$\pm$0.006	& 0.803$\pm$0.005	& 2558$\pm$8			& 0.577$\pm$0.011		& 53$\pm$4			& 1 \\
HD29587 	& 04:41:36.318 	& +42:07:06.49 	& 36.27$\pm$0.87 	& 0.633$\pm$0.010 	& 1.033$\pm$0.010	& 1481$\pm$22        	& 0.713$\pm$0.006 		& 55.2$\pm$9.2   	& 2, 4   \\
HD30246 	& 04:46:30.386	& +15:28:19.35	& 21.08$\pm$0.86	& 0.665$\pm$0.006	& 1.002$\pm$0.006	& 990.7$\pm$5.6			& 0.838$\pm$0.081		& 55$^{+20}_{-8}$	& 1 \\
HD33636	& 05:11:46.449 	& +04:24:12.76	& 35.25$\pm$1.02	& 0.588$\pm$0.016	& 1.079$\pm$0.017	& 2128	& 0.48	& 9.3 	& 10\\
HD38529 	& 05:46:34.913 	& +01:10:05.51	& 25.46$\pm$0.40	& 0.773$\pm$0.001	& 0.902$\pm$0.001	& 2136.14$\pm$0.29		& 0.362$\pm$0.002		& 13.99$\pm$0.59	& 1 \\
HD65430	& 07:59:33.937 	& +20:50:38.19	& 42.15$\pm$0.71	& 0.833$\pm$0.008	& 0.851$\pm$0.007	& 3138.0	& 0.32	& 67.8	& 10 \\
HD77065 	& 09:00:47.445 	& +21:27:13.37 	& 31.52$\pm$1.05 	& 0.839$\pm$0.010 	& 0.846$\pm$0.008	& 119.1135$\pm$0.0027	& 0.35$\pm$0.05 		& 41$\pm$2       	& 1, 2, 5   \\
HD72946 	& 08:35:51.266	& +06:37:21.97	& 38.11$\pm$0.85	& 0.710$\pm$0.015	& 0.959$\pm$0.014	& 5814$\pm$50			& 0.495$\pm$0.006		& 60.4$\pm$2.2		& 1 \\
HD92320 	& 10:40:56.909	& +59:20:33.01	& 23.79$\pm$0.78	& 0.679$\pm$0.015	& 0.988$\pm$0.014	& 145.402$\pm$0.013		& 0.3226$\pm$0.0014		& 59.4$\pm$4.1		& 1 \\
HD98230	& 11:18:10.836	& +31:31:44.82	& 114.49$\pm$0.43\tablefootmark{$c$}	& 0.65$\pm$0.02\tablefootmark{$c$}	& 1.016$\pm$0.020	& 3.98	& 0	& 35	& 10 \\
\hline 
\end{tabular}
\tablefoot{\\
(1) Wilson et al. (2016); (2) SB9; (3) Pike et al. (1978); (4) Mazeh et al.(1996); (5) Latham et al. (2002); (6) Nidever et al. (2002); (7) Griffin (2009); 
(8) Tokovinin \& Gorynya (2001); (9) Griffin et al. (1985); (10) Sozzetti \& Desidera (2010) \\
\tablefoottext{$a$}{From SIMBAD.} \\
\tablefoottext{$b$}{From Hipparcos-2 catalog.}\\
\tablefoottext{$c$}{Using Noyes et al. (1984) empirical formula $\log (M/M_\odot$)=$0.28-0.42\,(B-V)$, with precision of $\sim$0.01.} 
}
\end{table*}
\twocolumn

\onecolumn
\section{Figures}
\pagestyle{plain}

\newdimen\LFcapwidth \LFcapwidth=\linewidth
\begin{longfigure}{c}
\caption{\label{fig:solutions_BD} Orbital solutions of the radial velocity variations of the 11 binary systems with a brown-dwarf candidate as secondary. RV vs time are presented
on the left panel, with O-C residuals below, and RV vs phase on the right panel. Data points colours are fixed with respect to the order given 
on the upper-right corner of each figure, with alternatively red, then blue, then yellow, then pink. Diminutives 'SP' stands for SOPHIE-plus, 'Soph' for SOPHIE and 'Elo' for Elodie.
'Keck' and 'SB9' are self-explanatory and references of the corresponding data can be found in Table~\ref{tab:pub}.} 
\endLFfirsthead
\caption{Continued.}
\endLFhead
\includegraphics[height=58mm, clip=true, trim=0 -12 0 7]{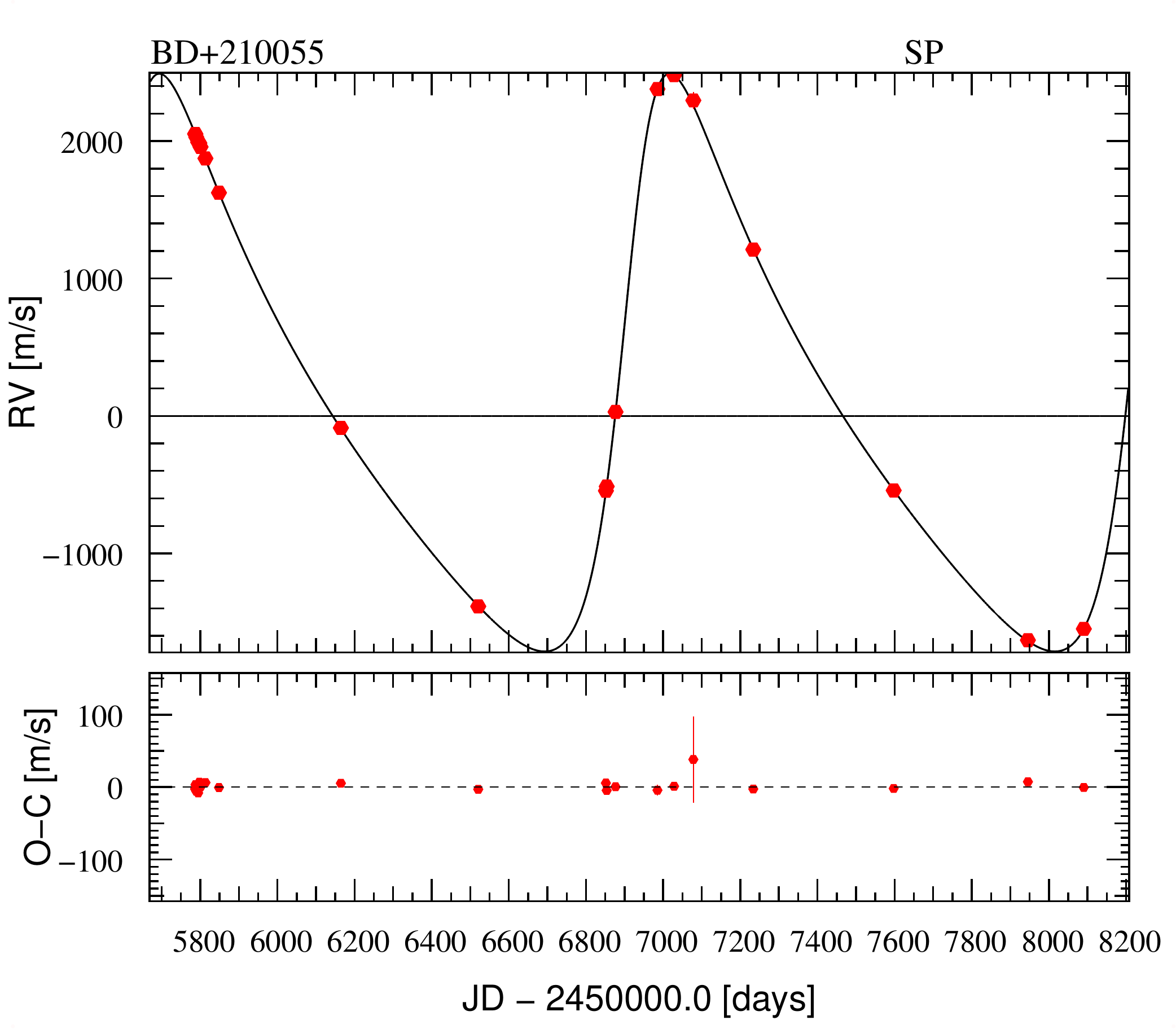}
\includegraphics[height=57mm, clip=true, trim=0  25 0 0]{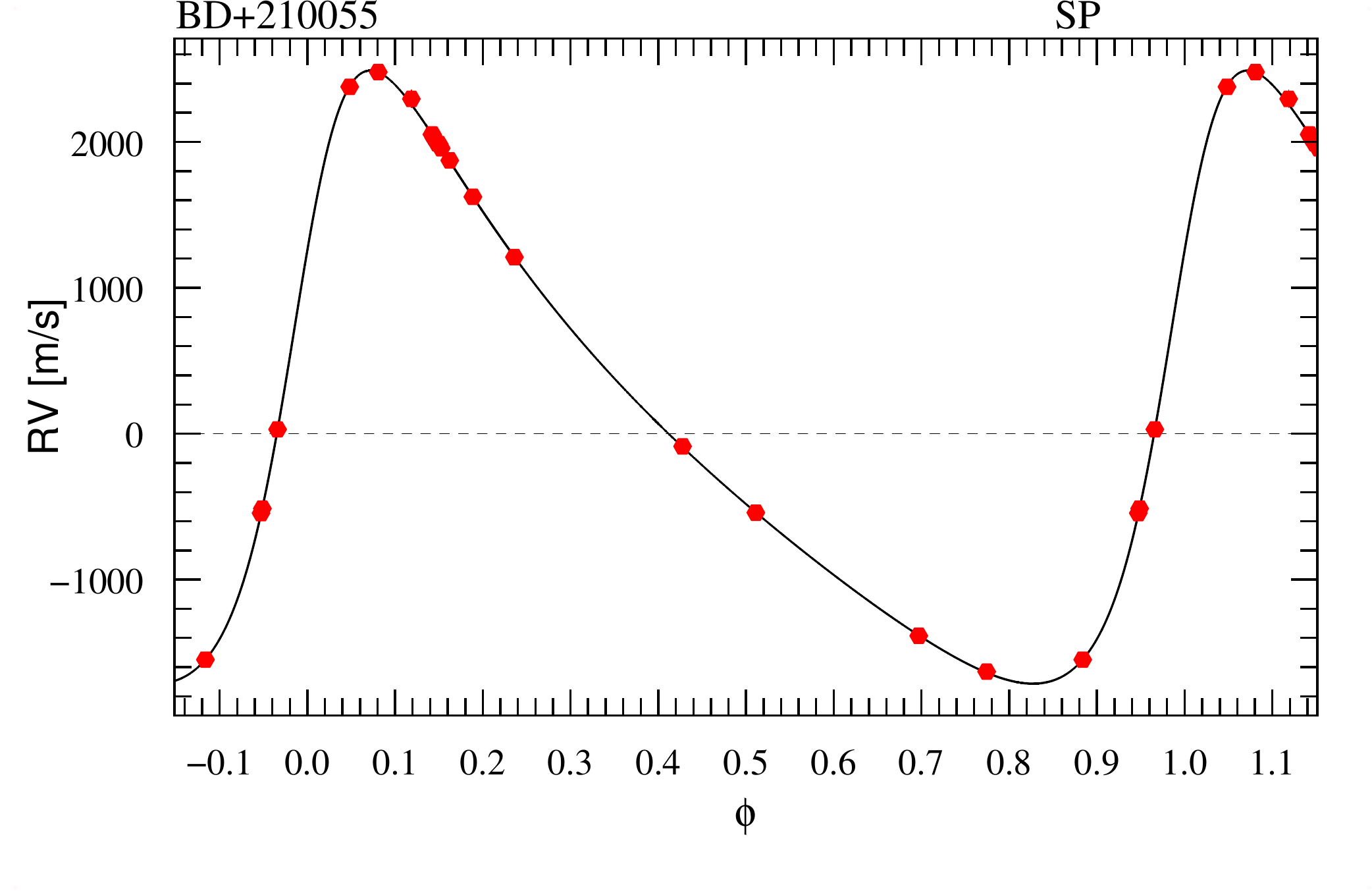} \\
\includegraphics[height=58mm, clip=true, trim=0 -12 0 7]{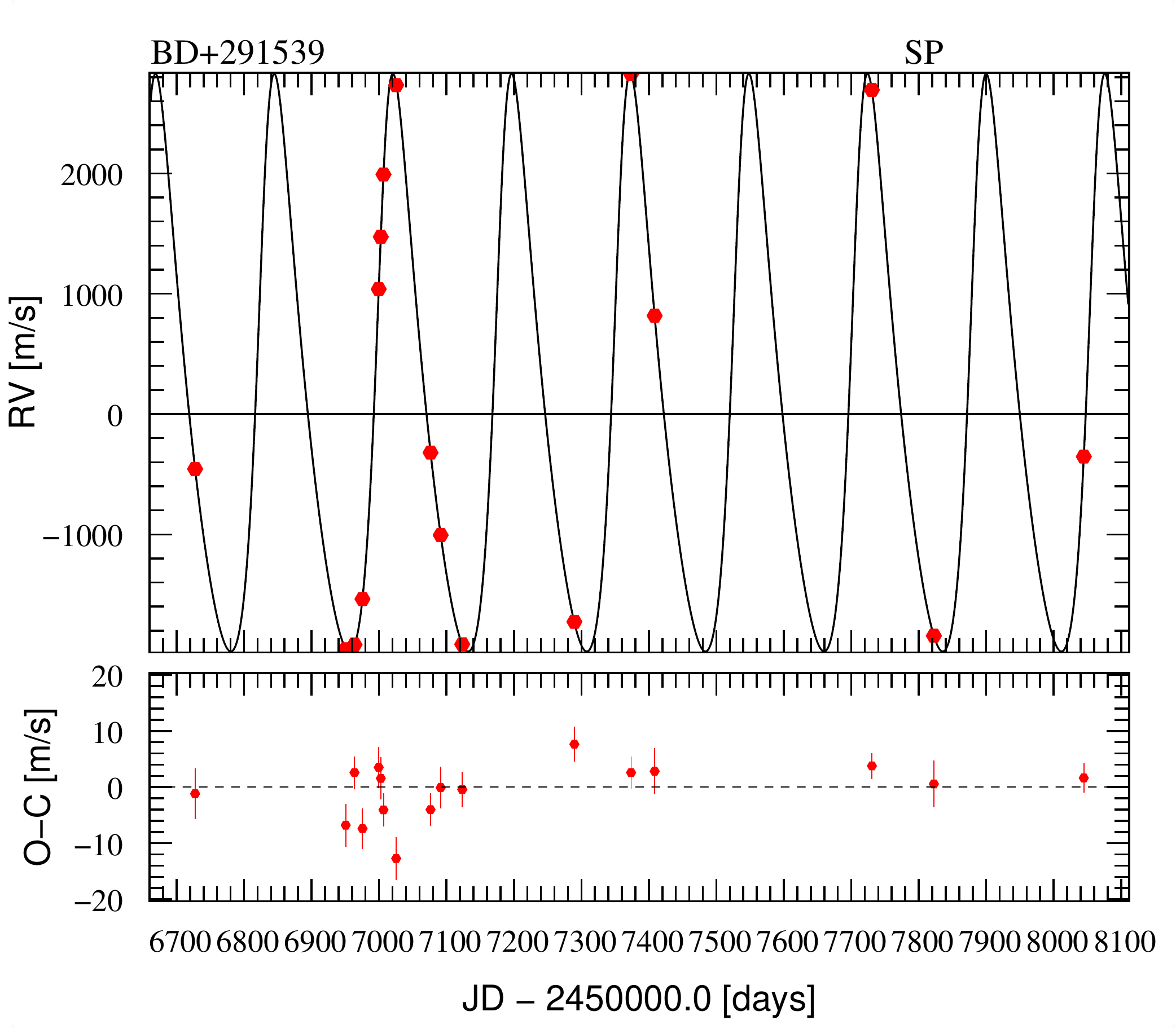} 
\includegraphics[height=57mm, clip=true, trim=0  25 0 0]{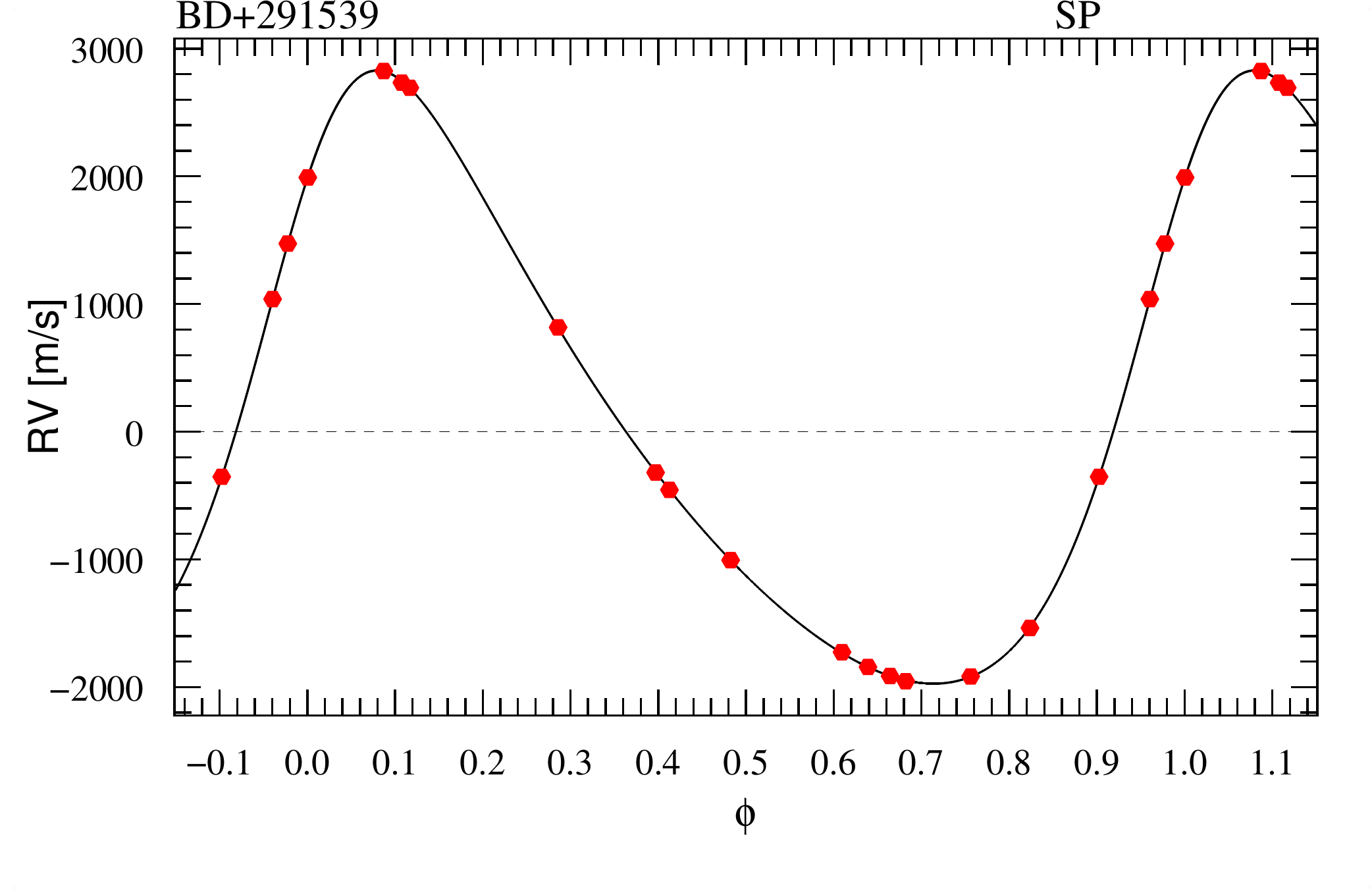} \\
\includegraphics[height=58mm, clip=true, trim=0 -12 0 7]{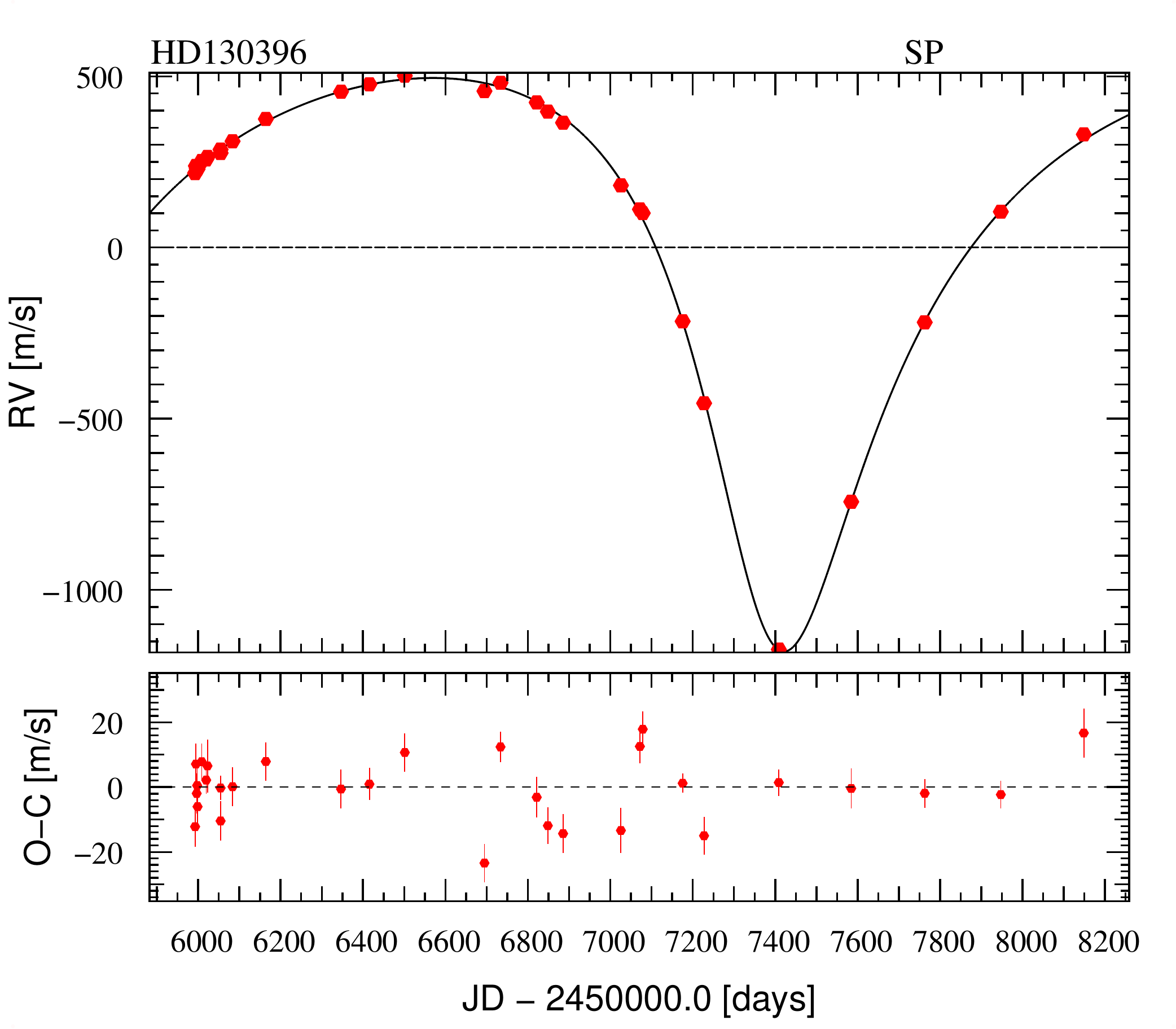}
\includegraphics[height=57mm, clip=true, trim=0  25 0 0]{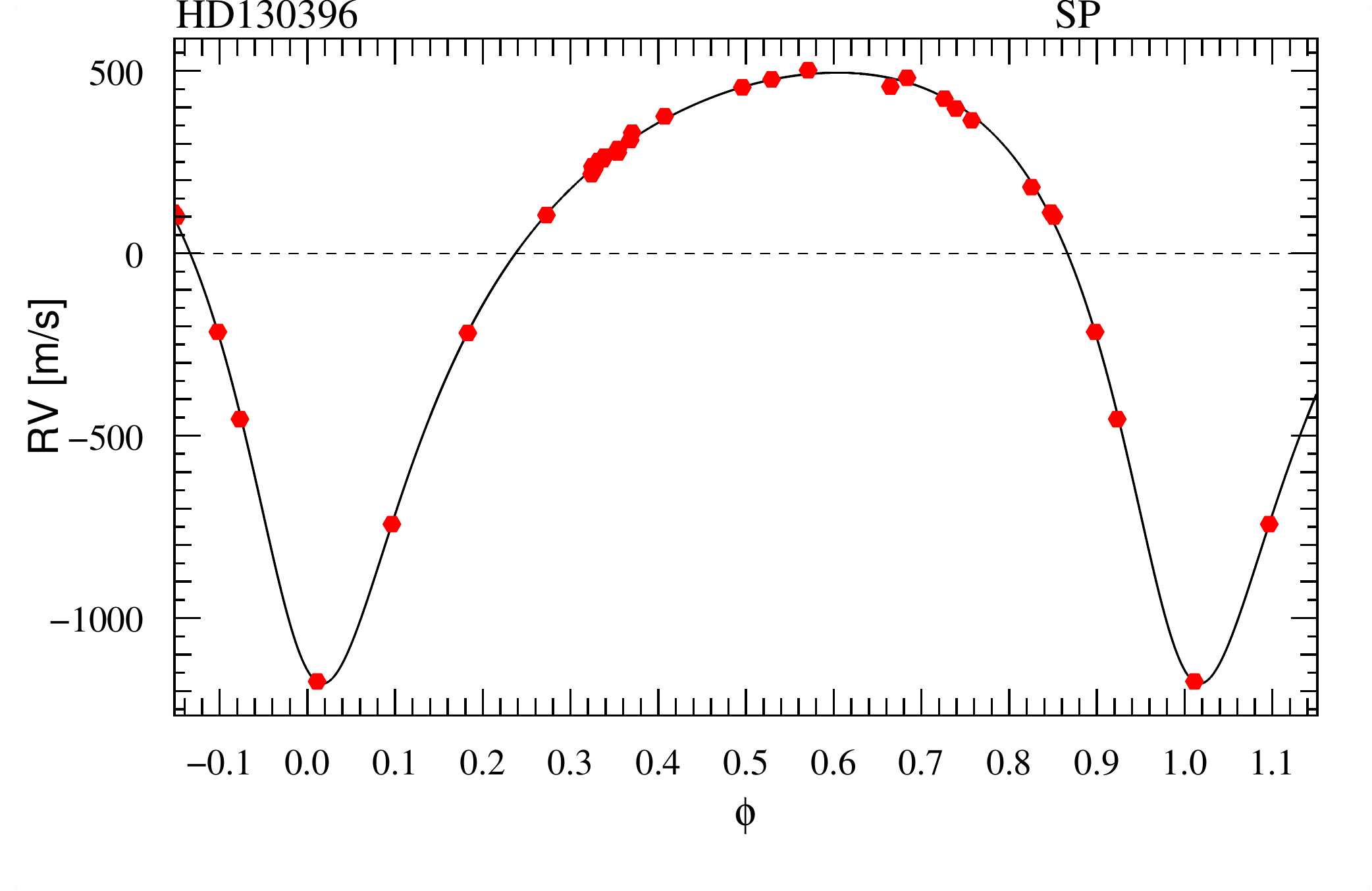} \\
\includegraphics[height=58mm, clip=true, trim=0 -12 0 7]{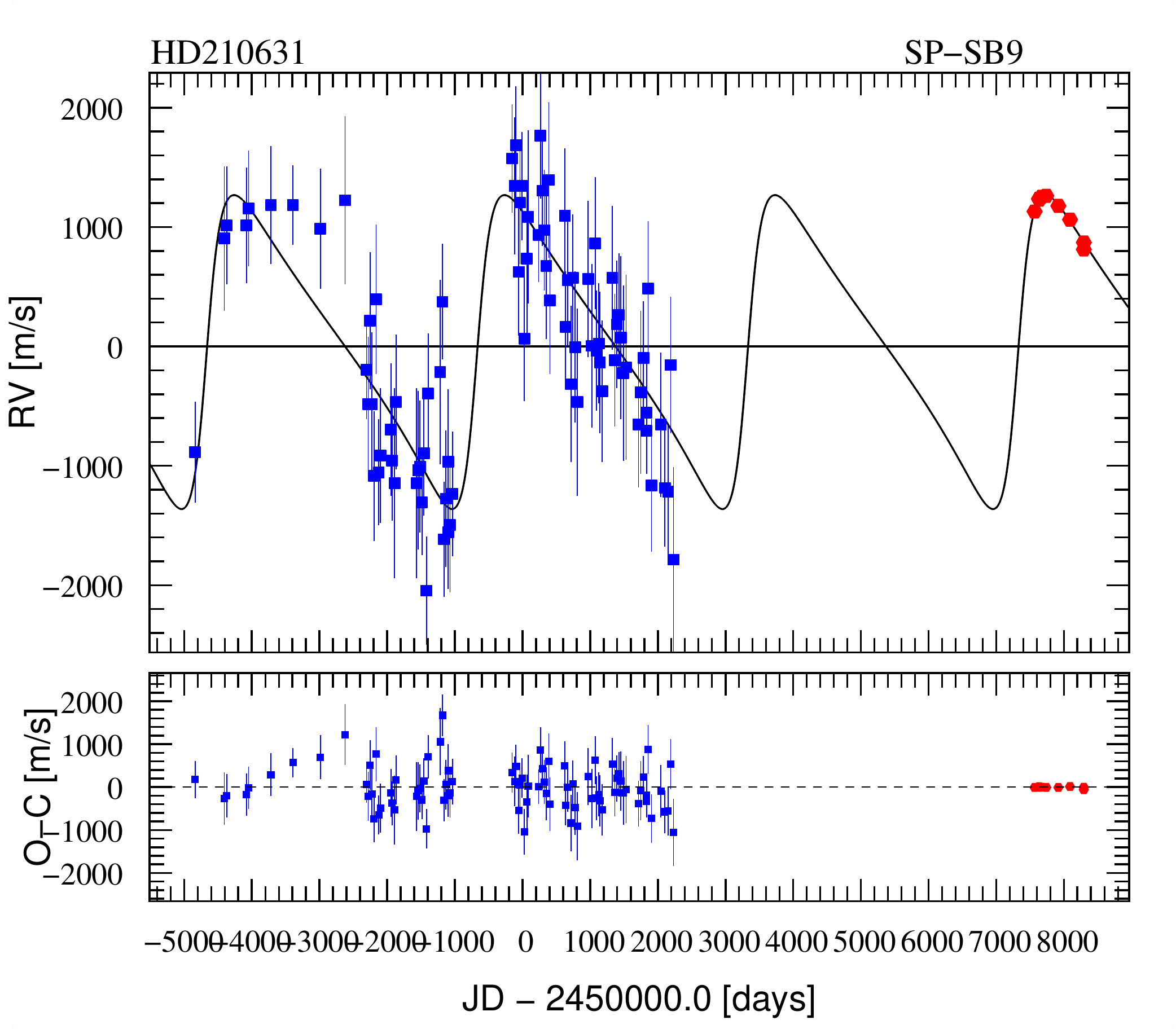} 
\includegraphics[height=57mm, clip=true, trim=0  25 0 0]{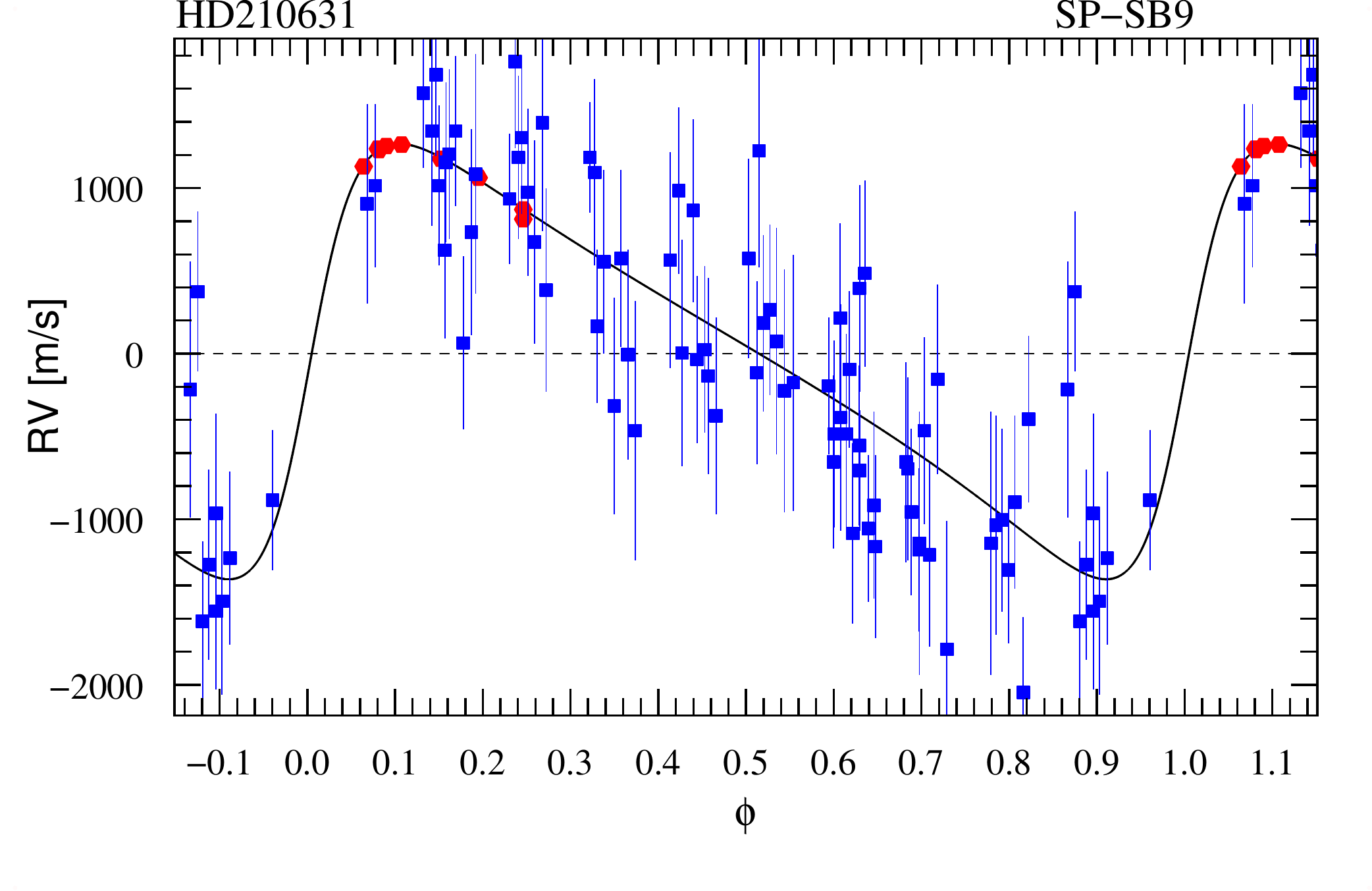} \\
\includegraphics[height=58mm, clip=true, trim=0 -12 0 7]{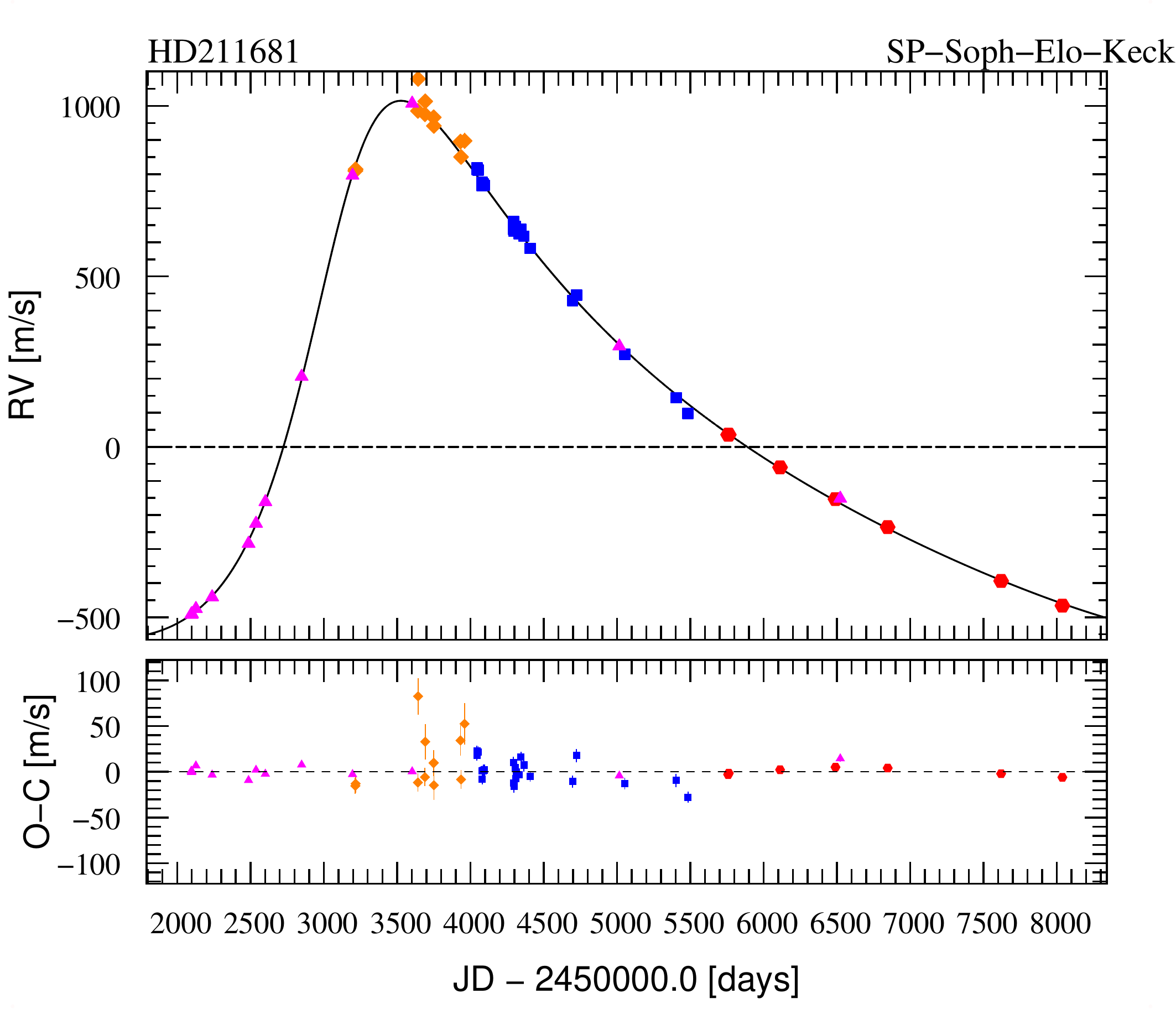}
\includegraphics[height=57mm, clip=true, trim=0  25 0 0]{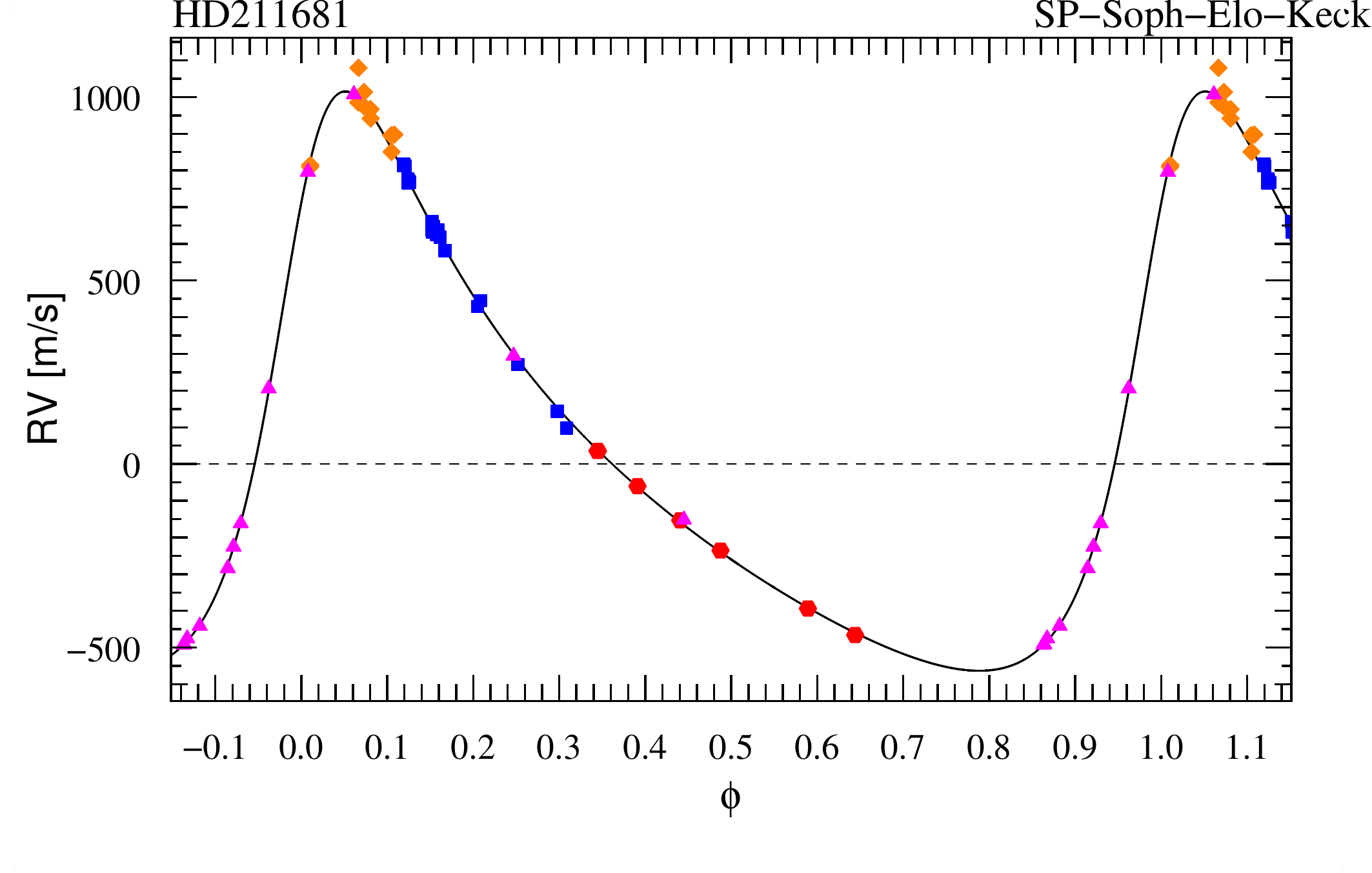} \\
\includegraphics[height=58mm, clip=true, trim=0 -12 0 7]{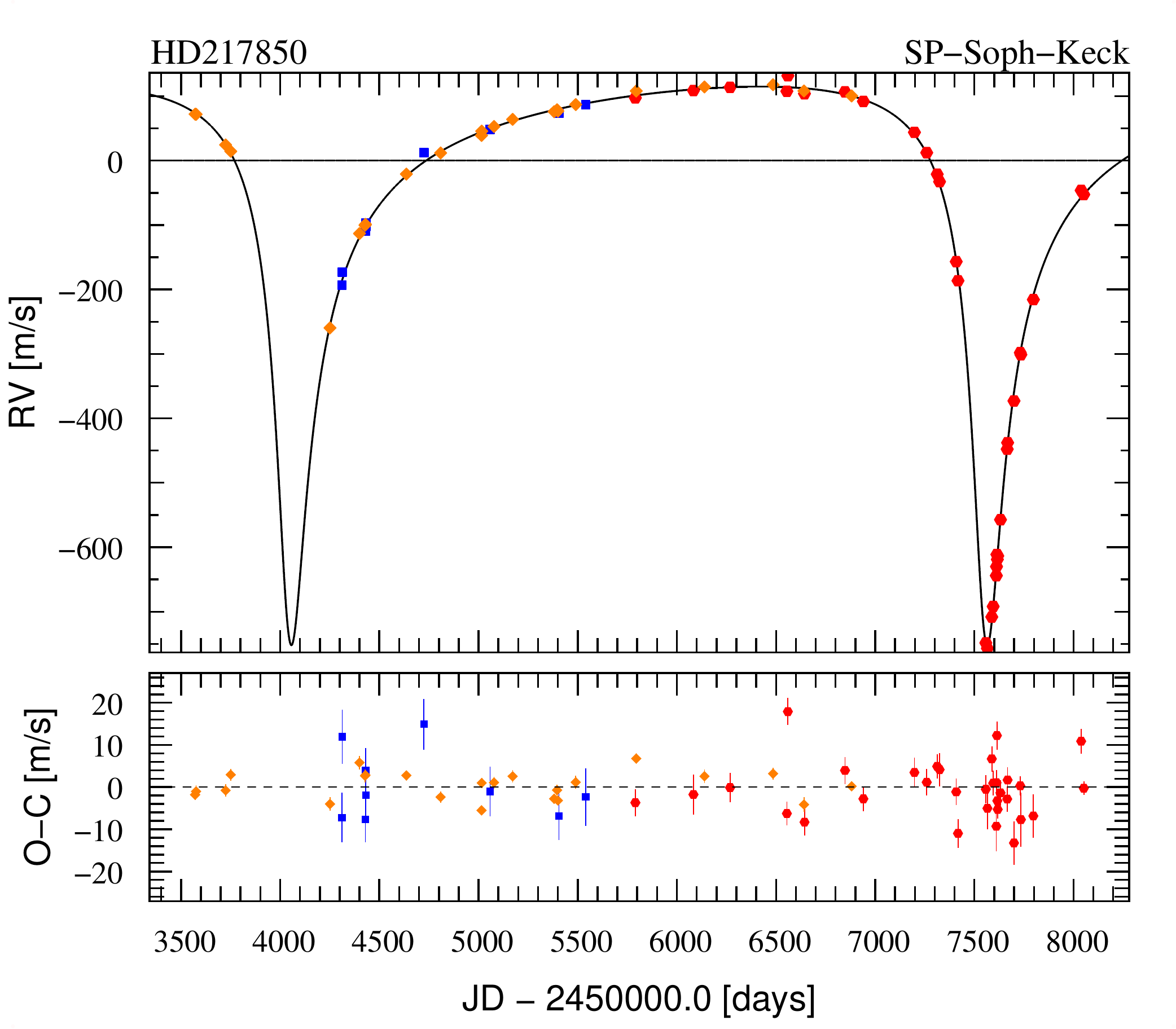} 
\includegraphics[height=57mm, clip=true, trim=0  25 0 0]{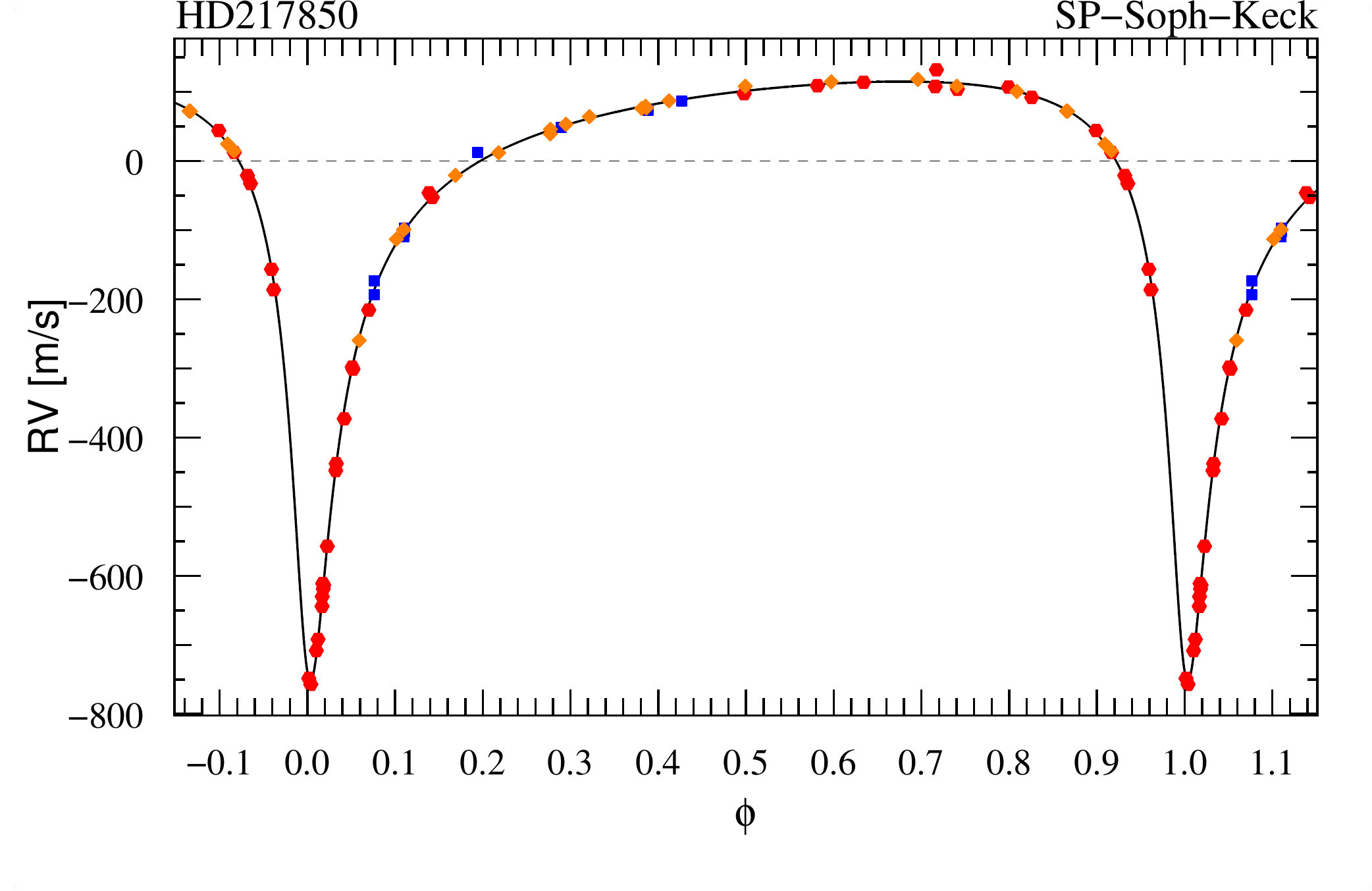}  \\
\includegraphics[height=58mm, clip=true, trim=0 -12 0 7]{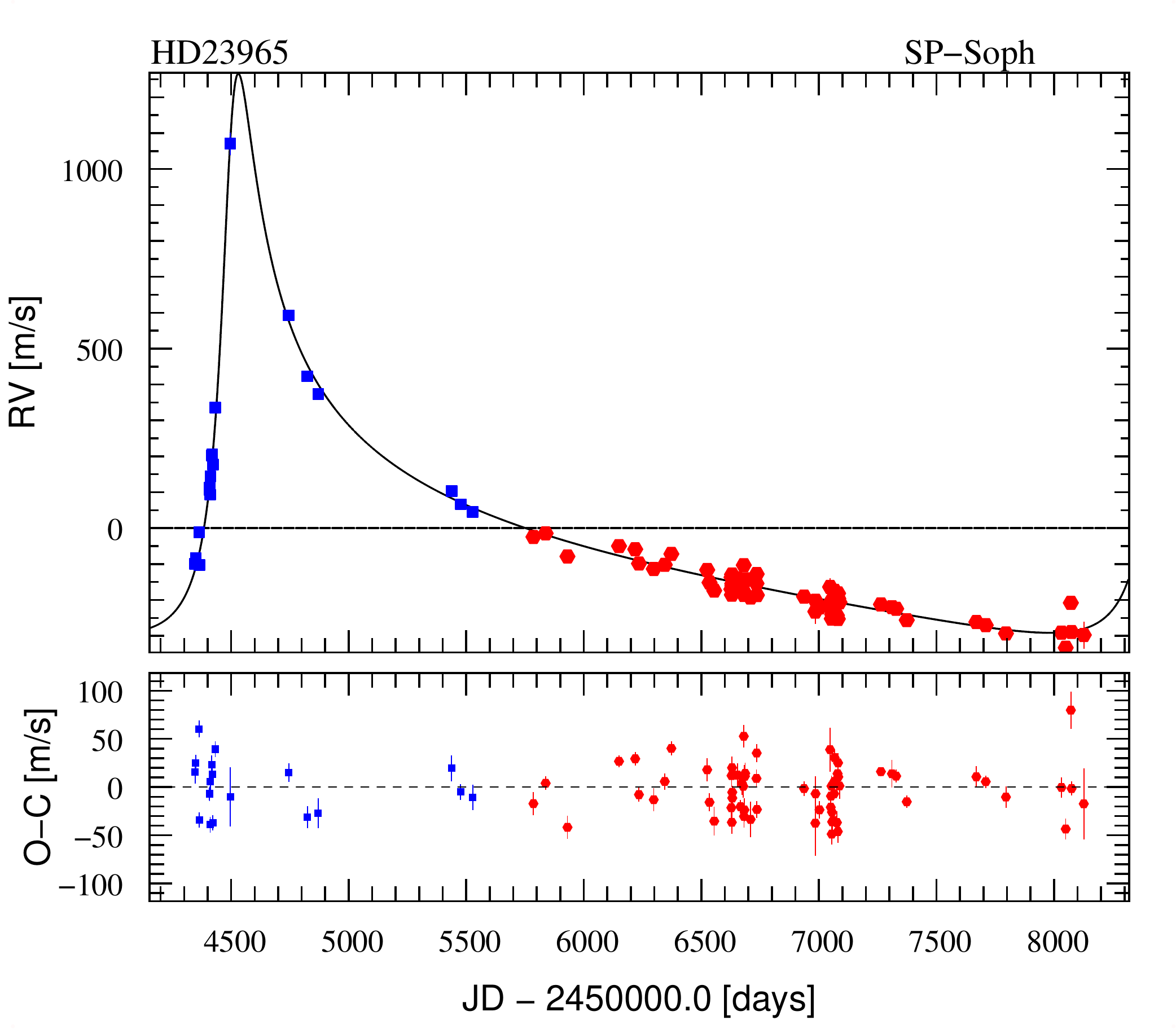} 
\includegraphics[height=57mm, clip=true, trim=0  25 0 0]{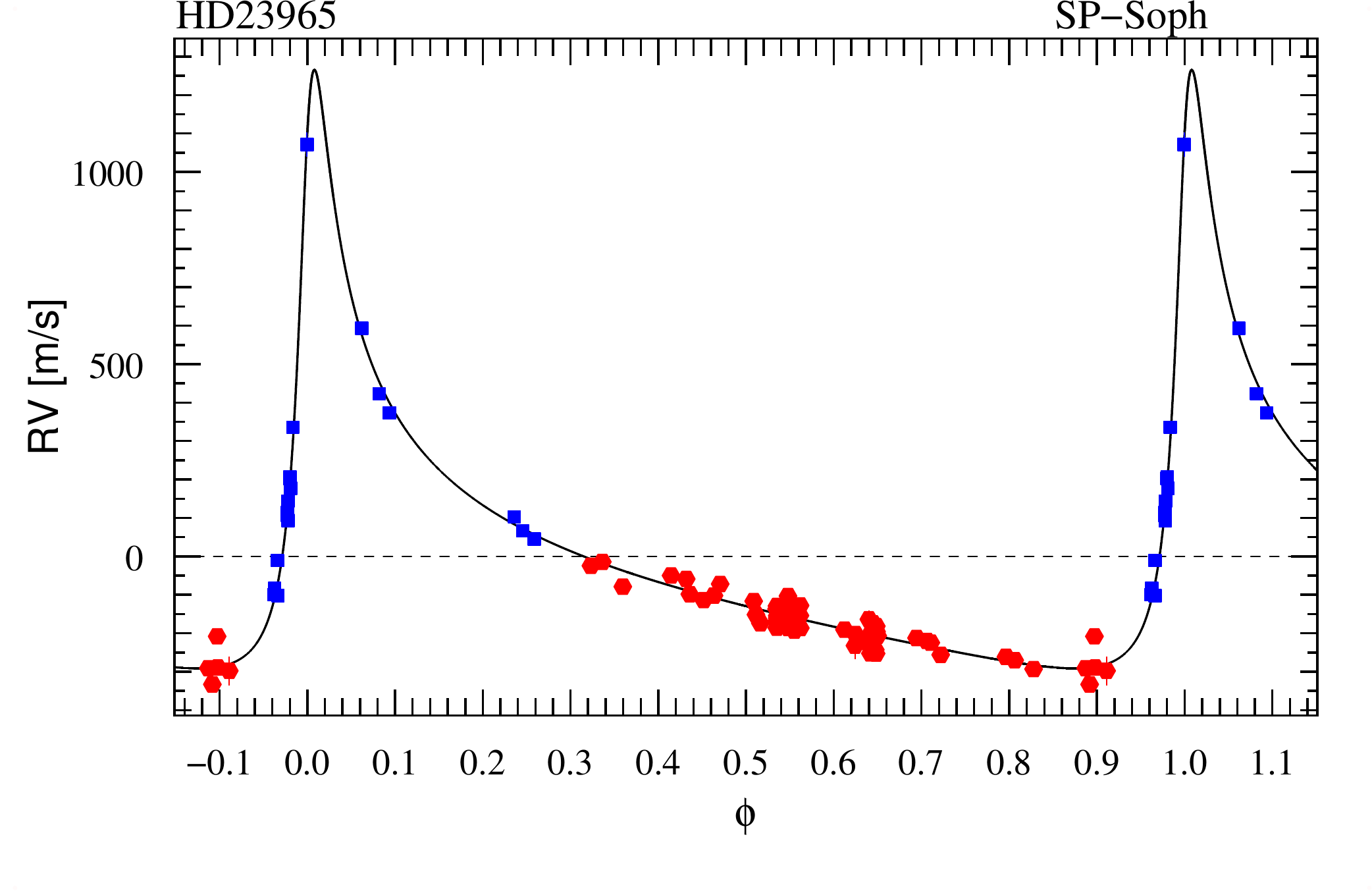}  \\
\includegraphics[height=58mm, clip=true, trim=0 -12 0 7]{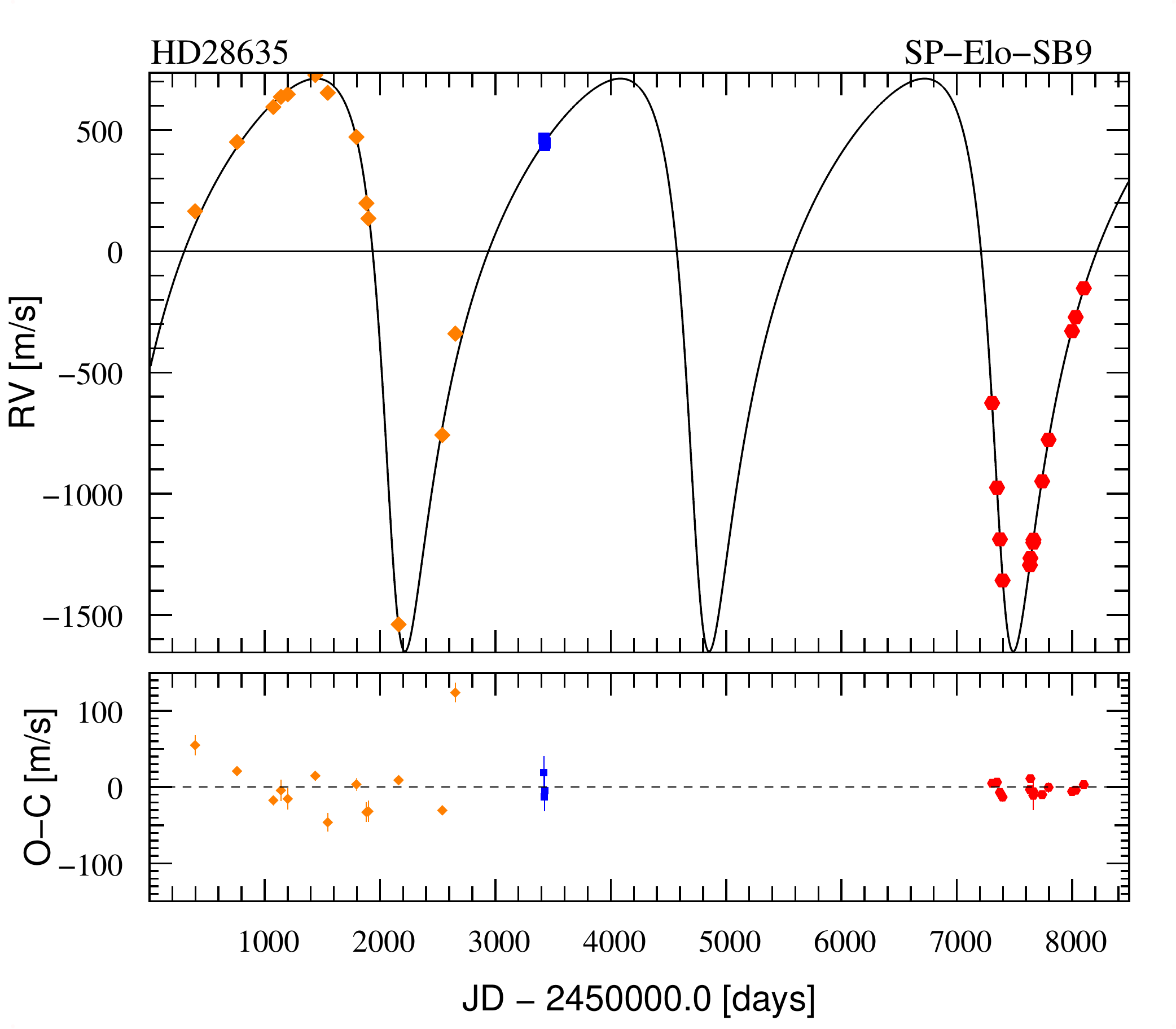} 
\includegraphics[height=57mm, clip=true, trim=0  25 0 0]{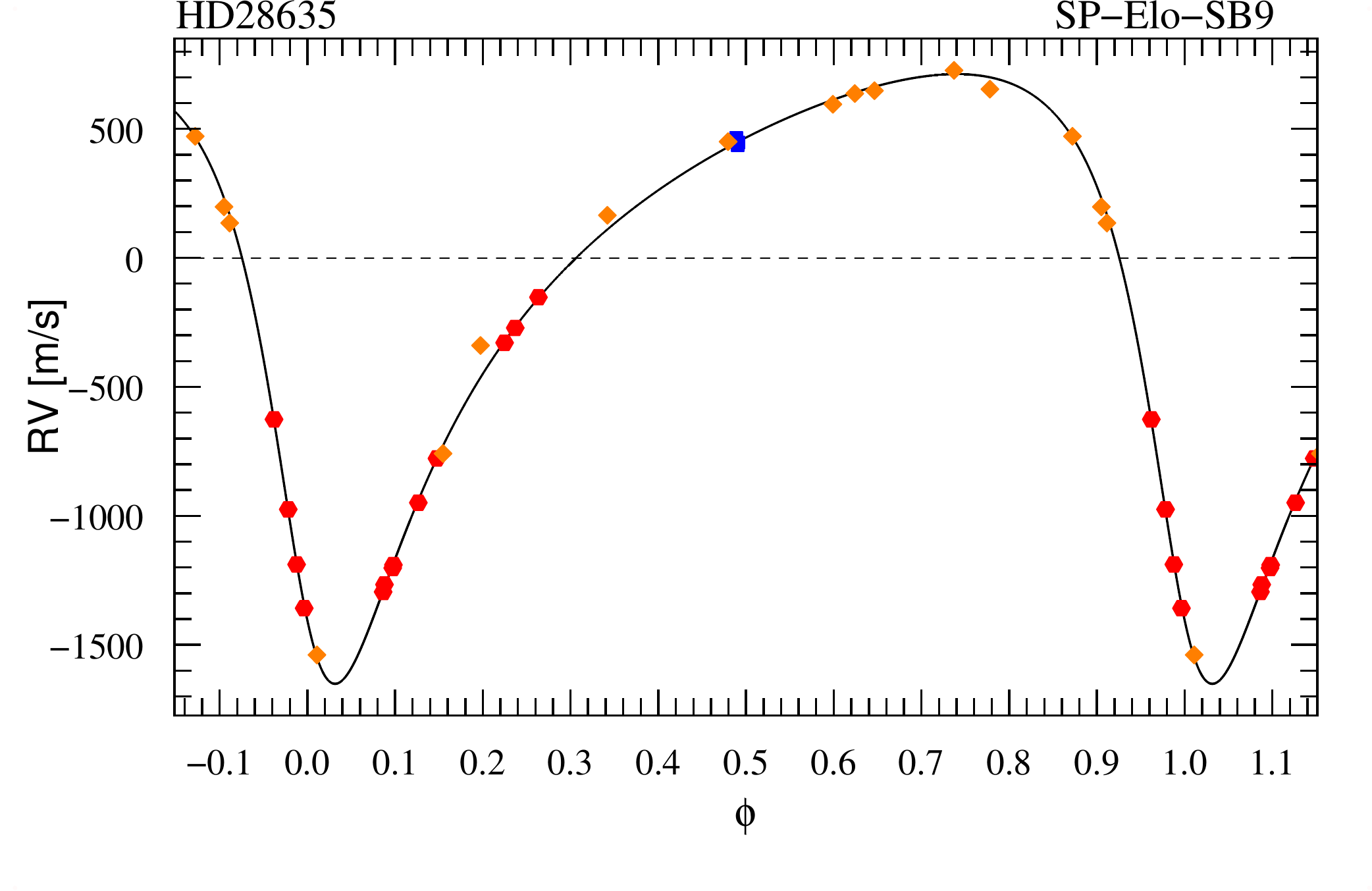} \\
\includegraphics[height=58mm, clip=true, trim=0 -12 0 7]{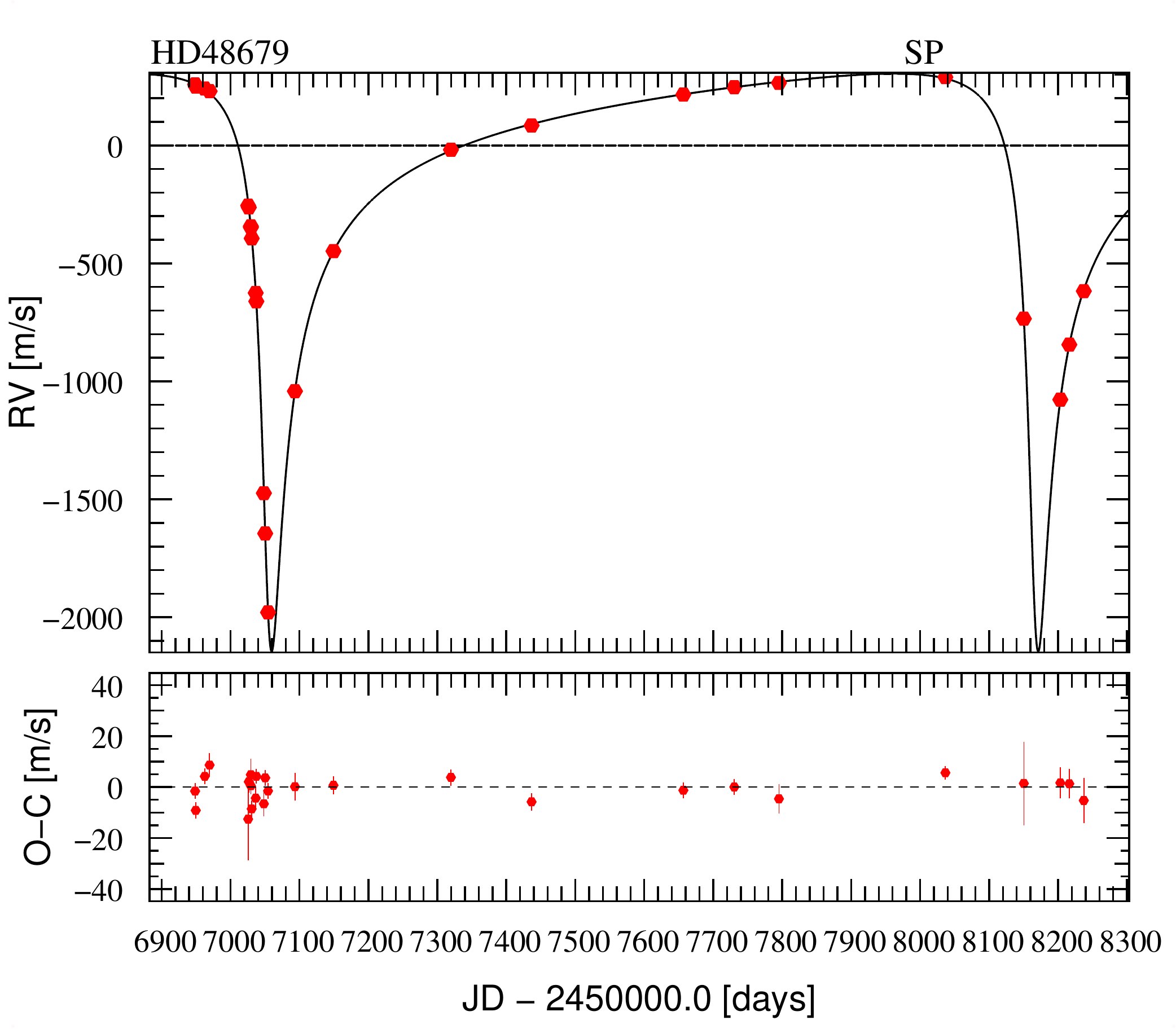} 
\includegraphics[height=57mm, clip=true, trim=0  25 0 0]{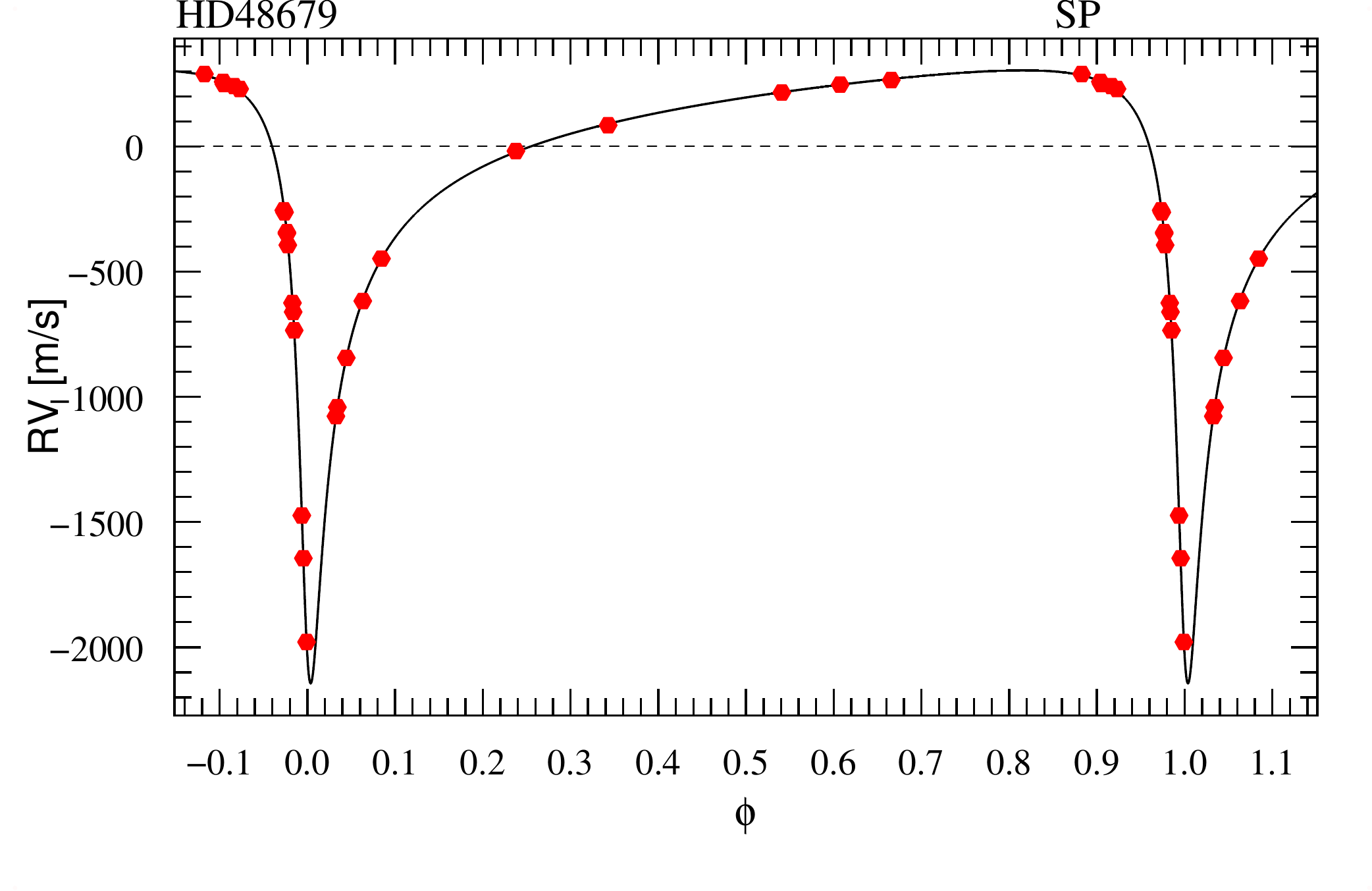} \\
\includegraphics[height=58mm, clip=true, trim=0 -12 0 7]{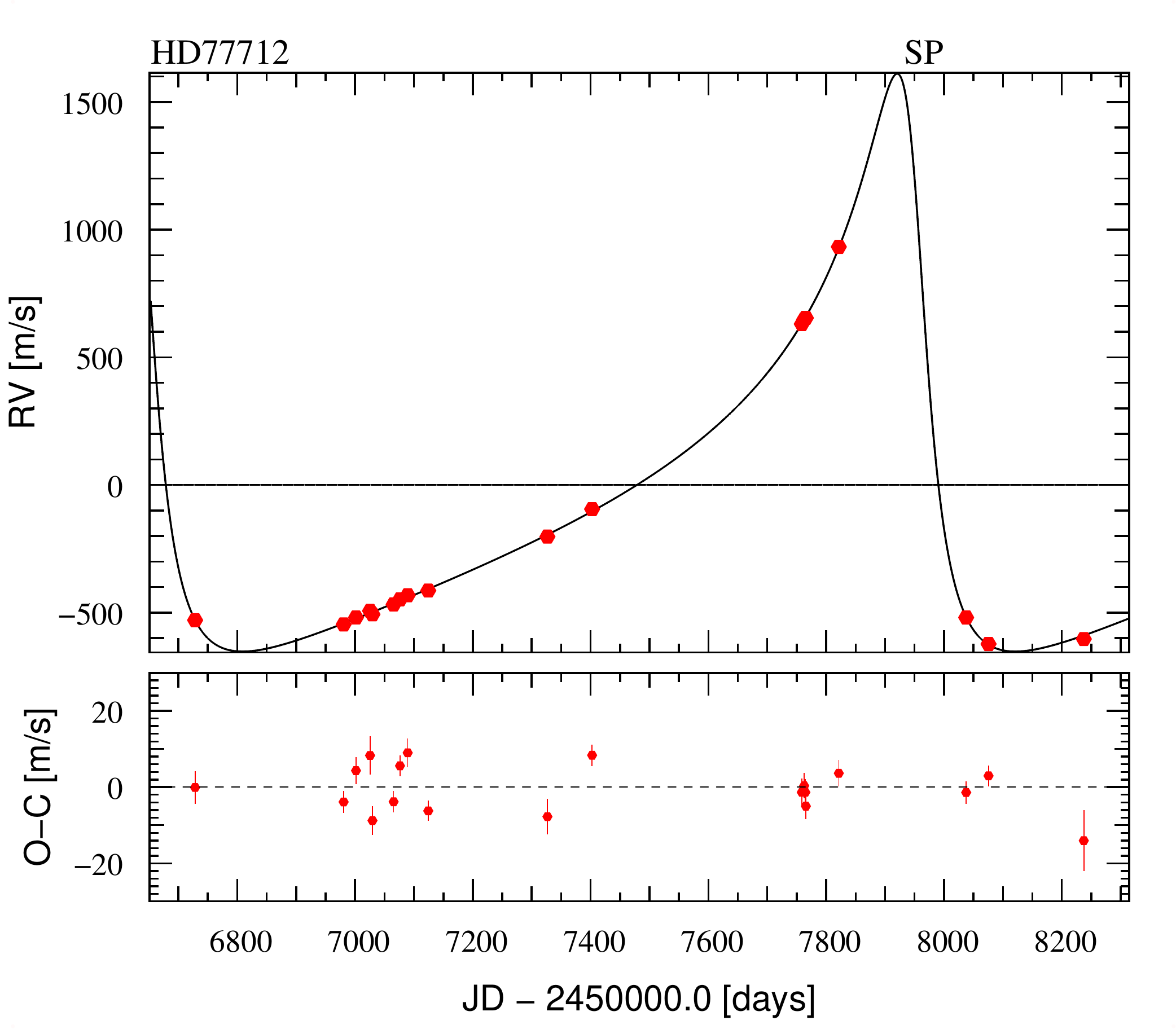} 
\includegraphics[height=57mm, clip=true, trim=0  25 0 0]{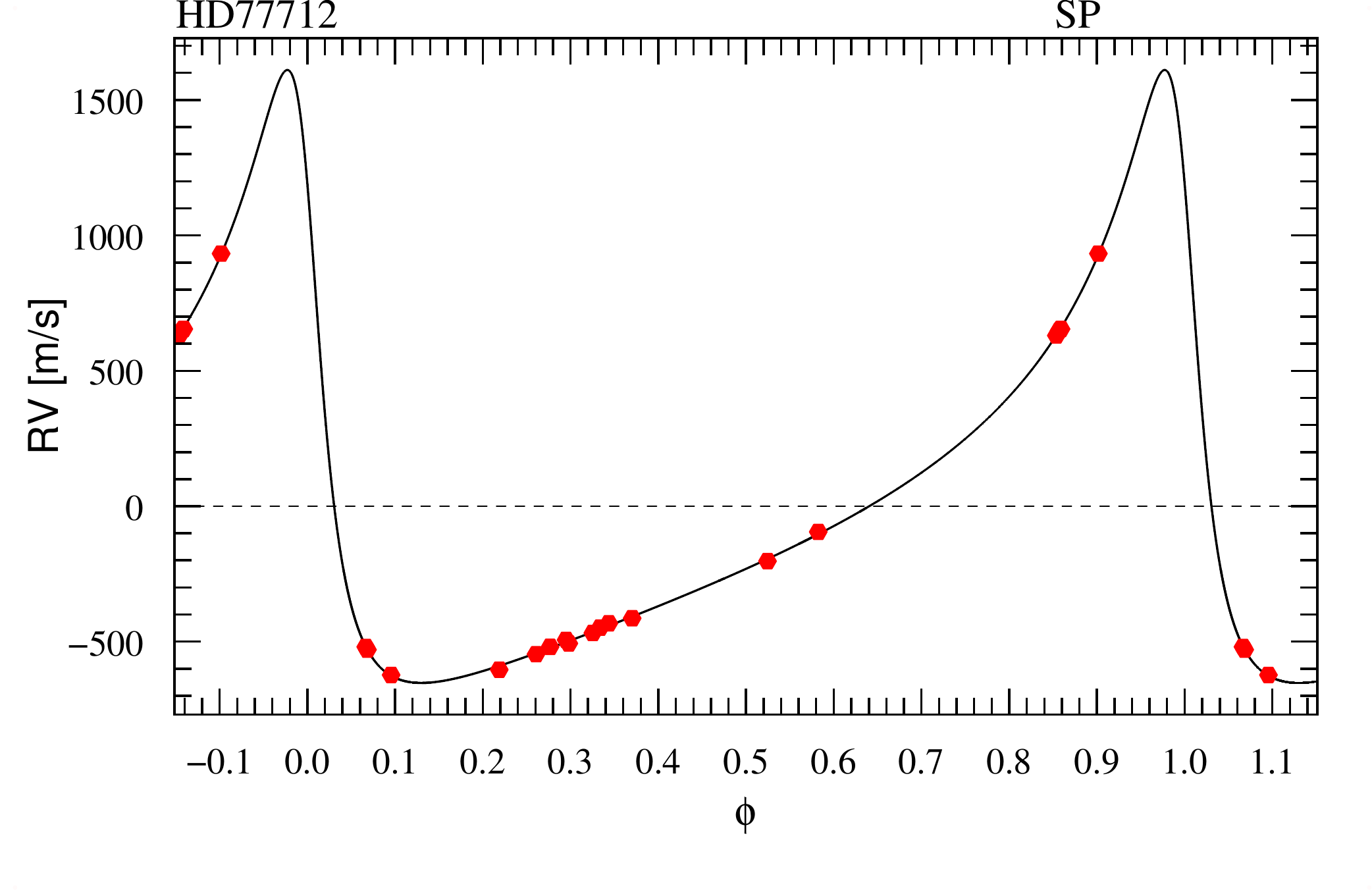} \\
\includegraphics[height=58mm, clip=true, trim=0 -12 0 7]{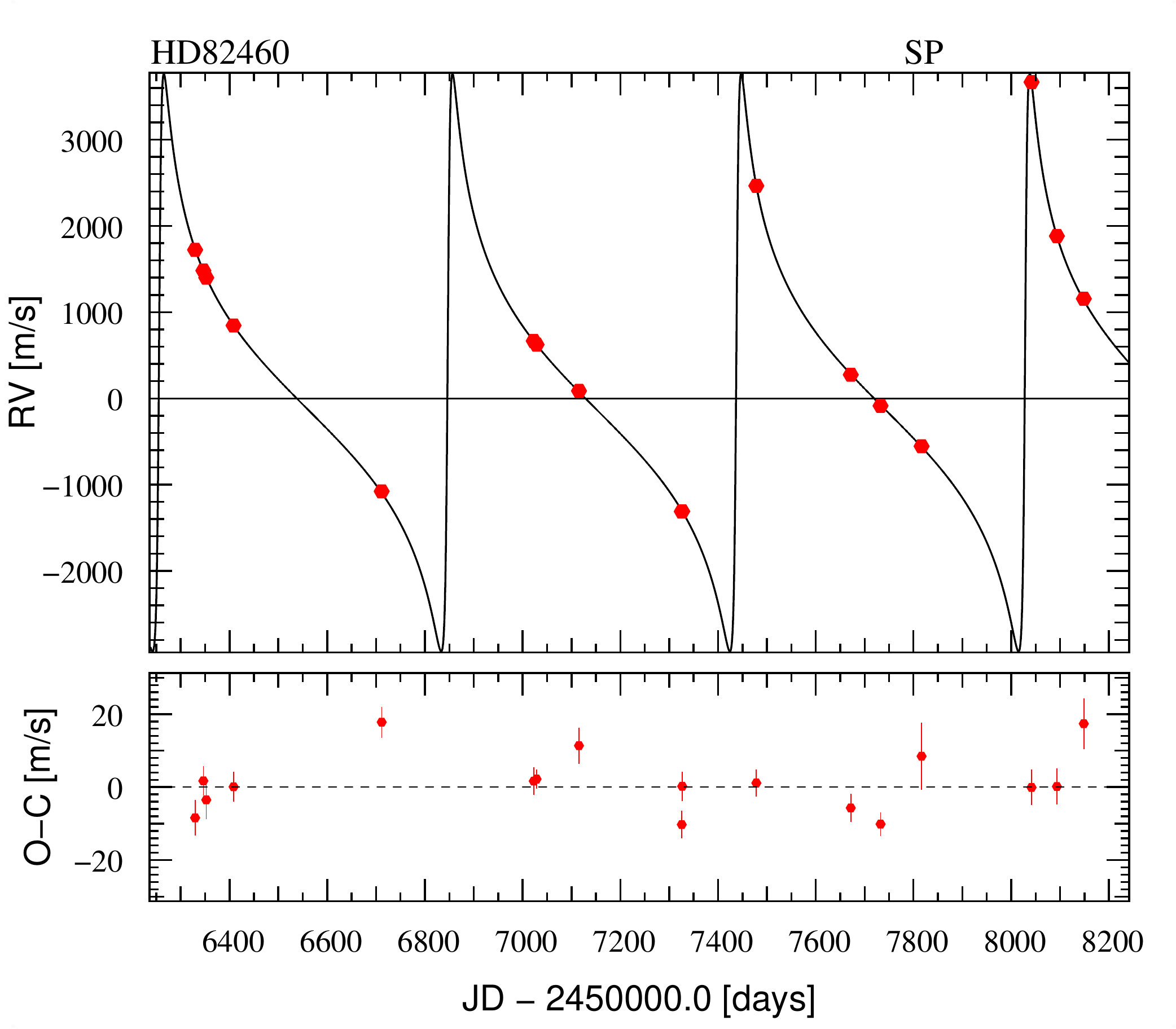}
\includegraphics[height=57mm, clip=true, trim=0  25 0 0]{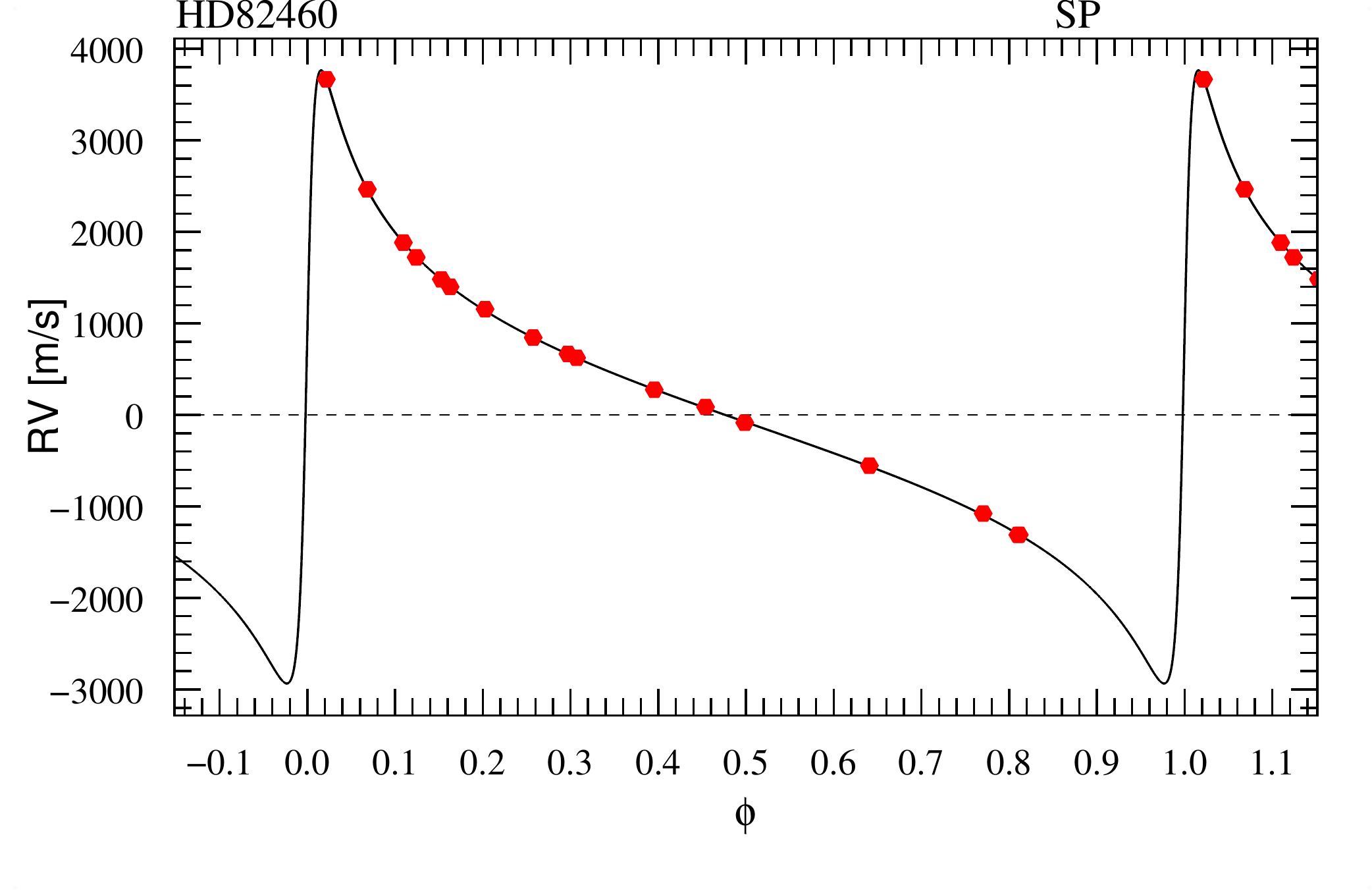} \\
\end{longfigure}

\begin{longfigure}{c}
\caption{\label{fig:solutions_triple} Drift+Keplerian orbital solutions of the radial velocity variations of the 3 triple systems. RV vs time are presented
on the left panel, with O-C residuals below, and RV vs phase for the Keplerian solution on the right panel. 'SPEloFix' means that the SOPHIE+ and Elodie datasets 
offset is fixed to 0\,m\,s$^{-1}$. 'SPSophFix' means that the SOPHIE+ and SOPHIE datasets offset is fixed to 0\,m\,s$^{-1}$. Details are given in Section~\ref{sec:multi}.} 
\endLFfirsthead
\caption{Continued.}
\endLFhead
\includegraphics[height=58mm, clip=true, trim=0 -12 0 7]{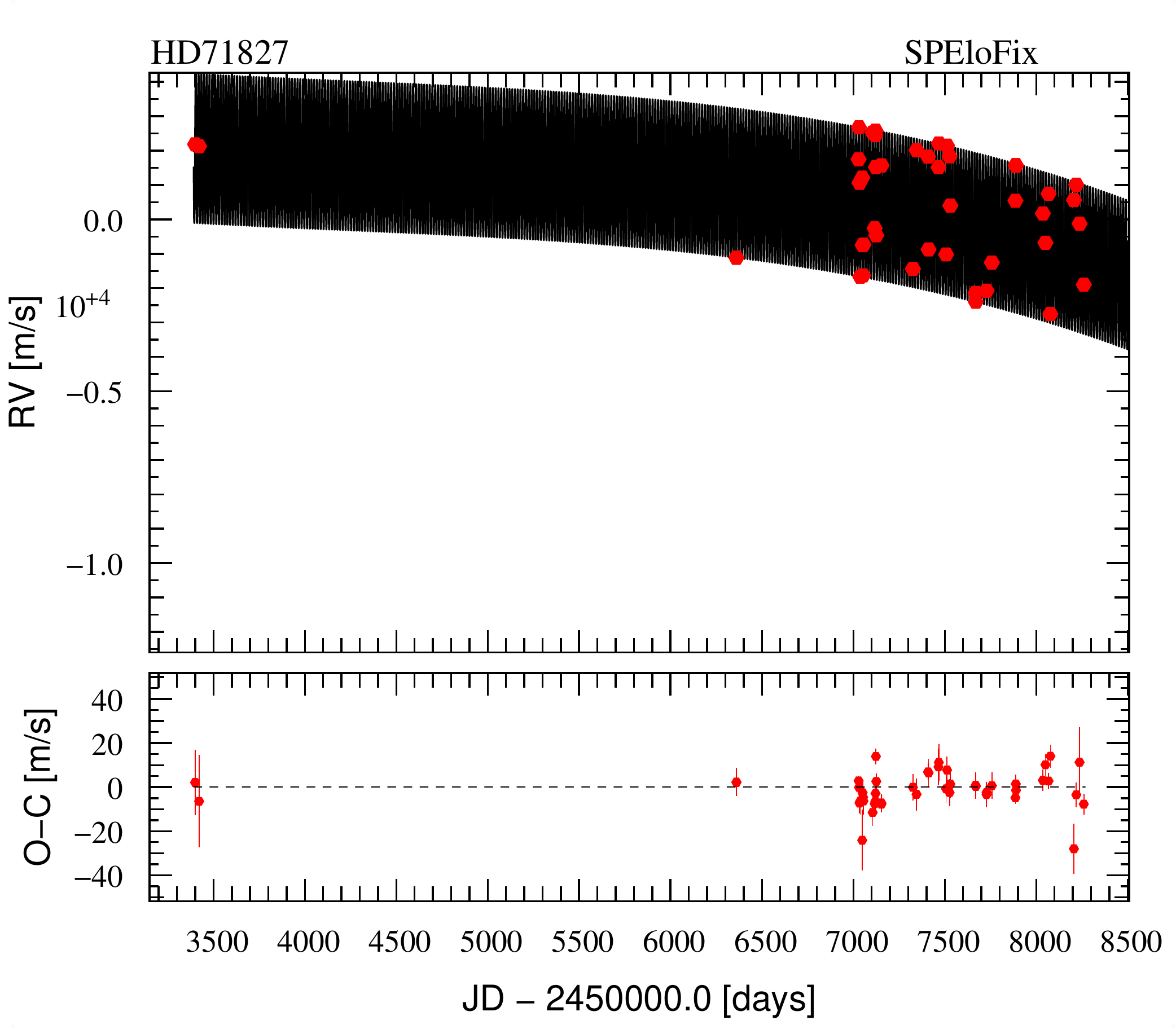}
\includegraphics[height=57mm, clip=true, trim=0  25 0 0]{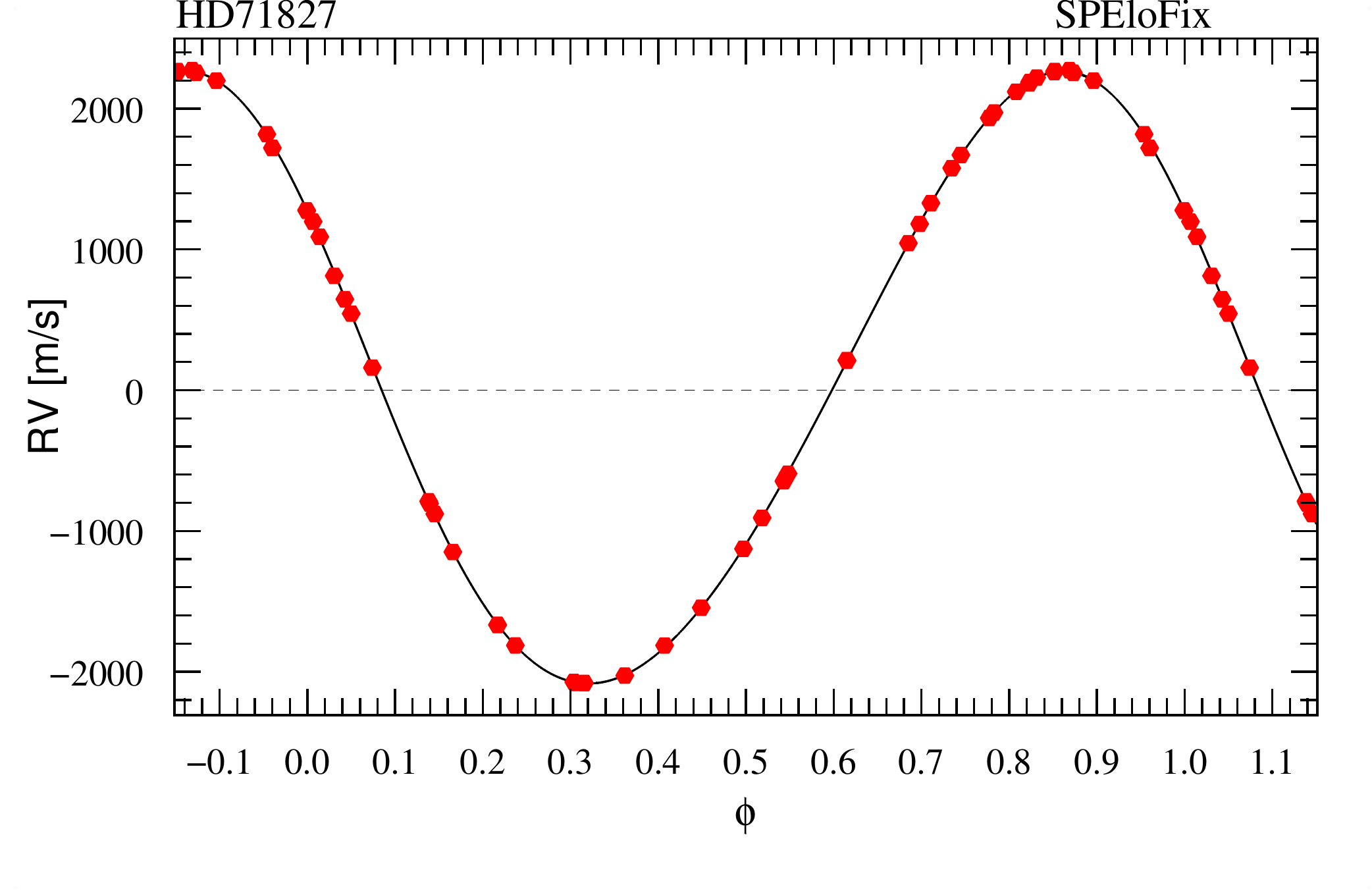} \\
\includegraphics[height=58mm, clip=true, trim=0 -12 0 7]{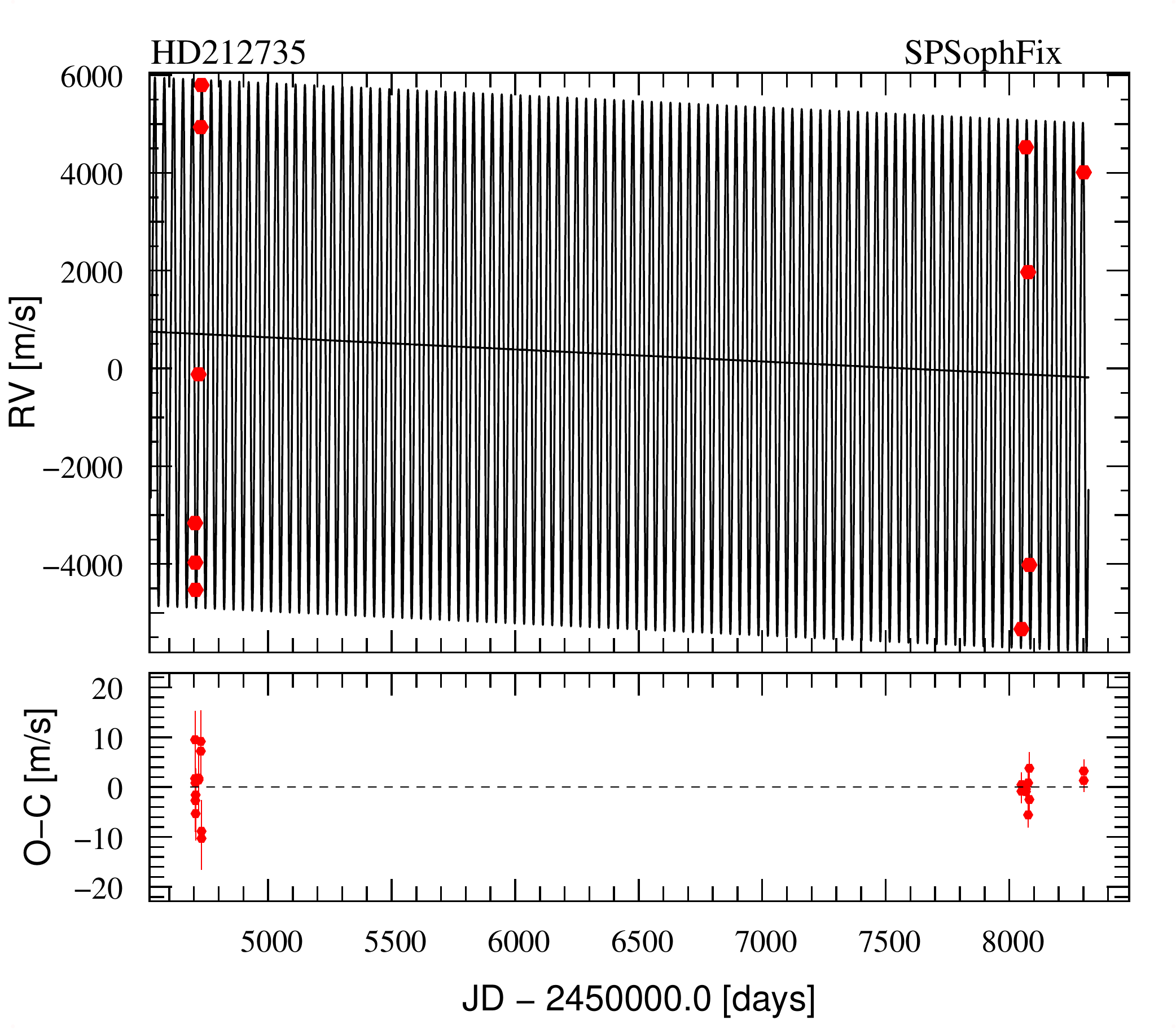} 
\includegraphics[height=57mm, clip=true, trim=0  25 0 0]{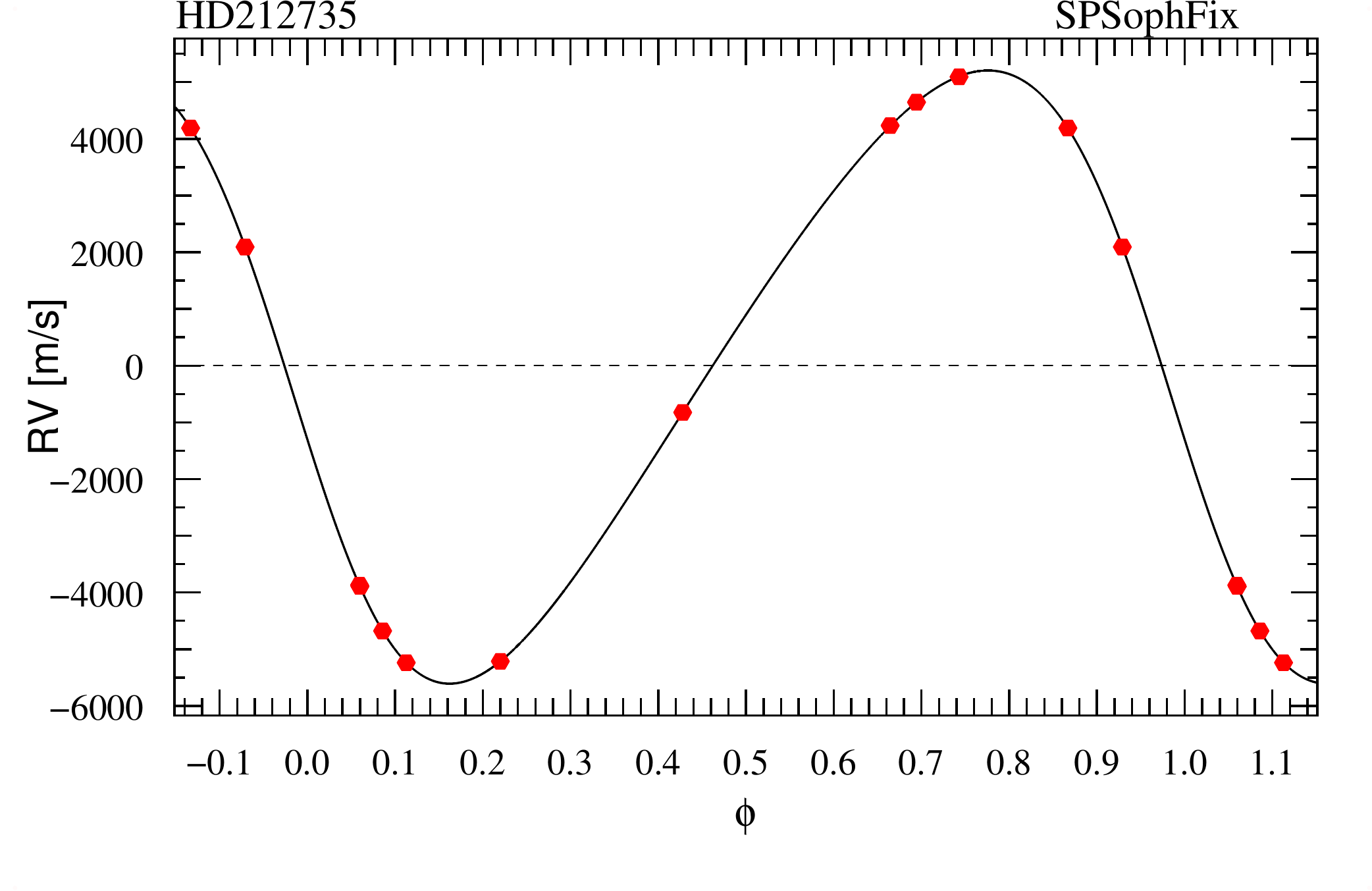} \\
\includegraphics[height=58mm, clip=true, trim=0 -12 0 7]{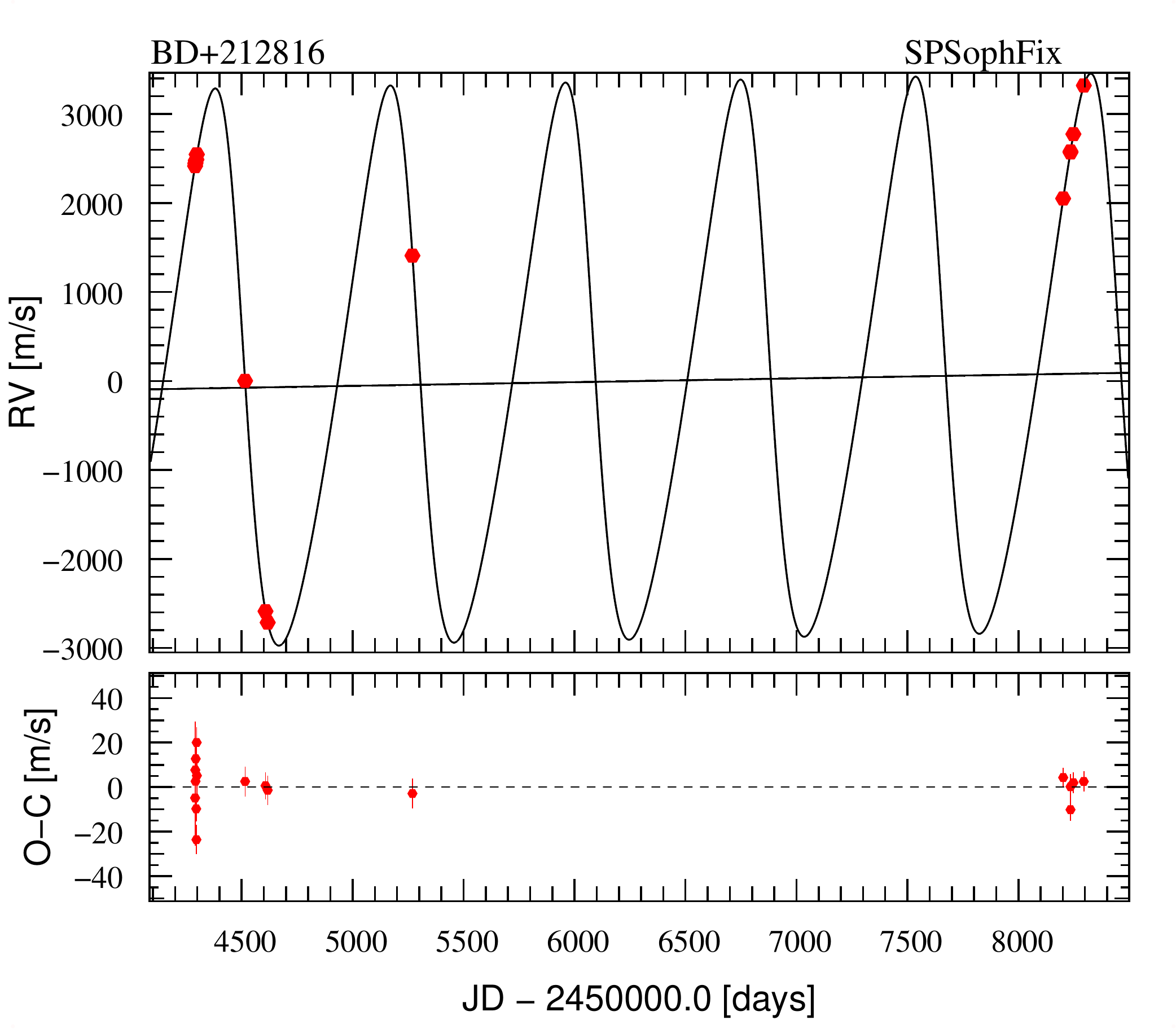} 
\includegraphics[height=57mm, clip=true, trim=0  25 0 0]{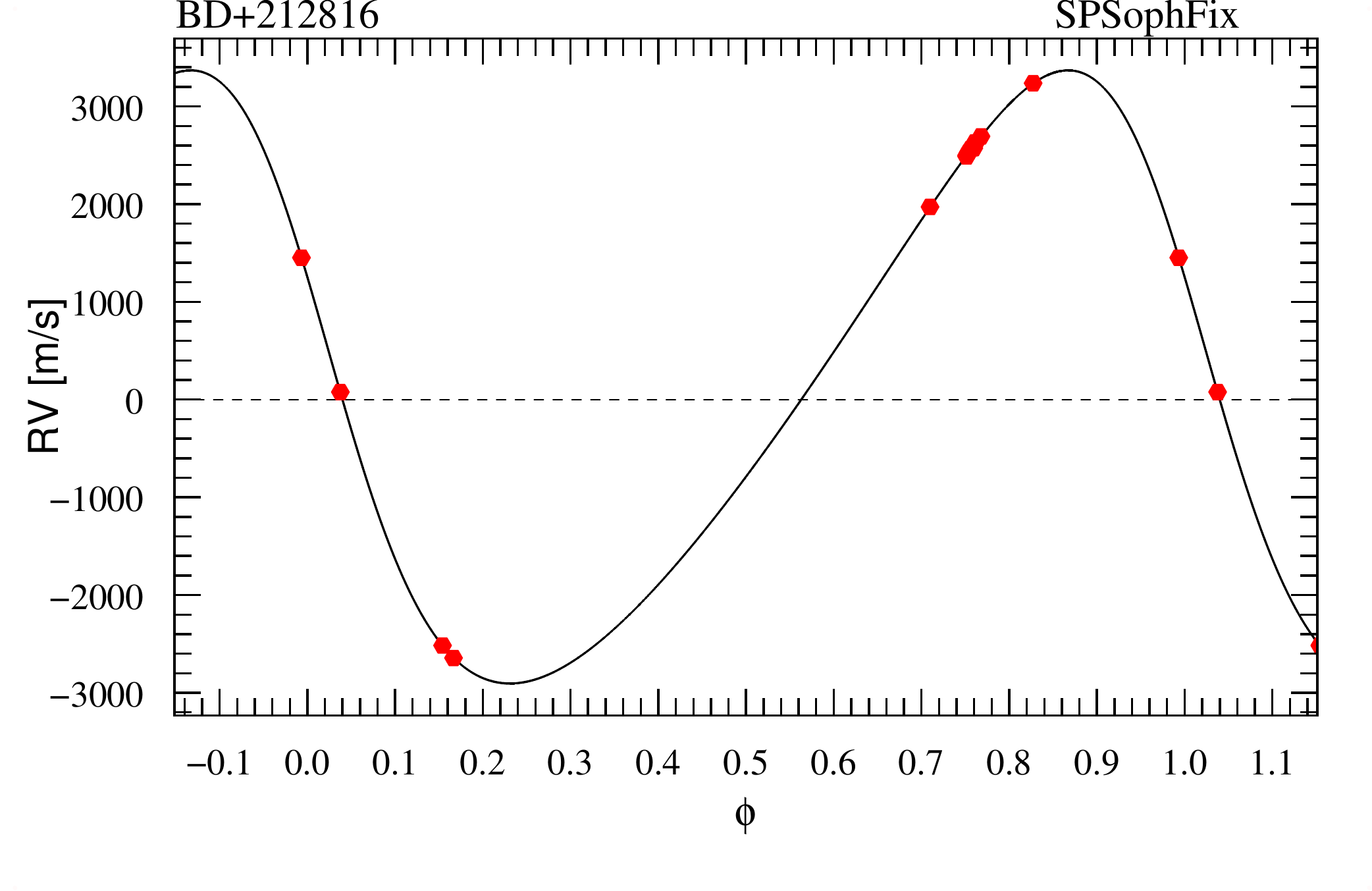} \\
\end{longfigure}

\begin{longfigure}{c}
\caption{\label{fig:solutions_dM}Orbital solutions of the radial velocity variations of the 39 stars with an M-dwarf companion. RV vs time are presented
on the left panel, with O-C residuals below, and RV vs phase on the right panel. The color code is explained in the caption of Figure~\ref{fig:solutions_BD}.
'SPSophFix' means that the SOPHIE+ and SOPHIE datasets offset is fixed to 0\,m\,s$^{-1}$.} 
\endLFfirsthead
\caption{Continued.}
\endLFhead
\includegraphics[height=58mm, clip=true, trim=0 -12 0 7]{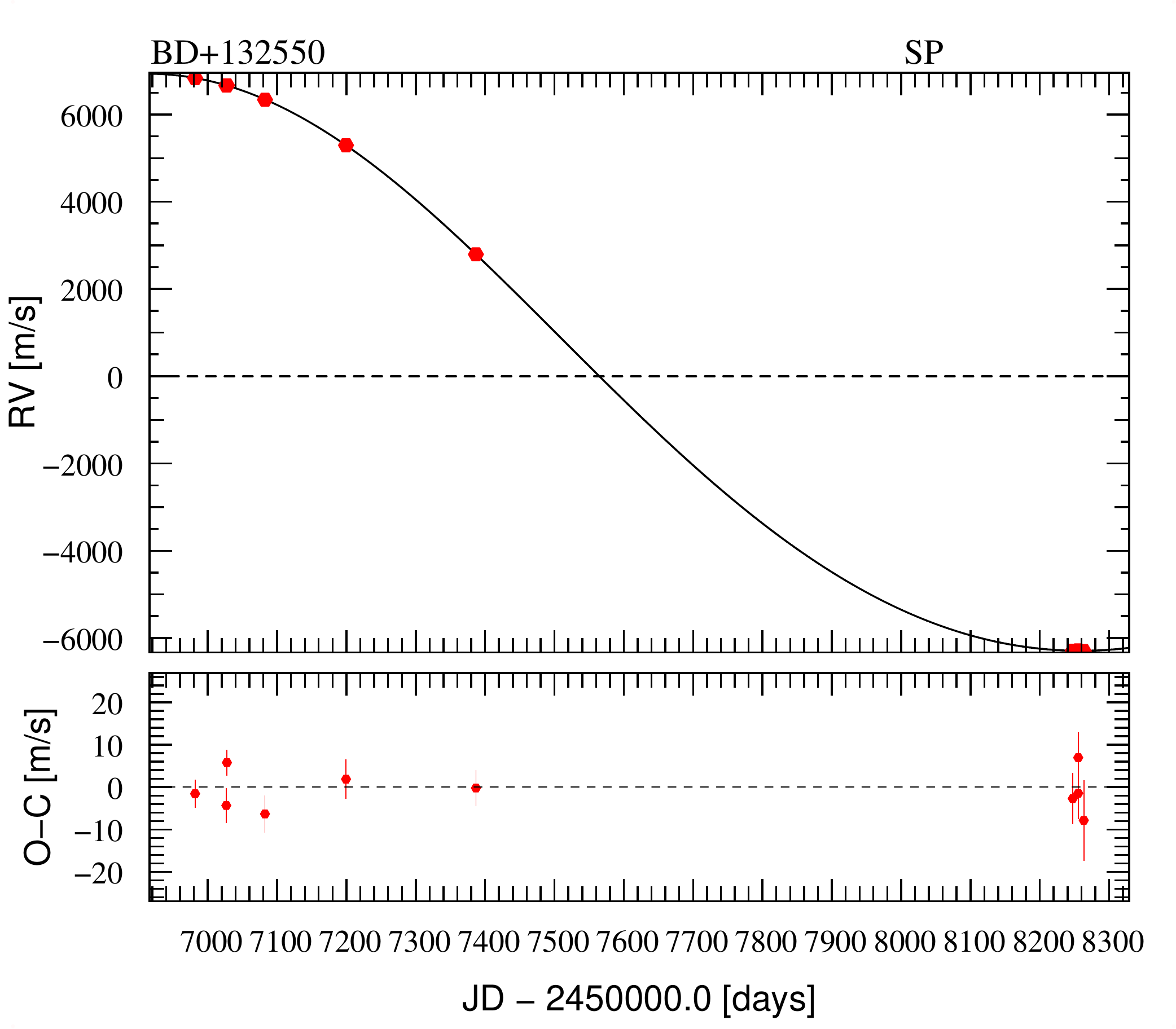}
\includegraphics[height=57mm, clip=true, trim=0  25 0 0]{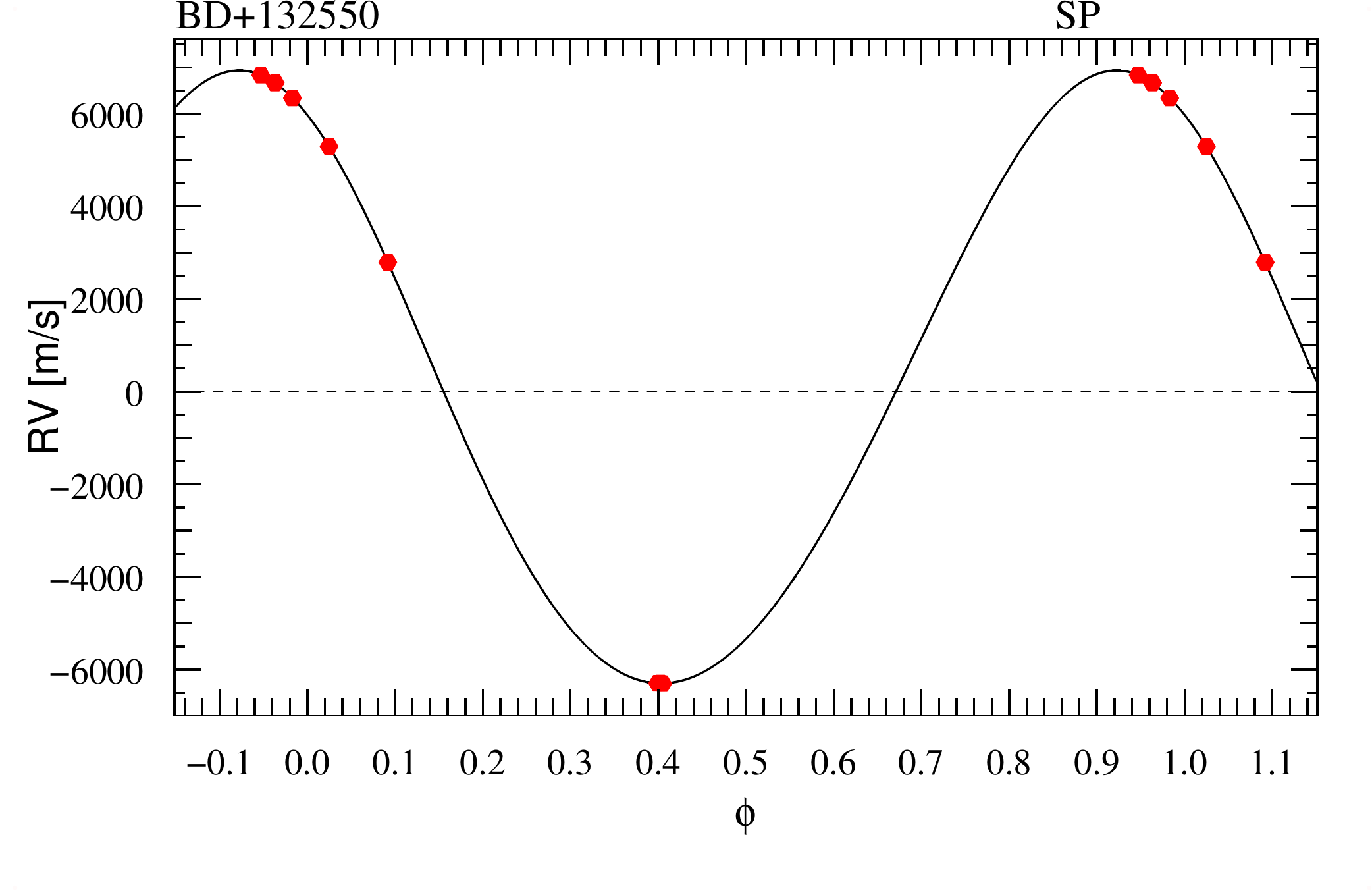}  \\
\includegraphics[height=58mm, clip=true, trim=0 -12 0 7]{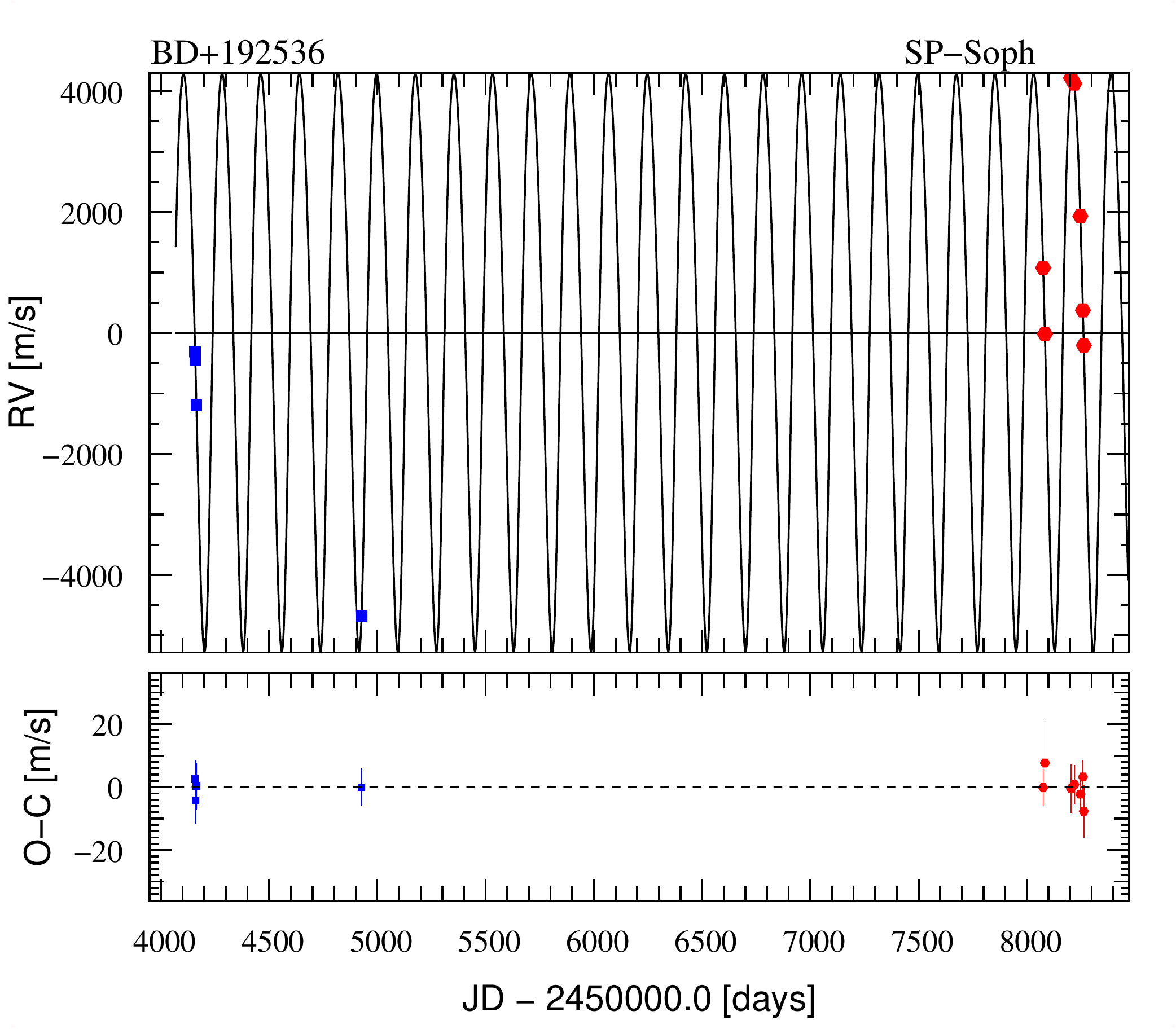}
\includegraphics[height=57mm, clip=true, trim=0  25 0 0]{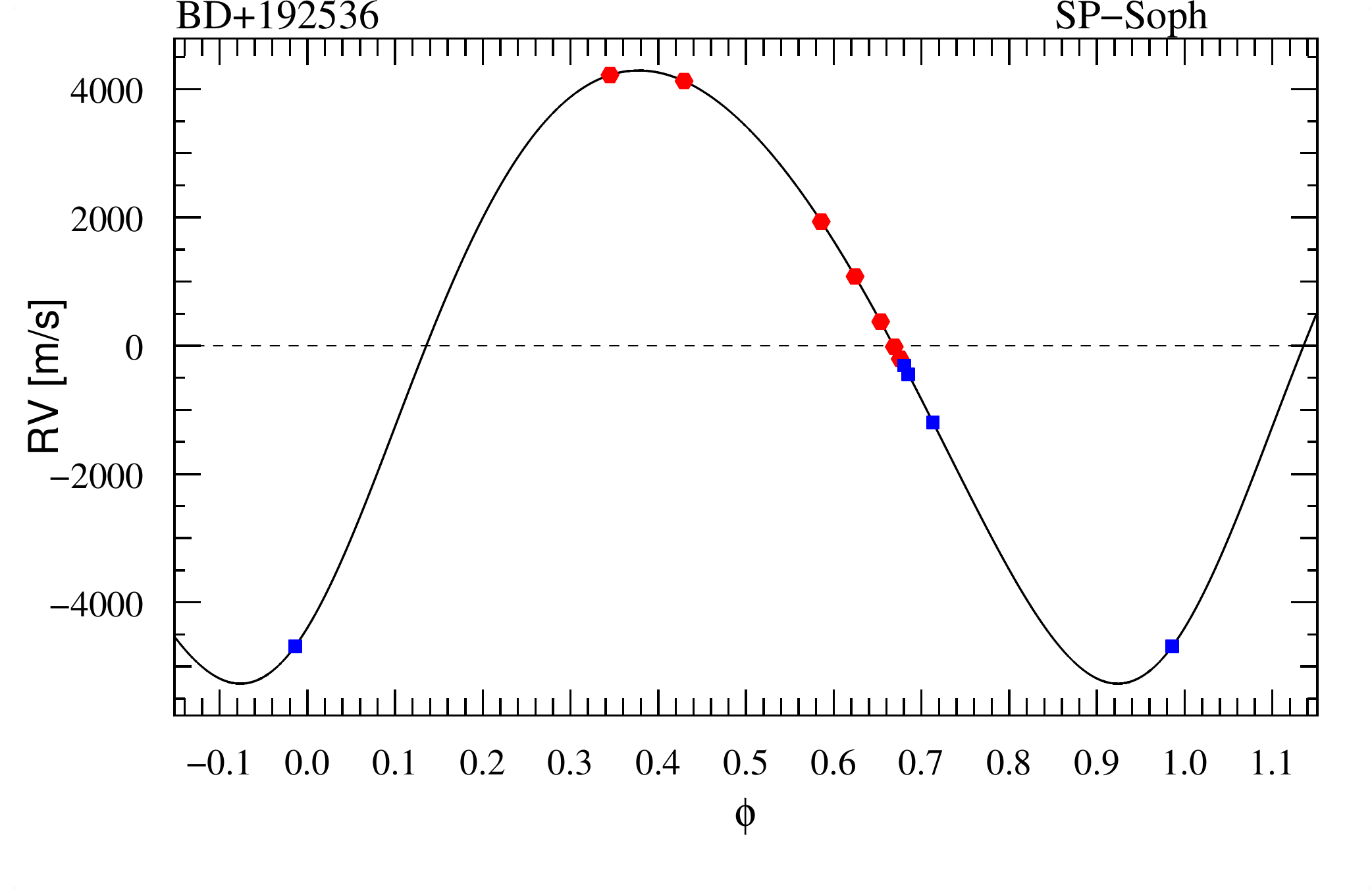} \\
\includegraphics[height=58mm, clip=true, trim=0 -12 0 7]{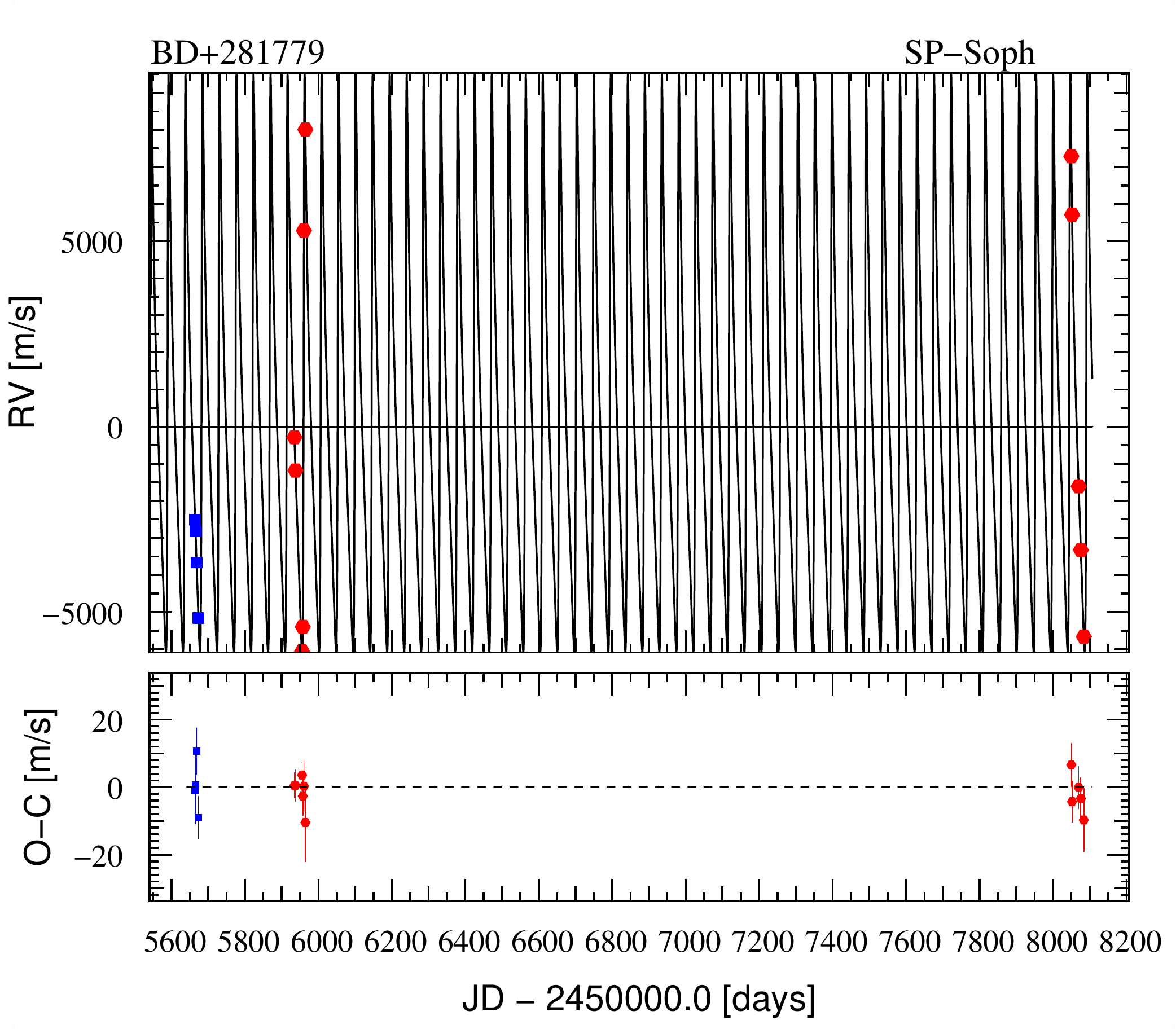} 
\includegraphics[height=57mm, clip=true, trim=0  25 0 0]{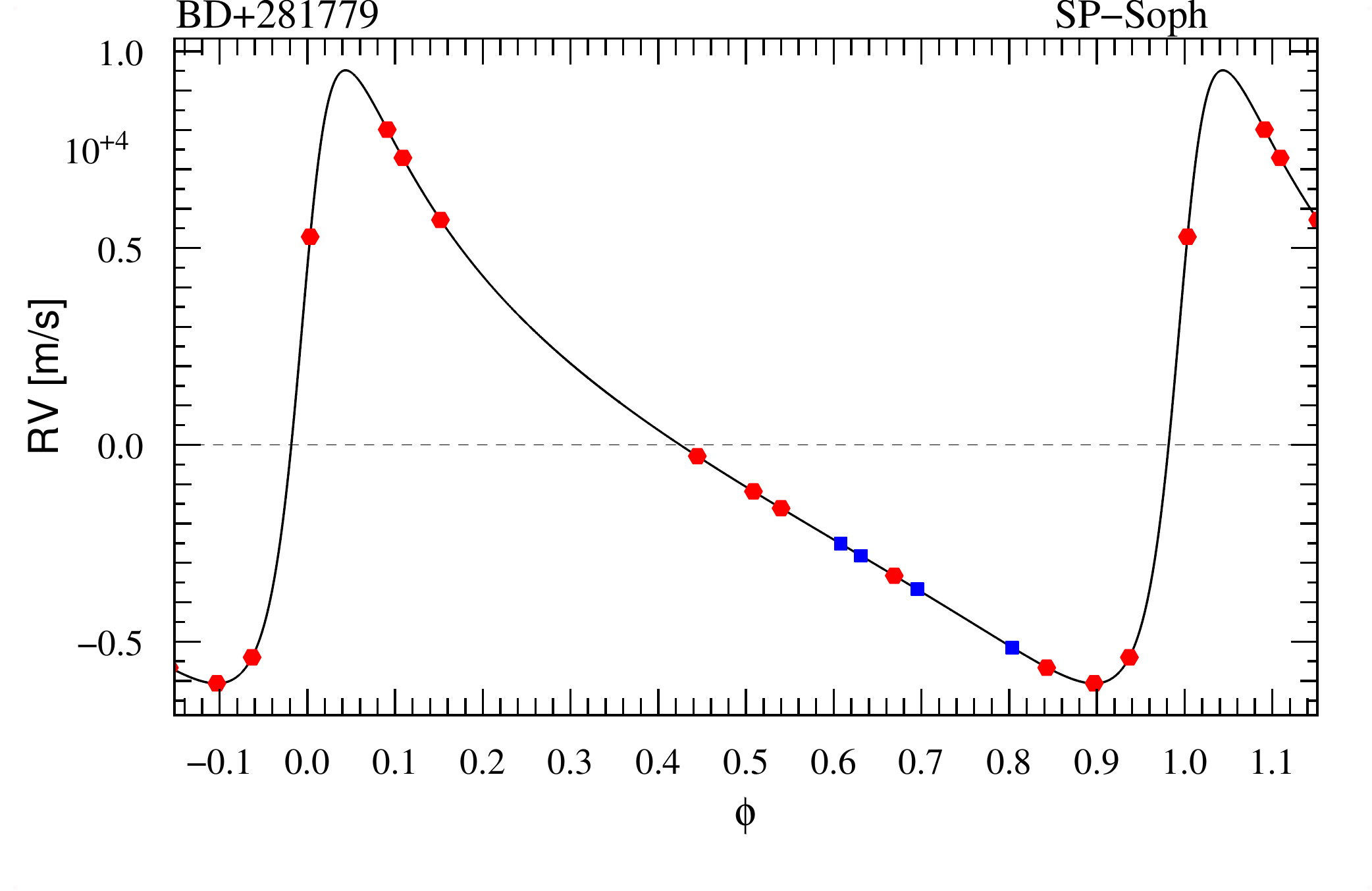} \\
\includegraphics[height=58mm, clip=true, trim=0 -12 0 7]{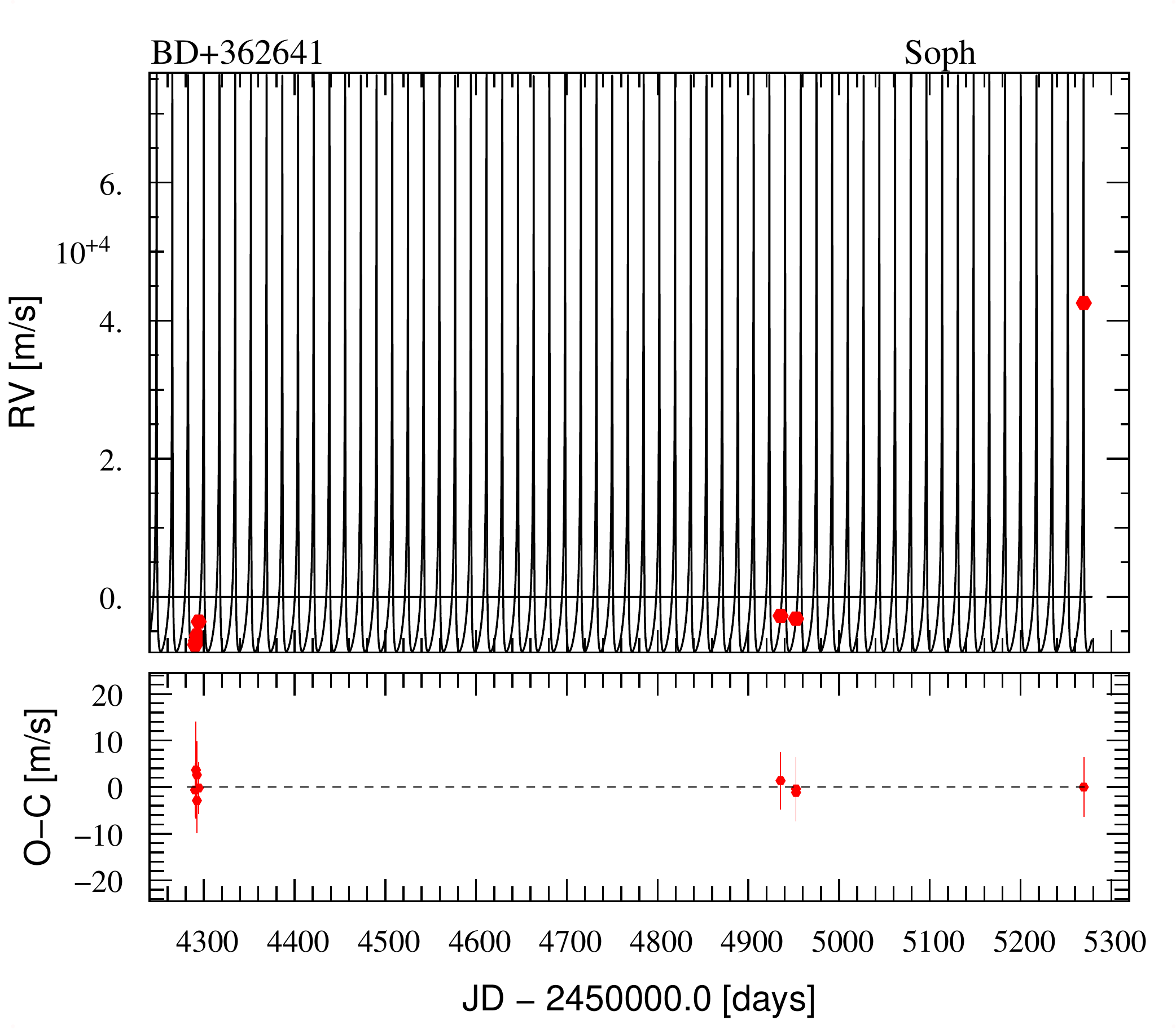}
\includegraphics[height=57mm, clip=true, trim=0  25 0 0]{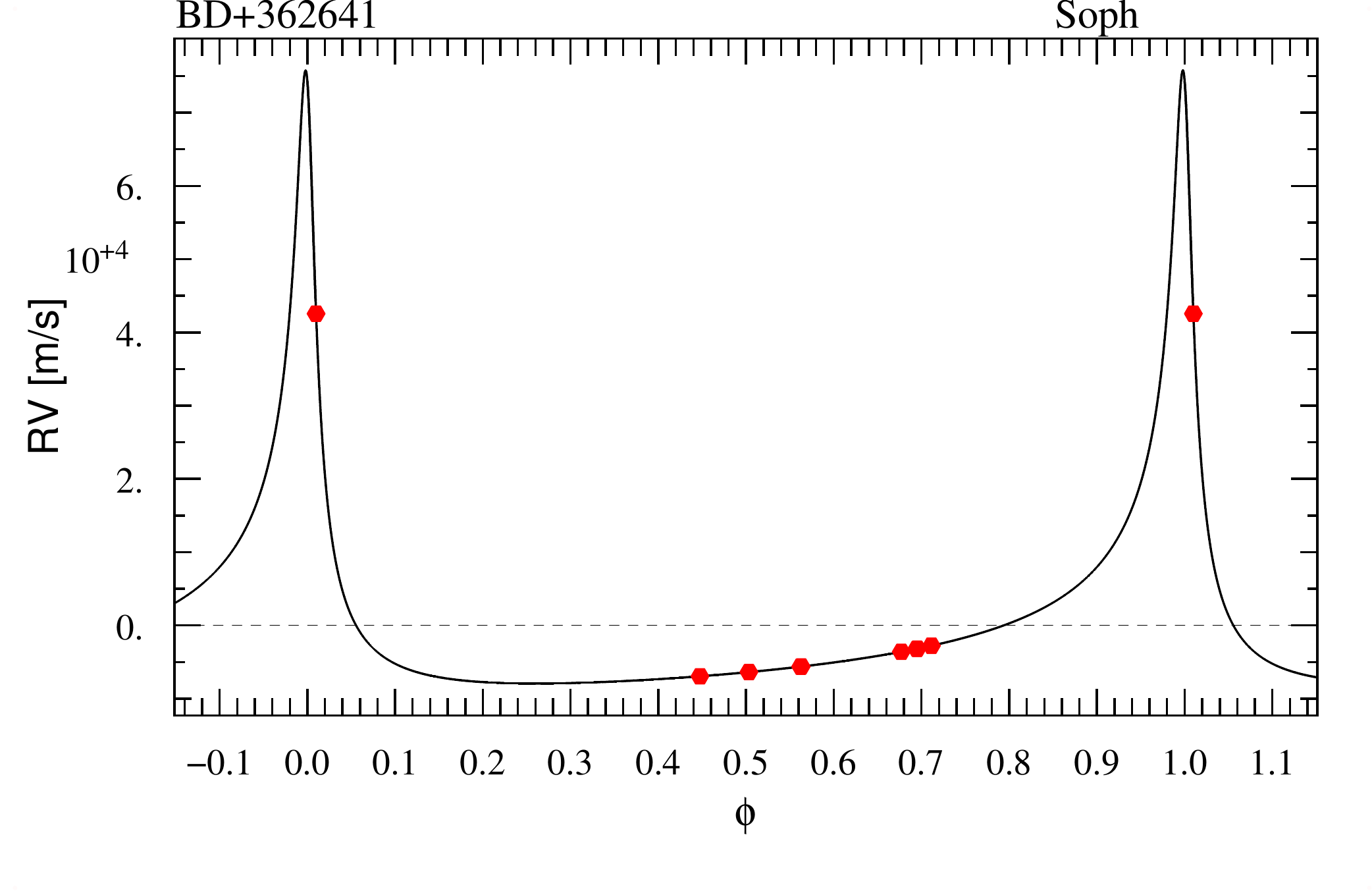} \\
\includegraphics[height=58mm, clip=true, trim=0 -12 0 7]{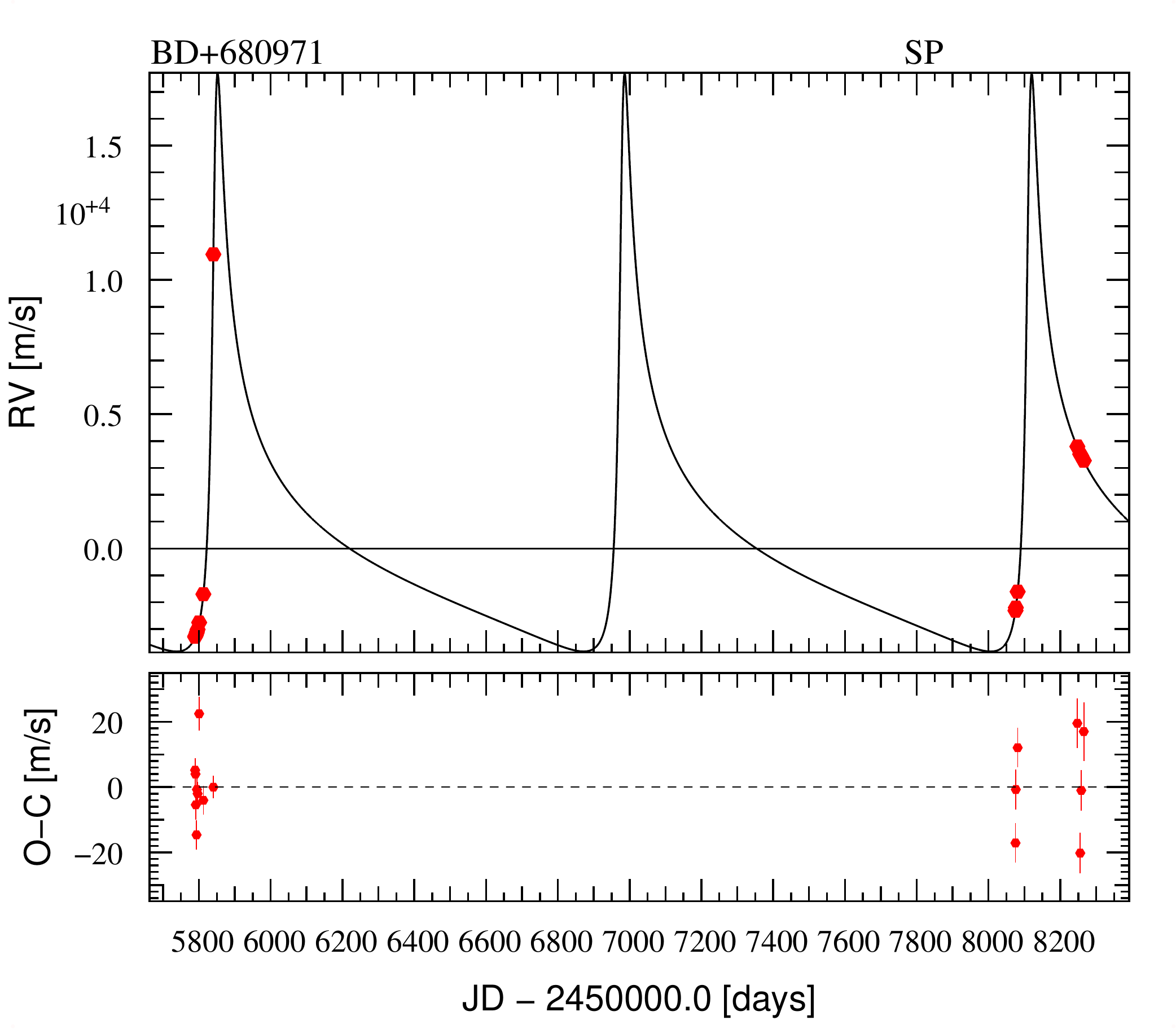}
\includegraphics[height=57mm, clip=true, trim=0  25 0 0]{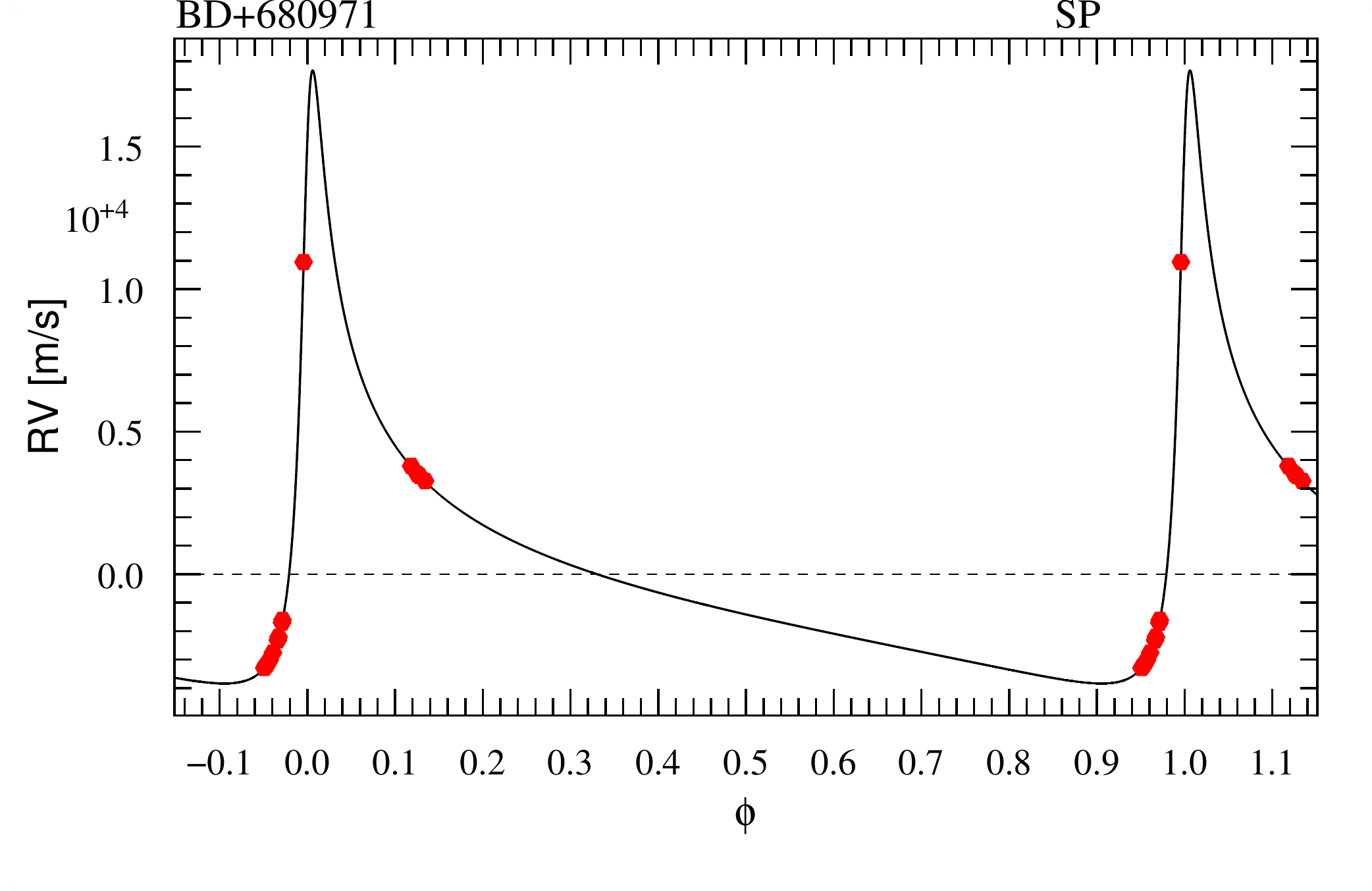} \\
\includegraphics[height=58mm, clip=true, trim=0 -12 0 7]{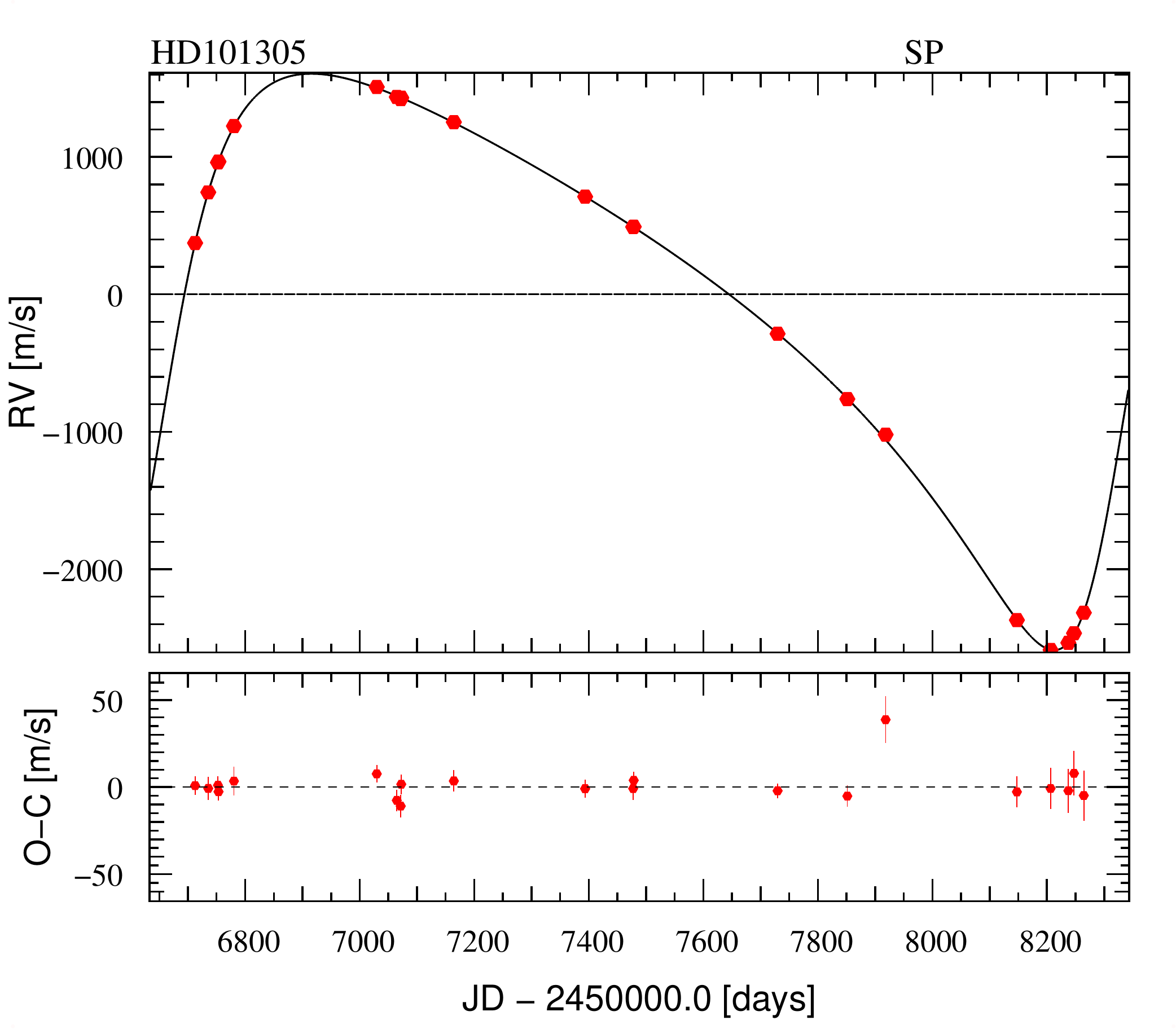}
\includegraphics[height=57mm, clip=true, trim=0  25 0 0]{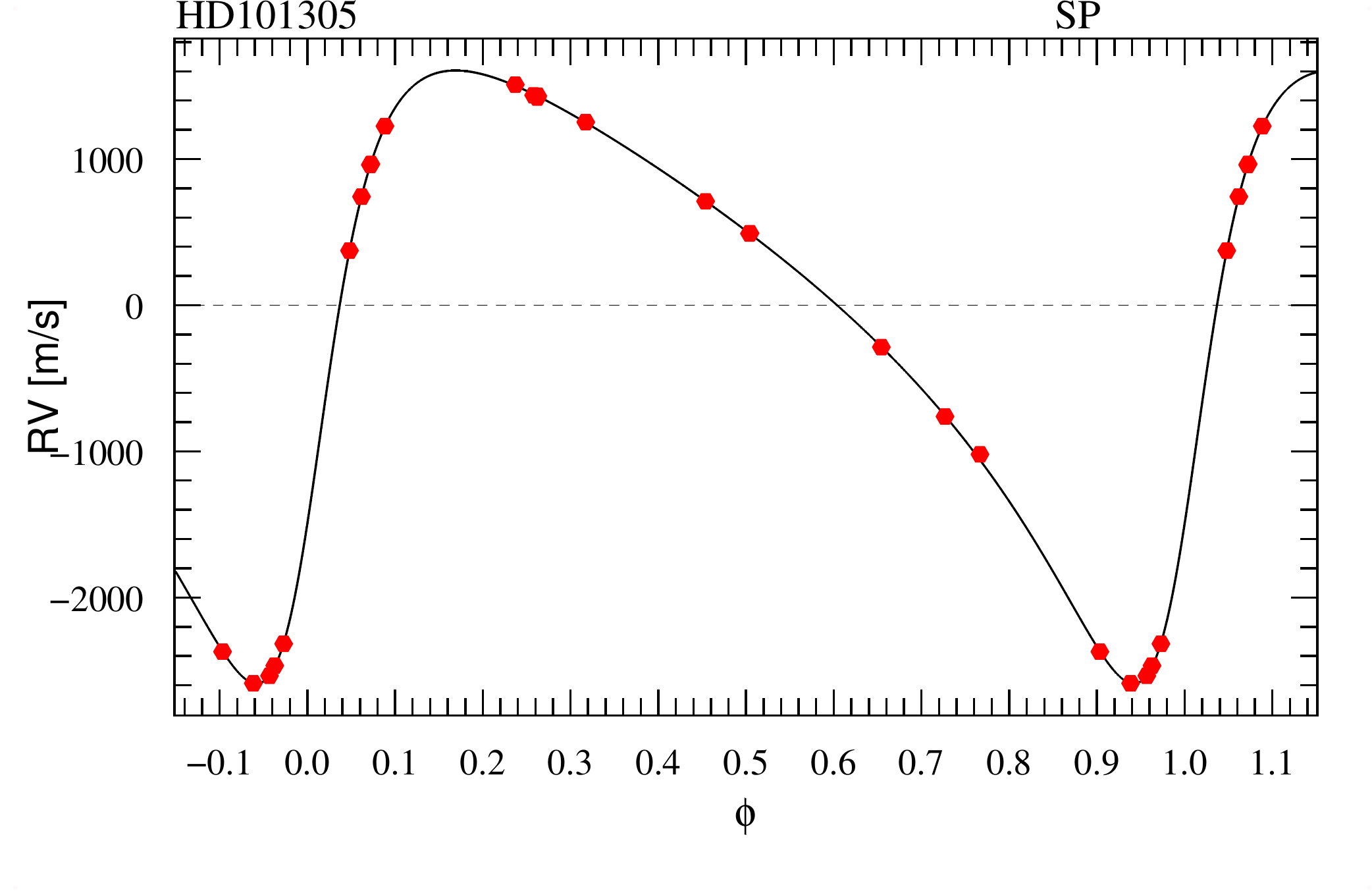} \\
\includegraphics[height=58mm, clip=true, trim=0 -12 0 7]{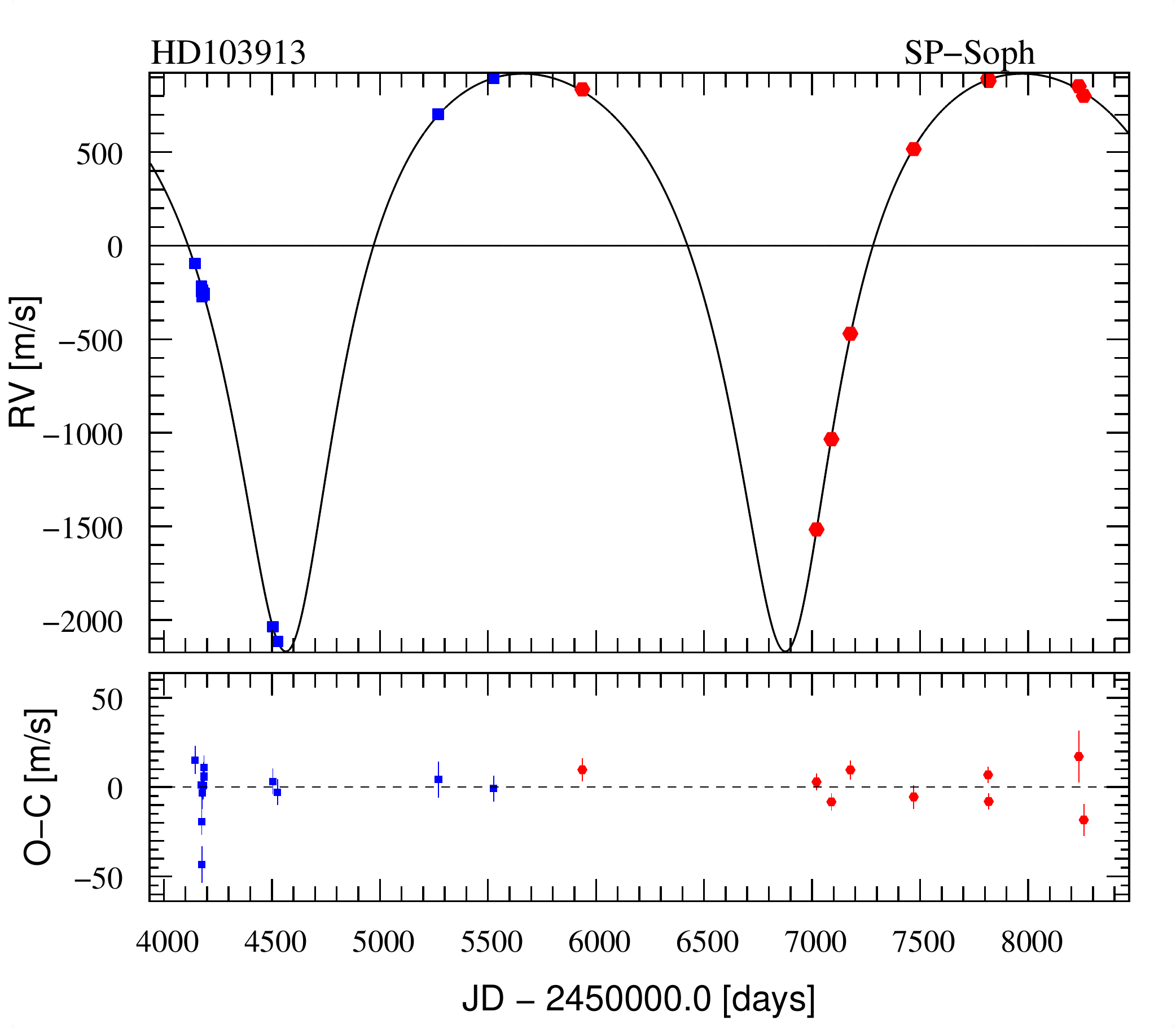}
\includegraphics[height=57mm, clip=true, trim=0  25 0 0]{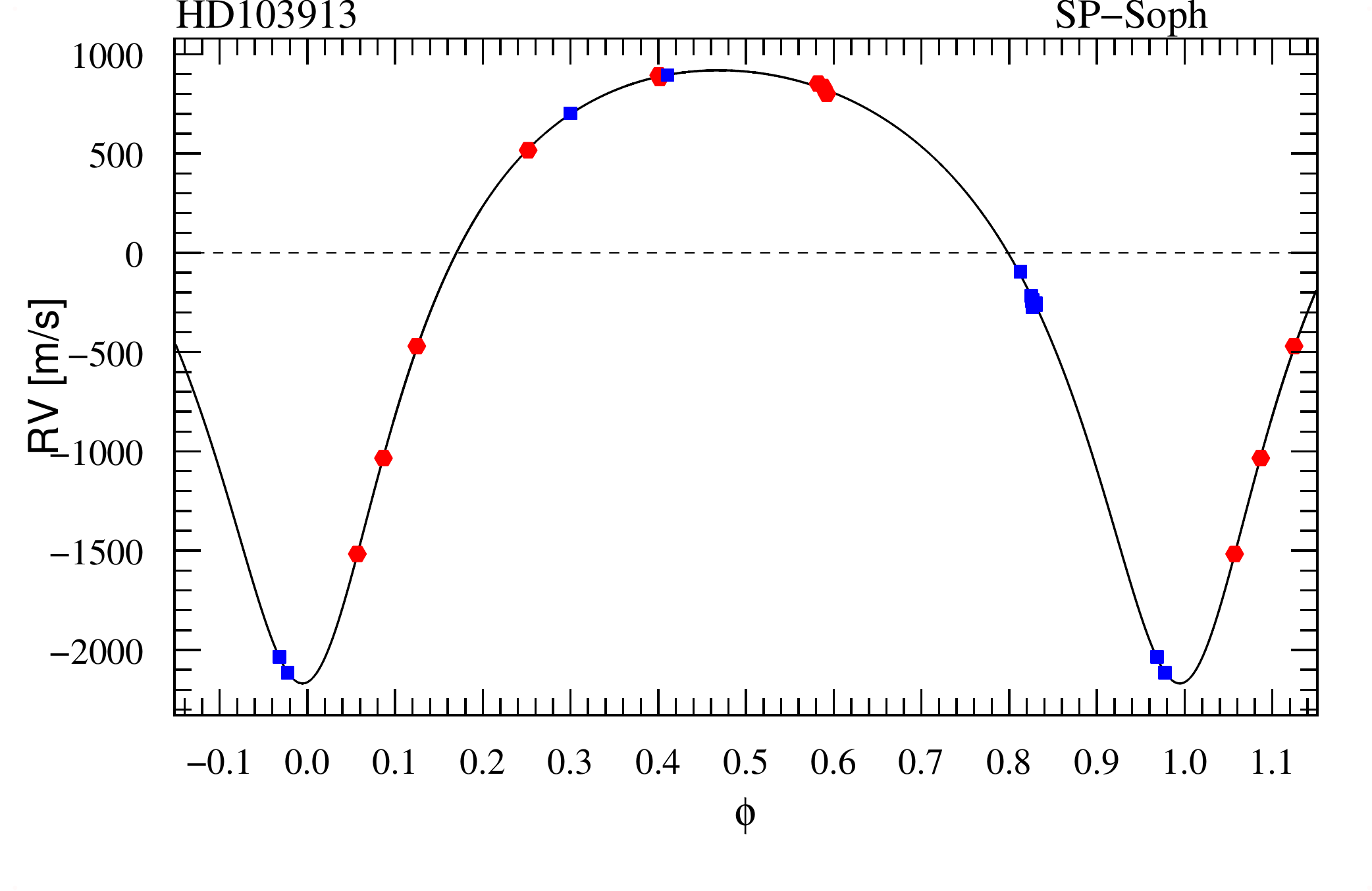} \\
\includegraphics[height=58mm, clip=true, trim=0 -12 0 7]{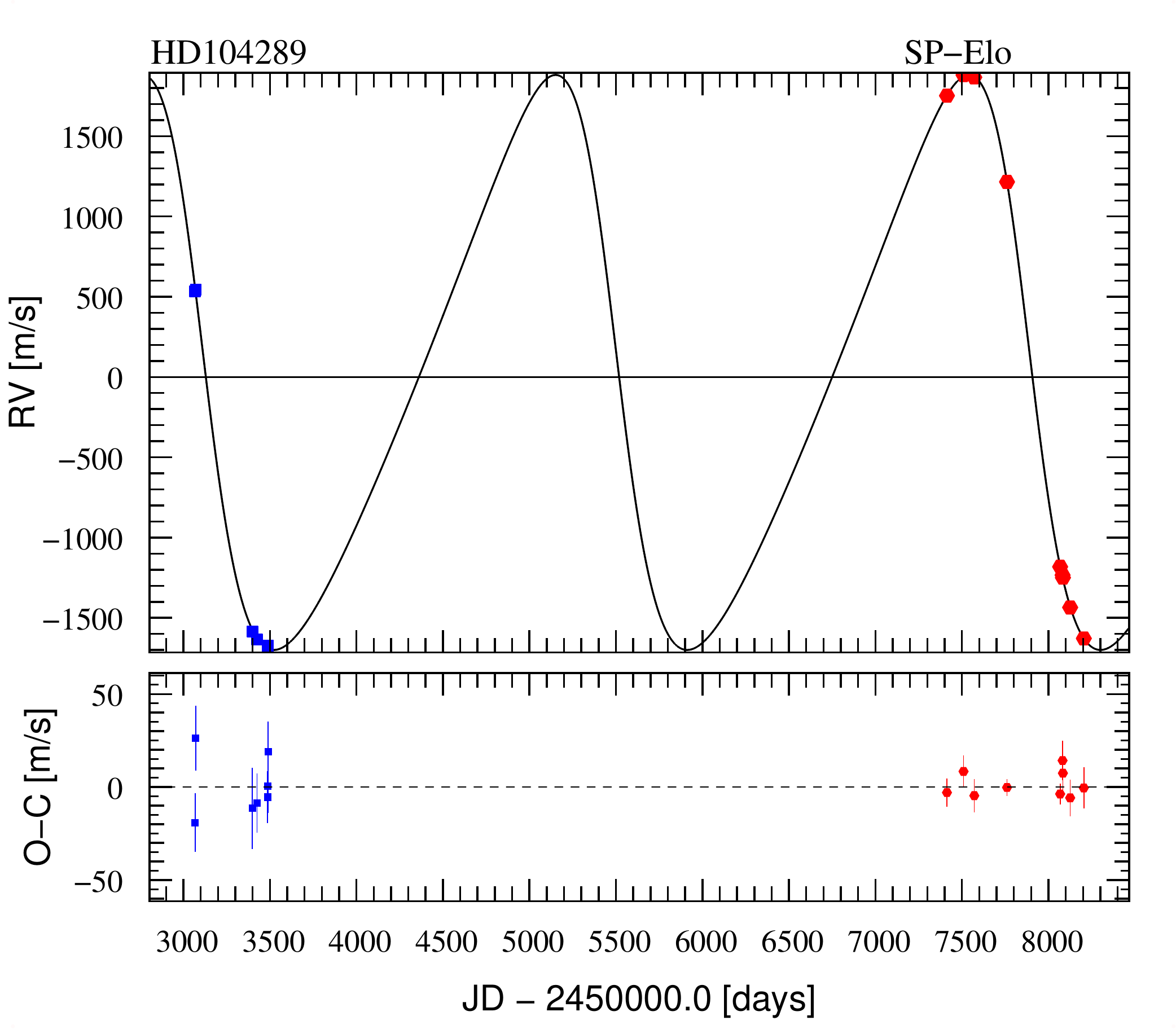}
\includegraphics[height=57mm, clip=true, trim=0  25 0 0]{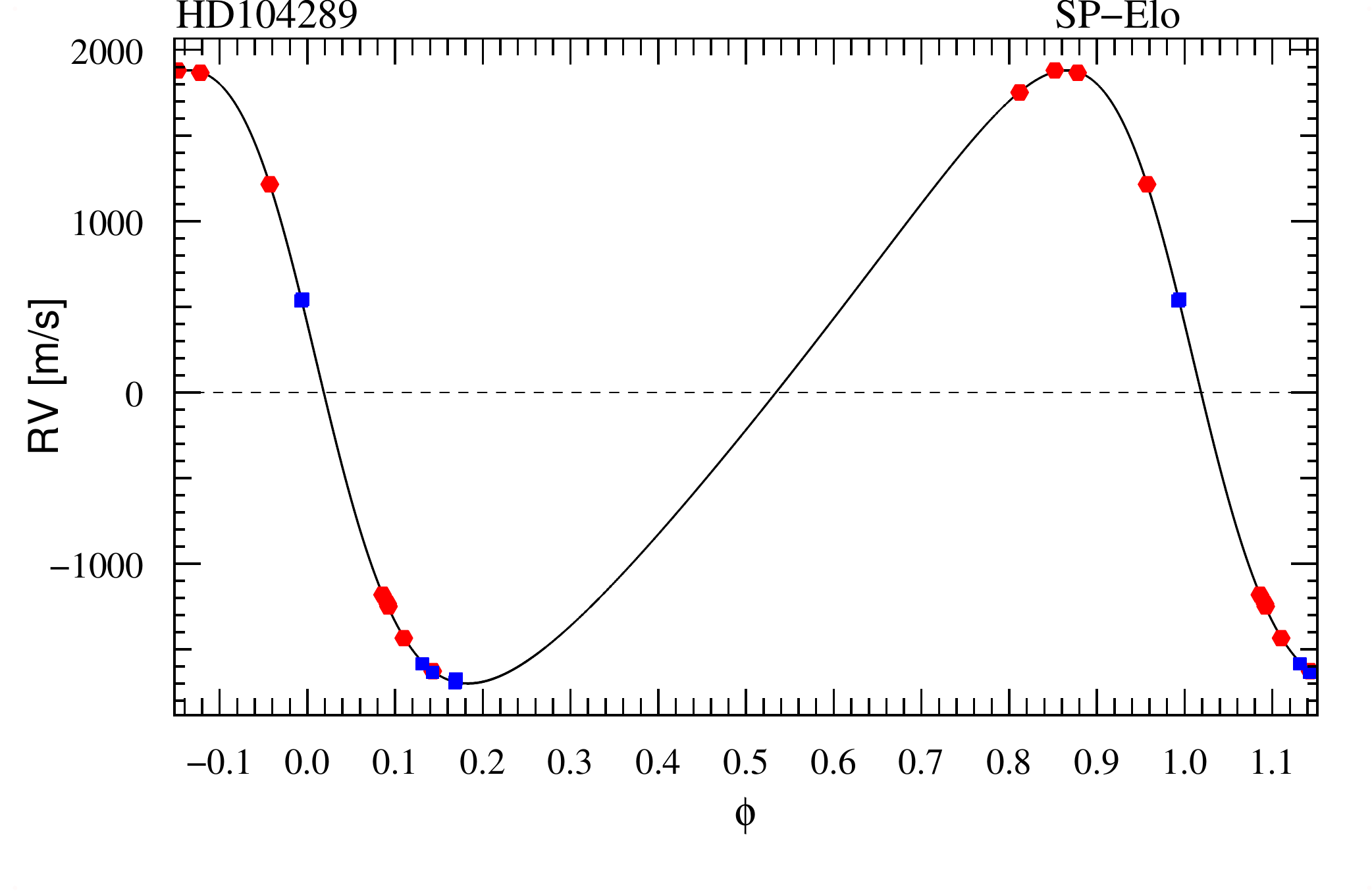} \\
\includegraphics[height=58mm, clip=true, trim=0 -12 0 7]{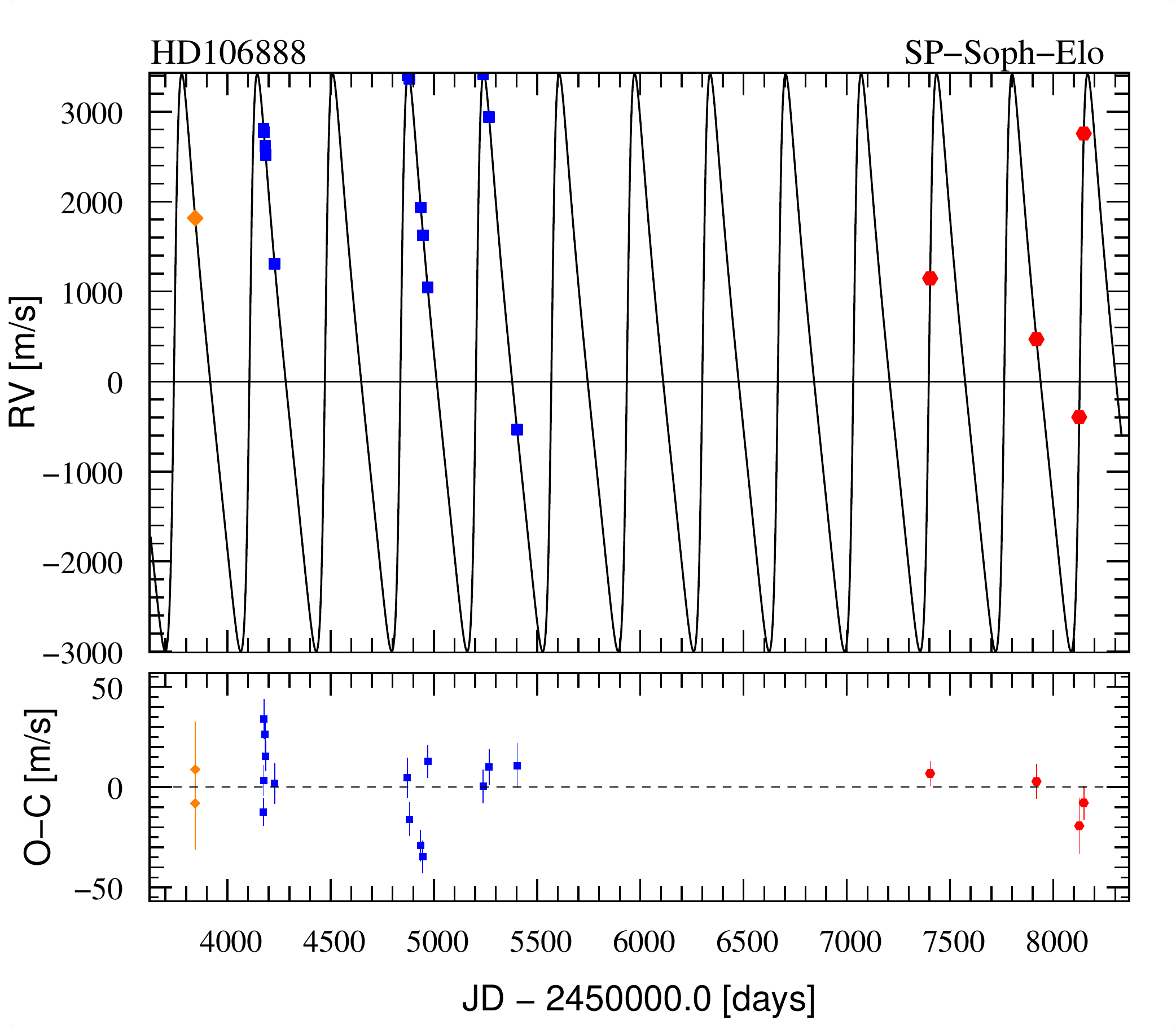} 
\includegraphics[height=57mm, clip=true, trim=0  25 0 0]{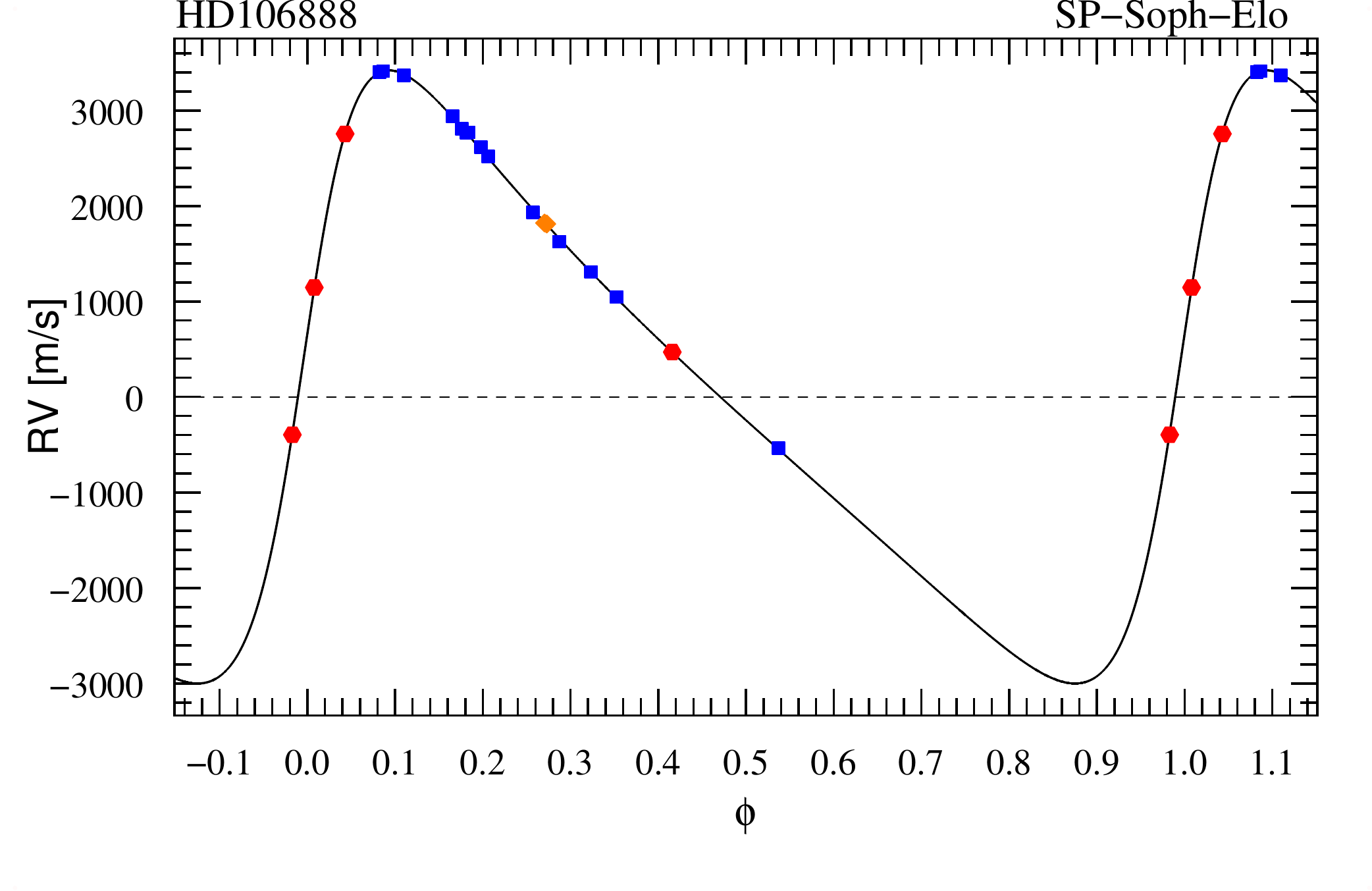} \\
\includegraphics[height=58mm, clip=true, trim=0 -12 0 7]{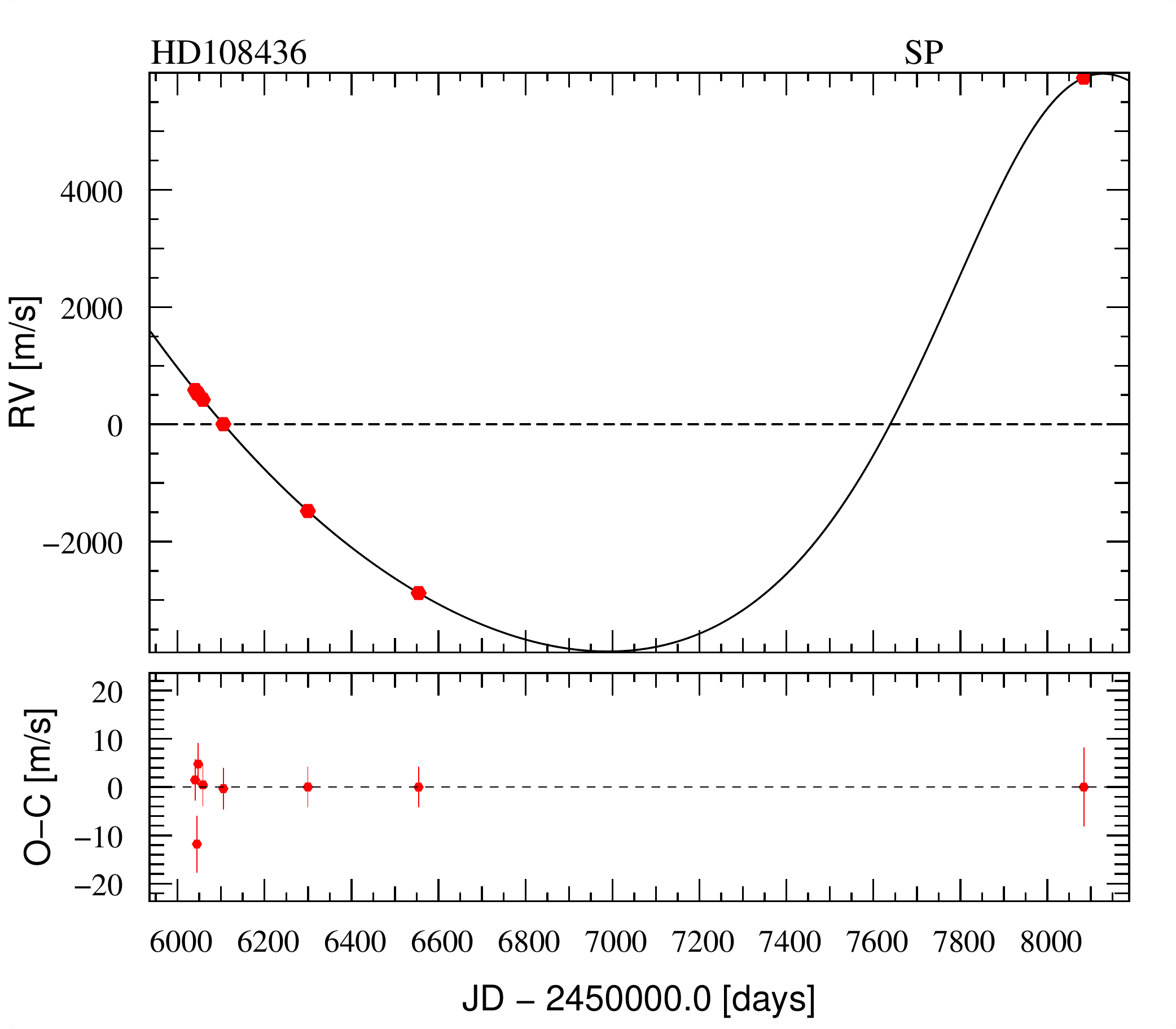}
\includegraphics[height=57mm, clip=true, trim=0  25 0 0]{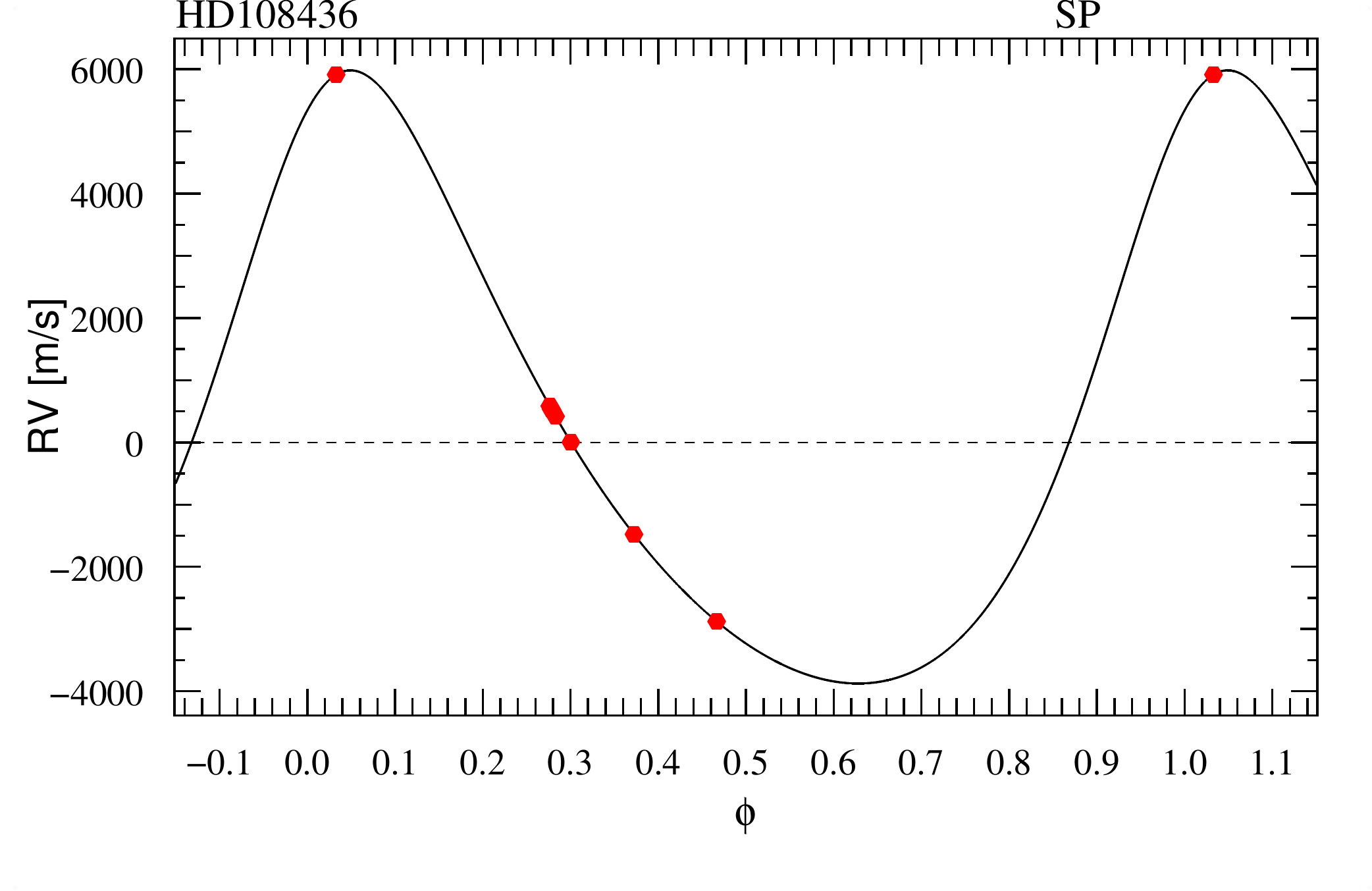} \\
\includegraphics[height=58mm, clip=true, trim=0 -12 0 7]{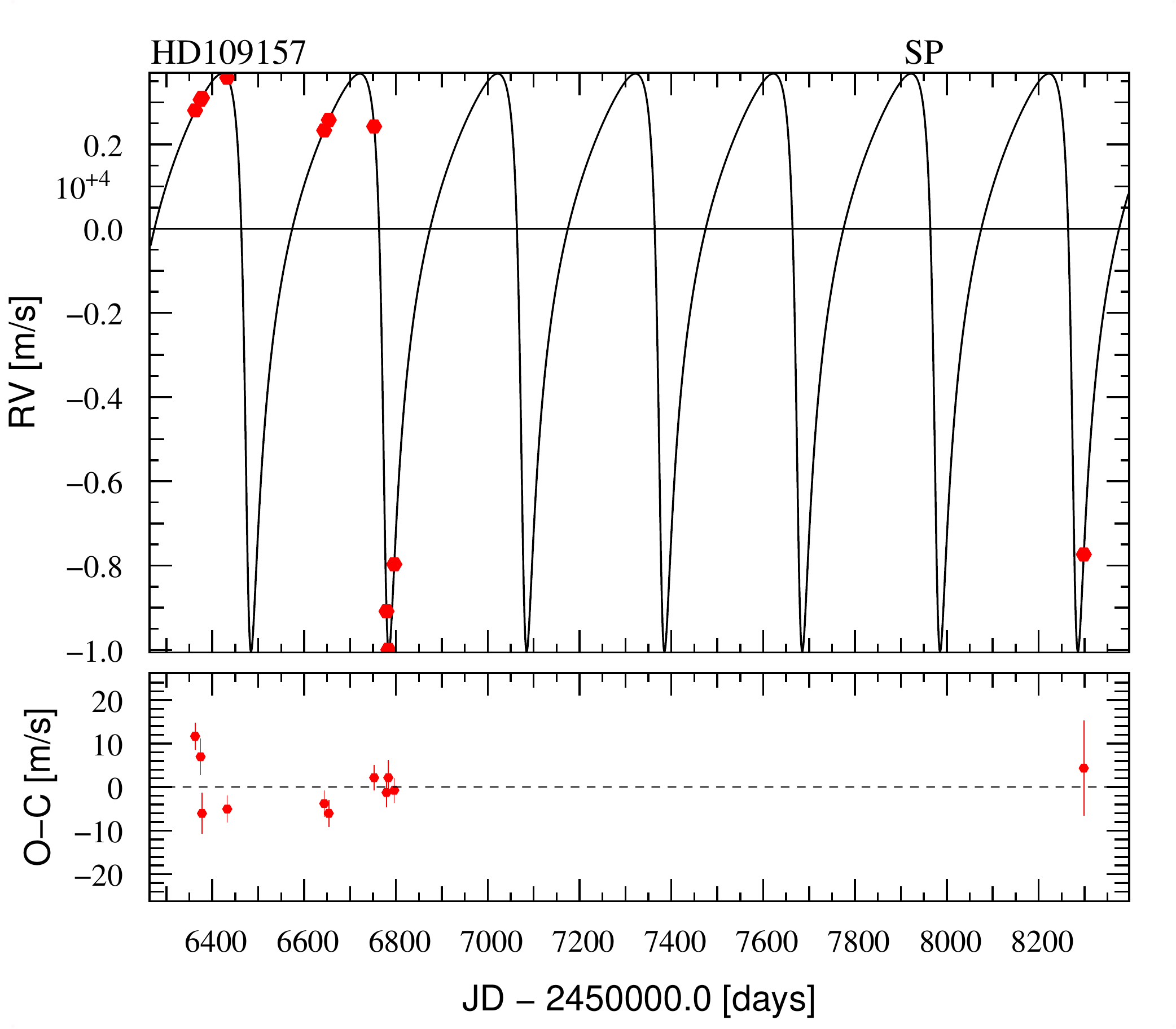}
\includegraphics[height=57mm, clip=true, trim=0  25 0 0]{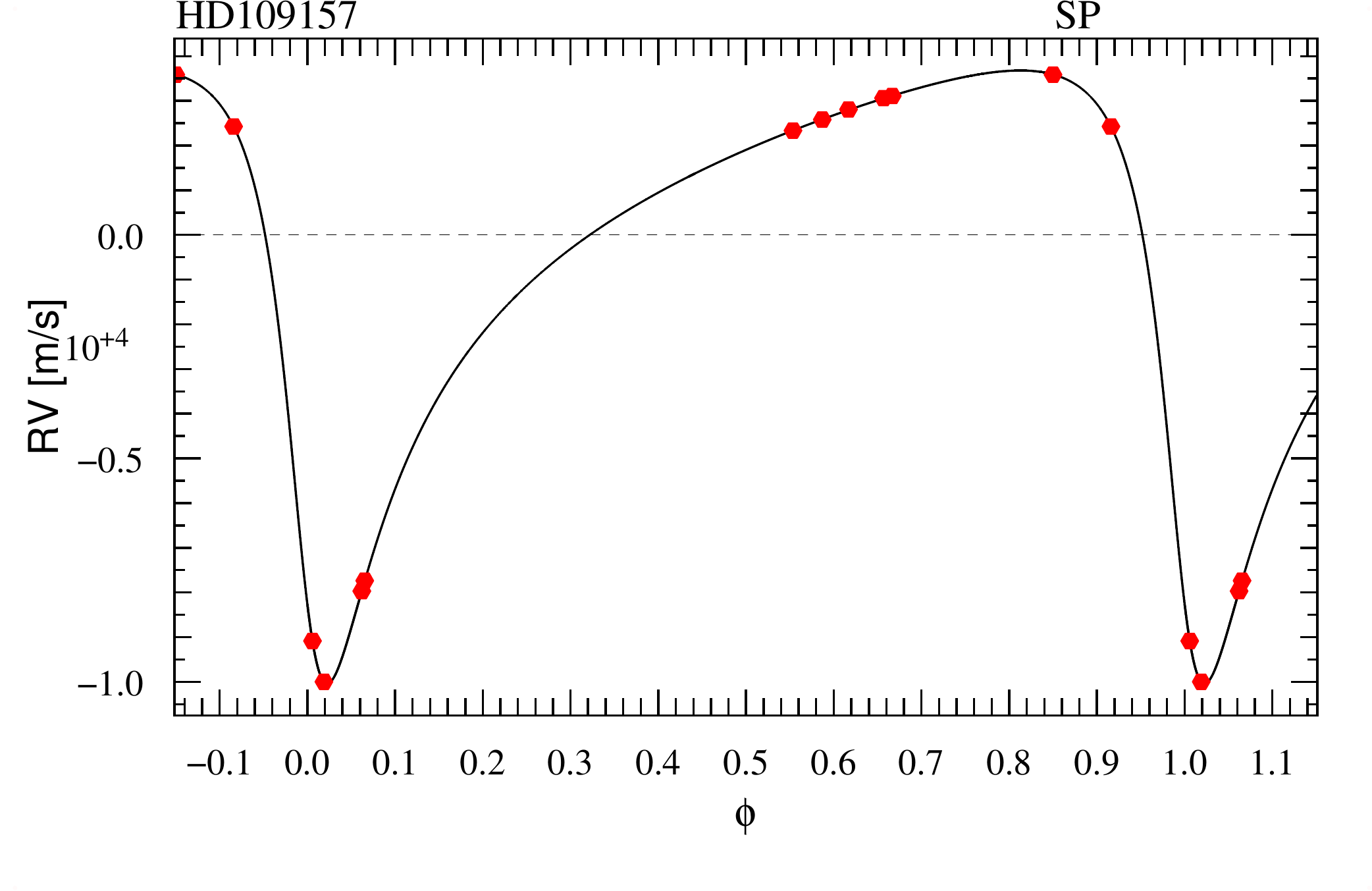} \\
\includegraphics[height=58mm, clip=true, trim=0 -12 0 7]{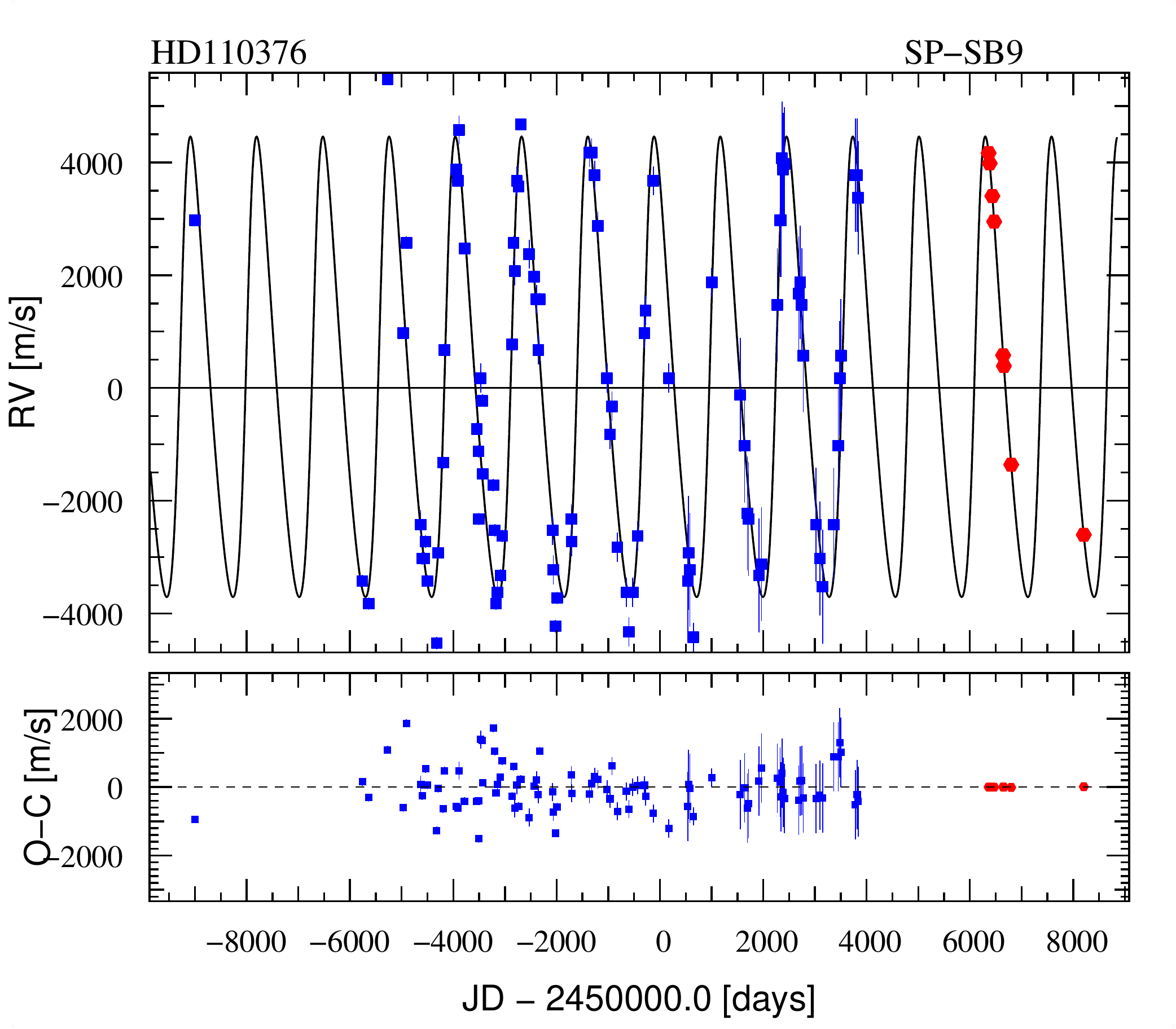}
\includegraphics[height=57mm, clip=true, trim=0  25 0 0]{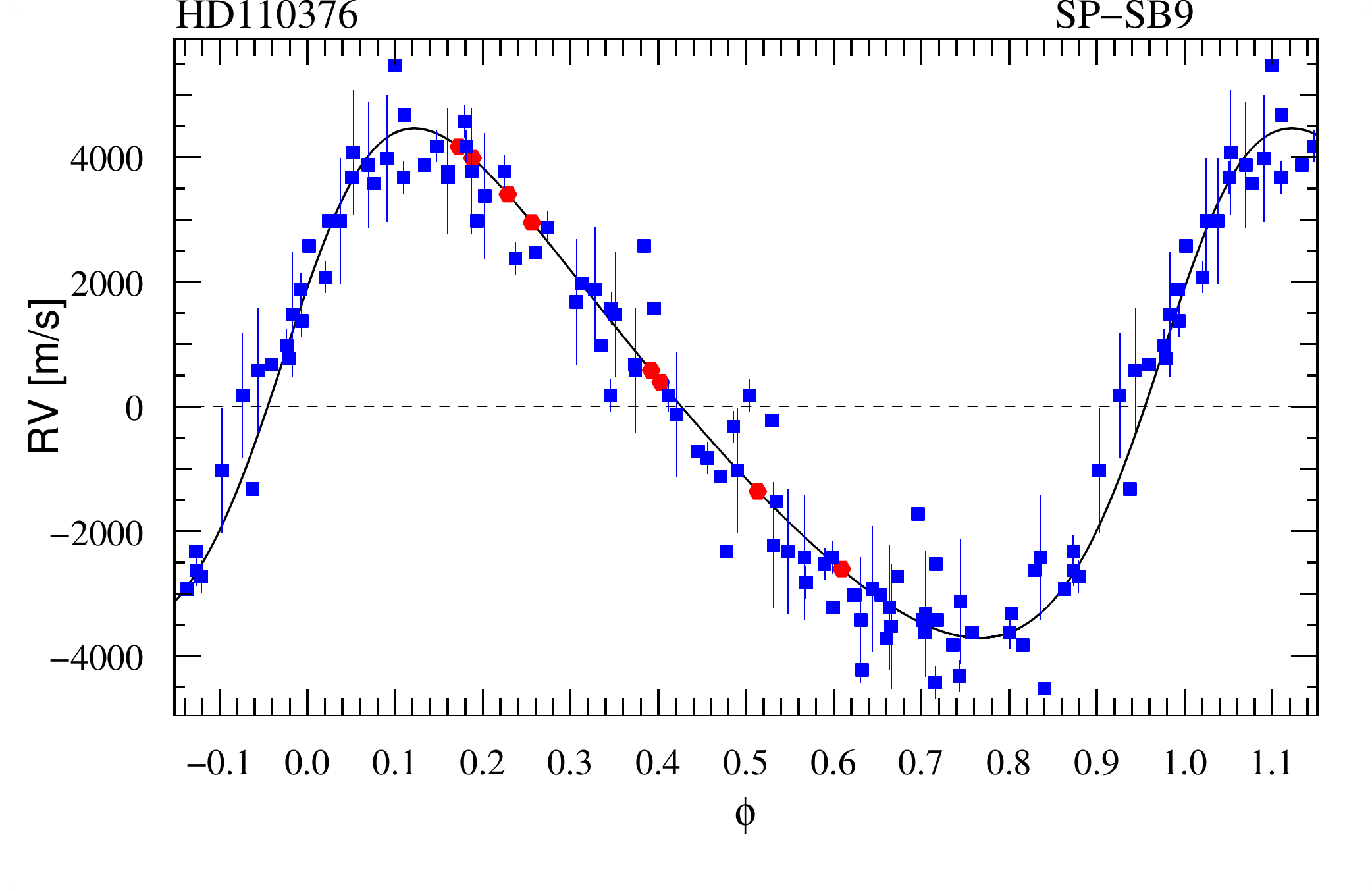} \\
\includegraphics[height=58mm, clip=true, trim=0 -12 0 7]{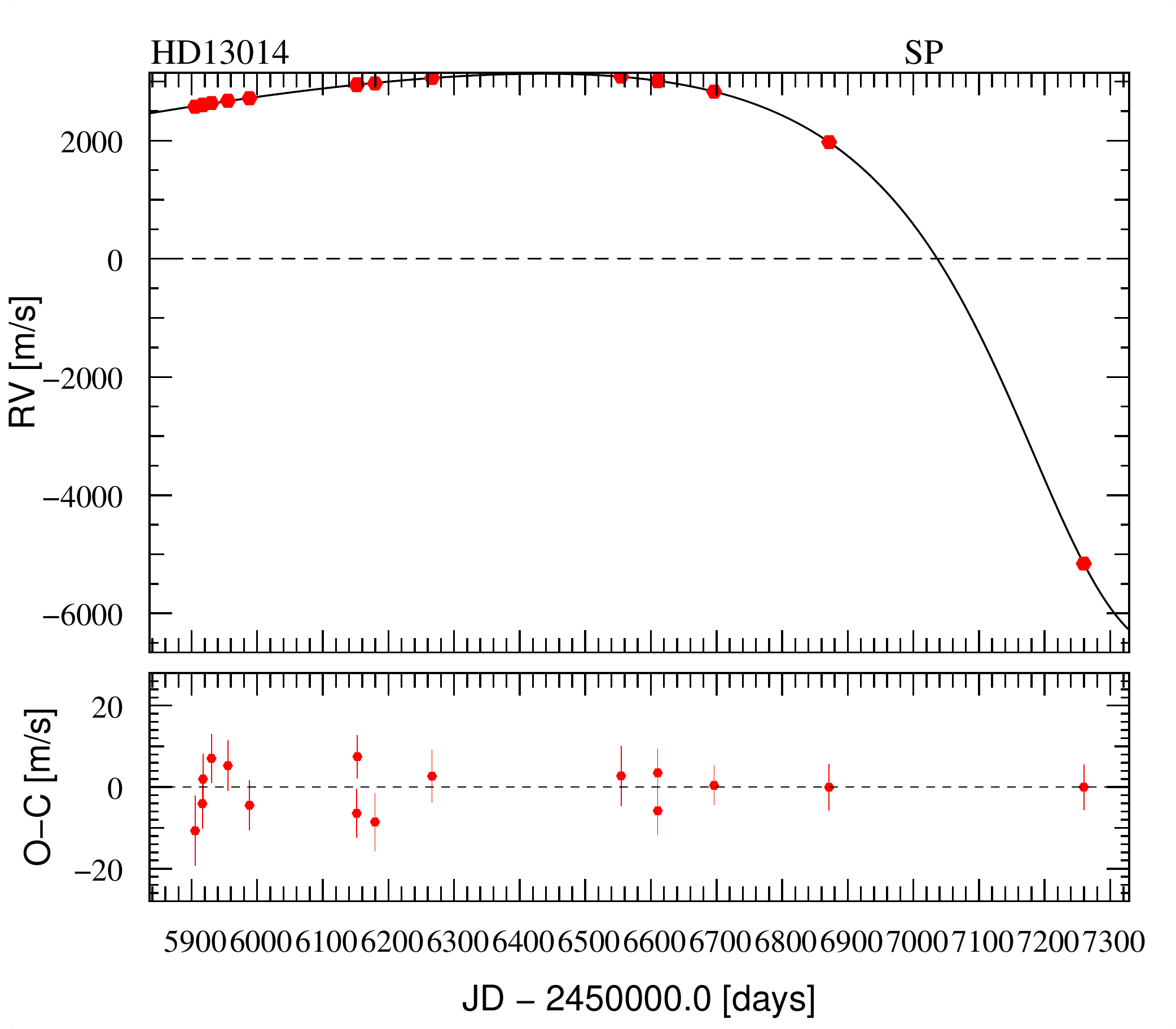}
\includegraphics[height=57mm, clip=true, trim=0  25 0 0]{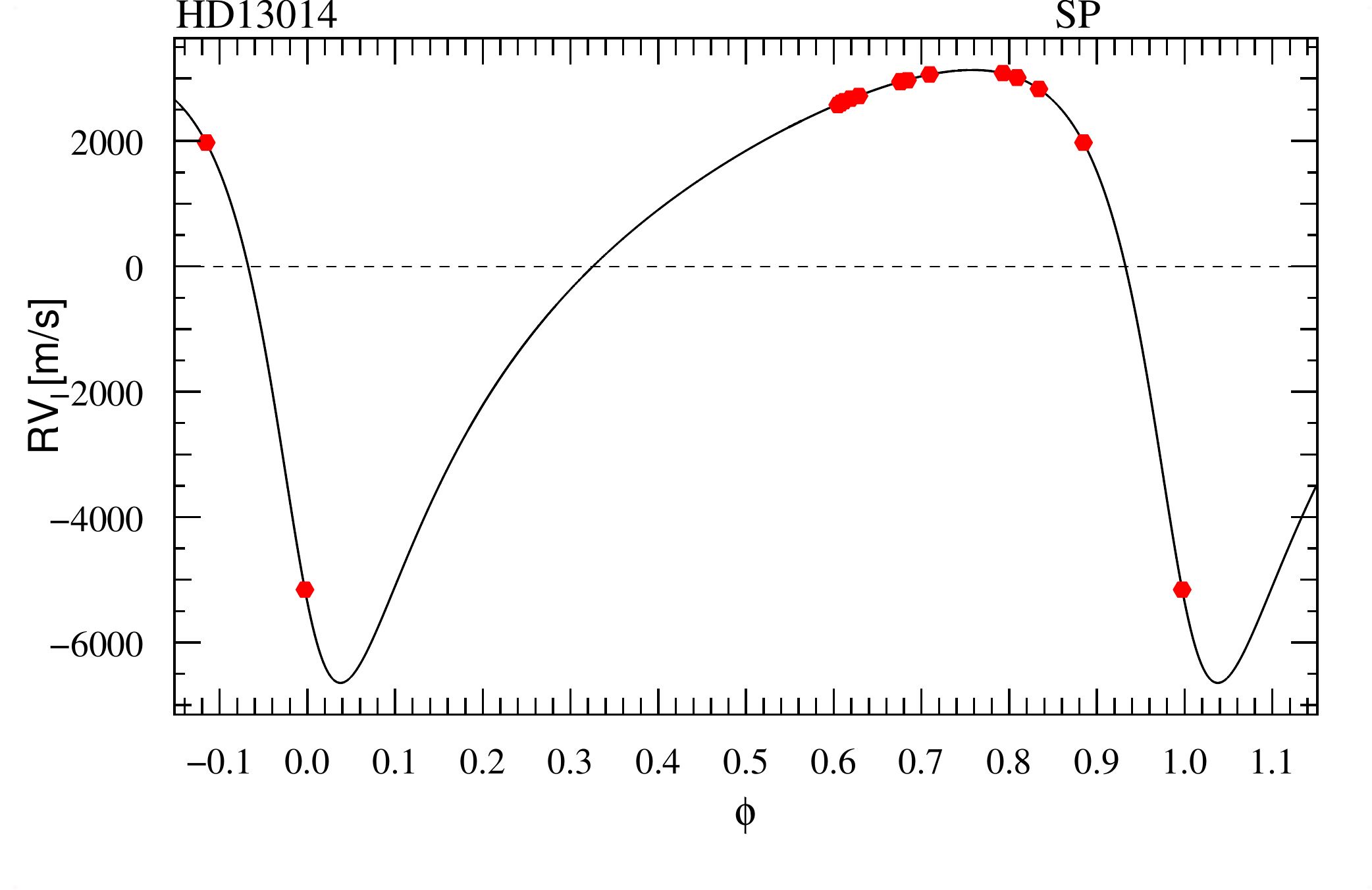} \\
\includegraphics[height=58mm, clip=true, trim=0 -12 0 7]{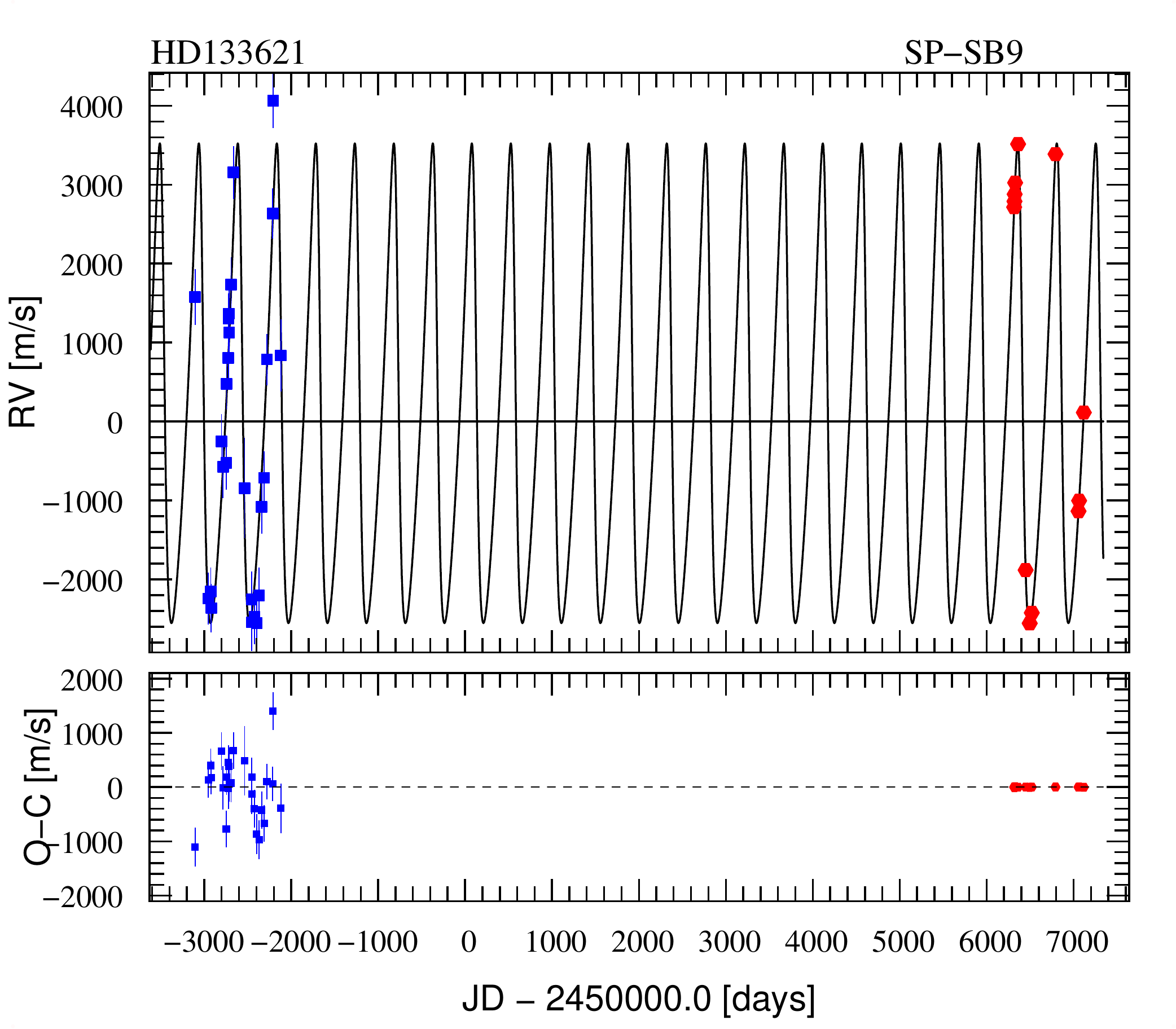}
\includegraphics[height=57mm, clip=true, trim=0  25 0 0]{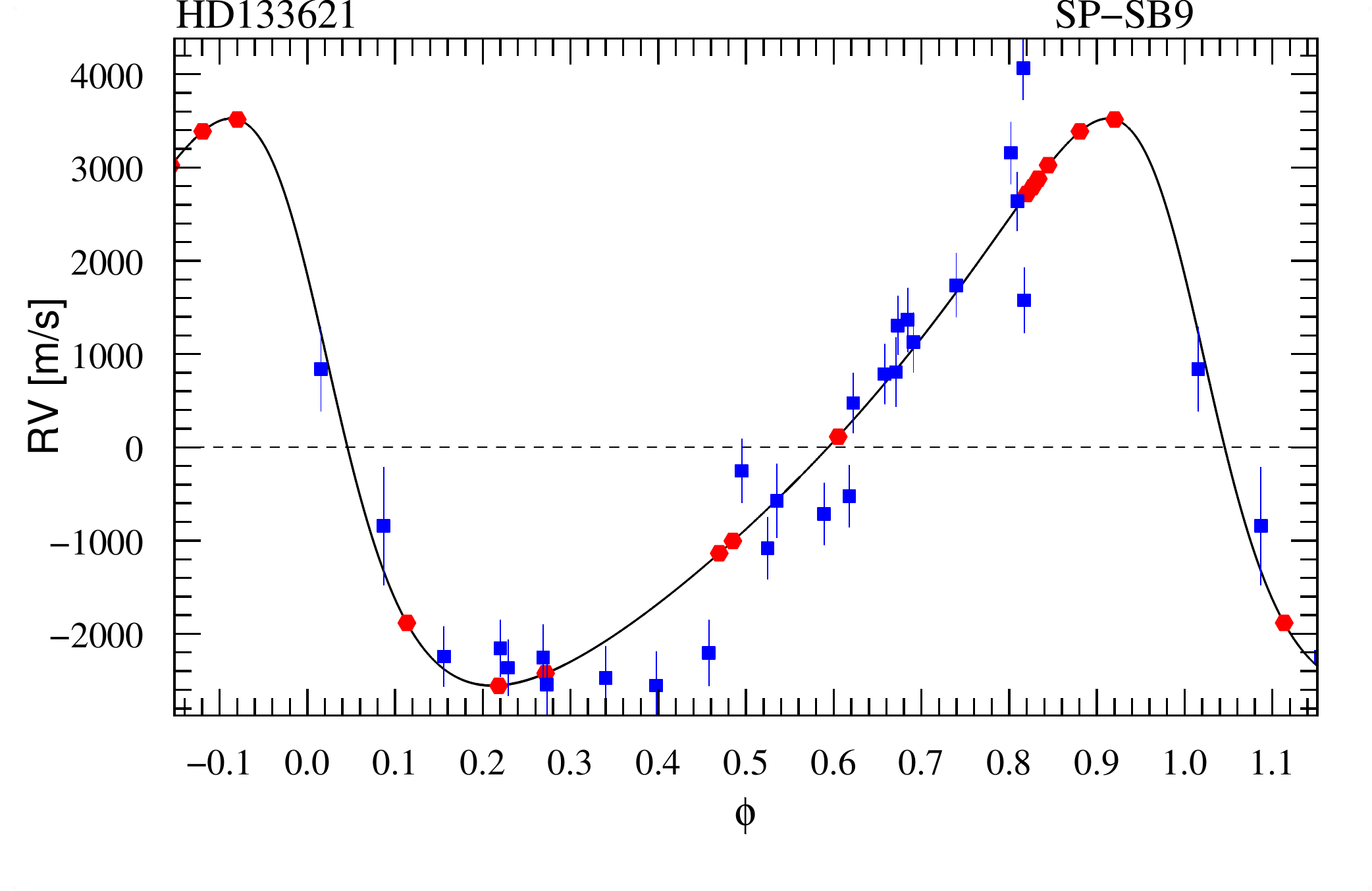} \\
\includegraphics[height=58mm, clip=true, trim=0 -12 0 7]{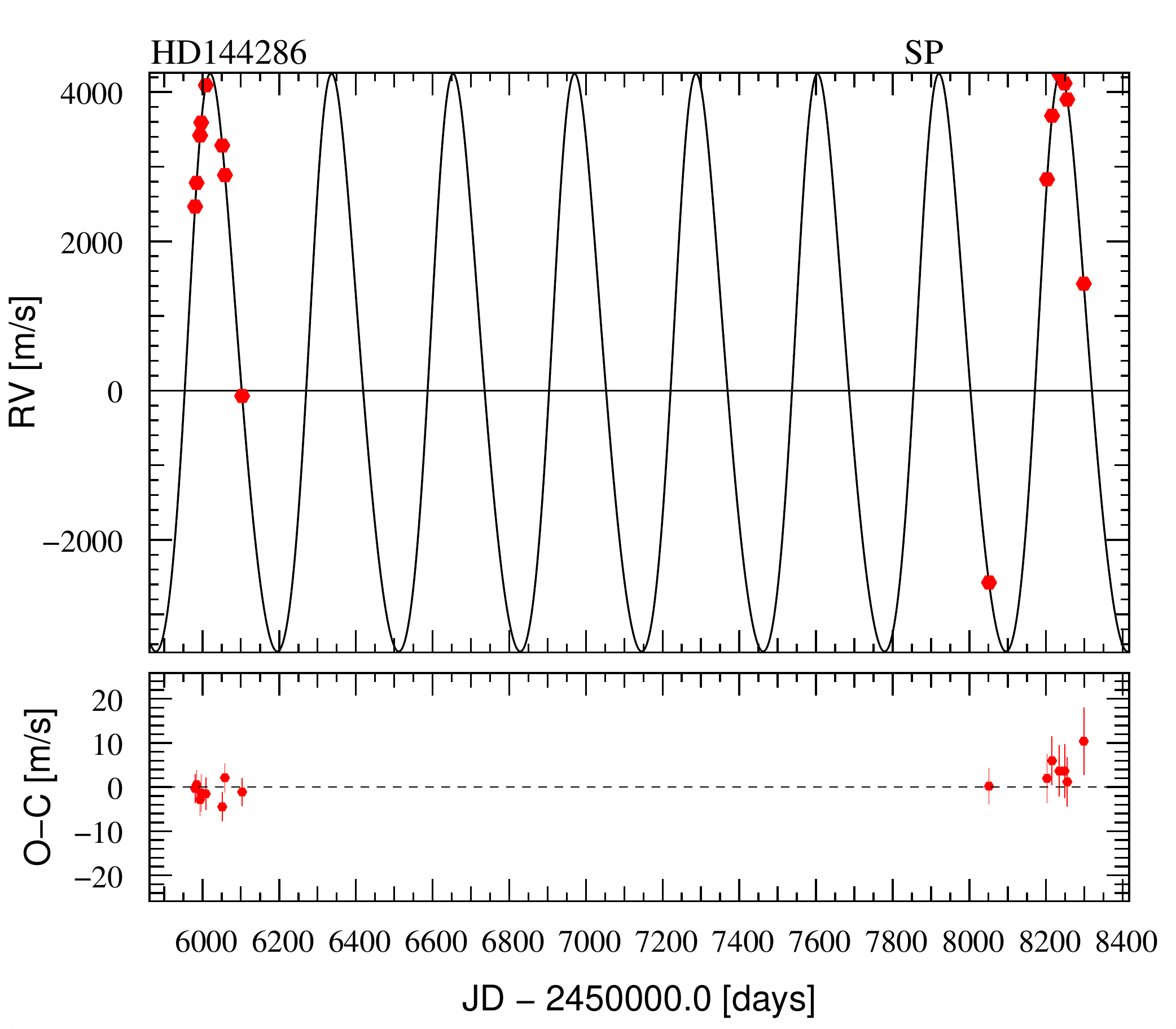} 
\includegraphics[height=57mm, clip=true, trim=0  25 0 0]{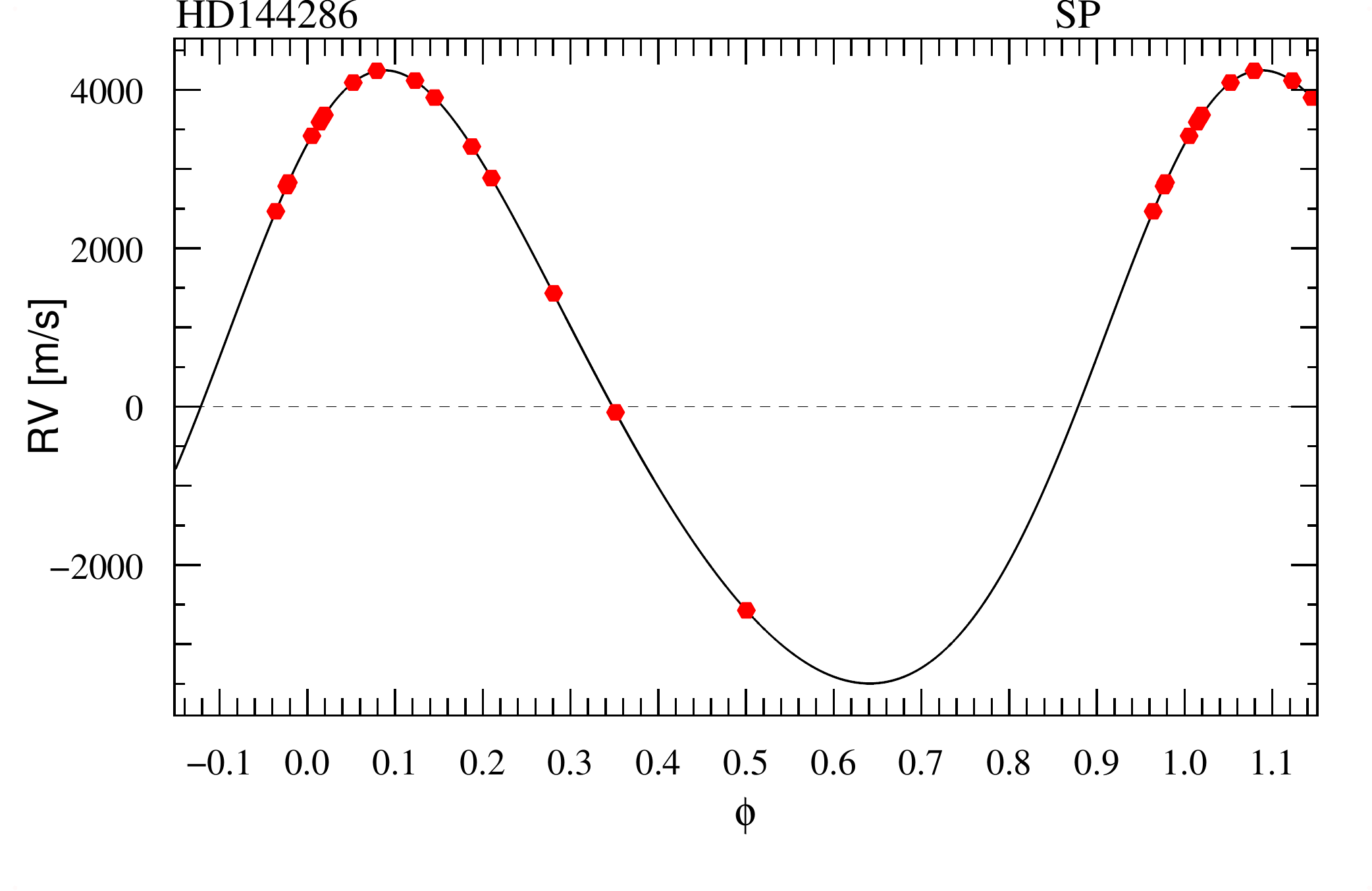} \\
\includegraphics[height=58mm, clip=true, trim=0 -12 0 7]{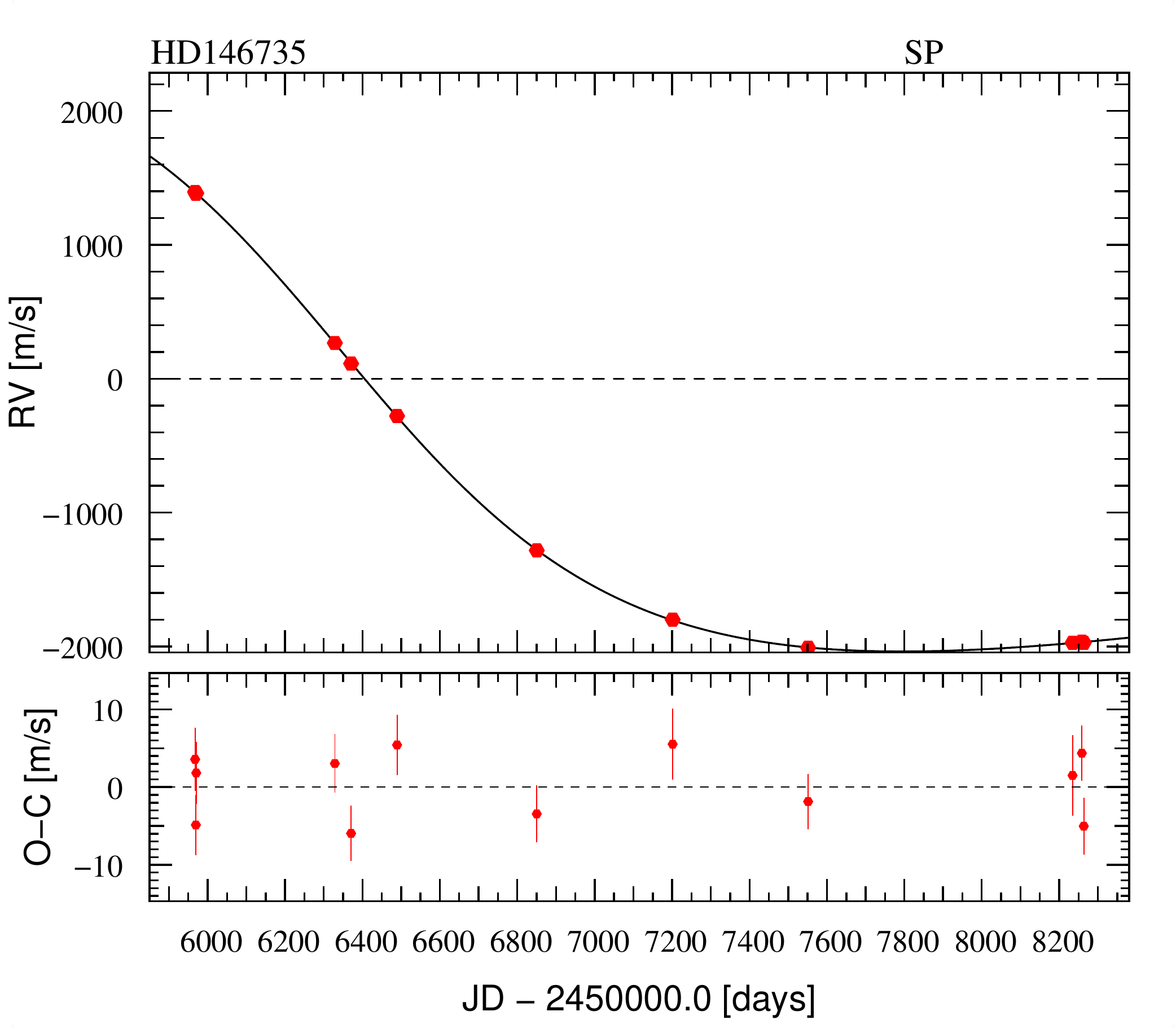}
\includegraphics[height=57mm, clip=true, trim=0  25 0 0]{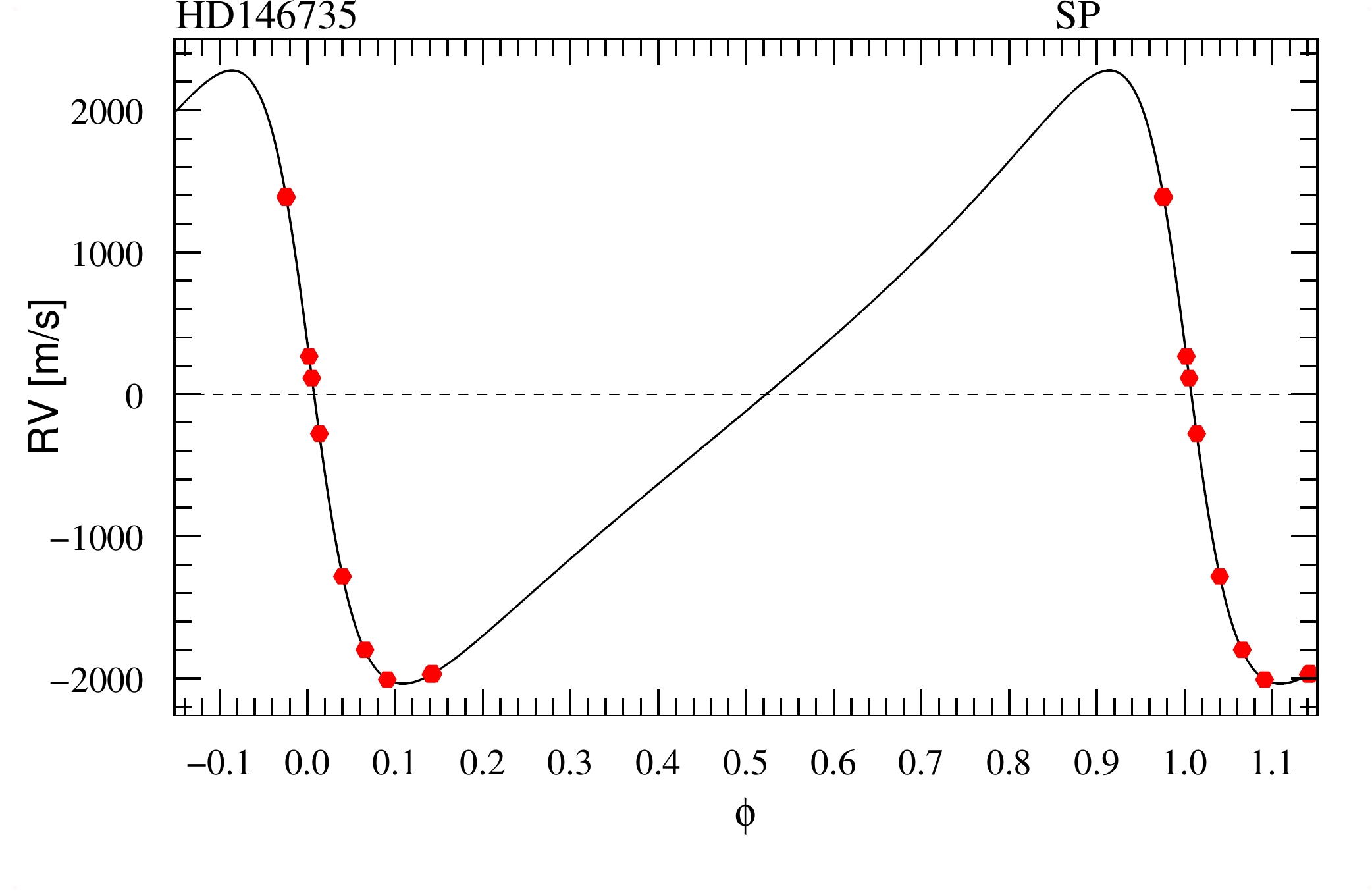} \\
\includegraphics[height=58mm, clip=true, trim=0 -12 0 7]{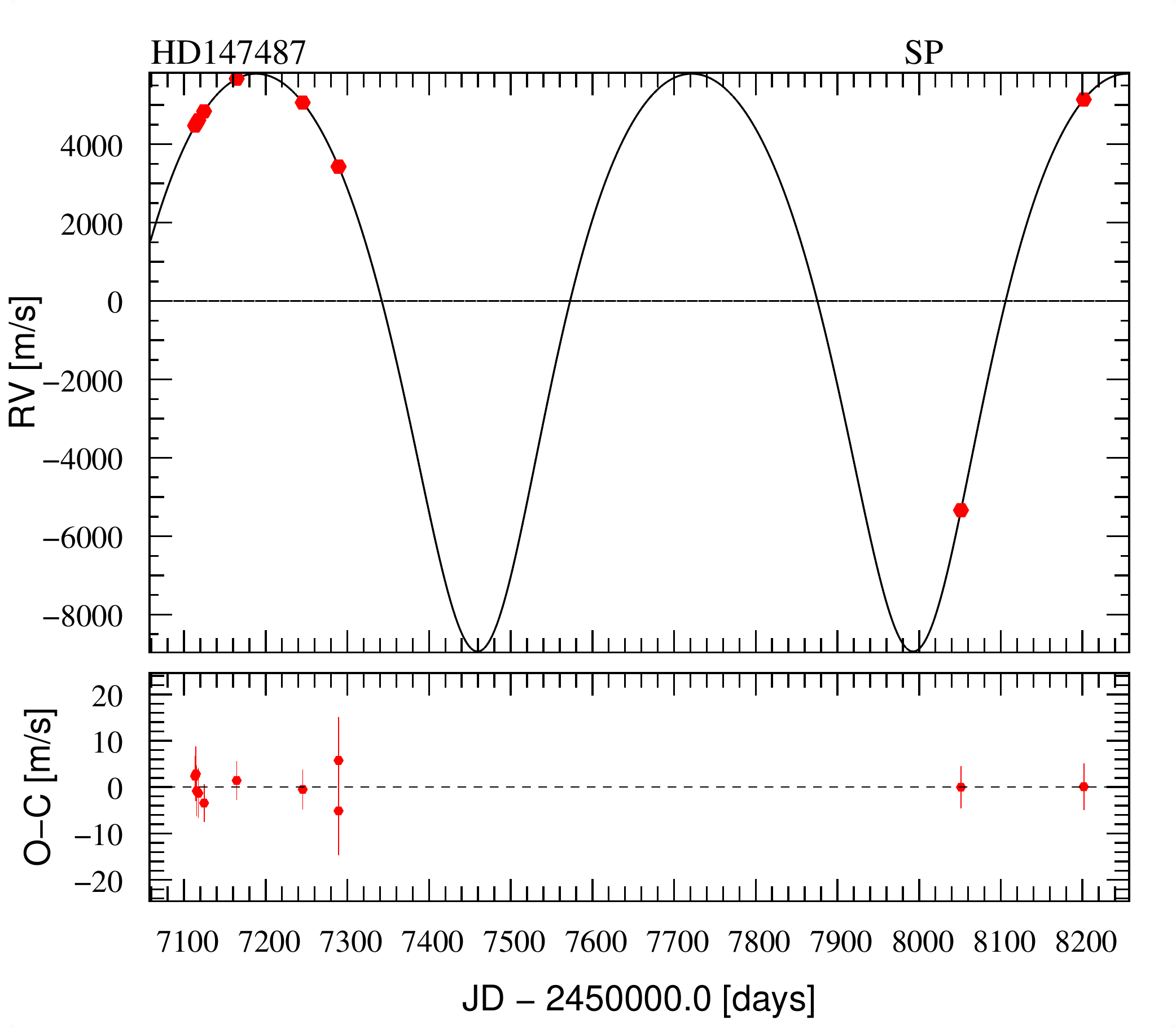}
\includegraphics[height=57mm, clip=true, trim=0  25 0 0]{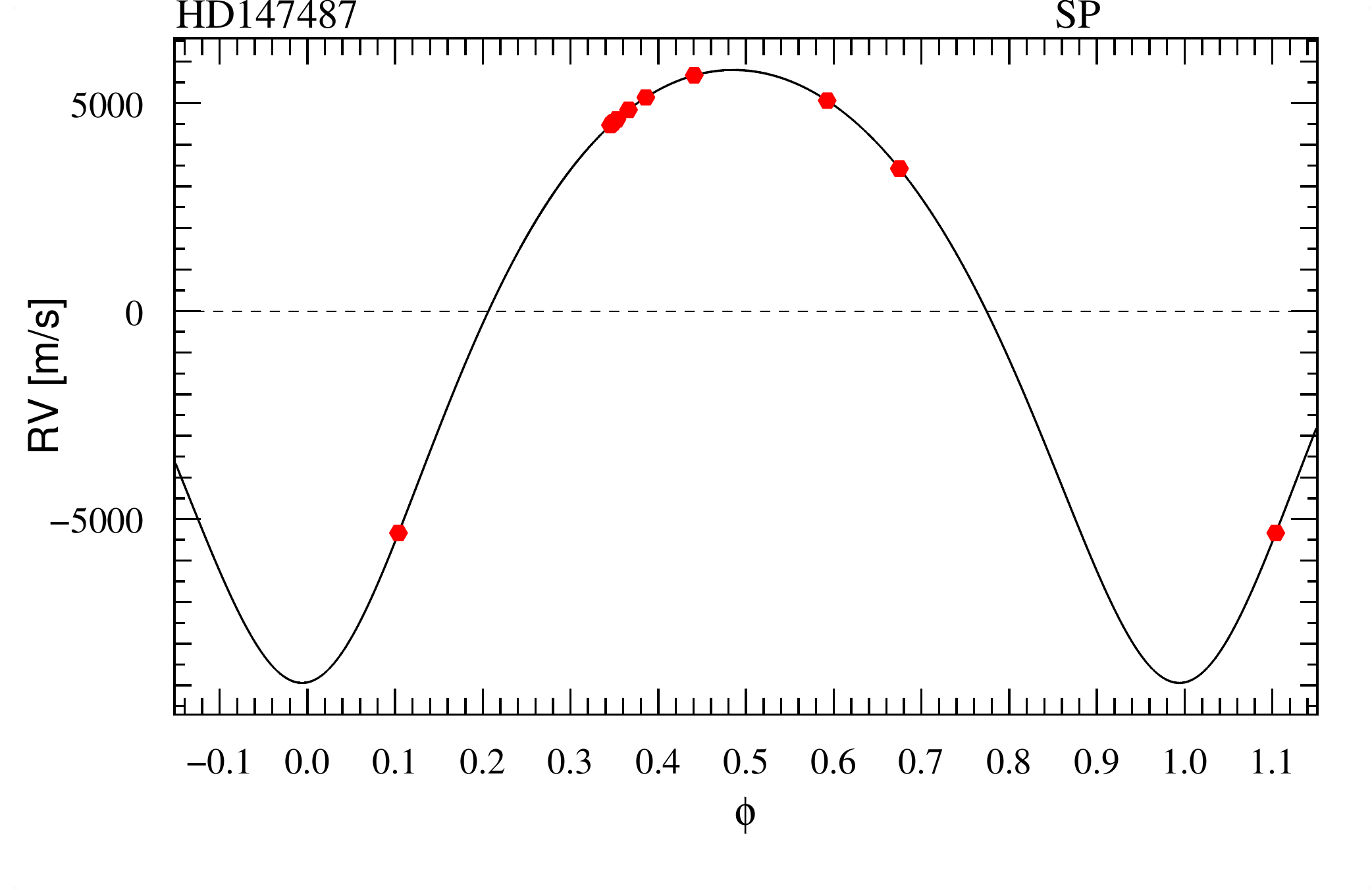} \\
\includegraphics[height=58mm, clip=true, trim=0 -12 0 7]{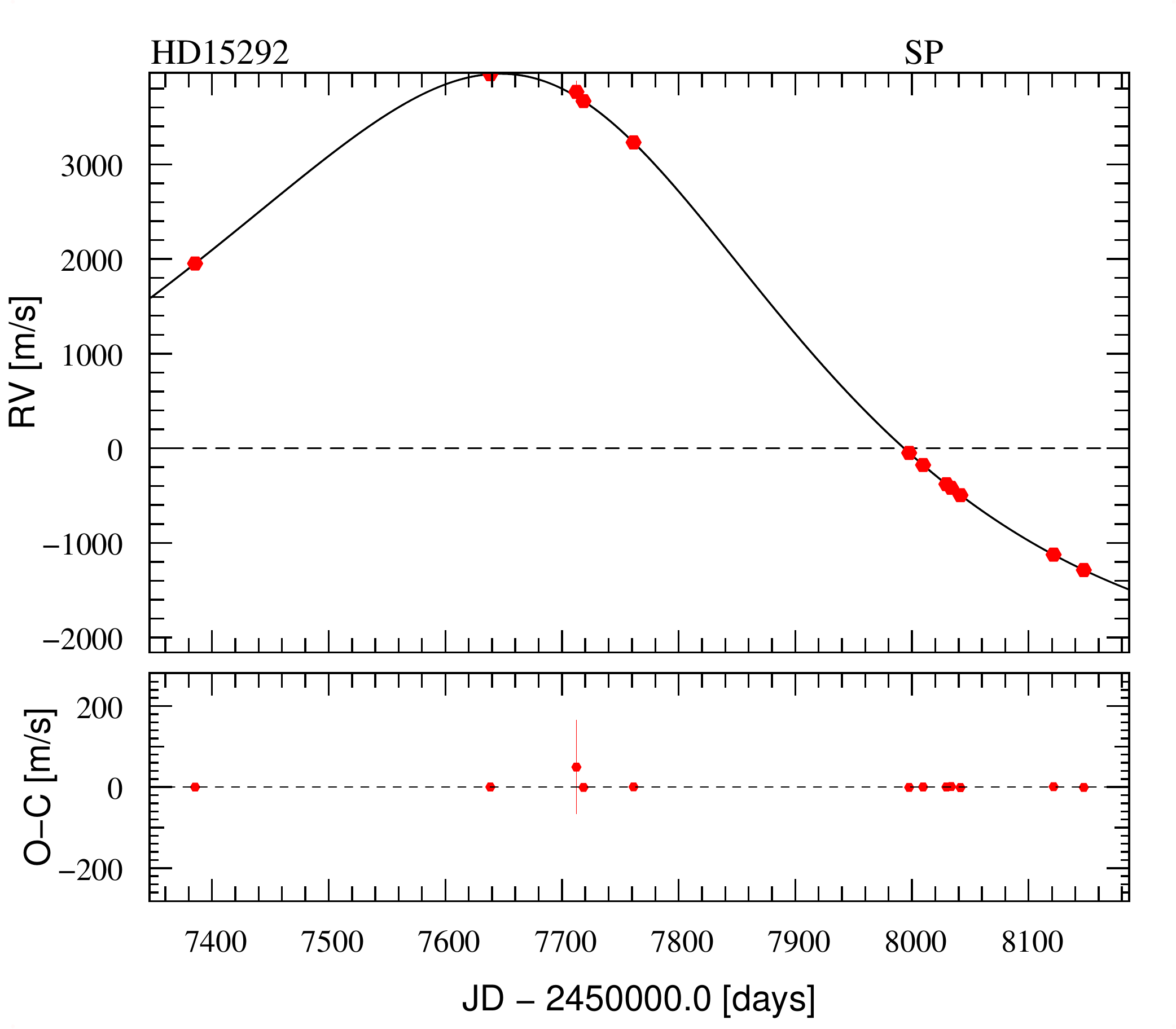}
\includegraphics[height=57mm, clip=true, trim=0  25 0 0]{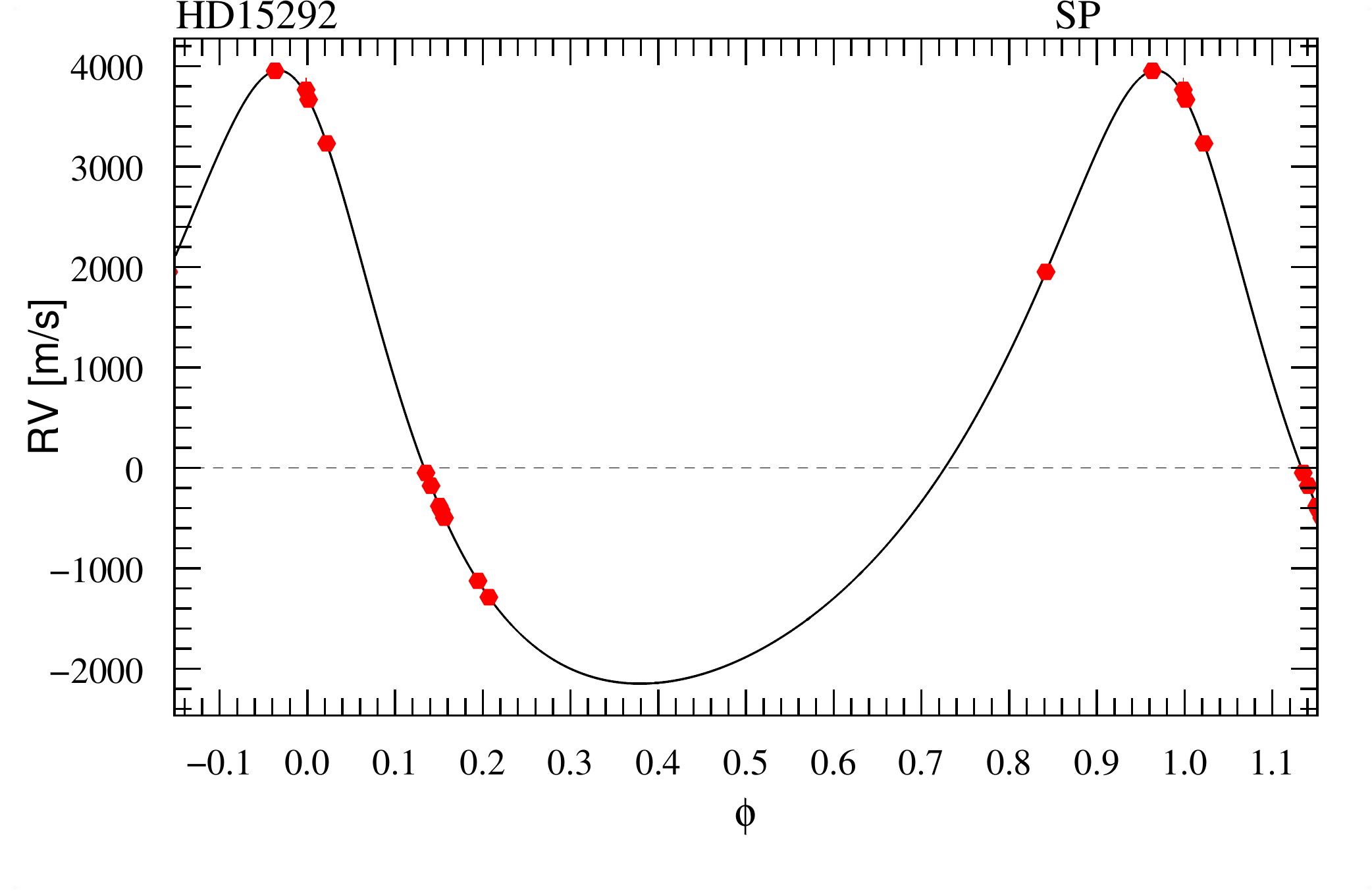} \\
\includegraphics[height=58mm, clip=true, trim=0 -12 0 7]{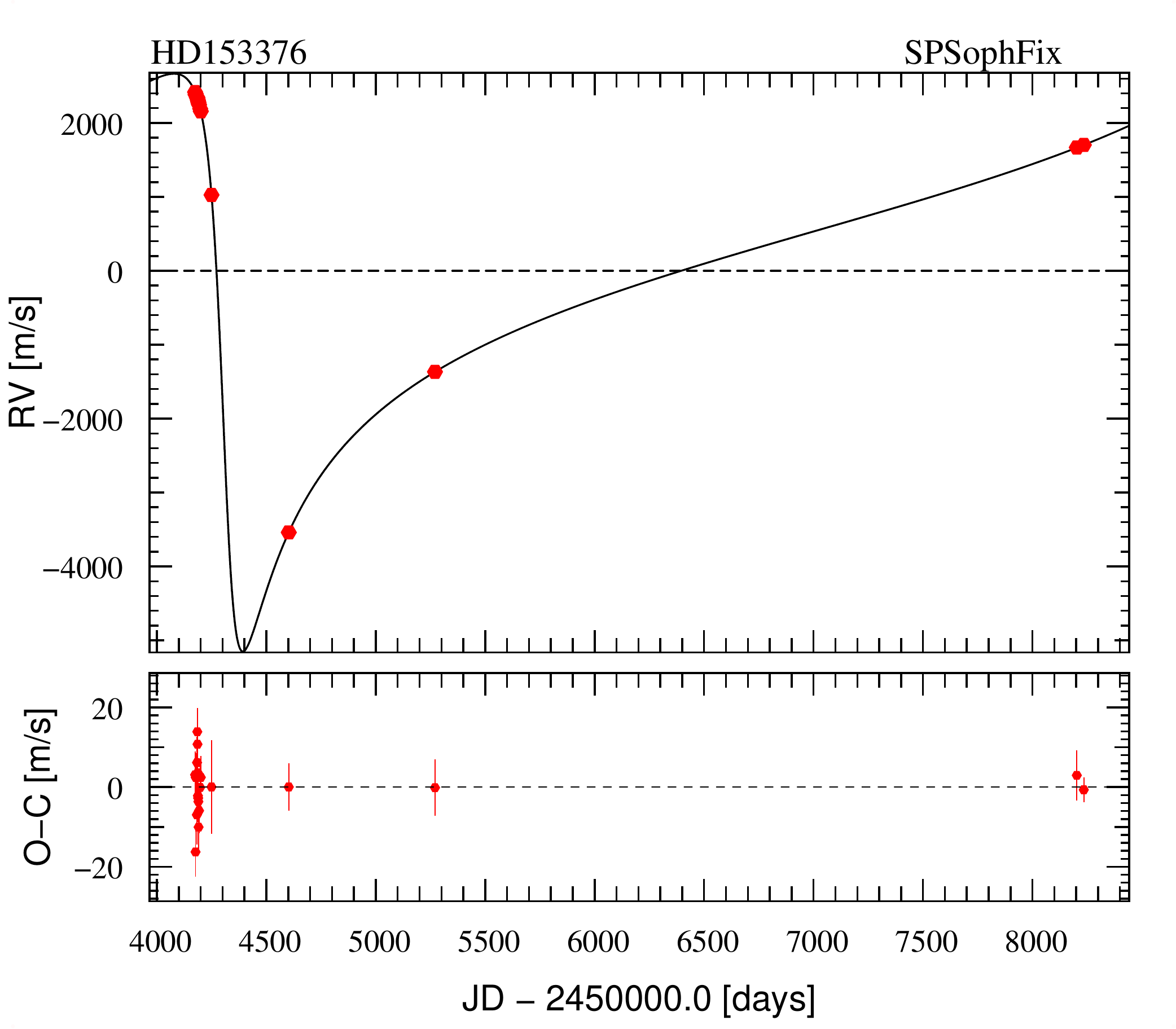}
\includegraphics[height=57mm, clip=true, trim=0  25 0 0]{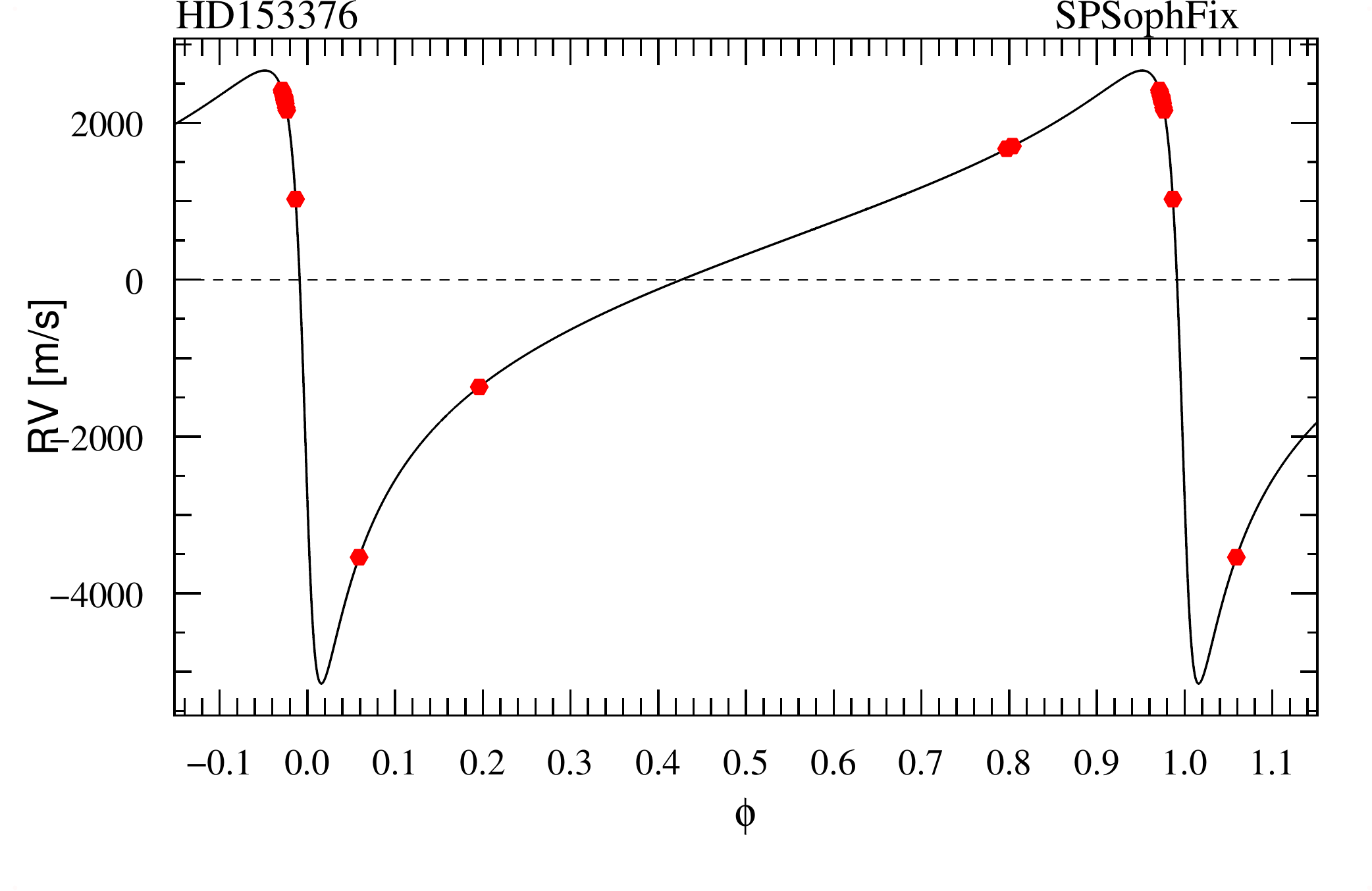} \\
\includegraphics[height=58mm, clip=true, trim=0 -12 0 7]{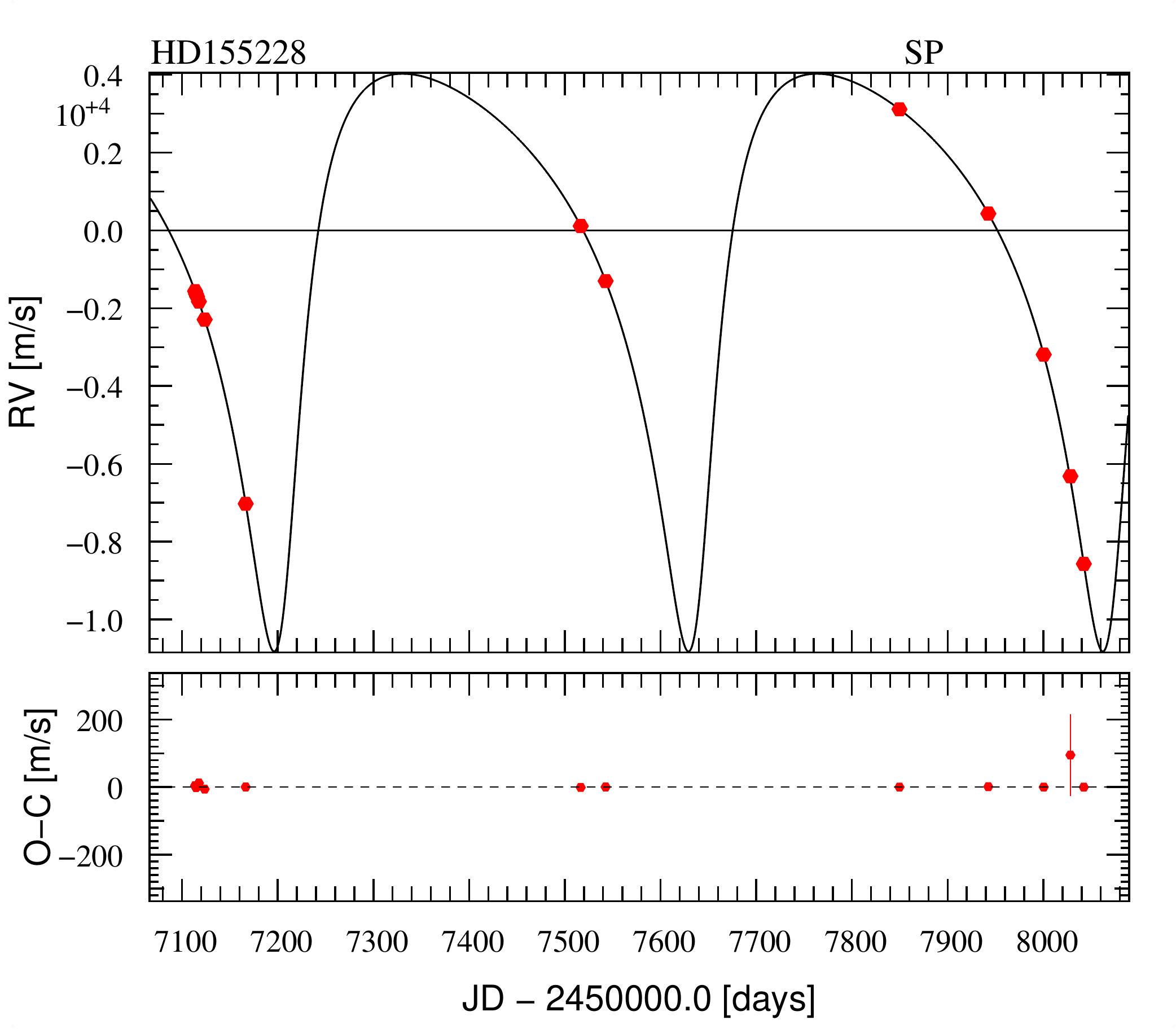}
\includegraphics[height=57mm, clip=true, trim=0  25 0 0]{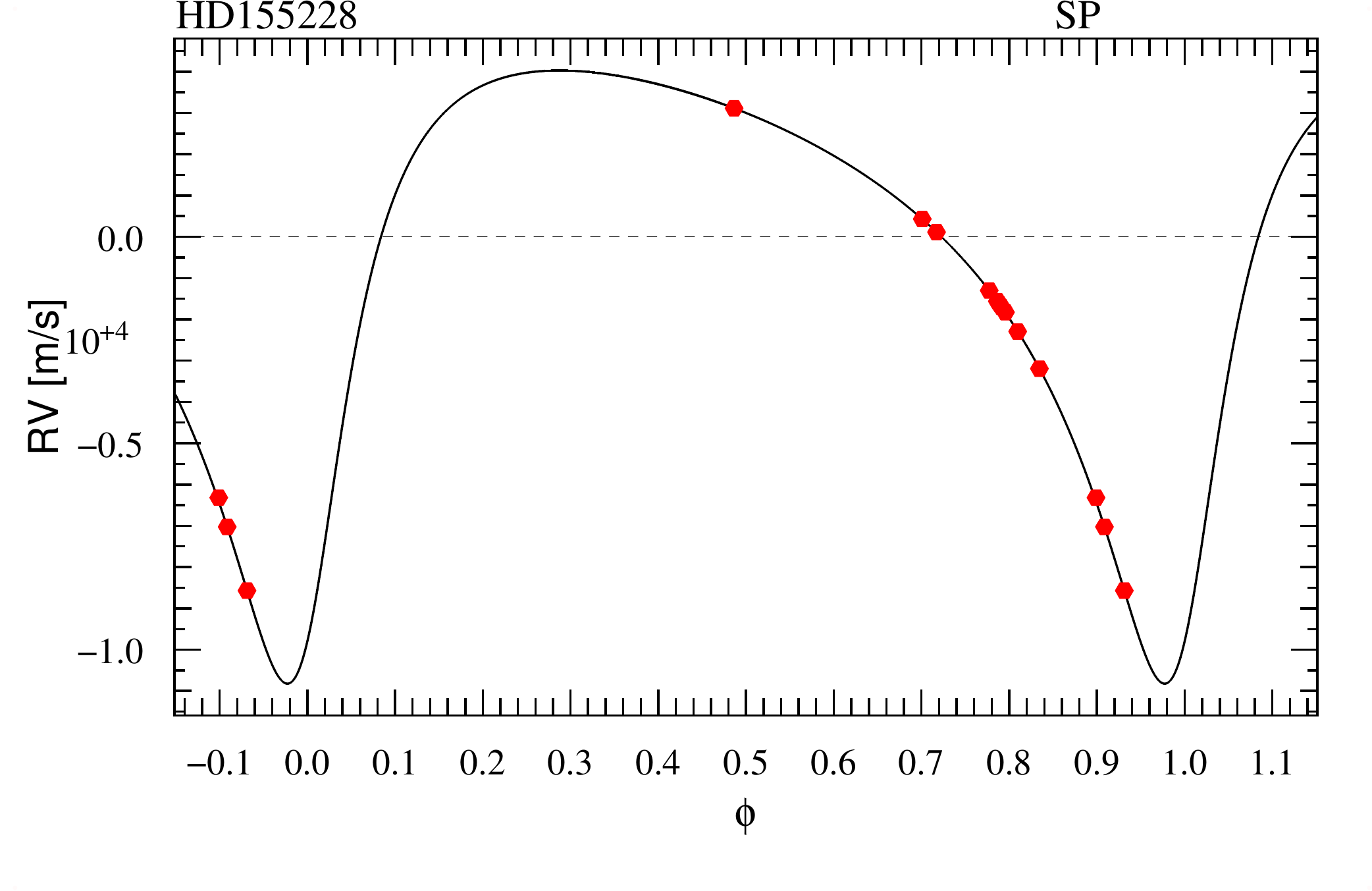} \\
\includegraphics[height=58mm, clip=true, trim=0 -12 0 7]{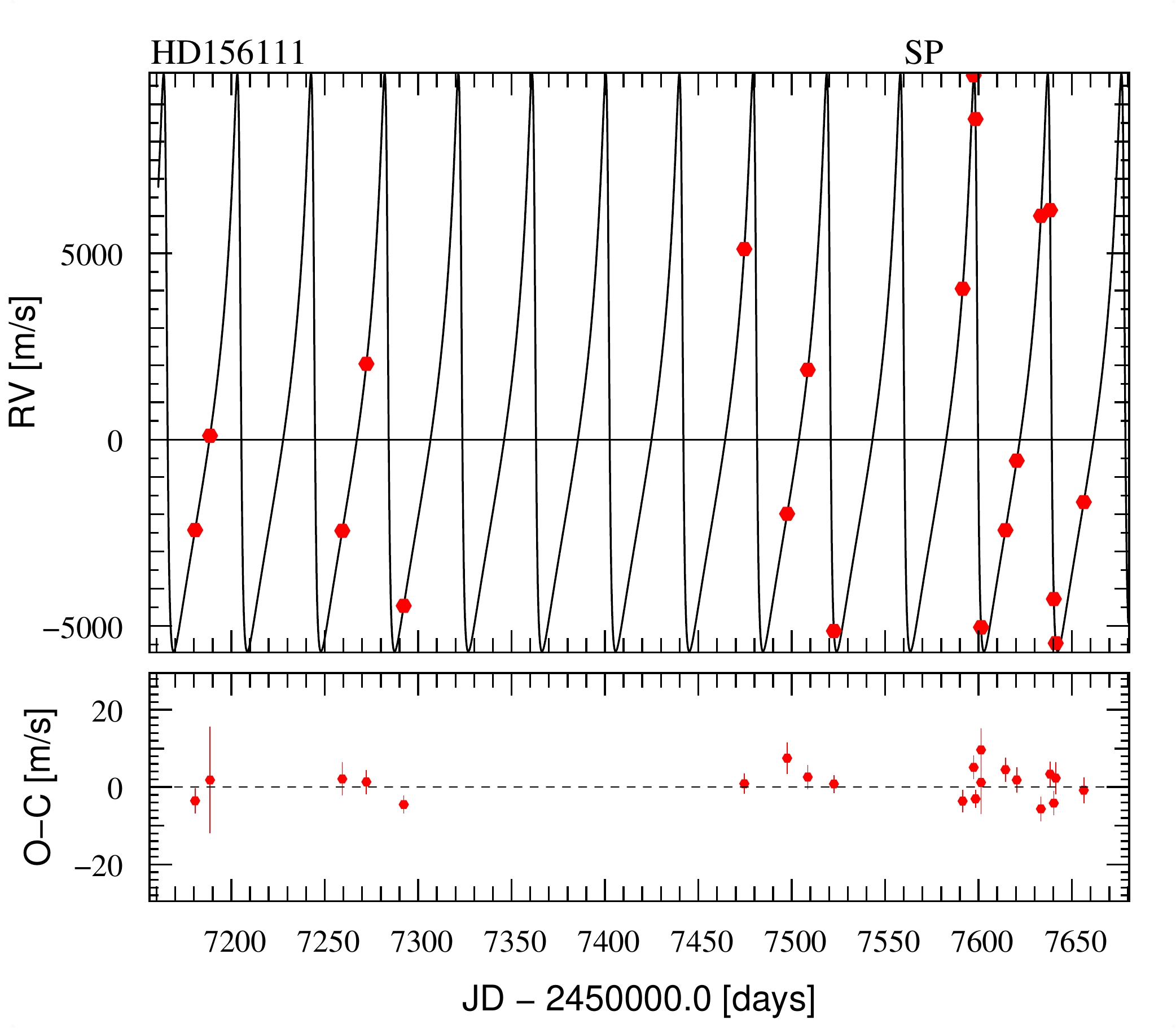}
\includegraphics[height=57mm, clip=true, trim=0  25 0 0]{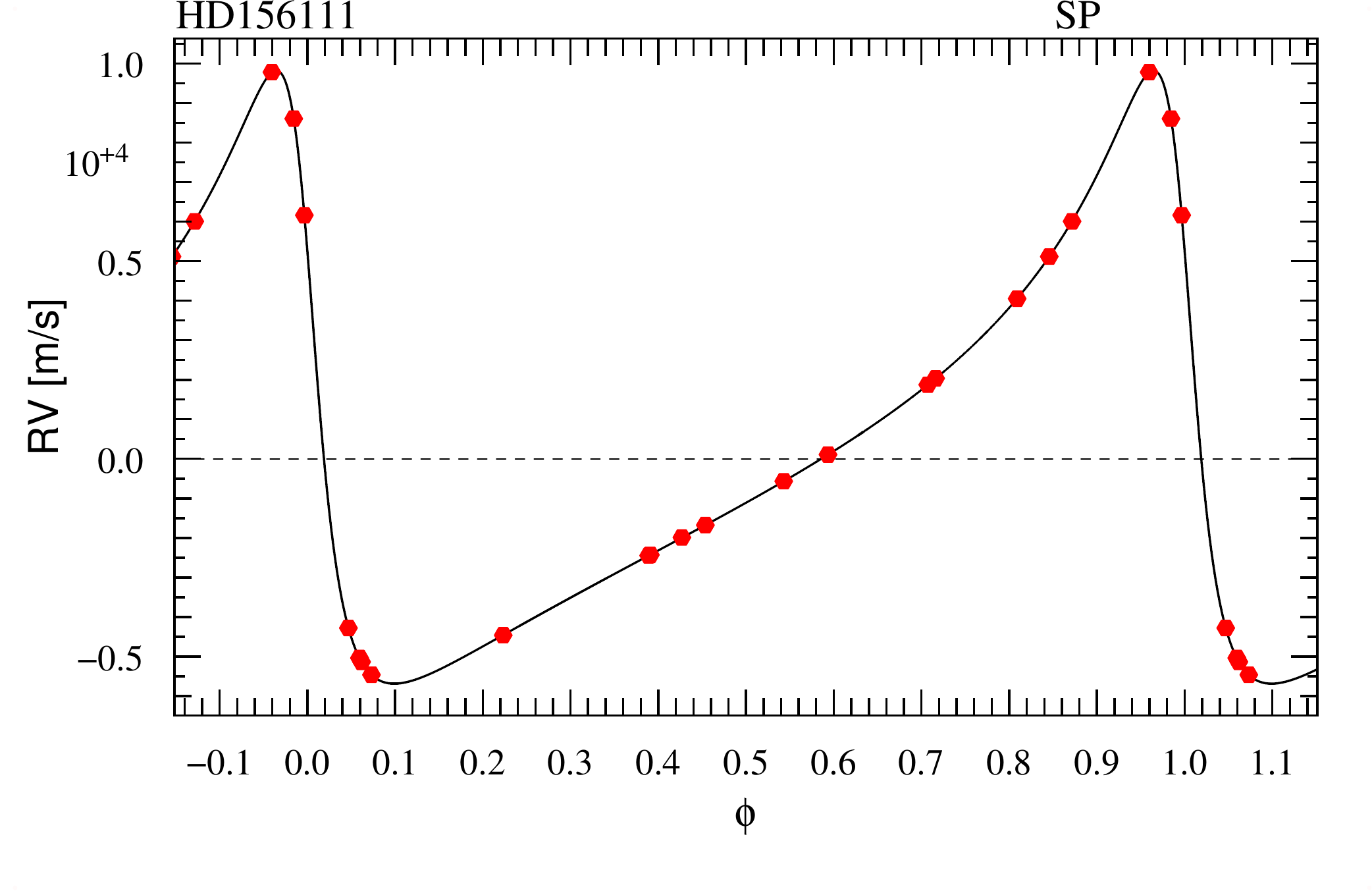} \\
\includegraphics[height=58mm, clip=true, trim=0 -12 0 7]{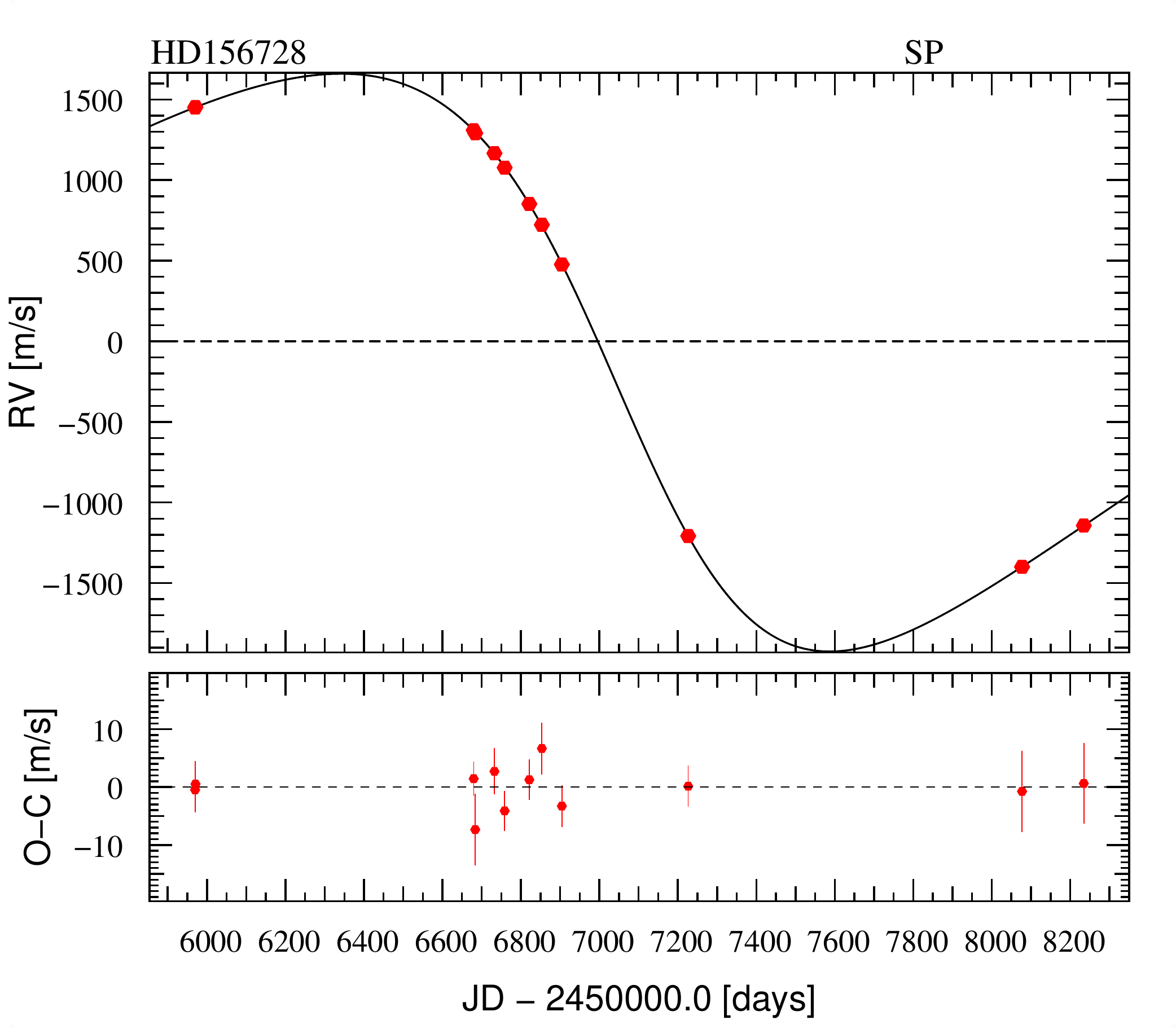} 
\includegraphics[height=57mm, clip=true, trim=0  25 0 0]{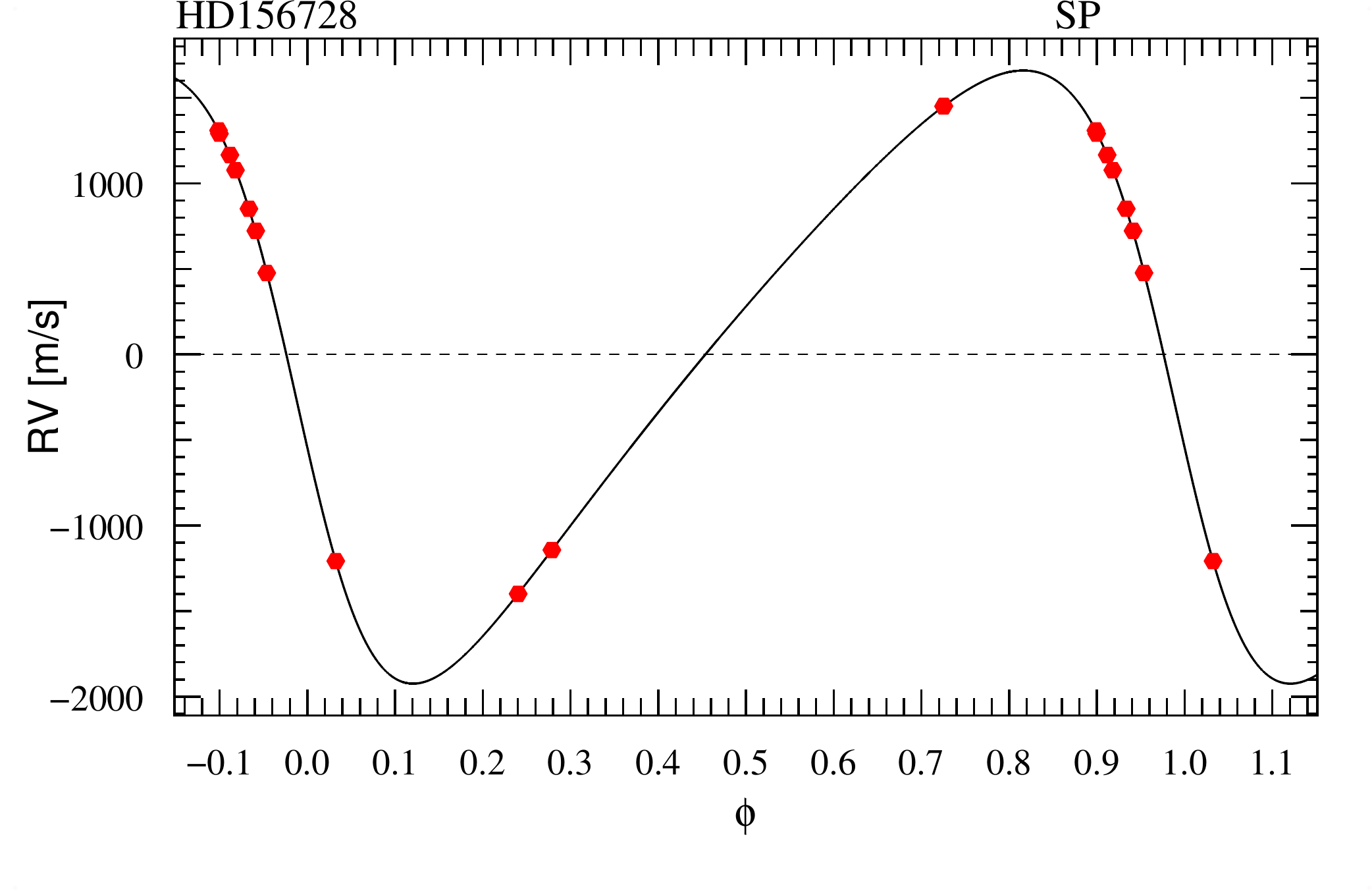} \\
\includegraphics[height=58mm, clip=true, trim=0 -12 0 7]{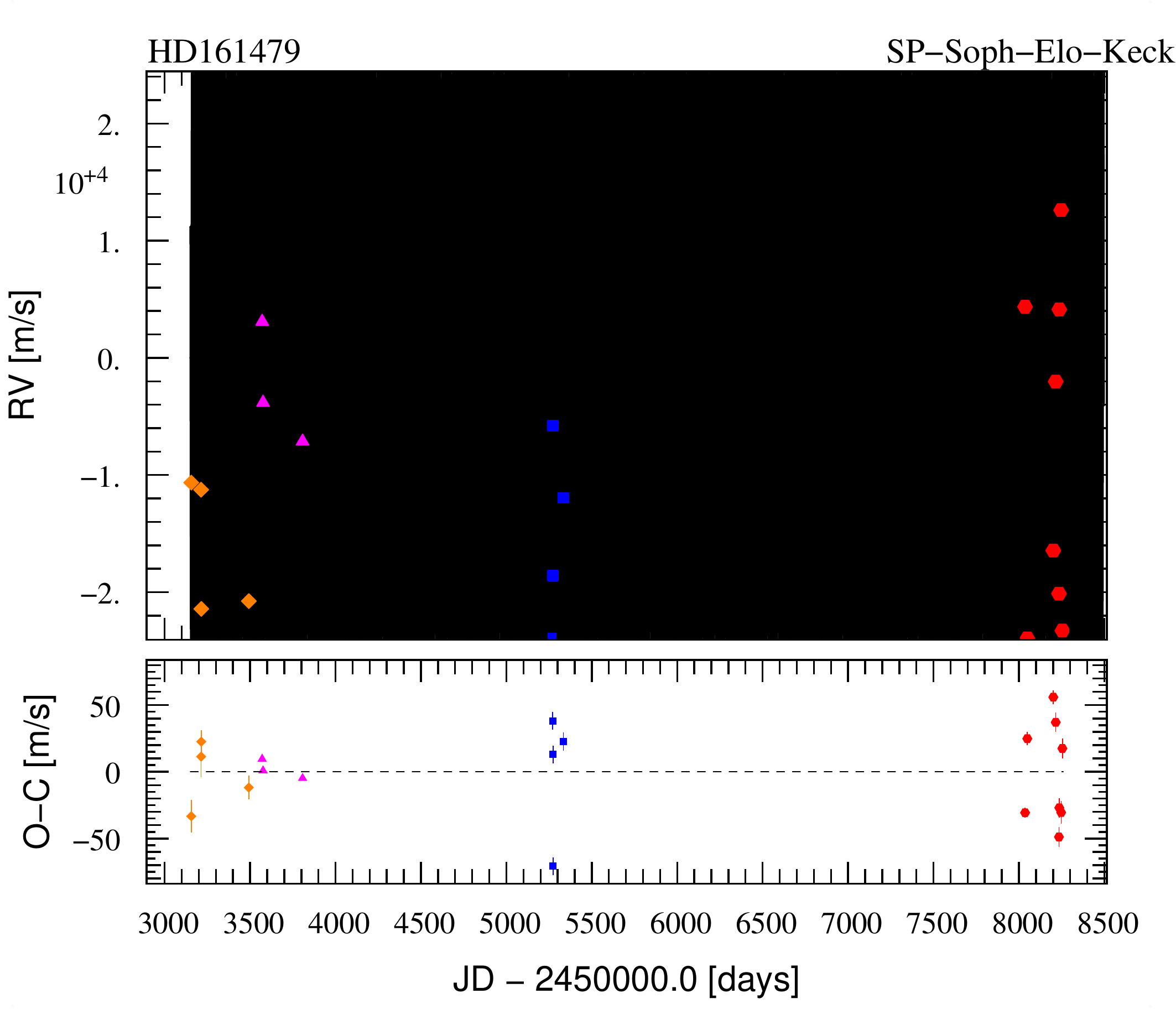} 
\includegraphics[height=57mm, clip=true, trim=0  25 0 0]{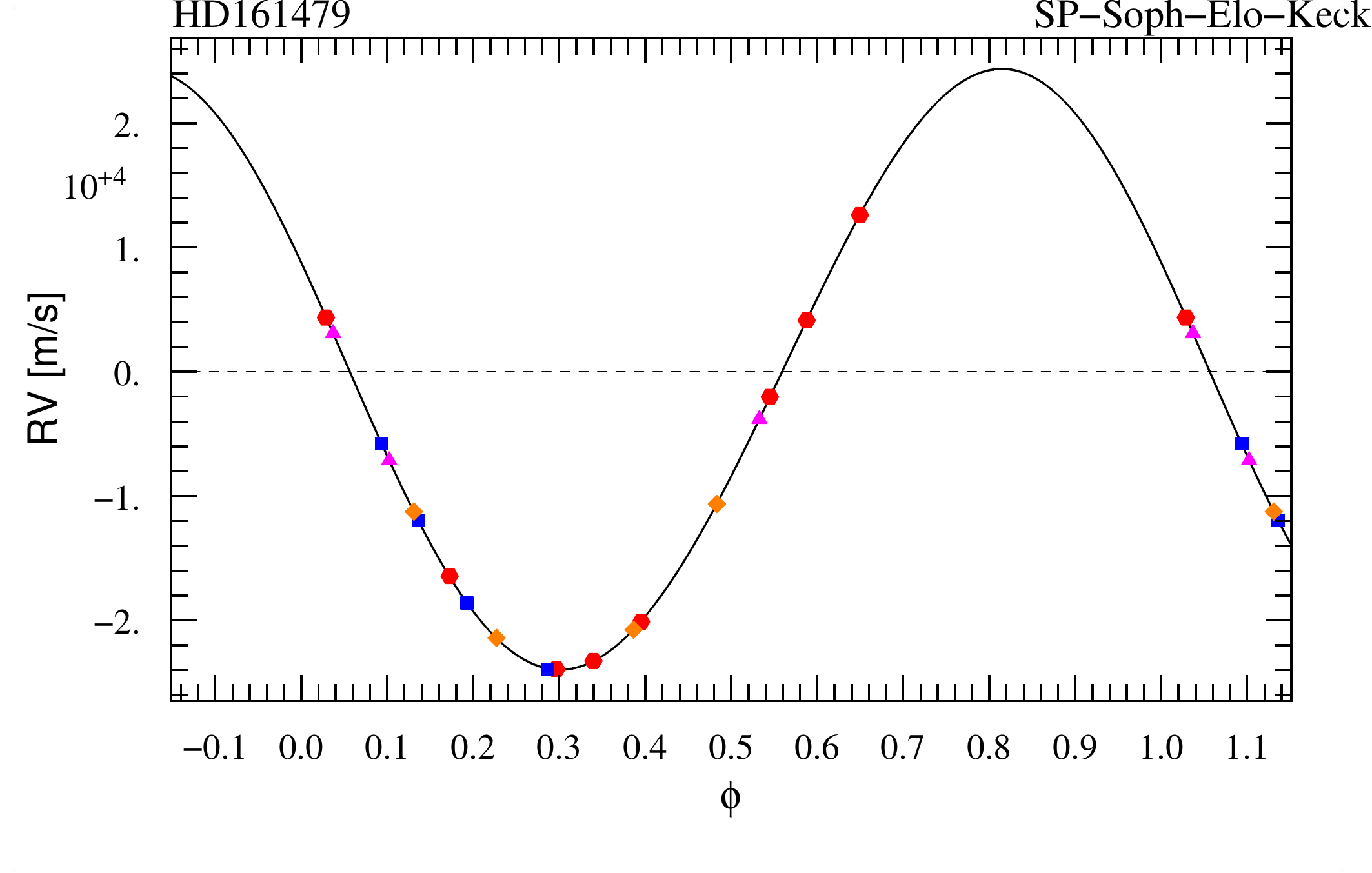}  \\
\includegraphics[height=58mm, clip=true, trim=0 -12 0 7]{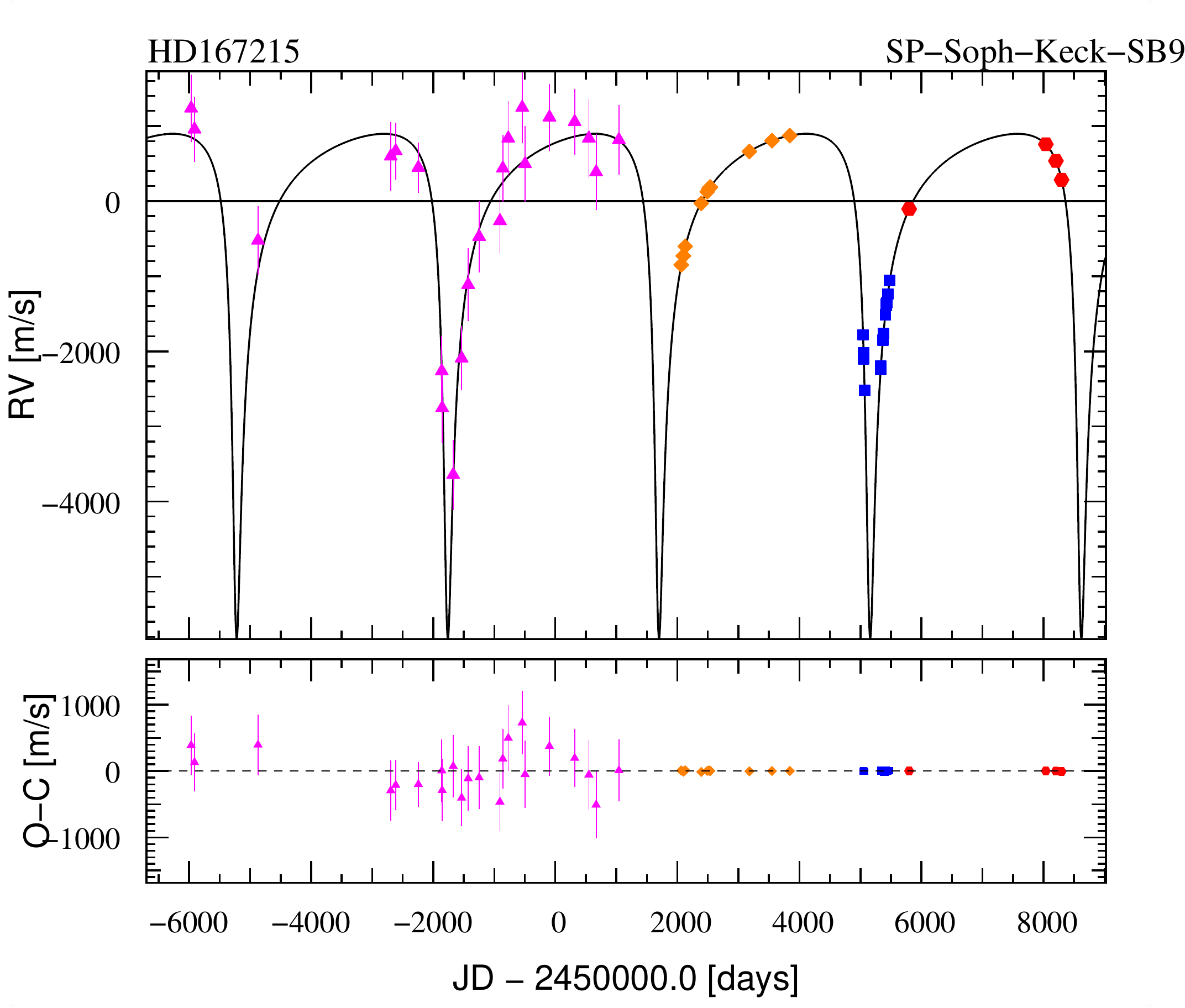}
\includegraphics[height=57mm, clip=true, trim=0  25 0 0]{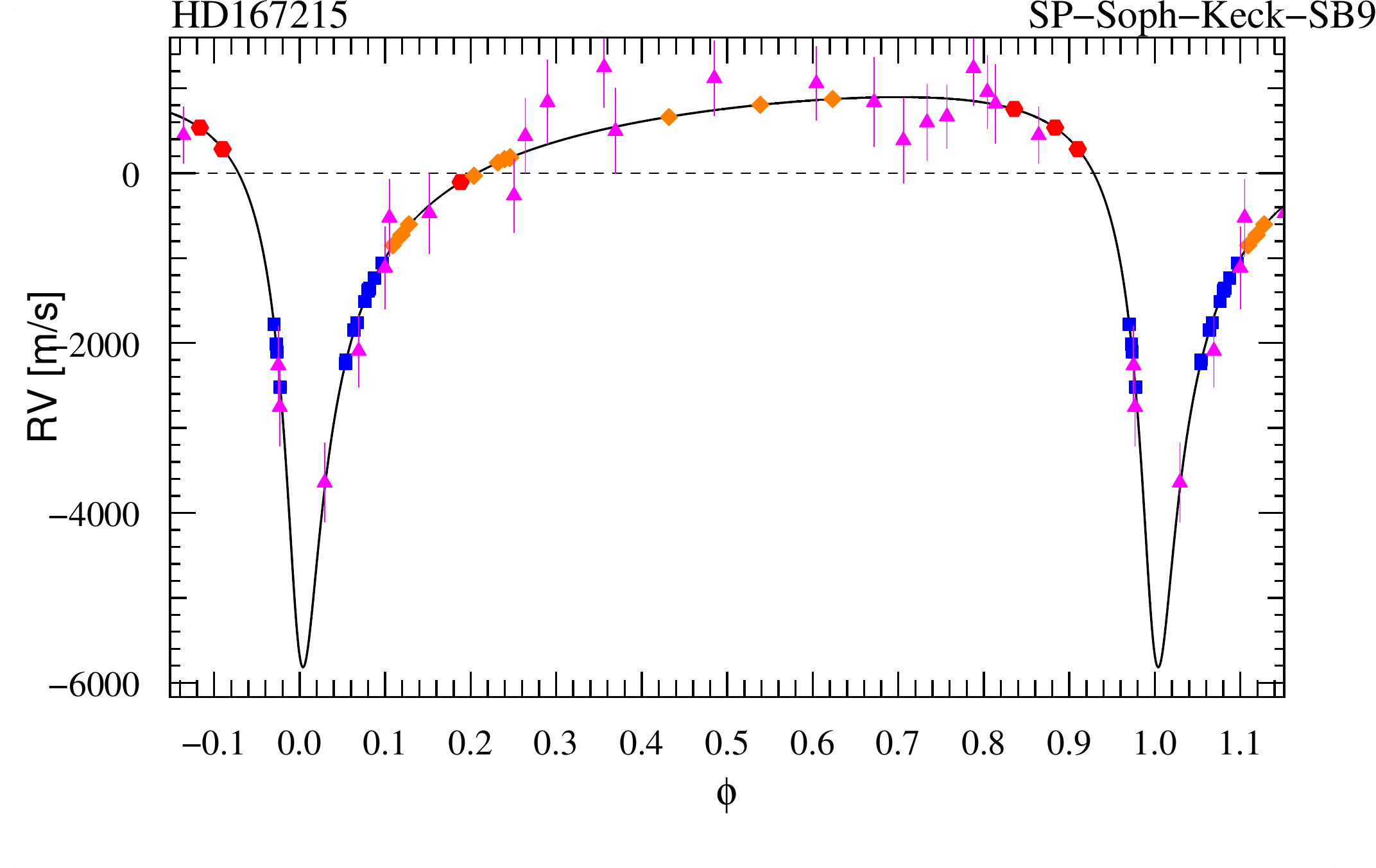} \\
\includegraphics[height=58mm, clip=true, trim=0 -12 0 7]{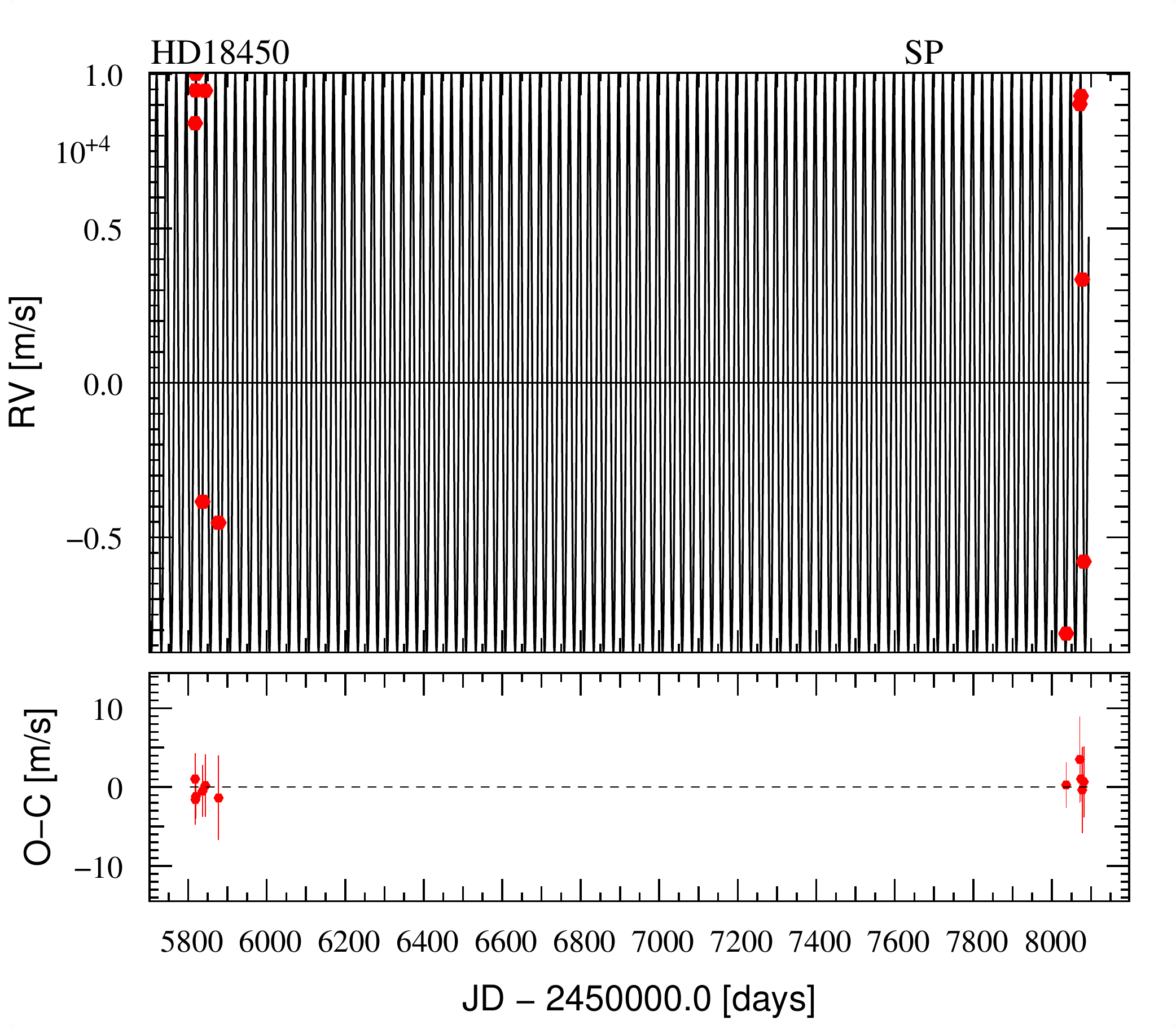}
\includegraphics[height=57mm, clip=true, trim=0  25 0 0]{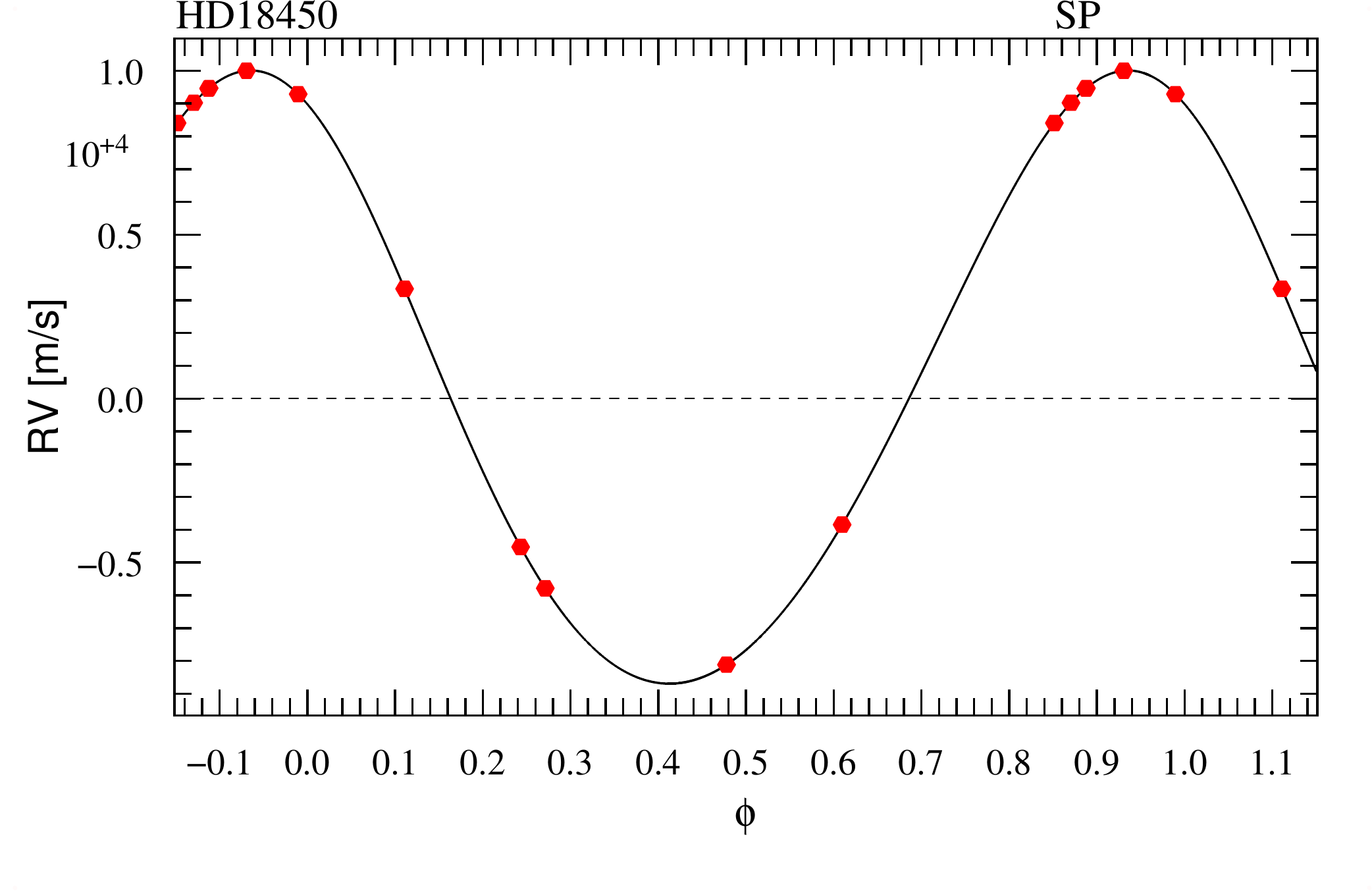} \\
\includegraphics[height=58mm, clip=true, trim=0 -12 0 7]{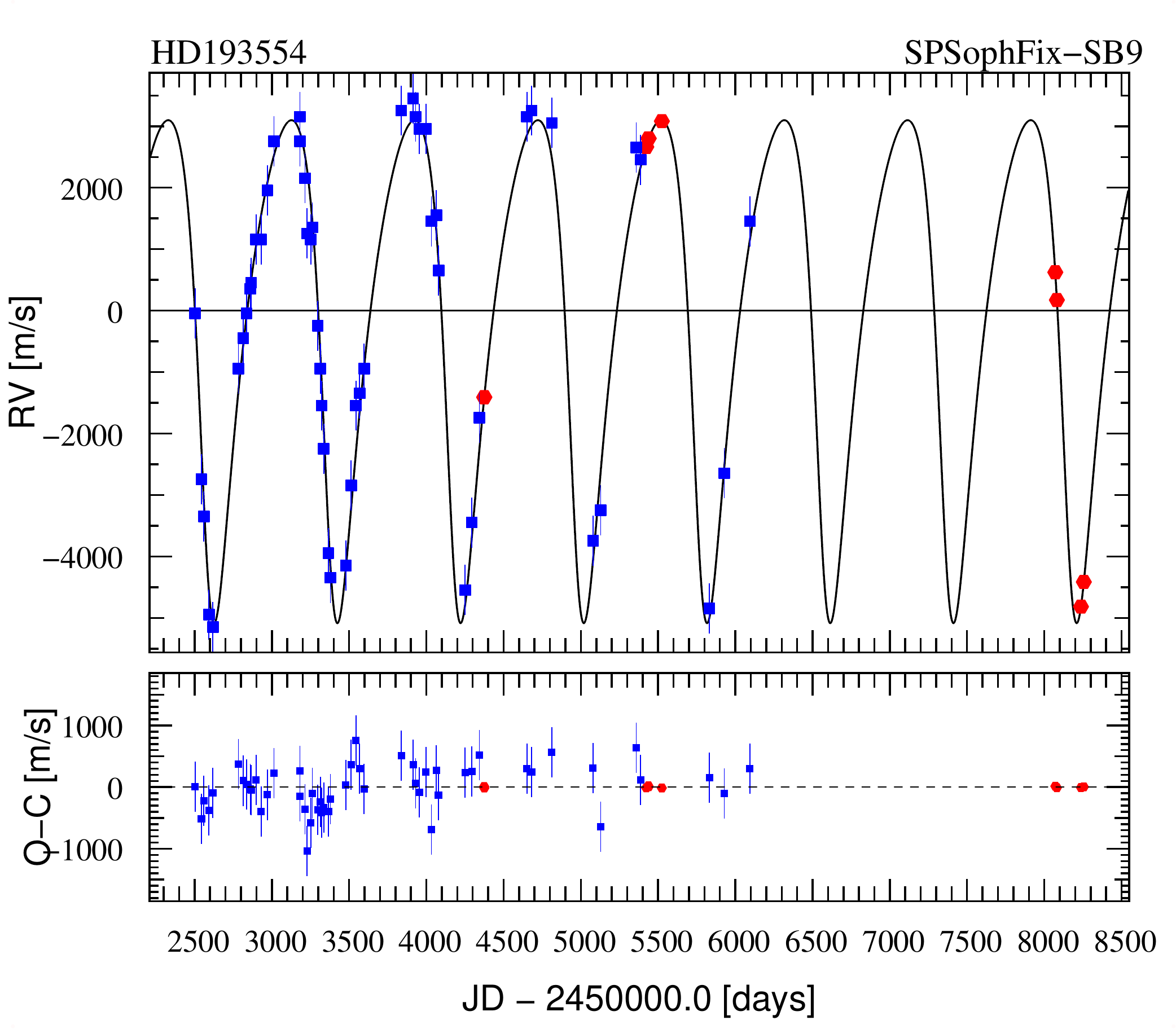} 
\includegraphics[height=57mm, clip=true, trim=0  25 0 0]{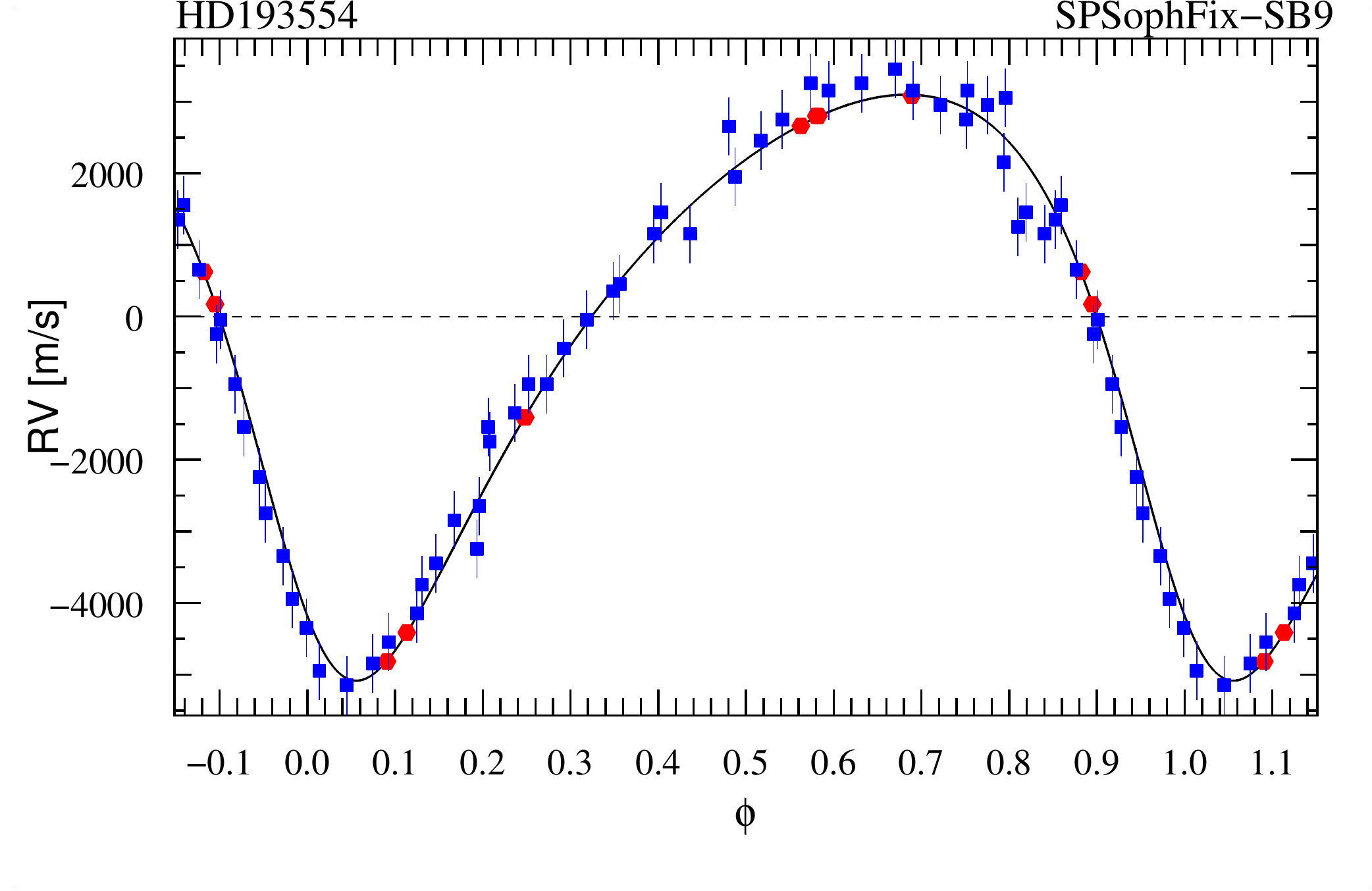} \\
\includegraphics[height=58mm, clip=true, trim=0 -12 0 7]{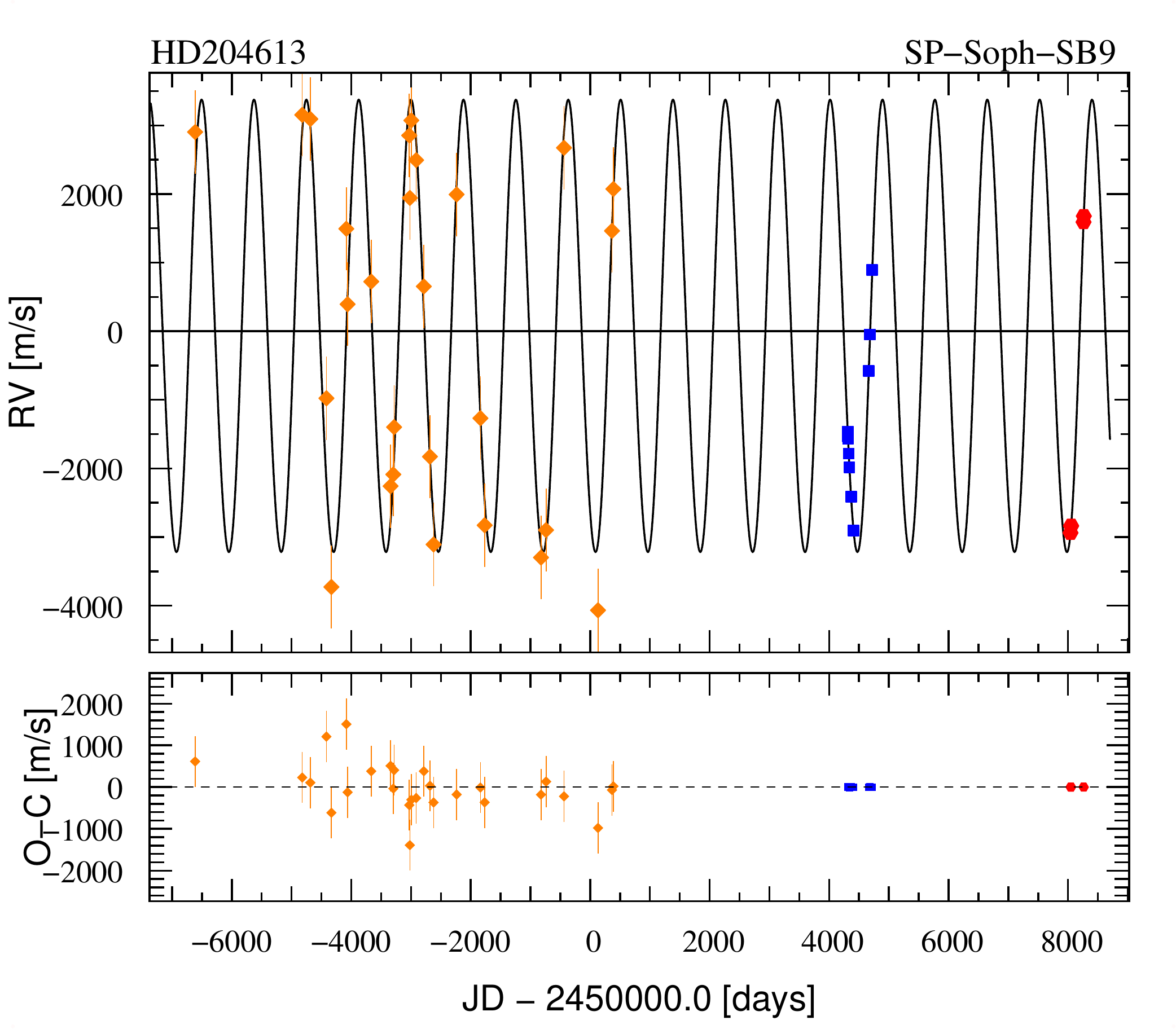}
\includegraphics[height=57mm, clip=true, trim=0  25 0 0]{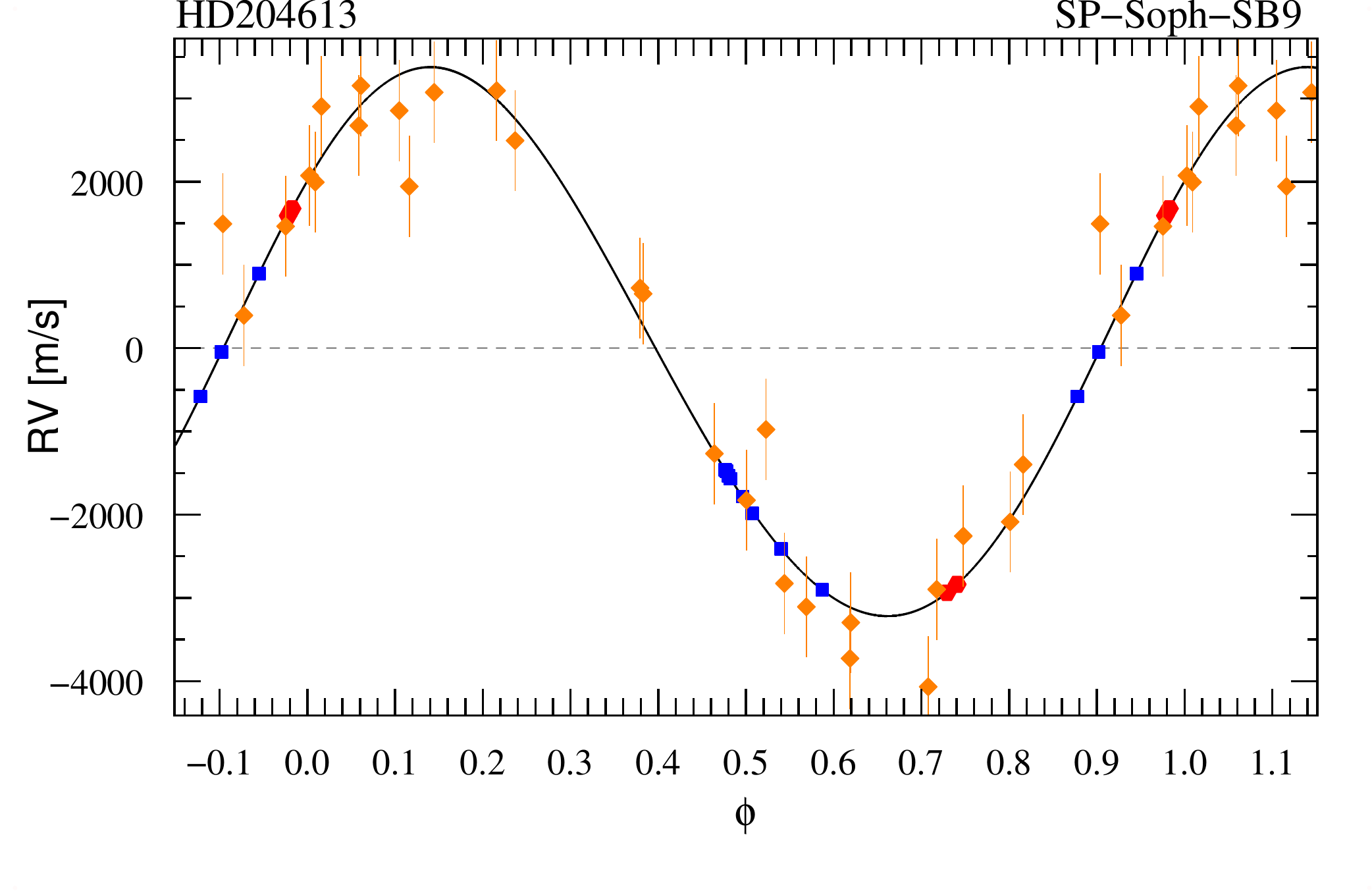} \\
\includegraphics[height=58mm, clip=true, trim=0 -12 0 7]{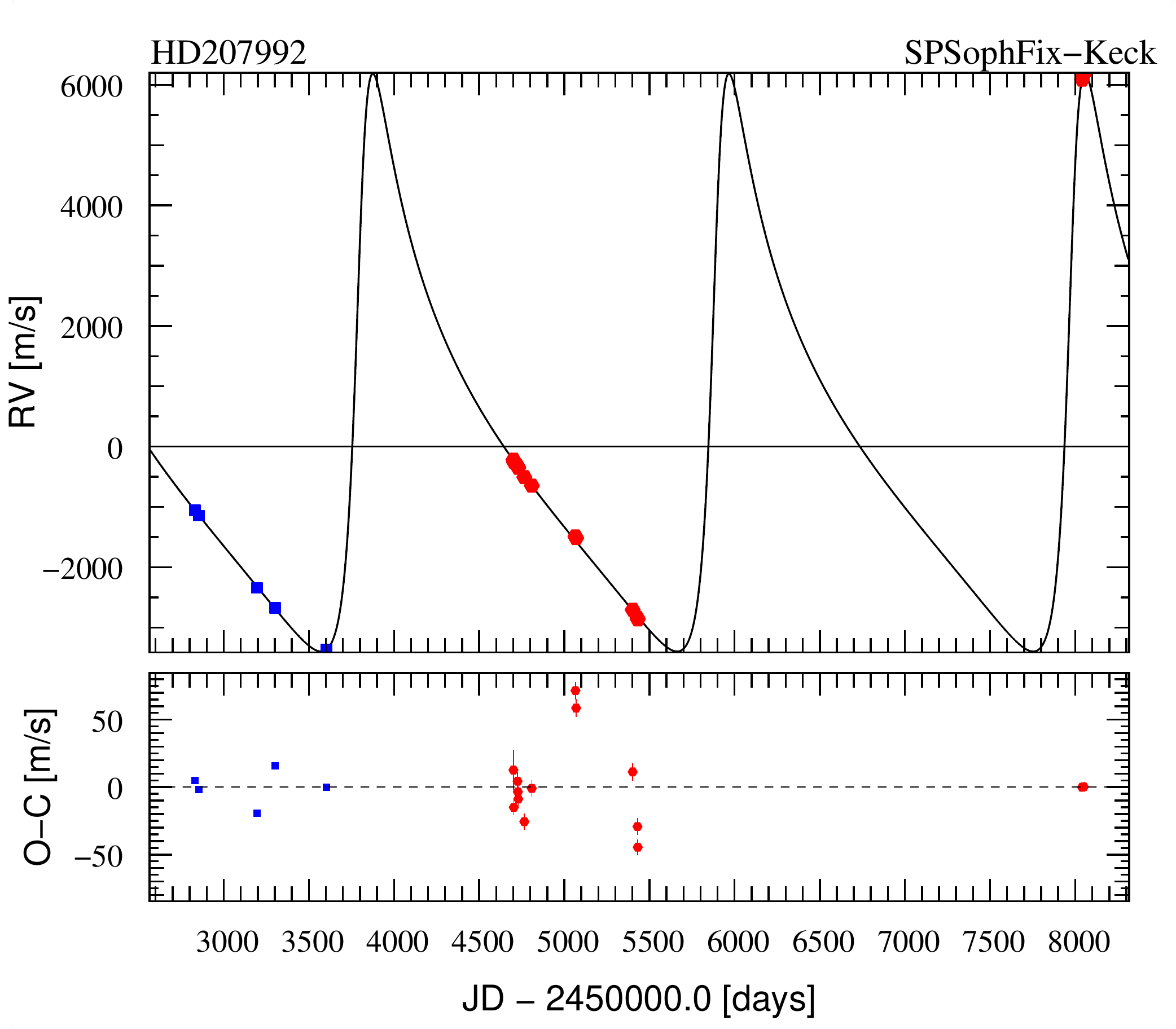}
\includegraphics[height=57mm, clip=true, trim=0  25 0 0]{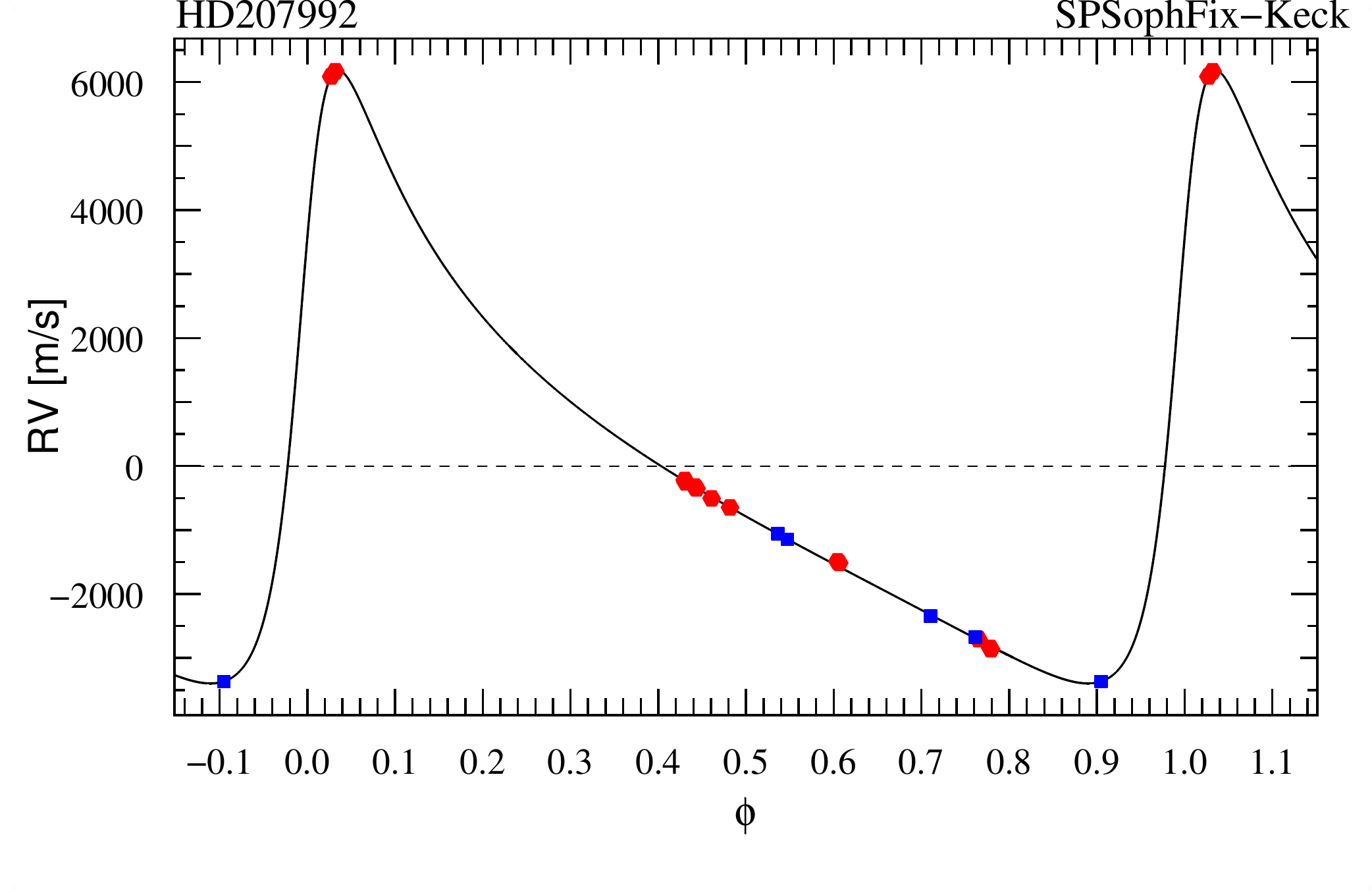} \\
\includegraphics[height=58mm, clip=true, trim=0 -12 0 7]{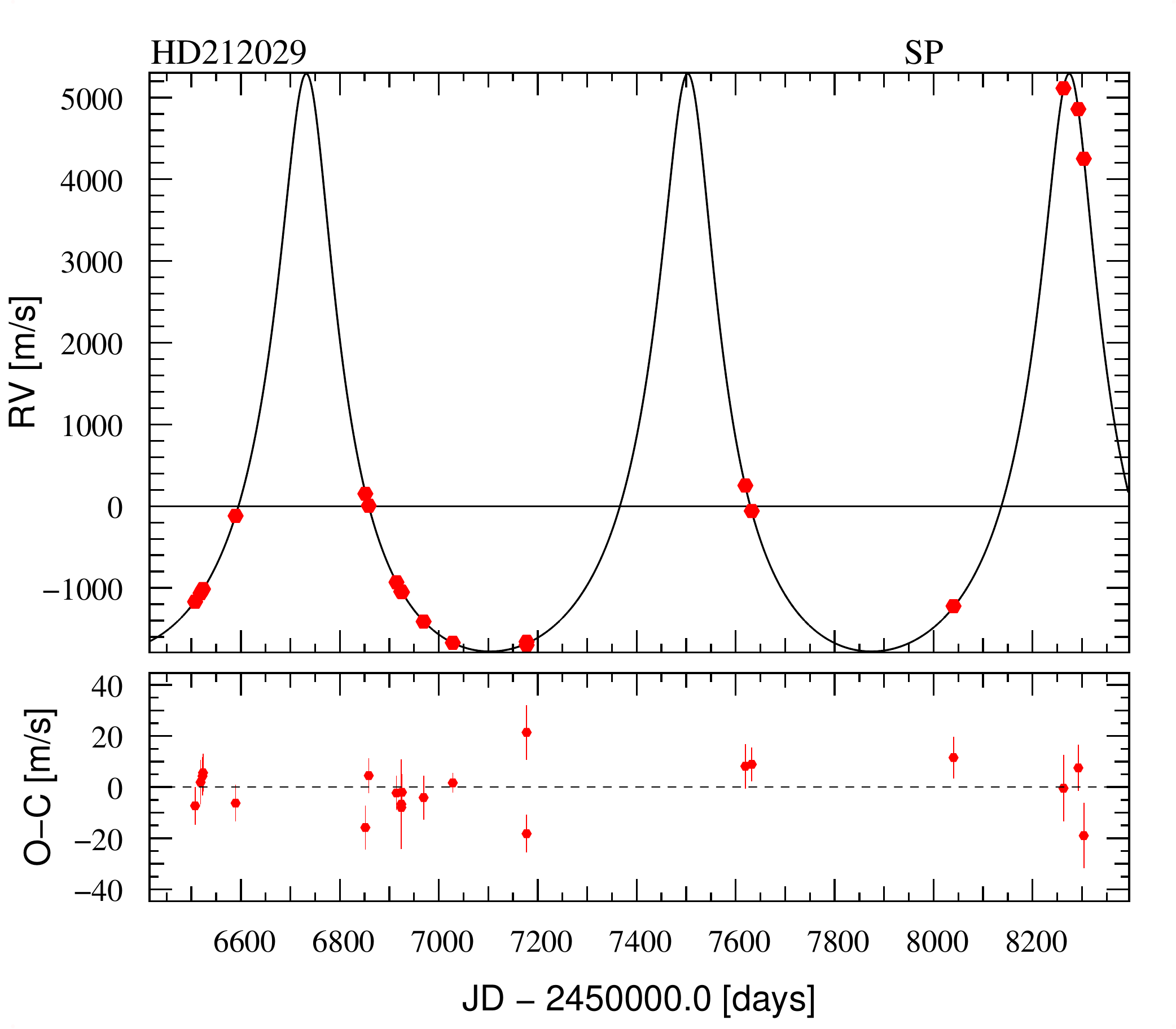} 
\includegraphics[height=57mm, clip=true, trim=0  25 0 0]{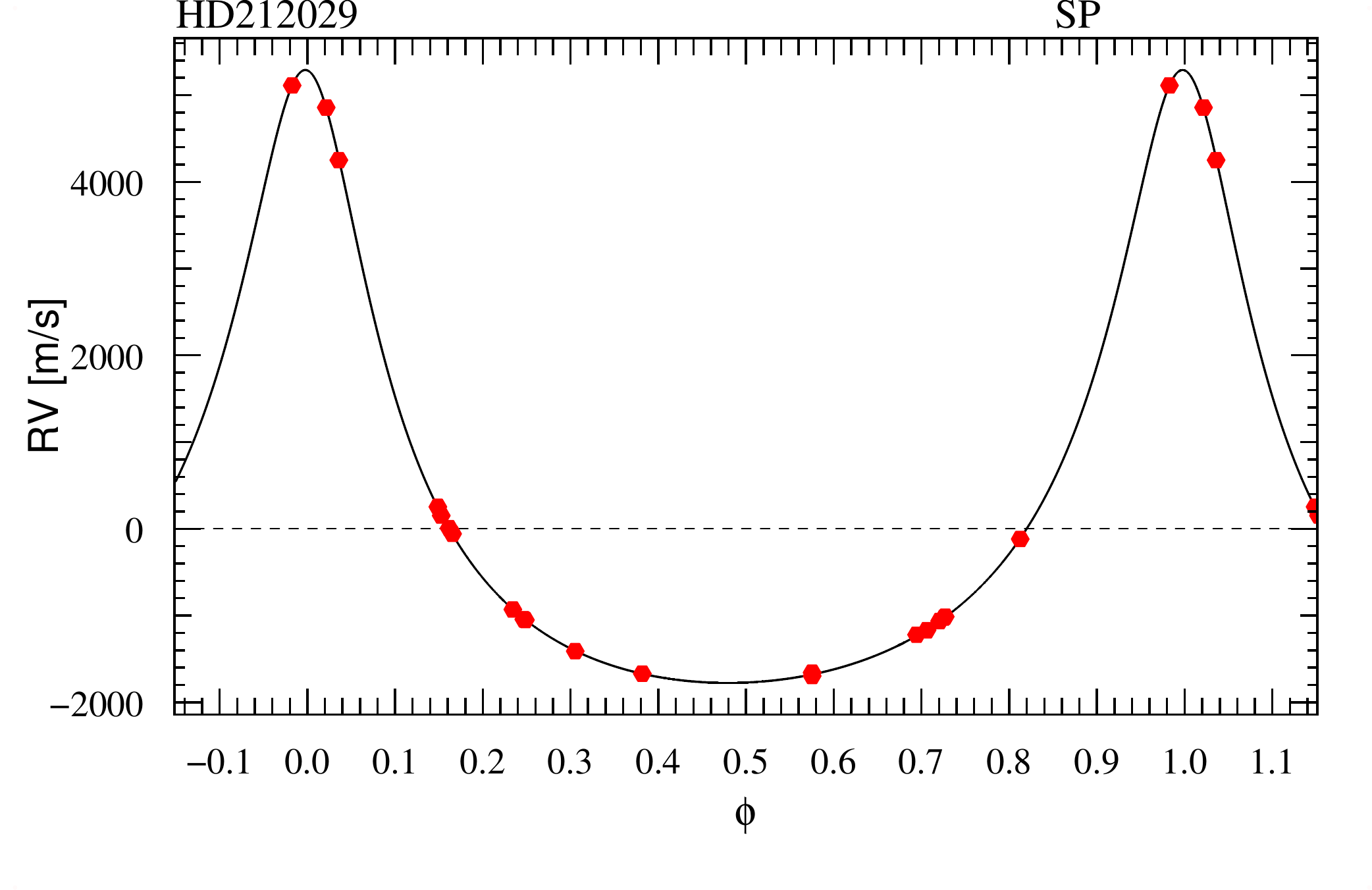} \\
\includegraphics[height=58mm, clip=true, trim=0 -12 0 7]{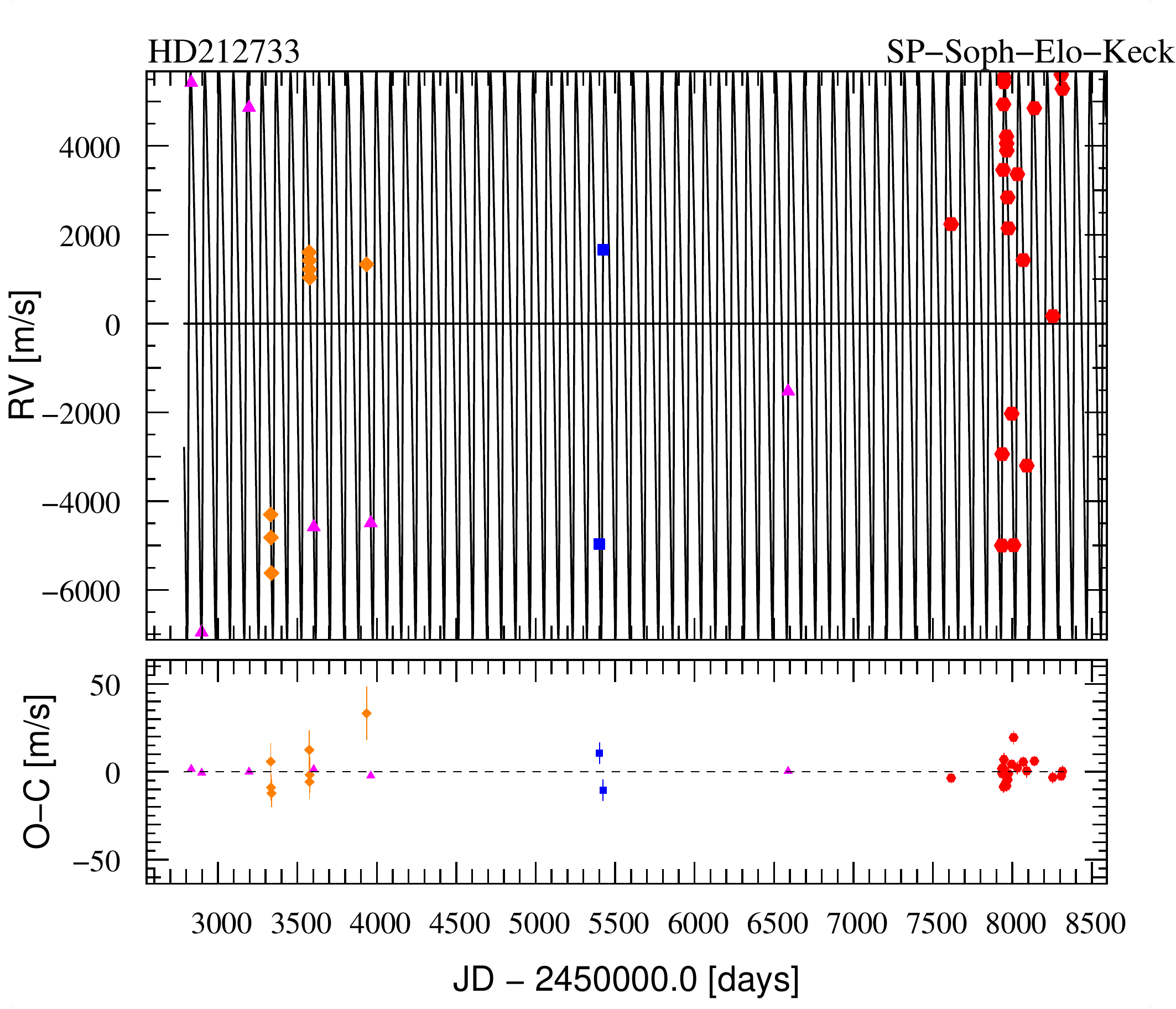} 
\includegraphics[height=57mm, clip=true, trim=0  25 0 0]{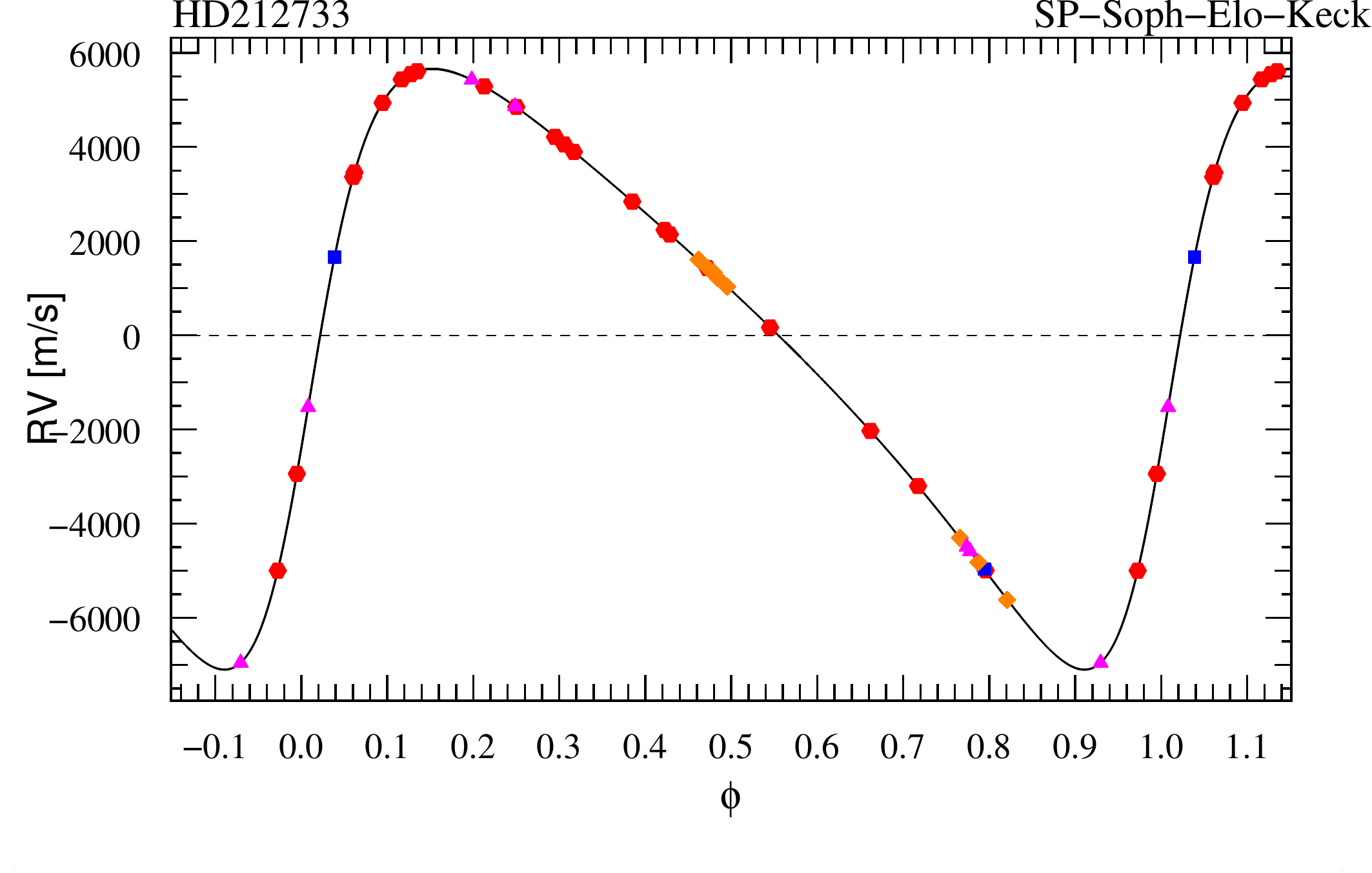} \\
\includegraphics[height=58mm, clip=true, trim=0 -12 0 7]{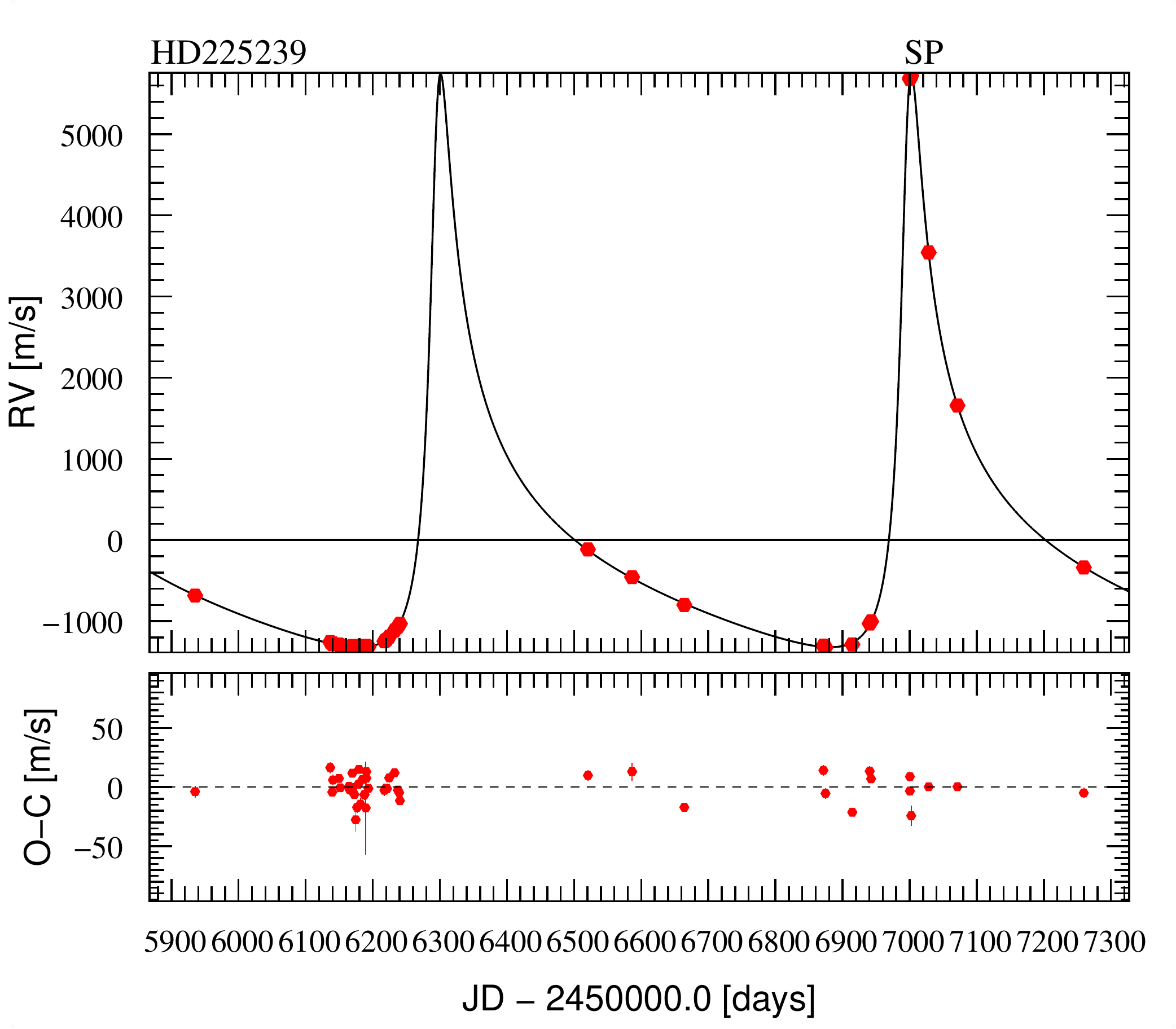} 
\includegraphics[height=57mm, clip=true, trim=0  25 0 0]{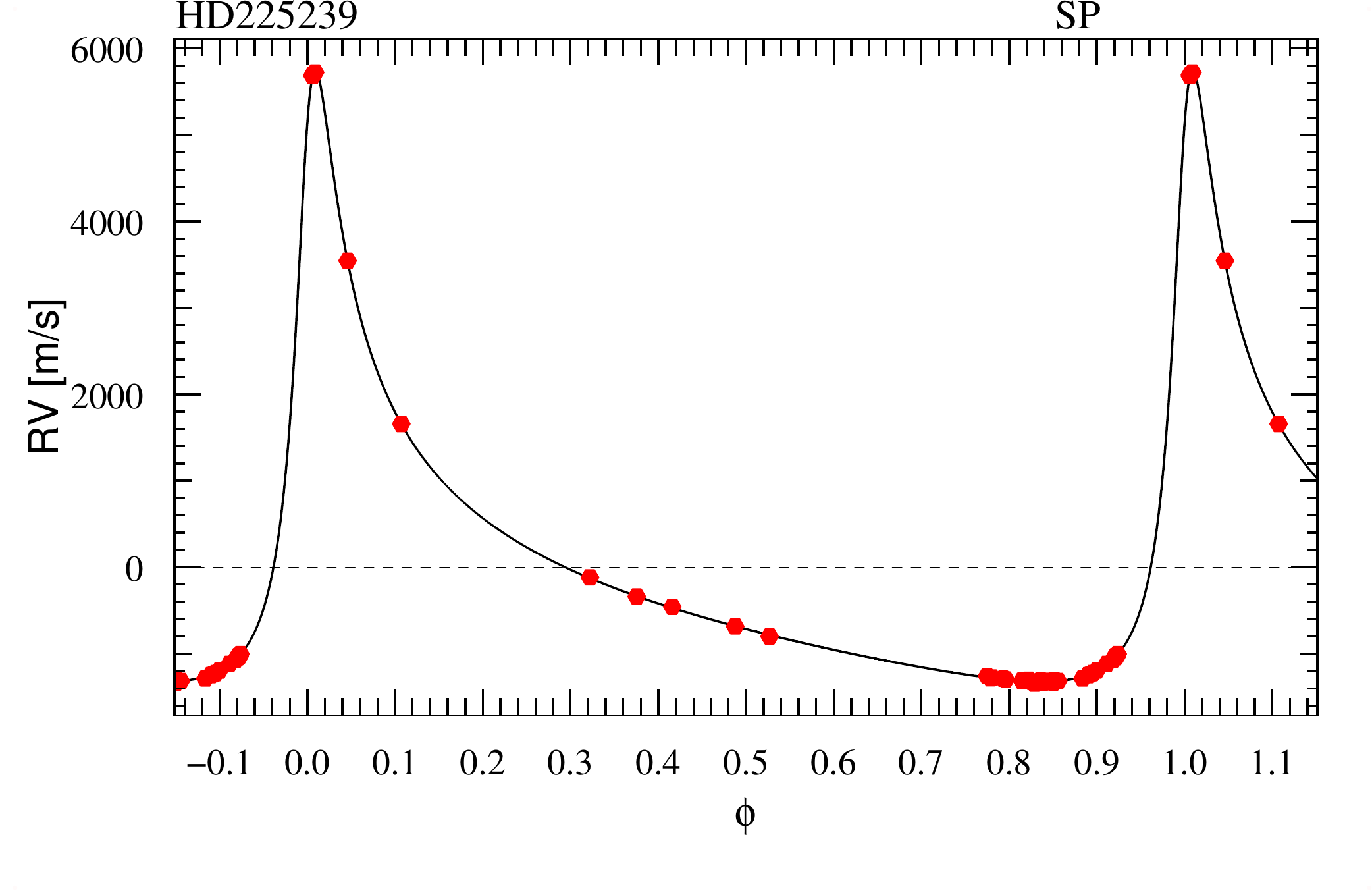} \\
\includegraphics[height=58mm, clip=true, trim=0 -12 0 7]{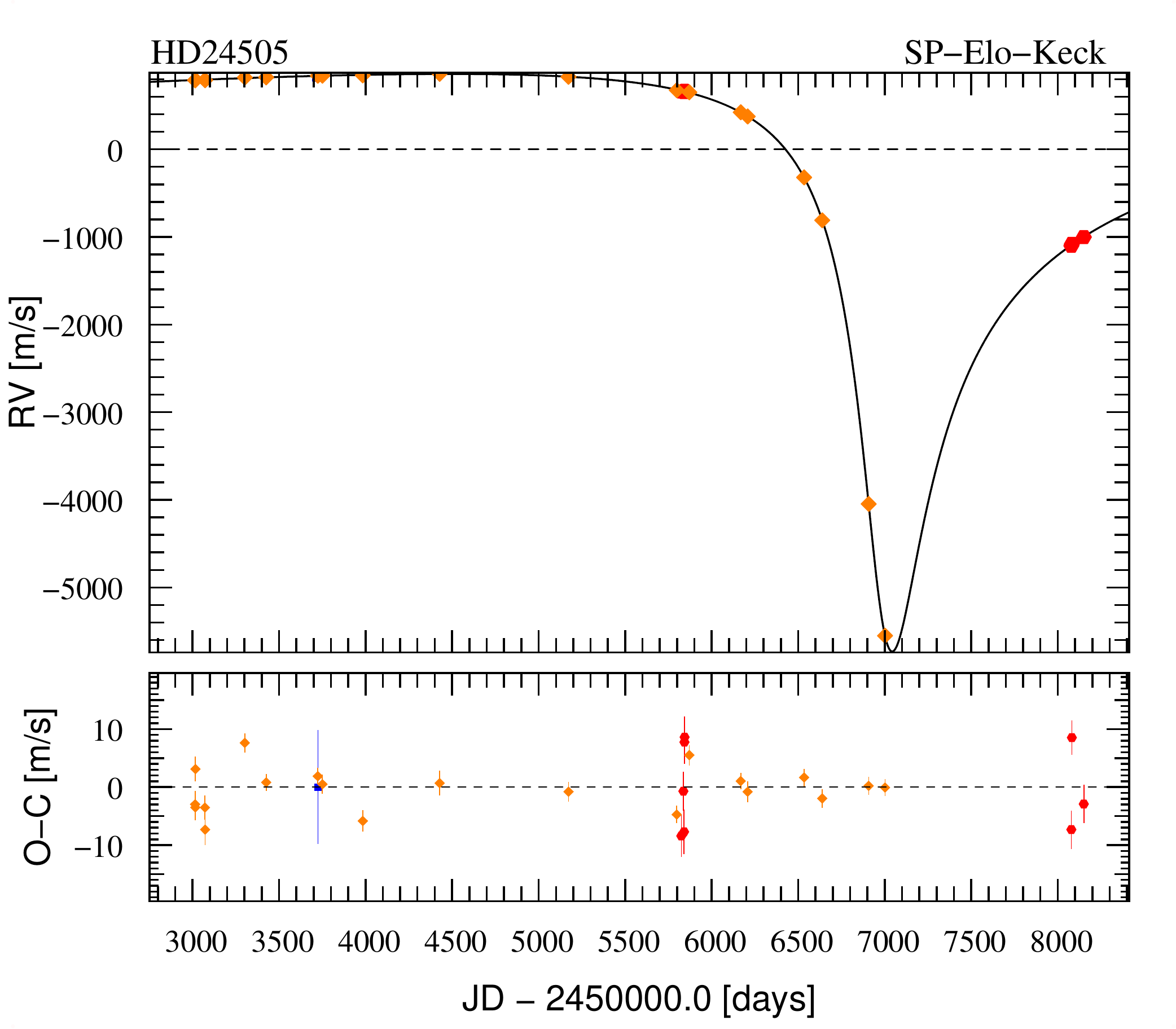}
\includegraphics[height=57mm, clip=true, trim=0  25 0 0]{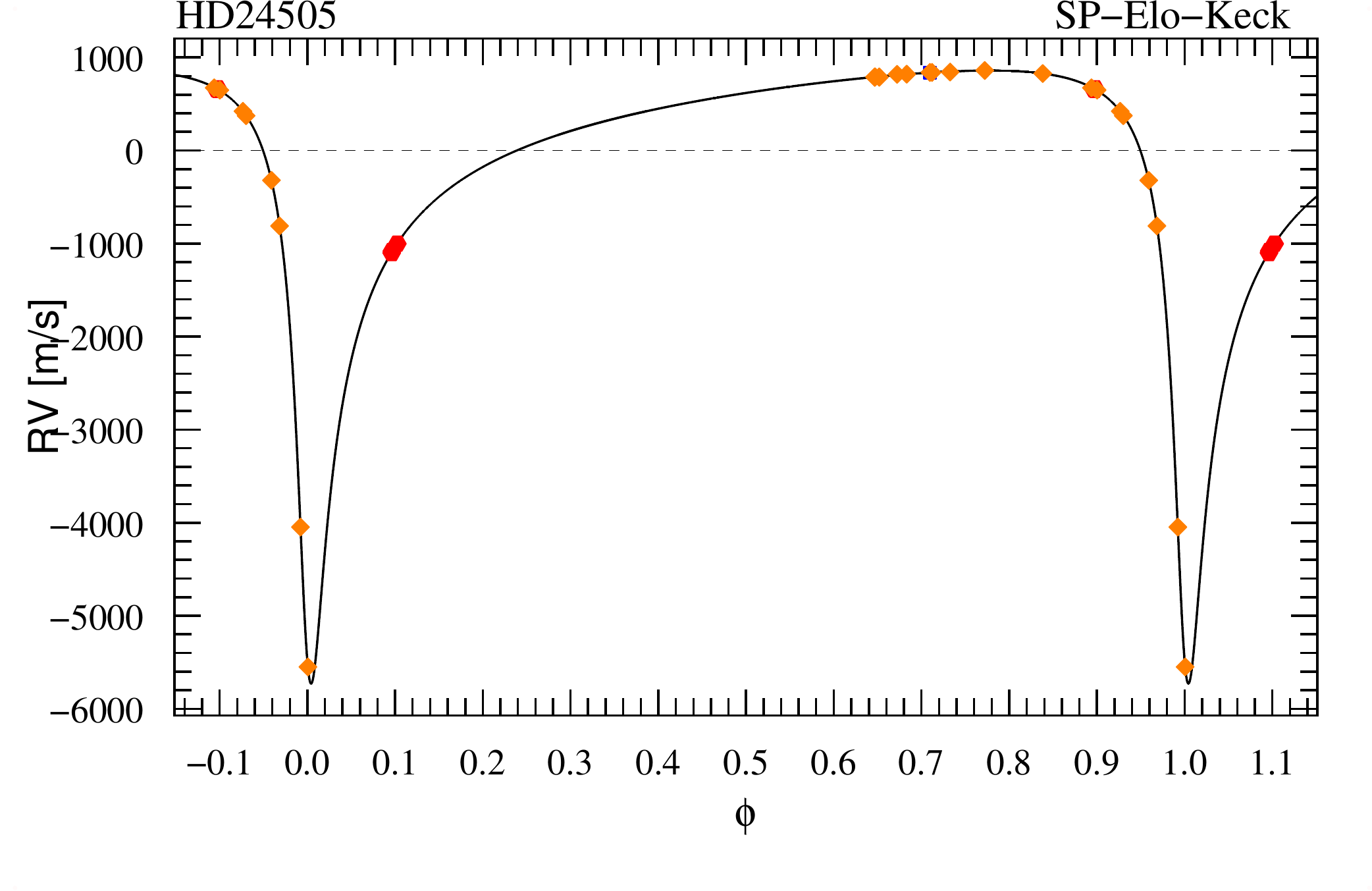} \\
\includegraphics[height=58mm, clip=true, trim=0 -12 0 7]{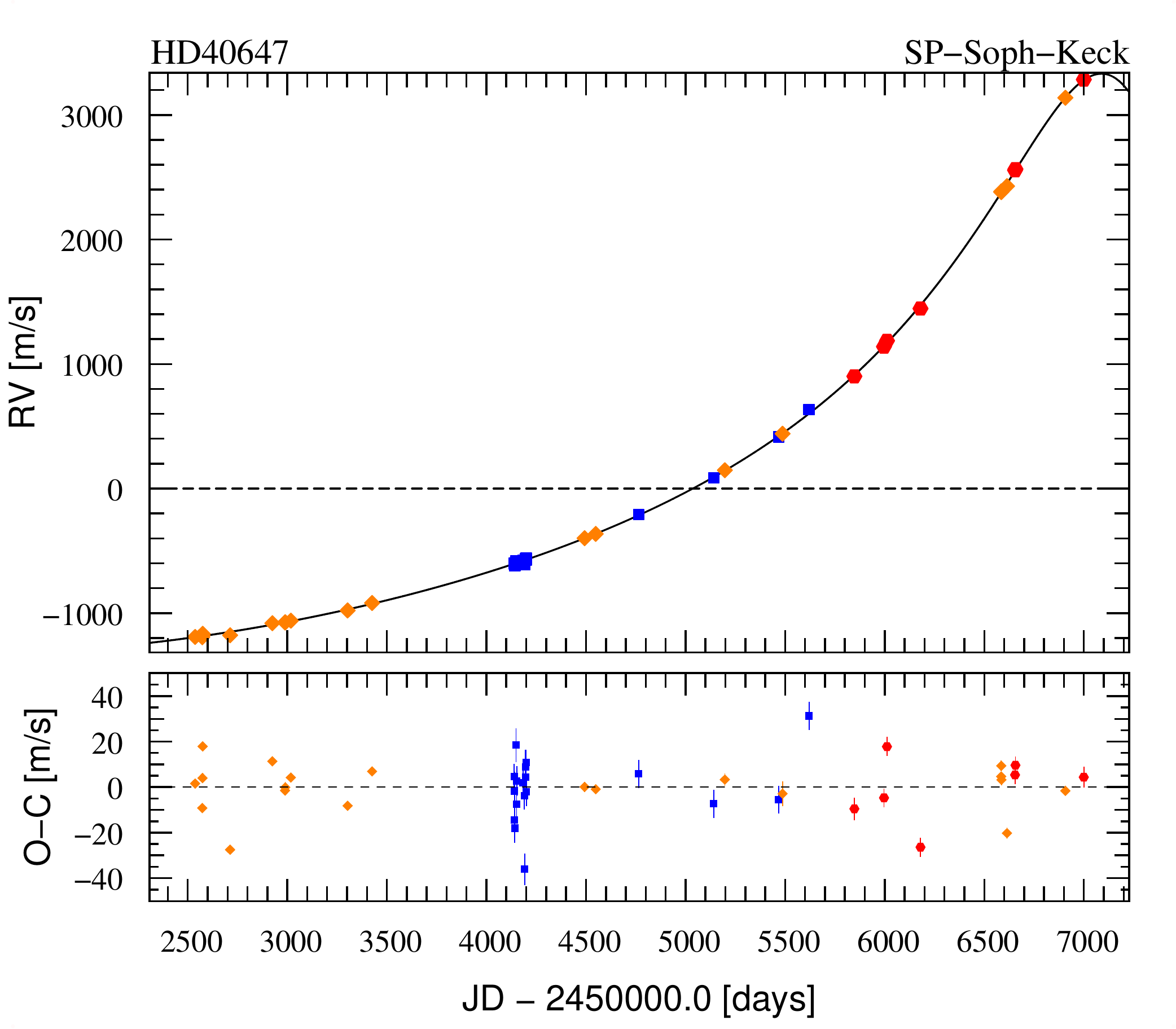}
\includegraphics[height=57mm, clip=true, trim=0  25 0 0]{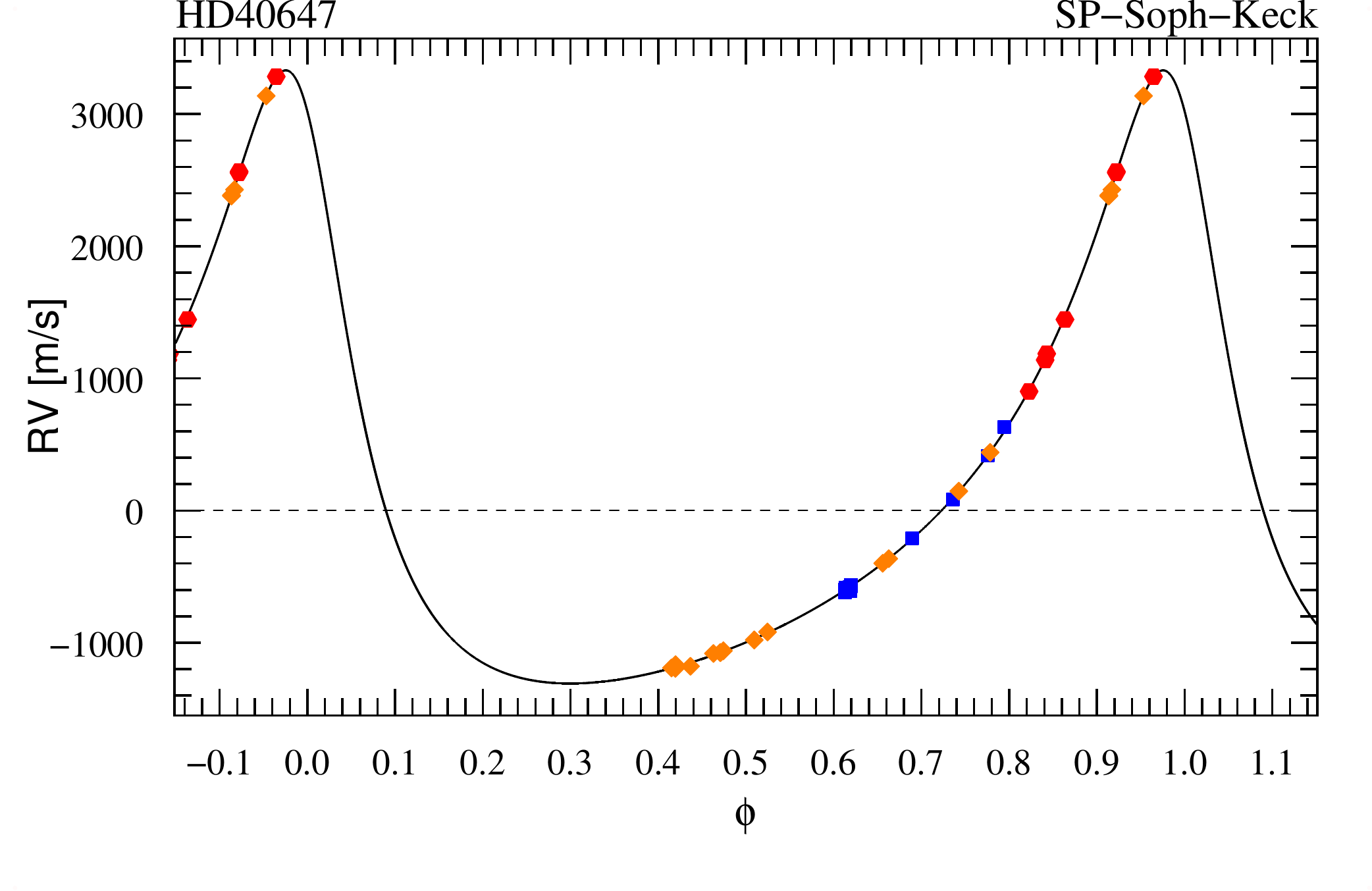} \\
\includegraphics[height=58mm, clip=true, trim=0 -12 0 7]{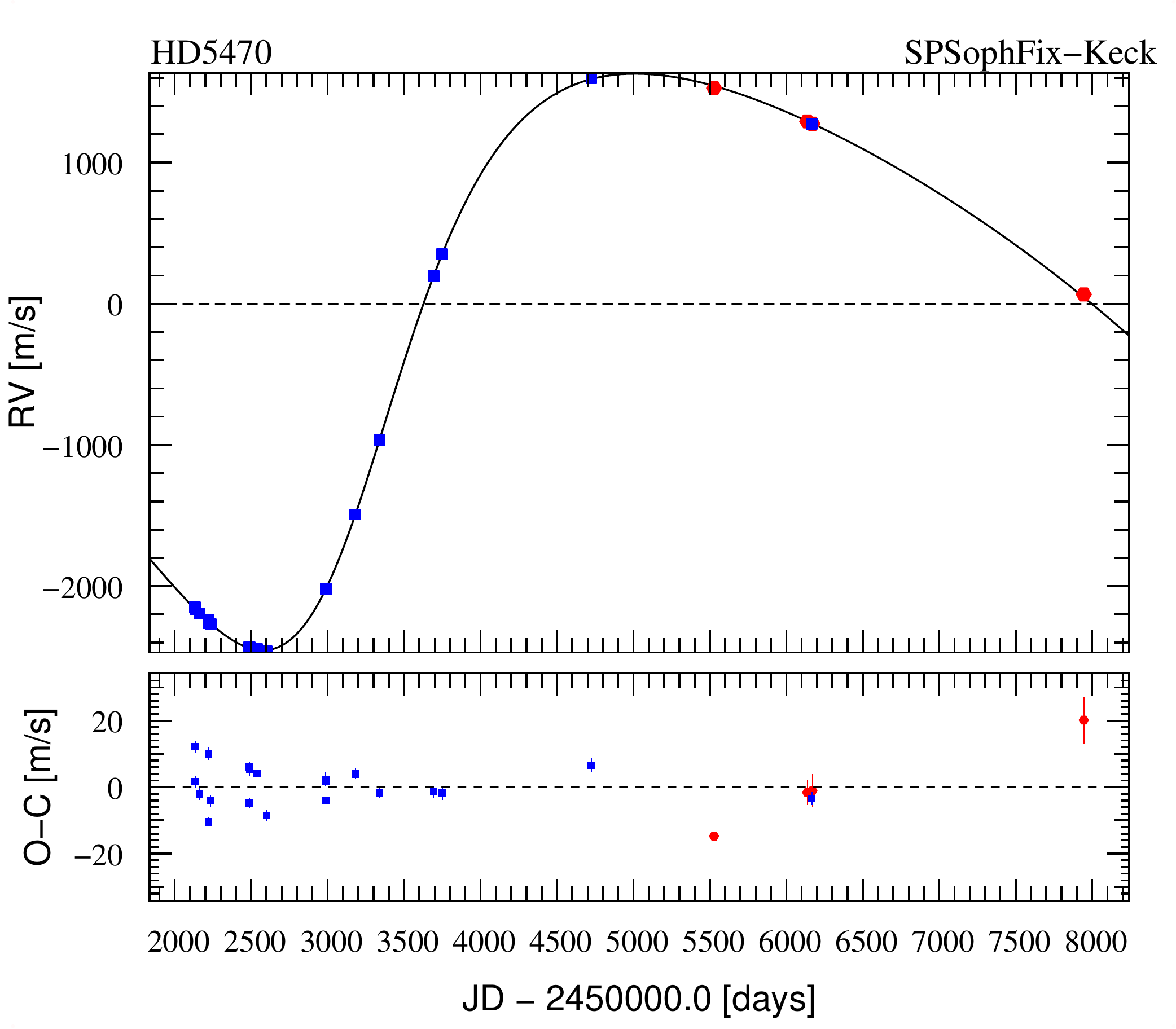}
\includegraphics[height=57mm, clip=true, trim=0  25 0 0]{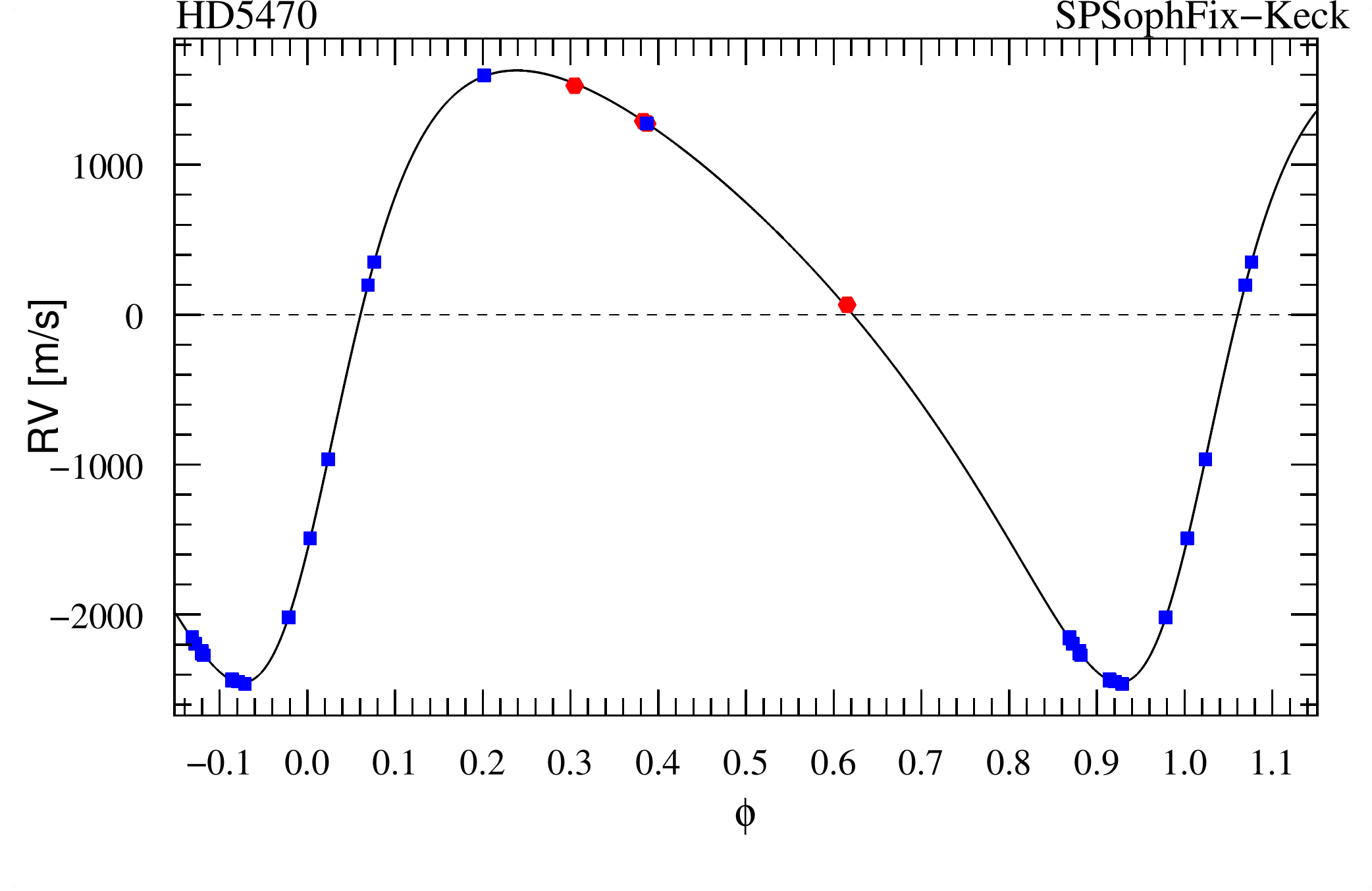} \\
\includegraphics[height=58mm, clip=true, trim=0 -12 0 7]{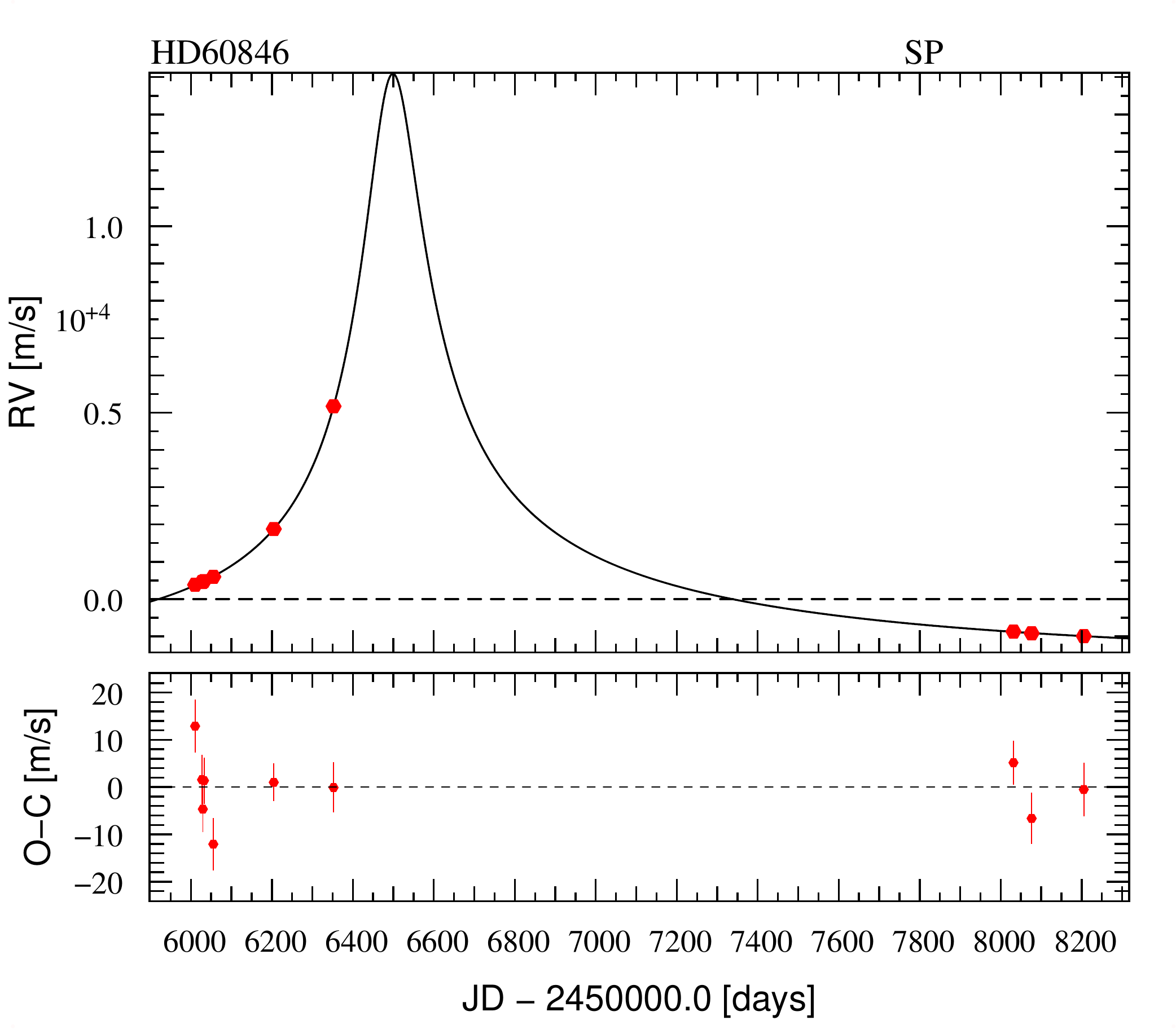}
\includegraphics[height=57mm, clip=true, trim=0  25 0 0]{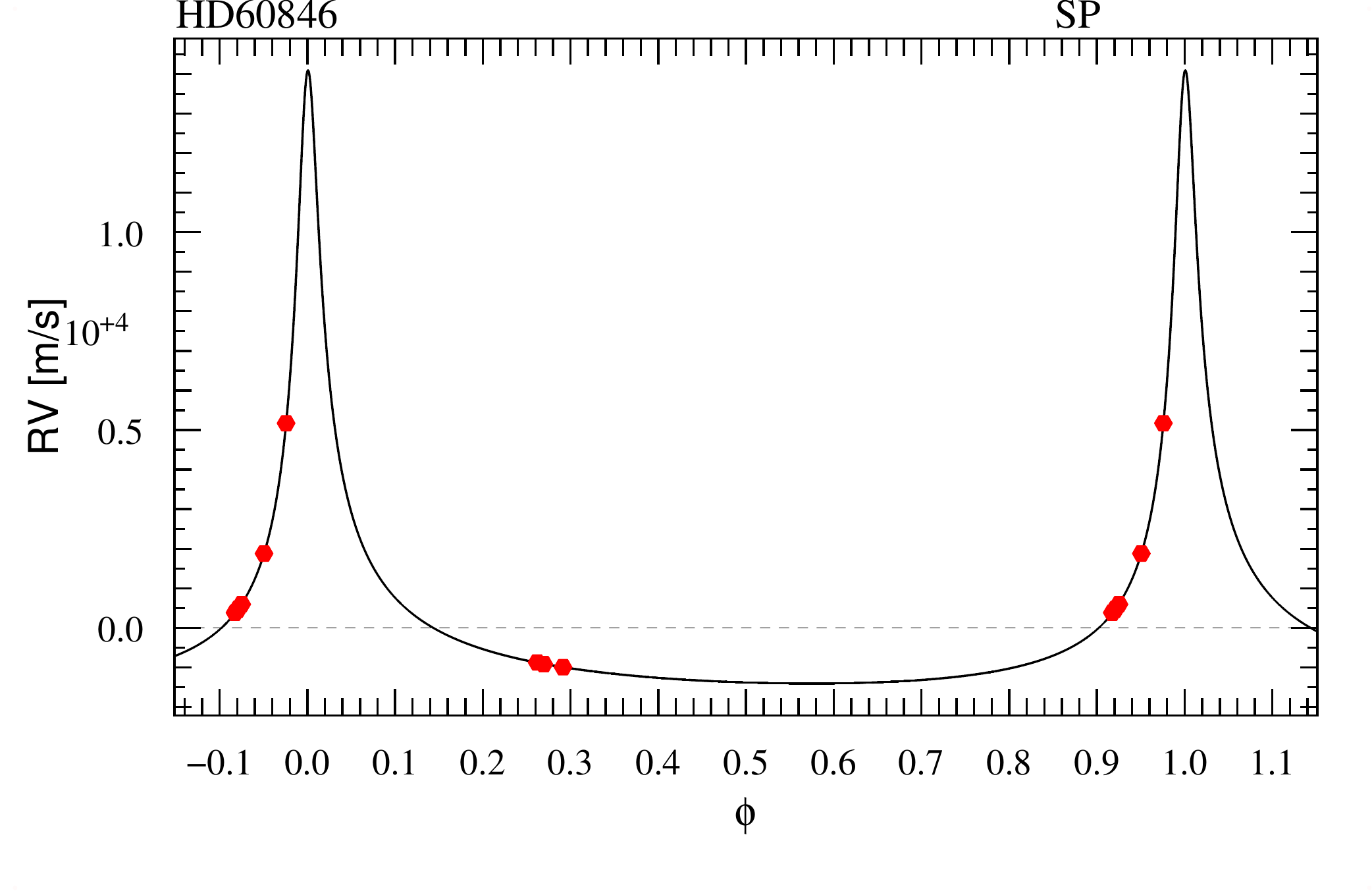} \\
\includegraphics[height=58mm, clip=true, trim=0 -12 0 7]{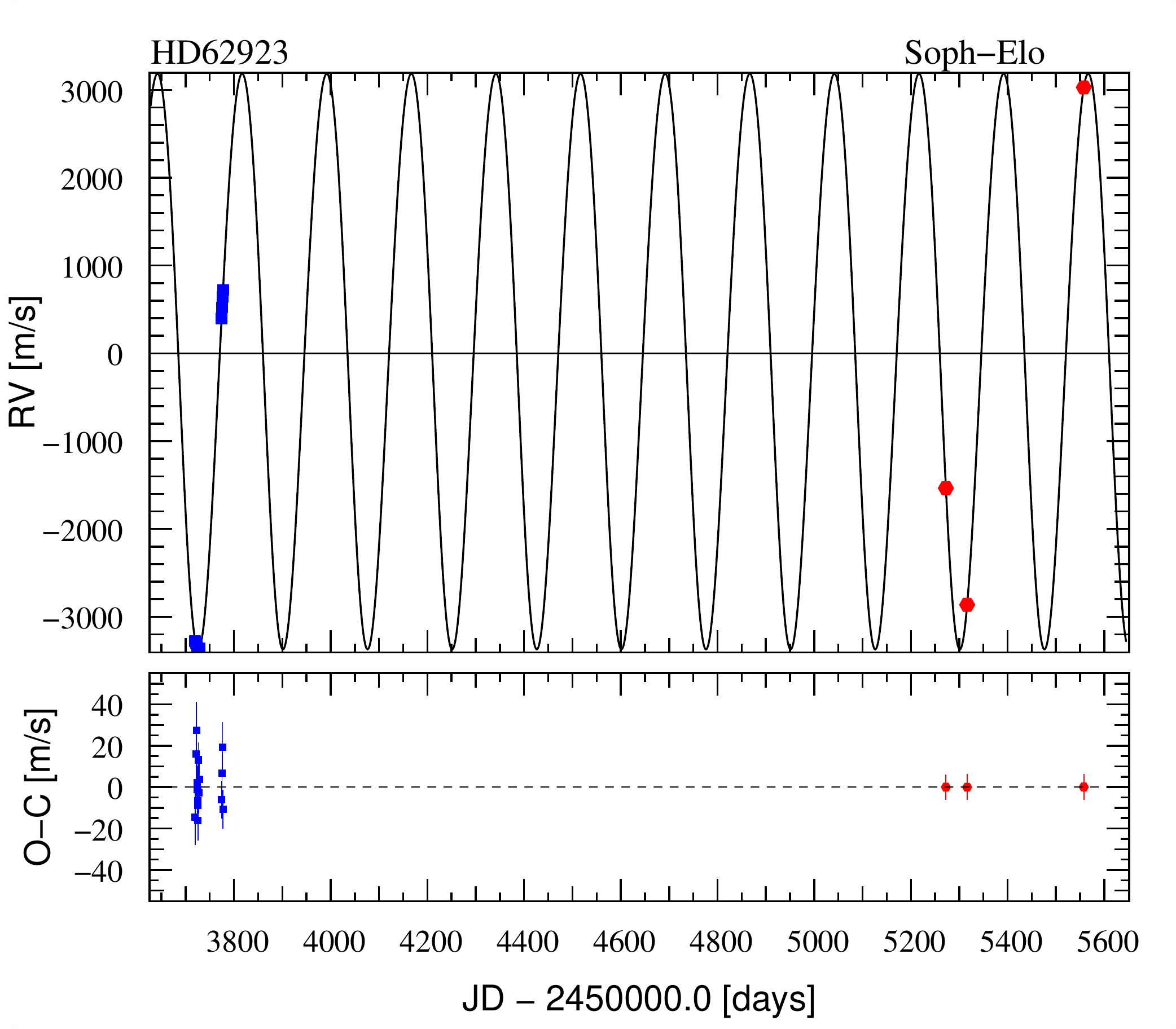} 
\includegraphics[height=57mm, clip=true, trim=0  25 0 0]{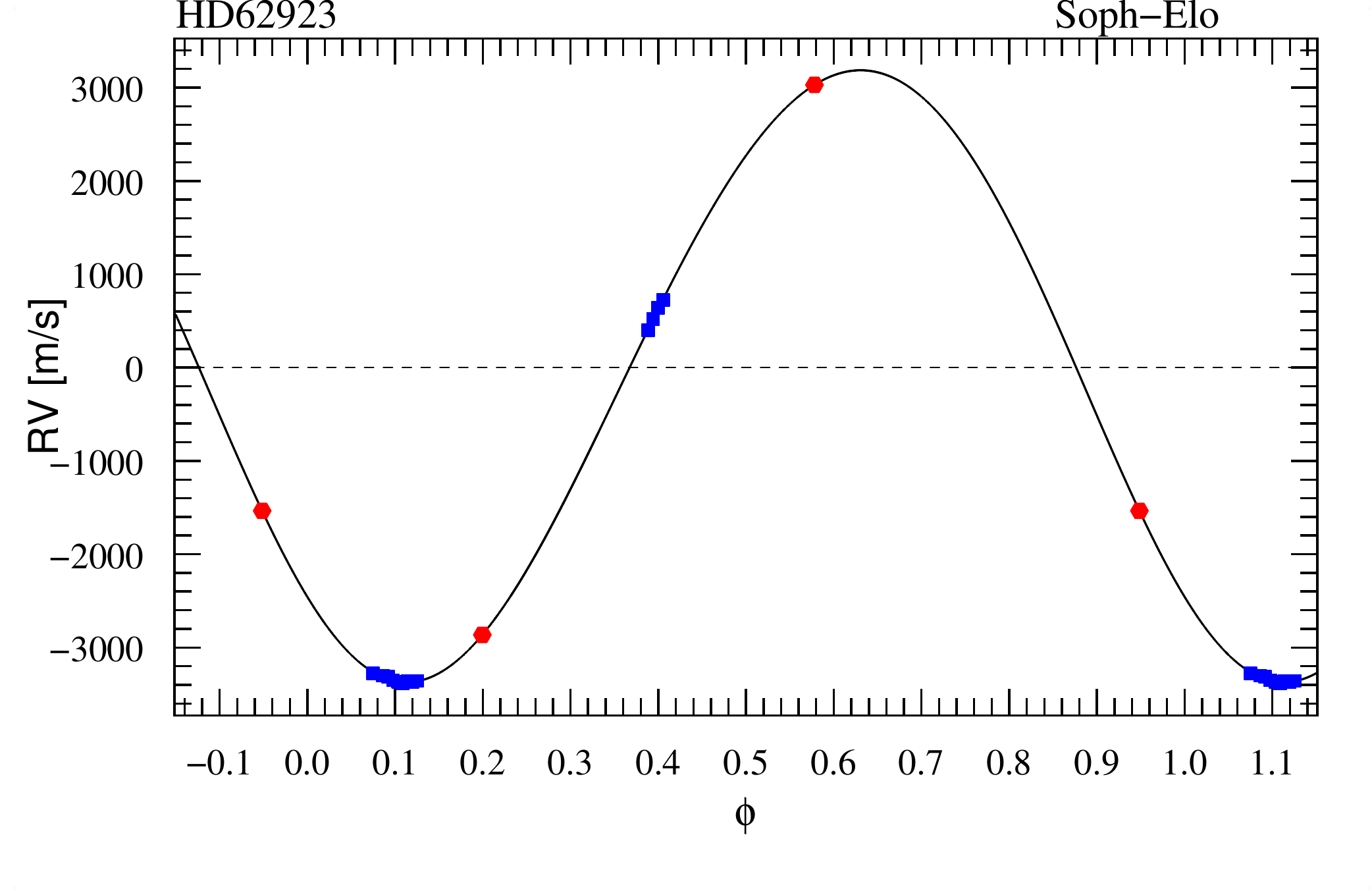} \\
\includegraphics[height=58mm, clip=true, trim=0 -12 0 7]{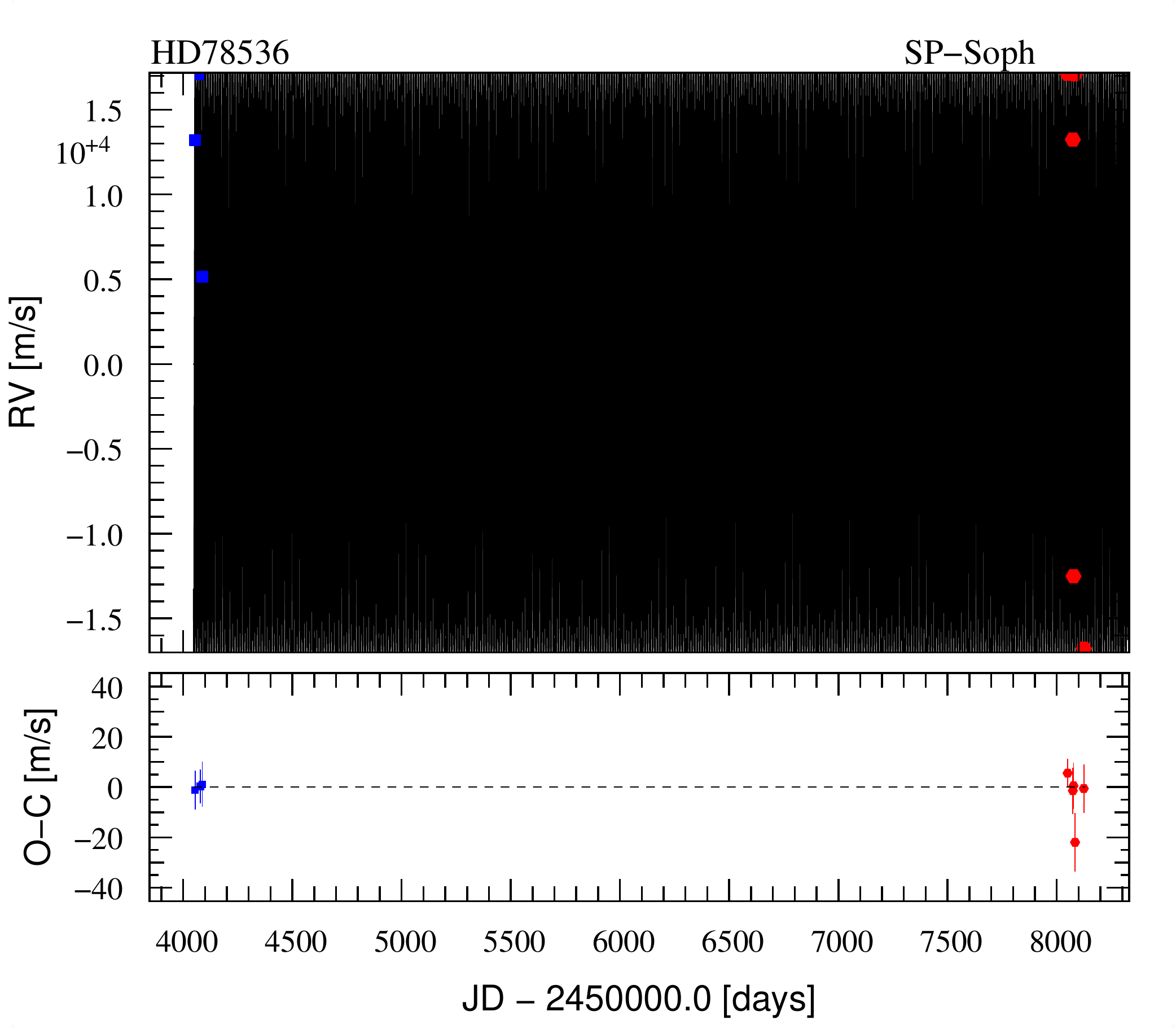}
\includegraphics[height=57mm, clip=true, trim=0  25 0 0]{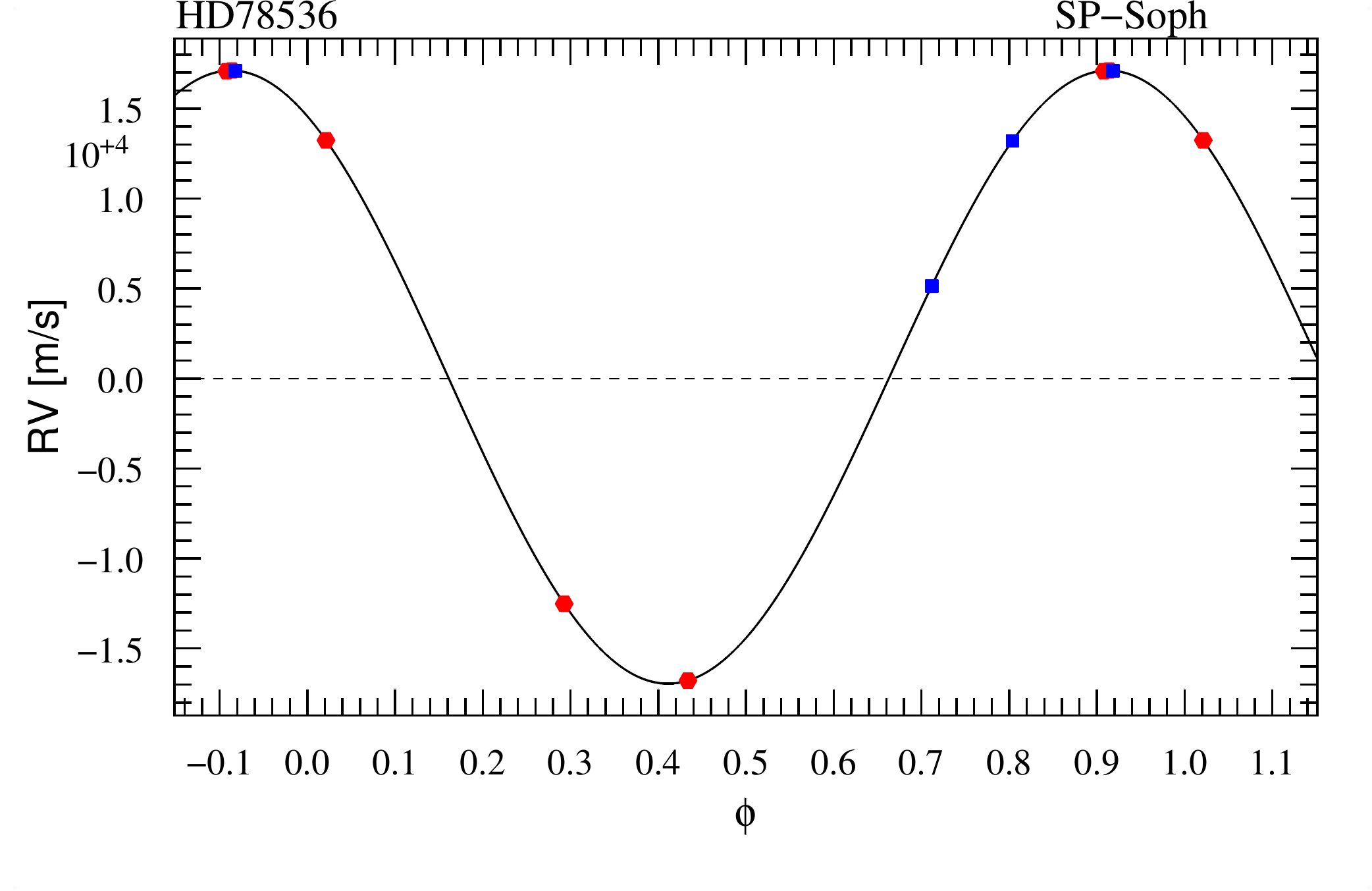} \\
\includegraphics[height=58mm, clip=true, trim=0 -12 0 7]{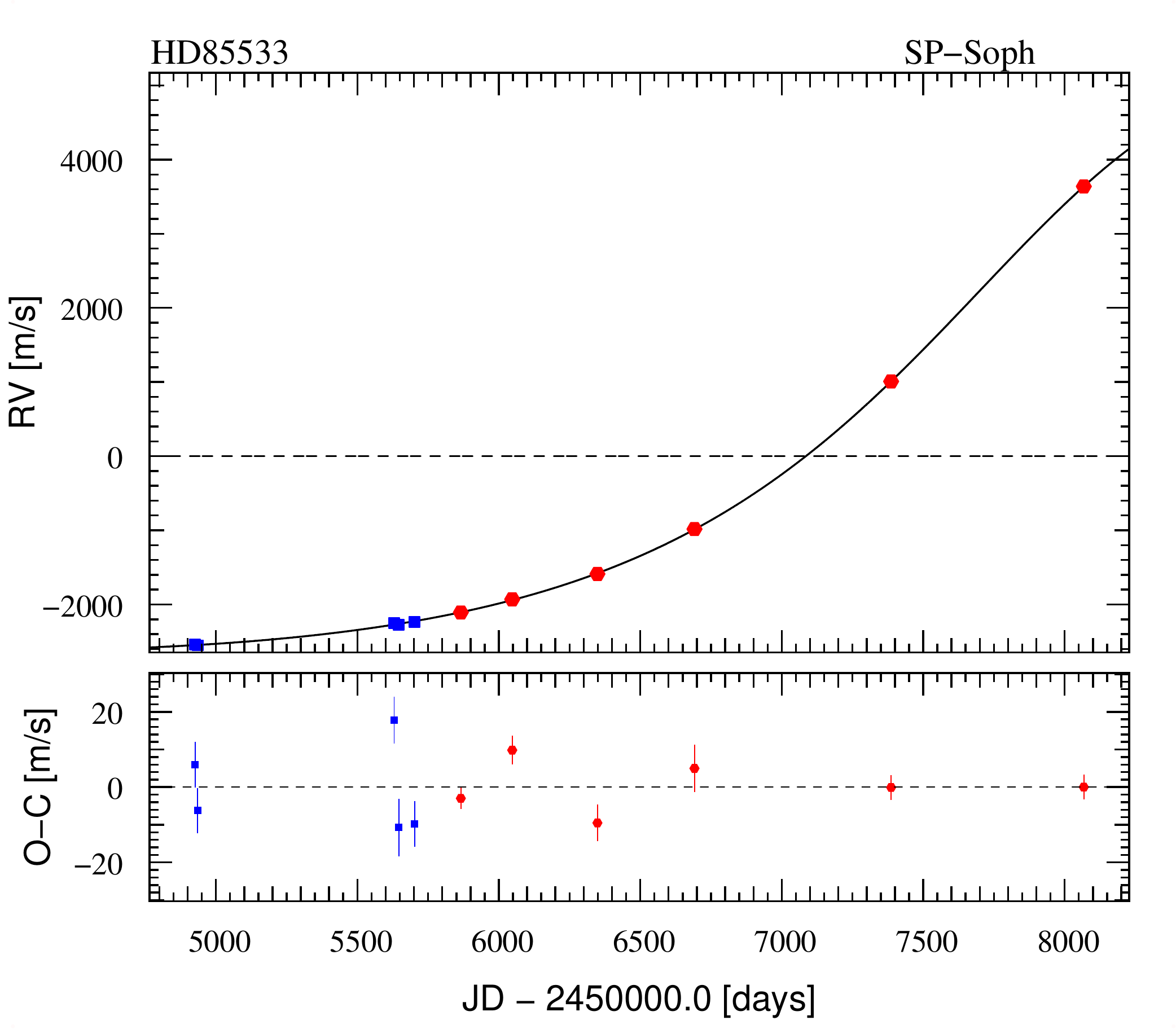}
\includegraphics[height=57mm, clip=true, trim=0  25 0 0]{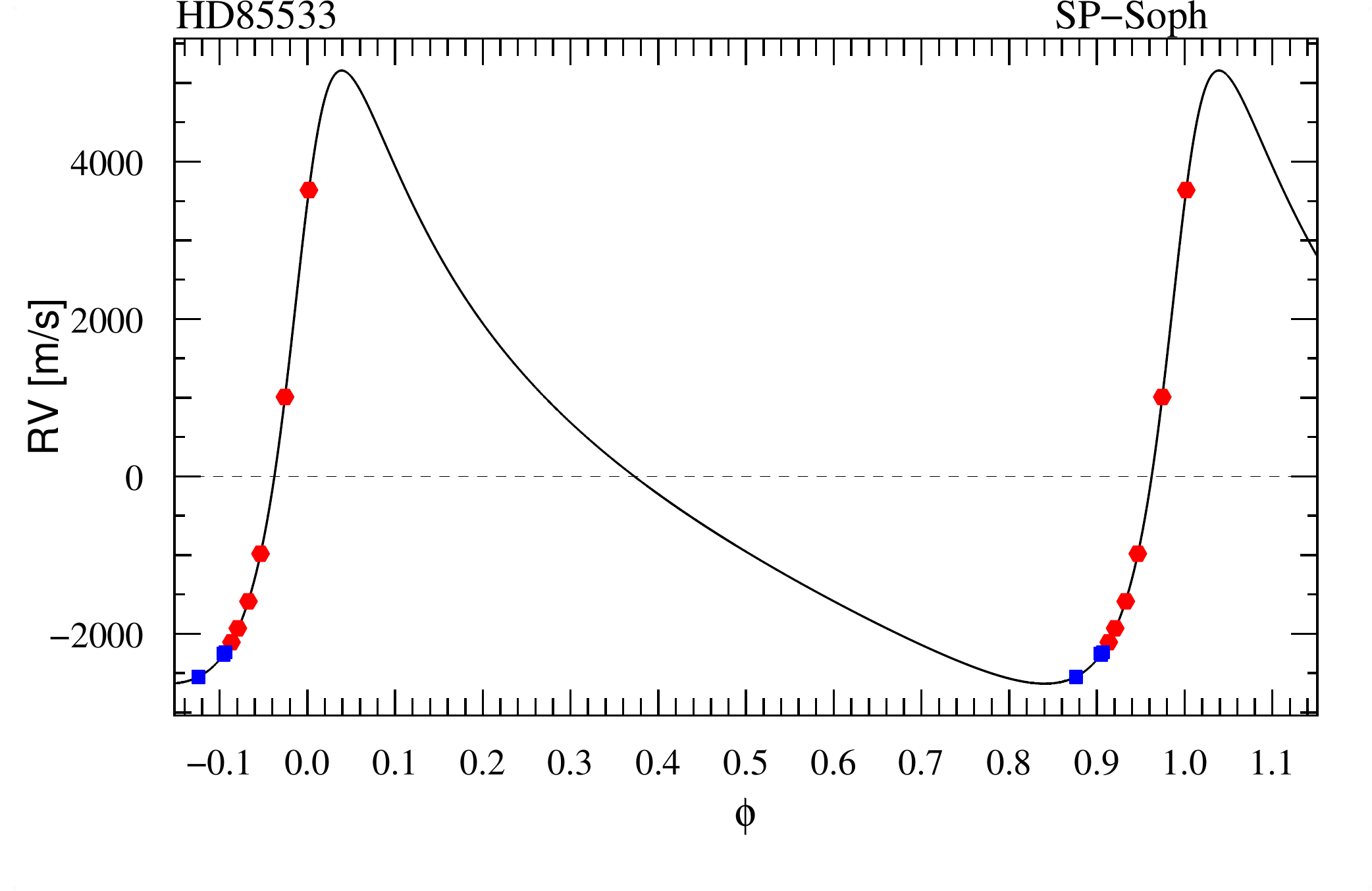} \\
\includegraphics[height=58mm, clip=true, trim=0 -12 0 7]{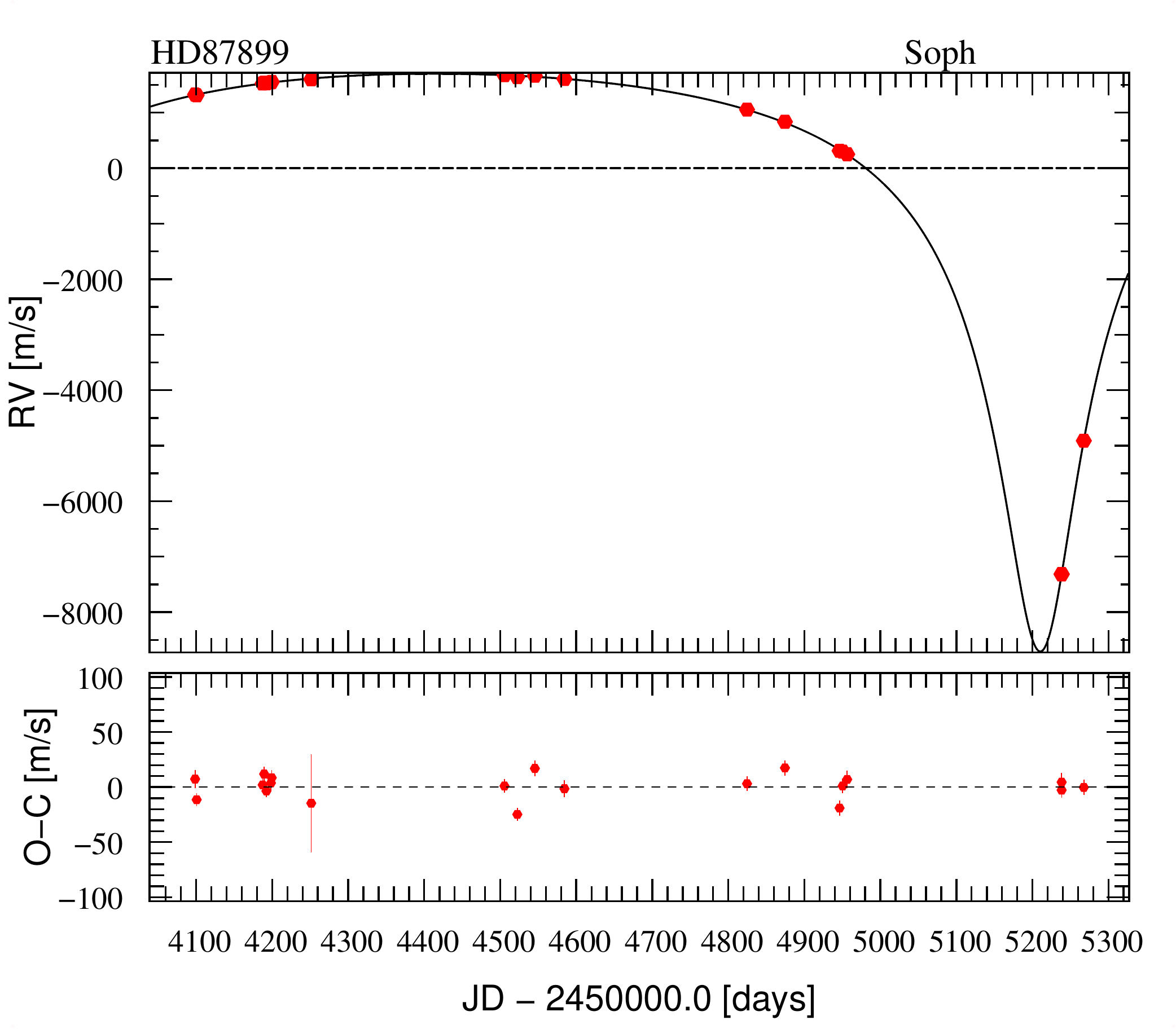}
\includegraphics[height=57mm, clip=true, trim=0  25 0 0]{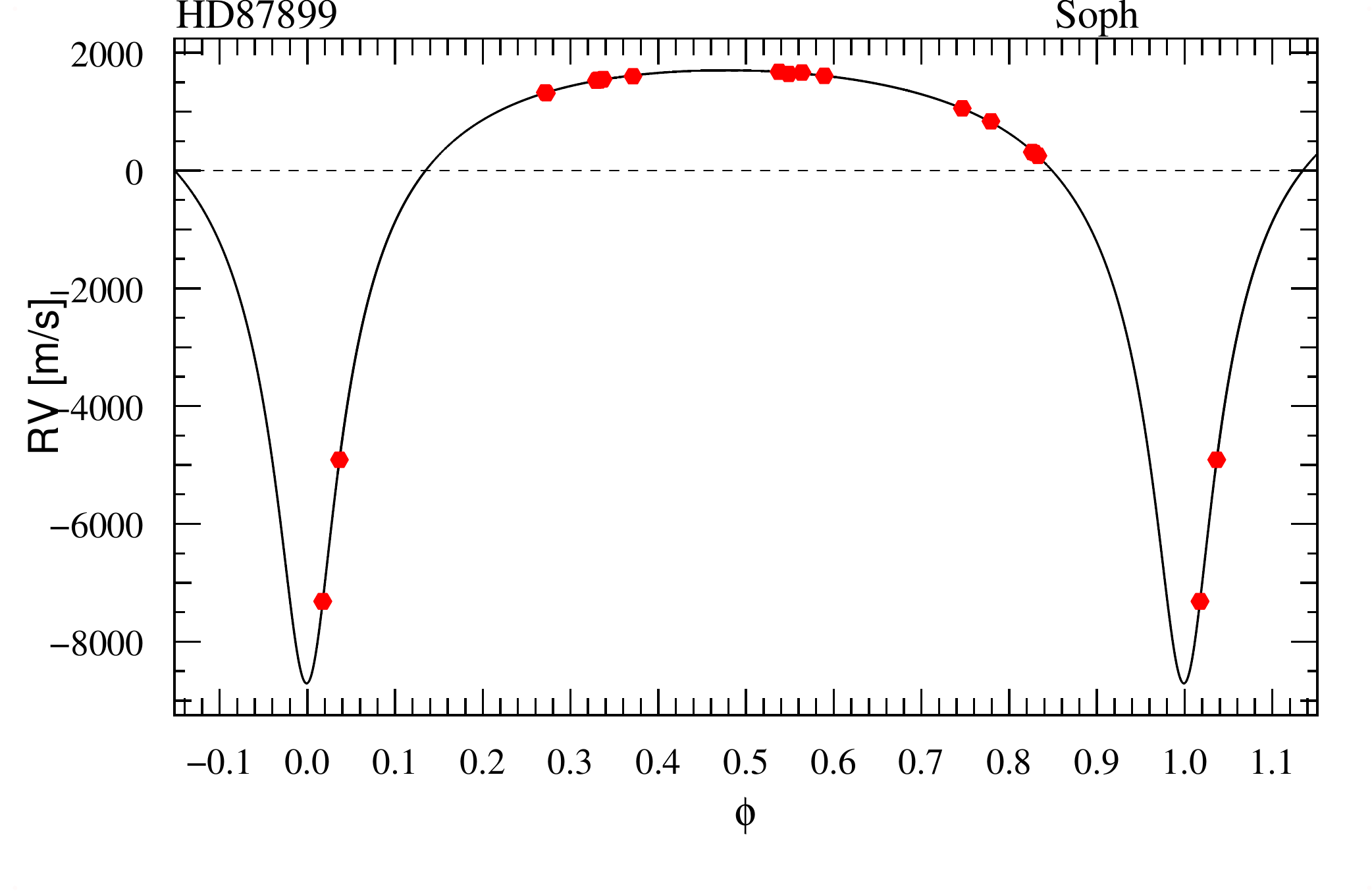} 
\end{longfigure}

\pagebreak
 \begin{figure}[ht]
 \begin{center} 
\includegraphics[width= 0.265\linewidth]{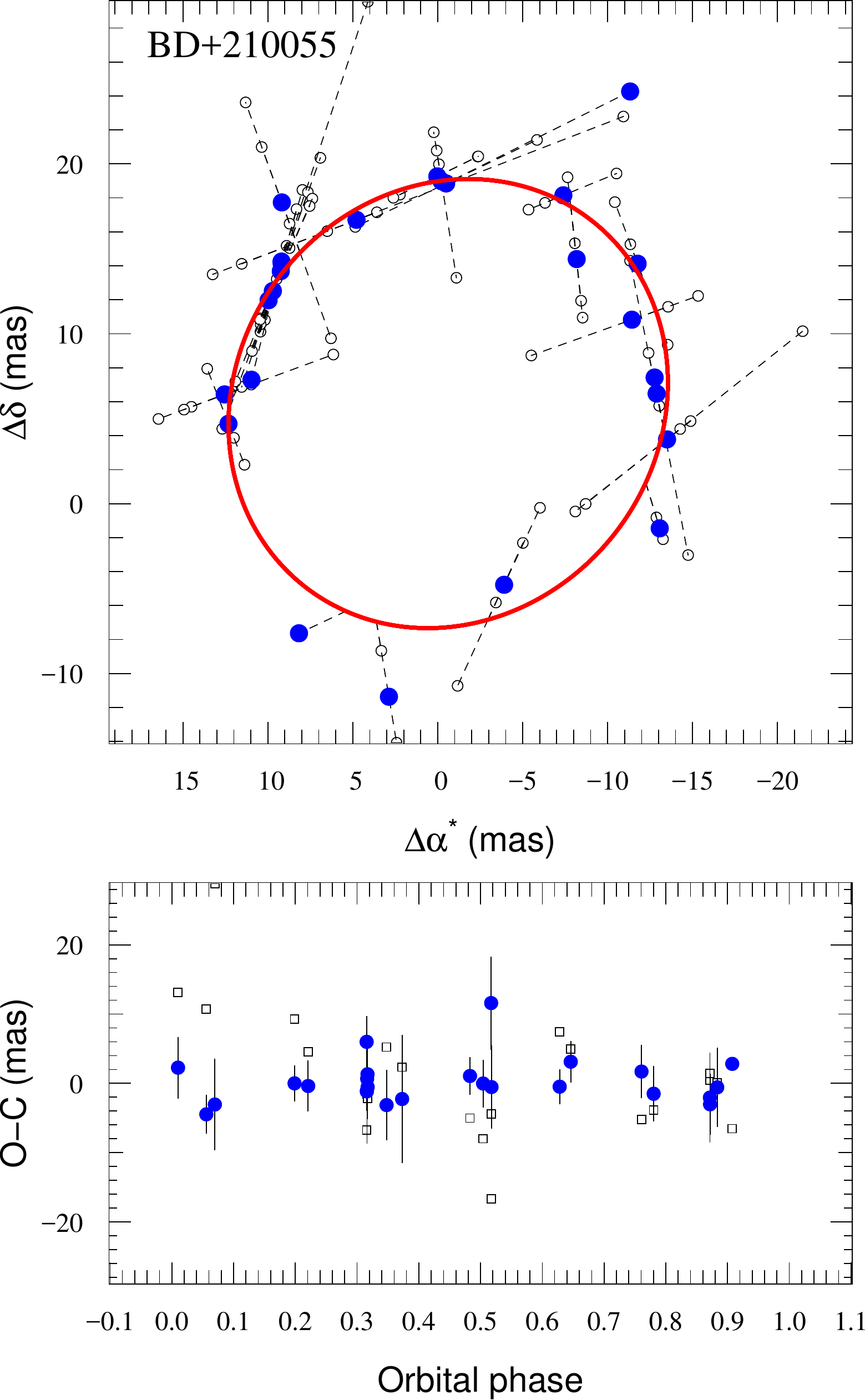} 
\includegraphics[width= 0.265\linewidth]{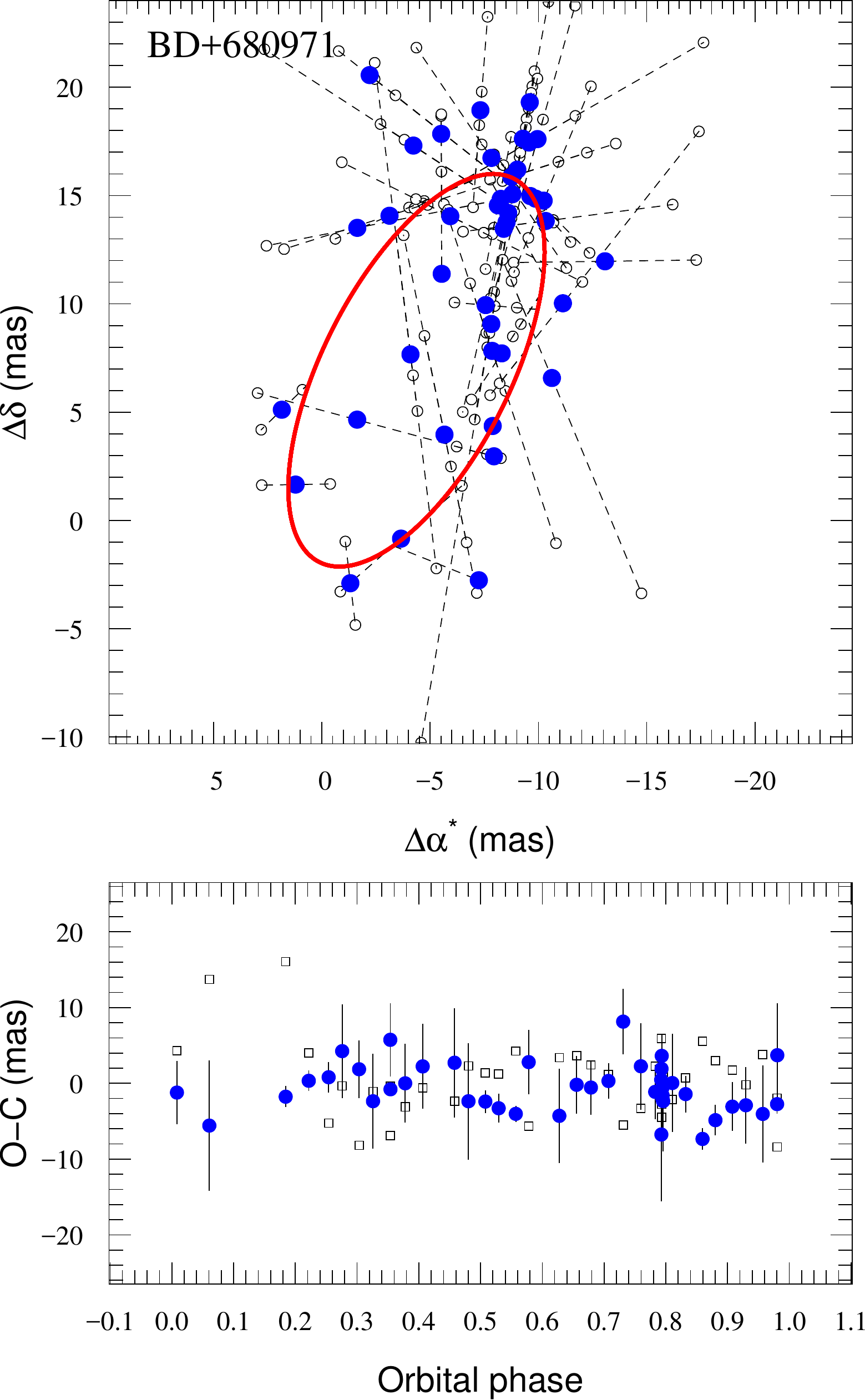} 
\includegraphics[width= 0.265\linewidth]{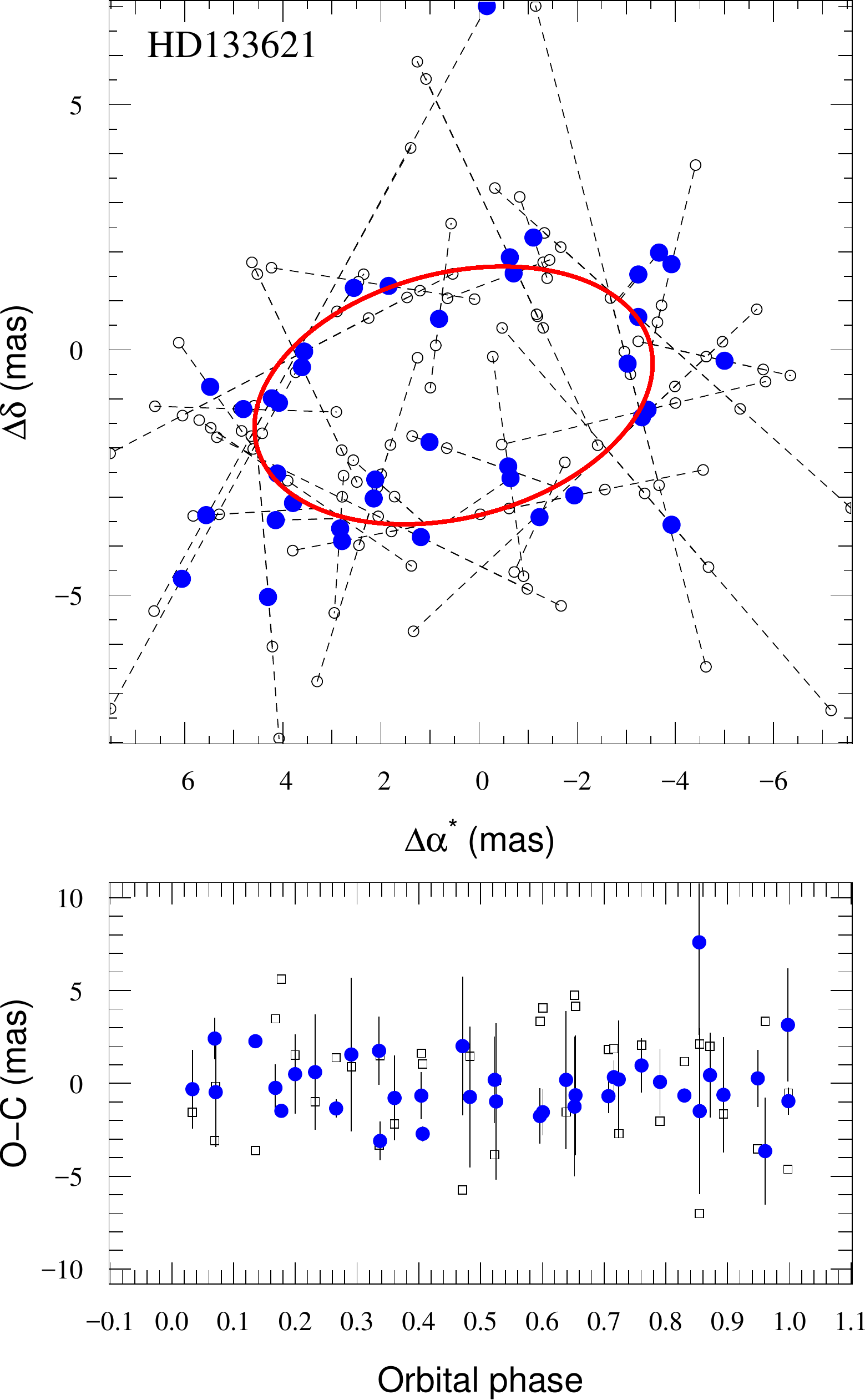} 
\includegraphics[width= 0.265\linewidth]{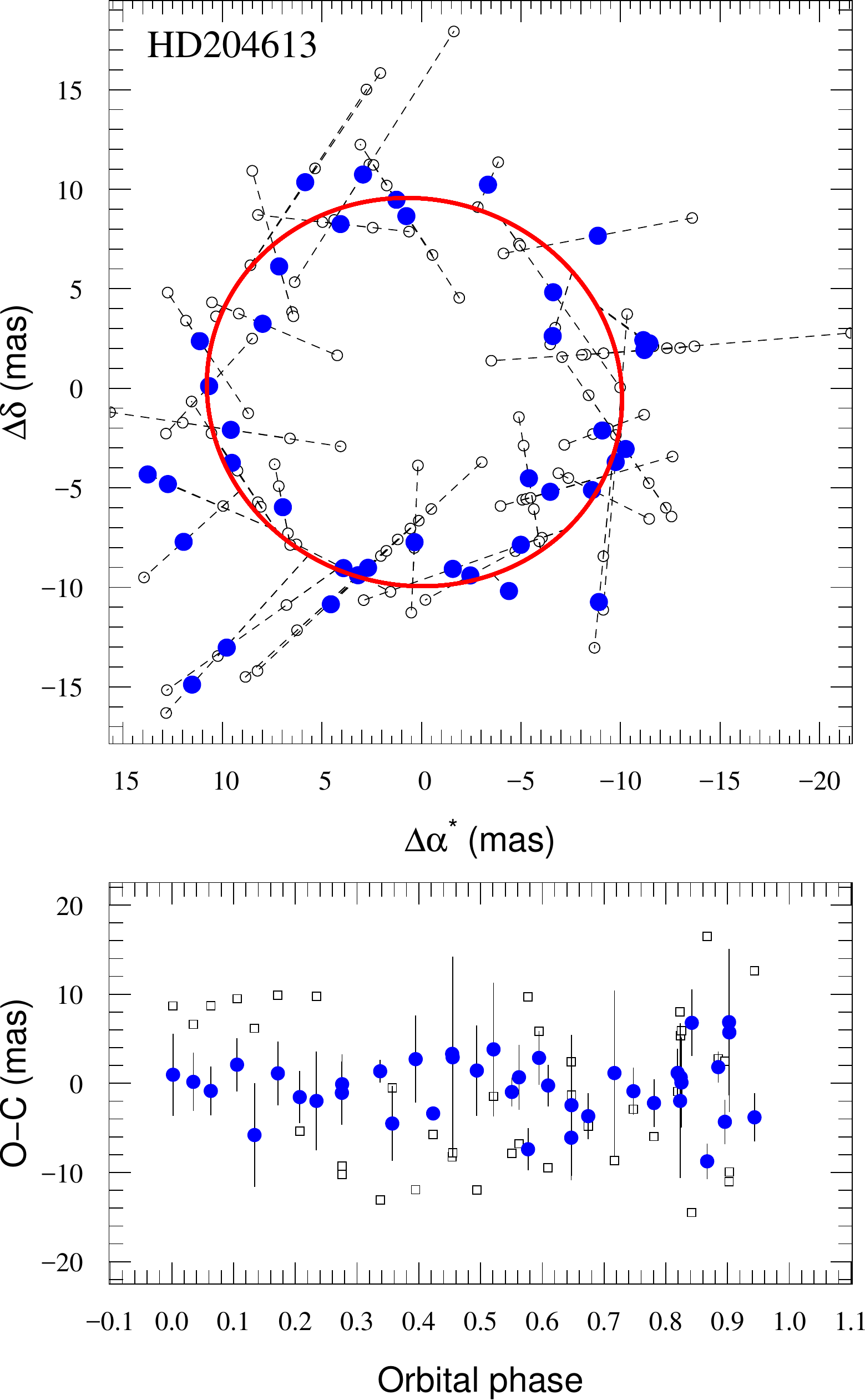} 
\includegraphics[width= 0.265\linewidth]{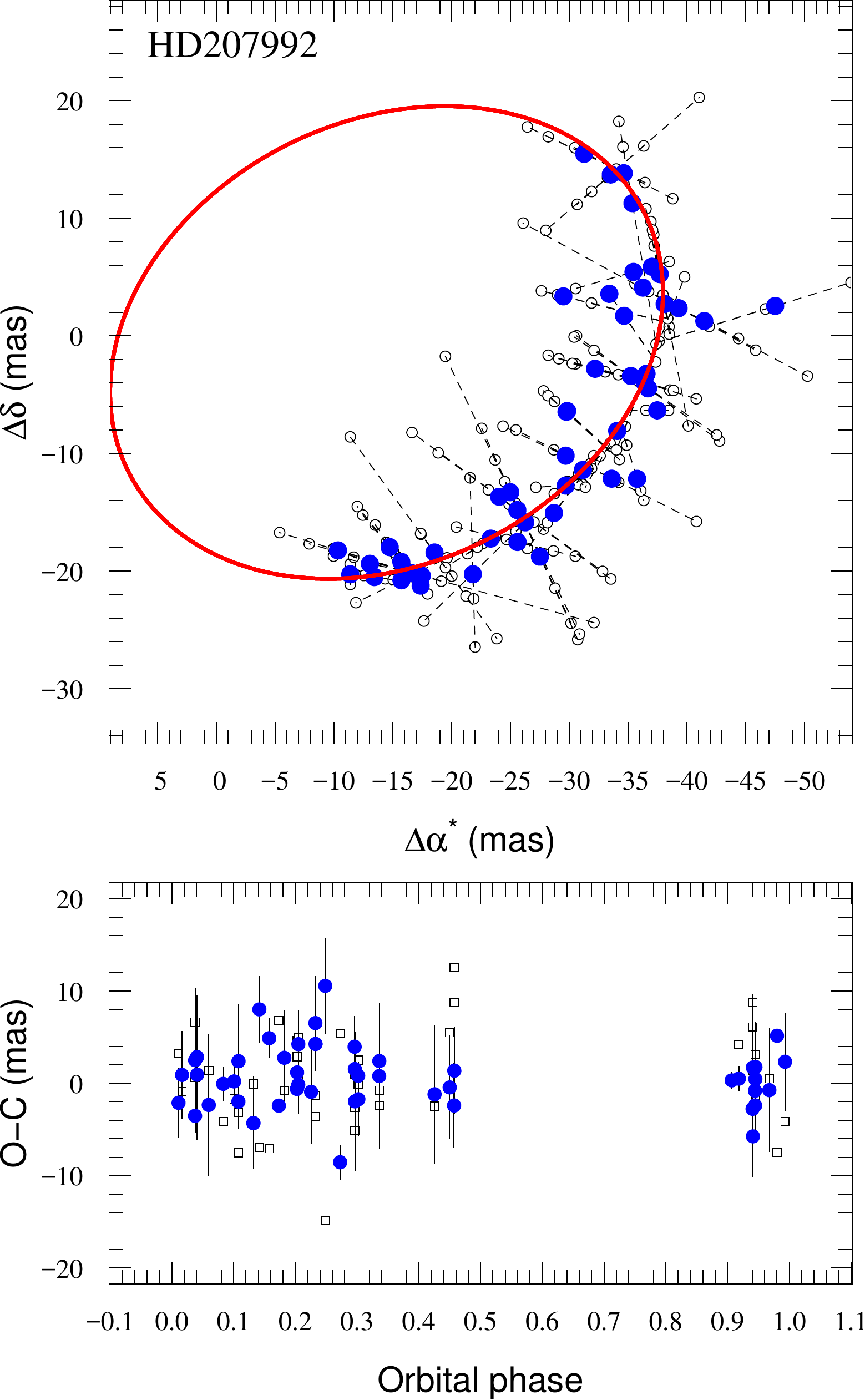} 
\includegraphics[width= 0.265\linewidth]{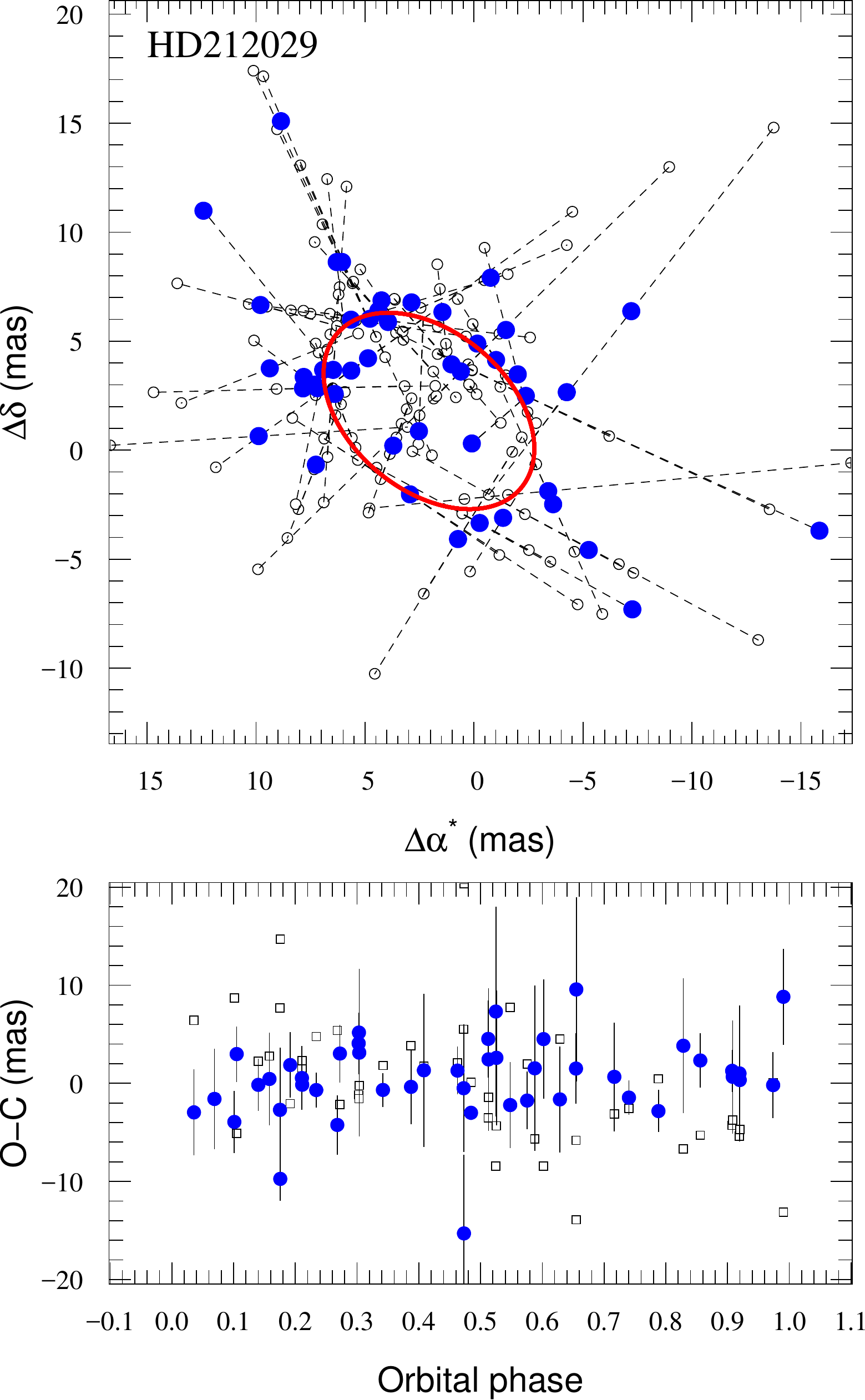} 
\includegraphics[width= 0.265\linewidth]{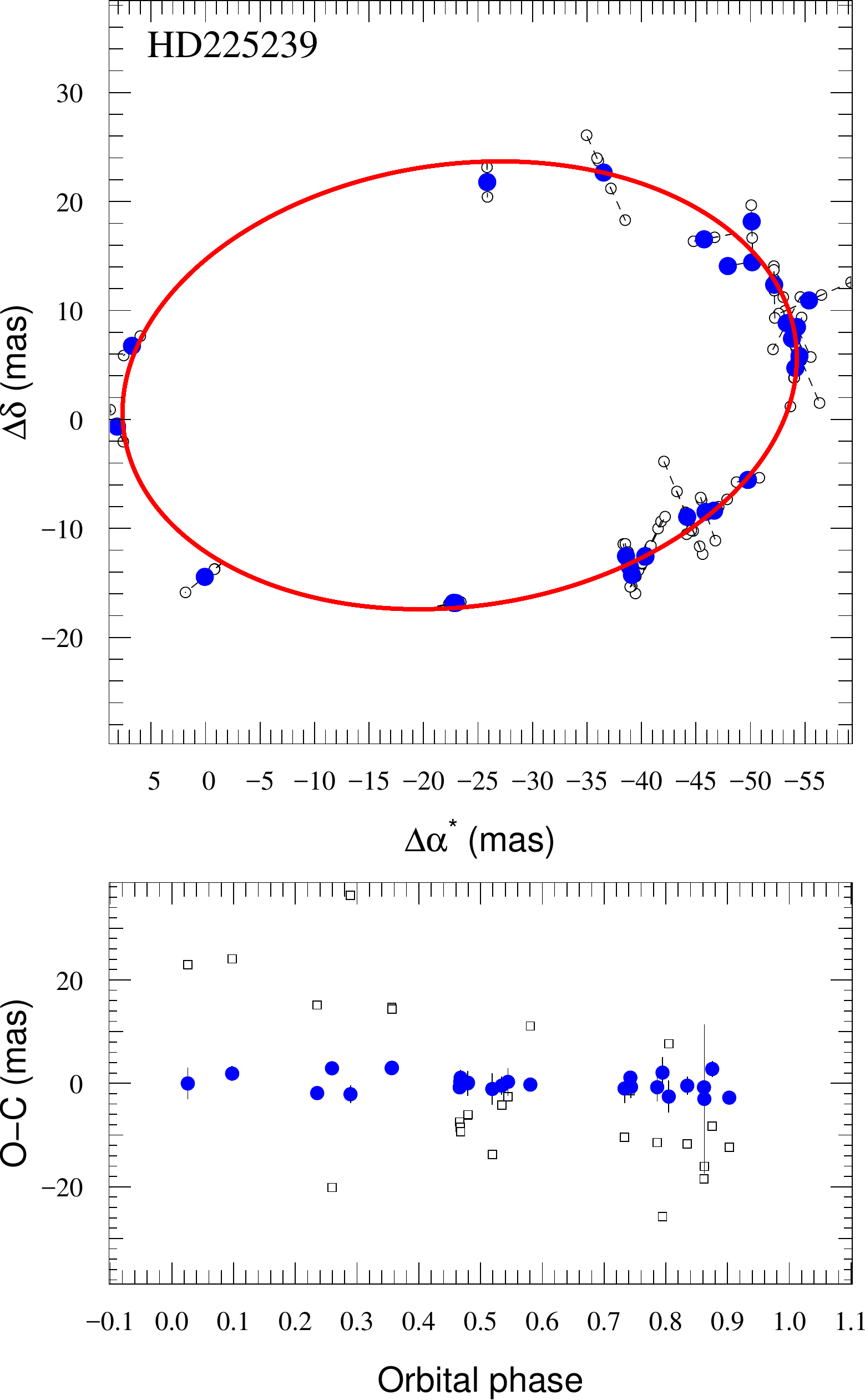} 
\includegraphics[width= 0.265\linewidth]{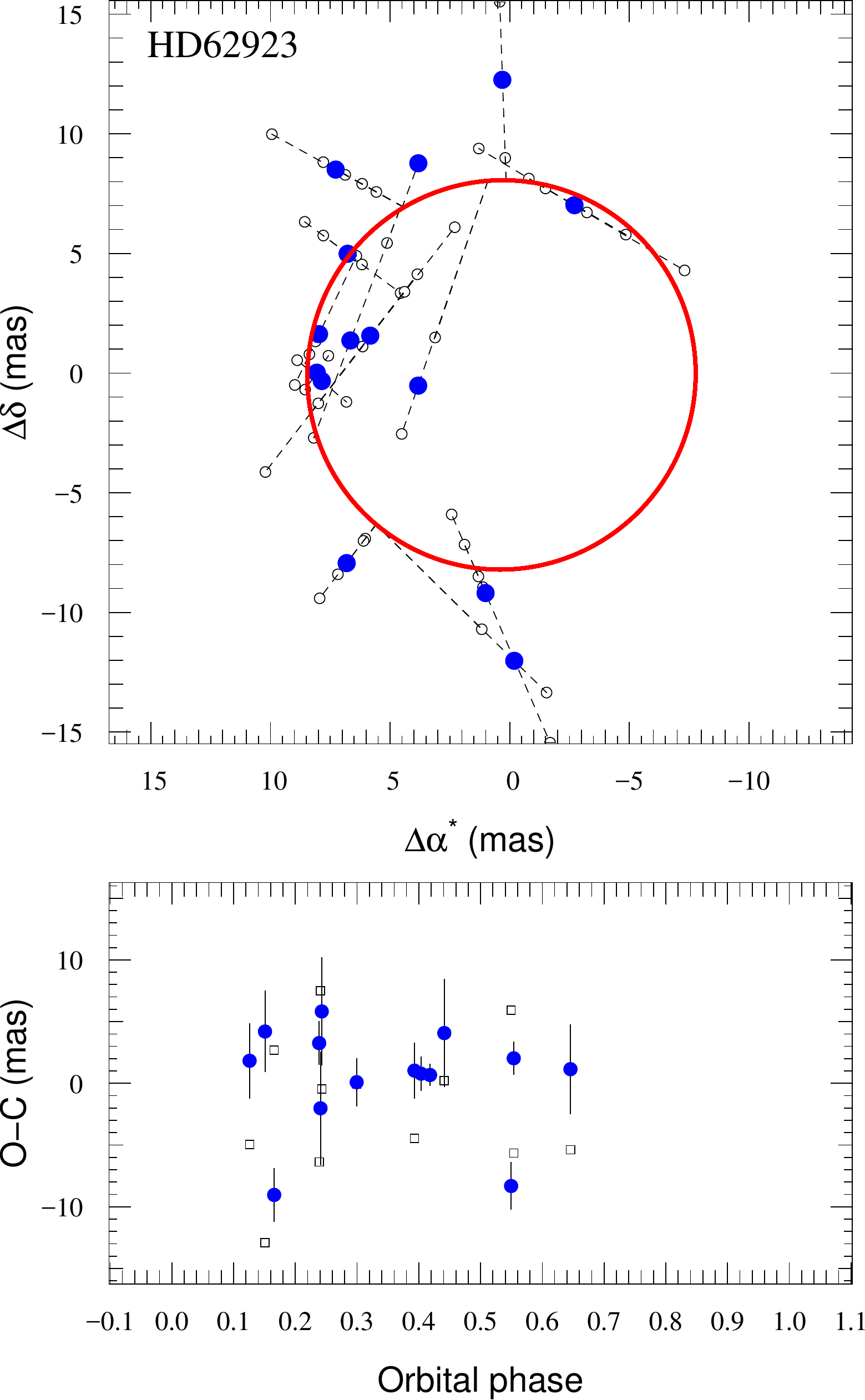} 
\includegraphics[width= 0.265\linewidth]{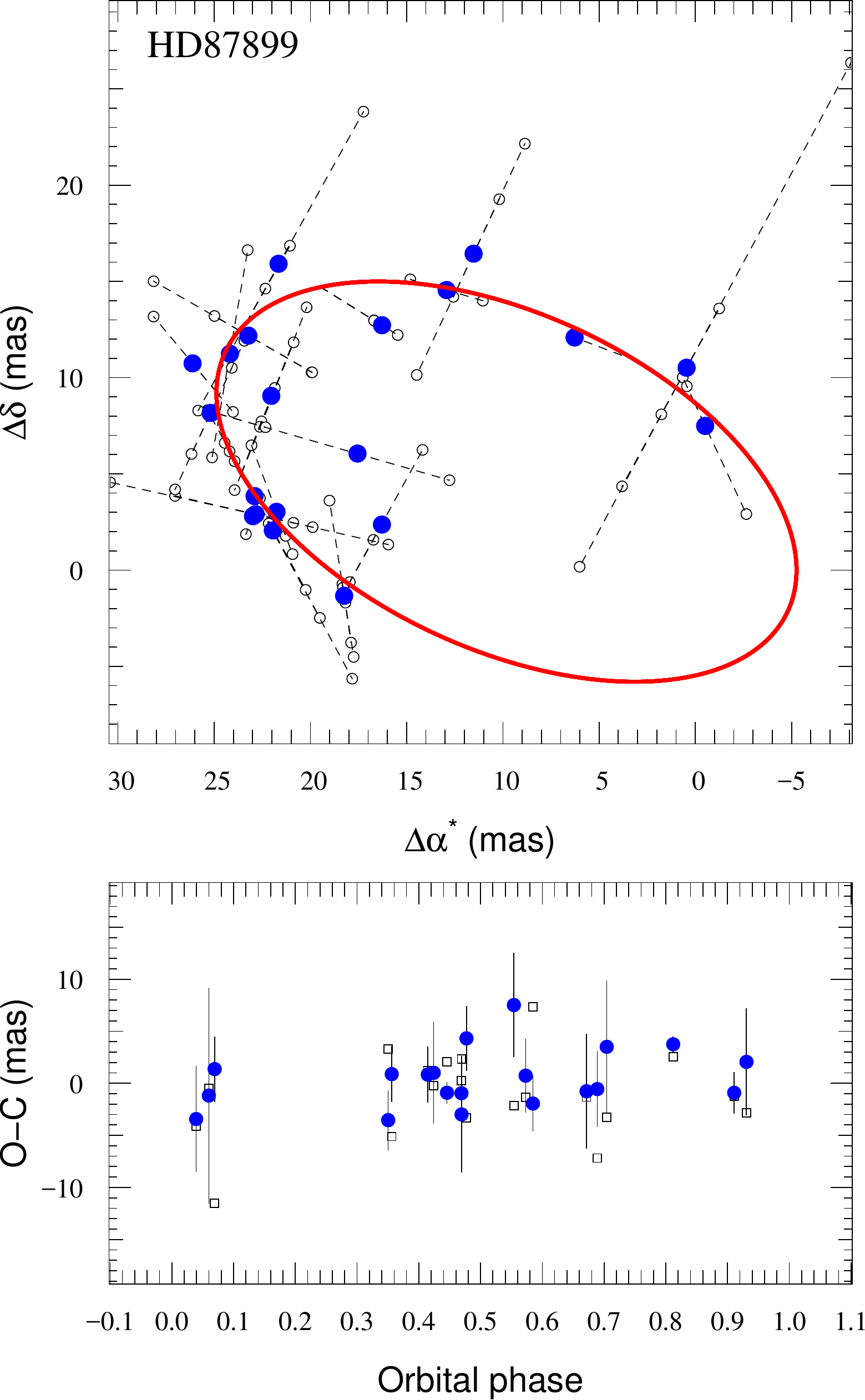} 
 \caption{Hipparcos astrometry 3-$\sigma$ detections in 9 systems. \emph{Top panels:} Modelled astrometric orbits projected on the sky. North is up and east is left. 
 The solid red line shows the model orbit and open circles mark the individual Hipparcos measurements. \emph{Bottom panels:} O--C residuals for the normal points of the orbital 
 solution (filled blue circles) and of the five-parameter model without companion (open squares).} 
\label{fig:orbits_3sigma}
 \end{center} 
 \end{figure}

 \begin{figure}[ht]
 \begin{center} 
\includegraphics[width= 0.3\linewidth]{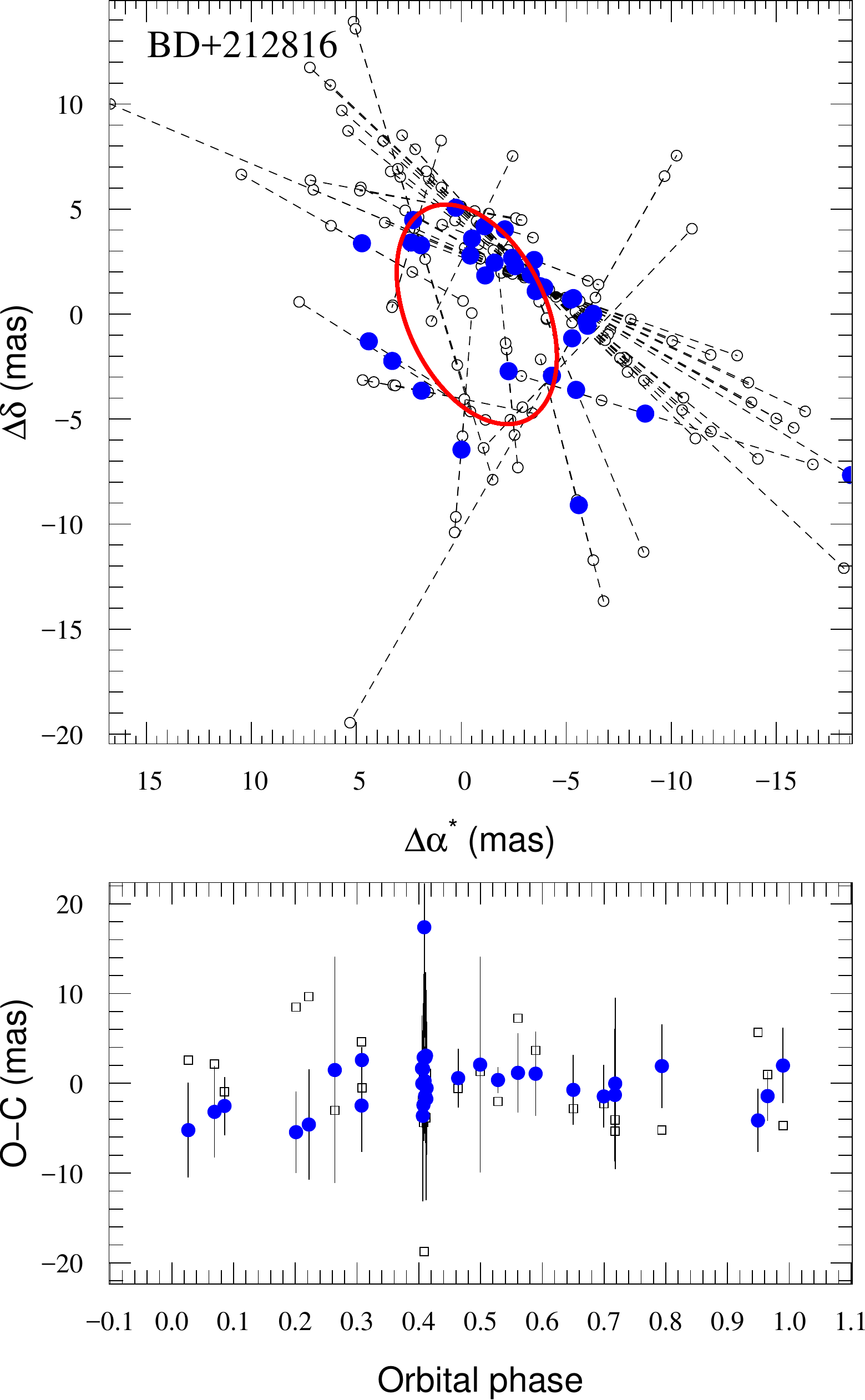} 
\includegraphics[width= 0.3\linewidth]{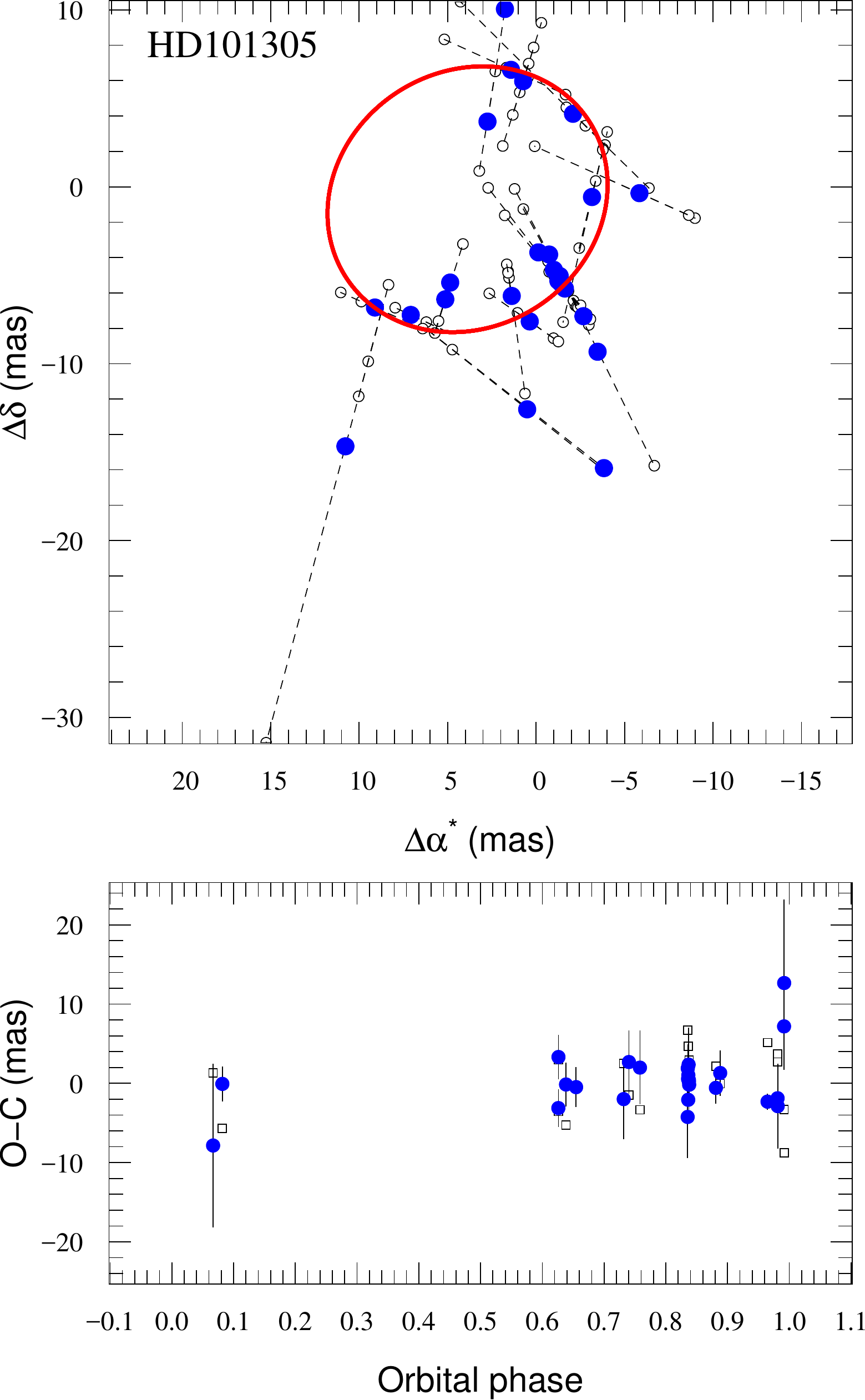} 
\includegraphics[width= 0.3\linewidth]{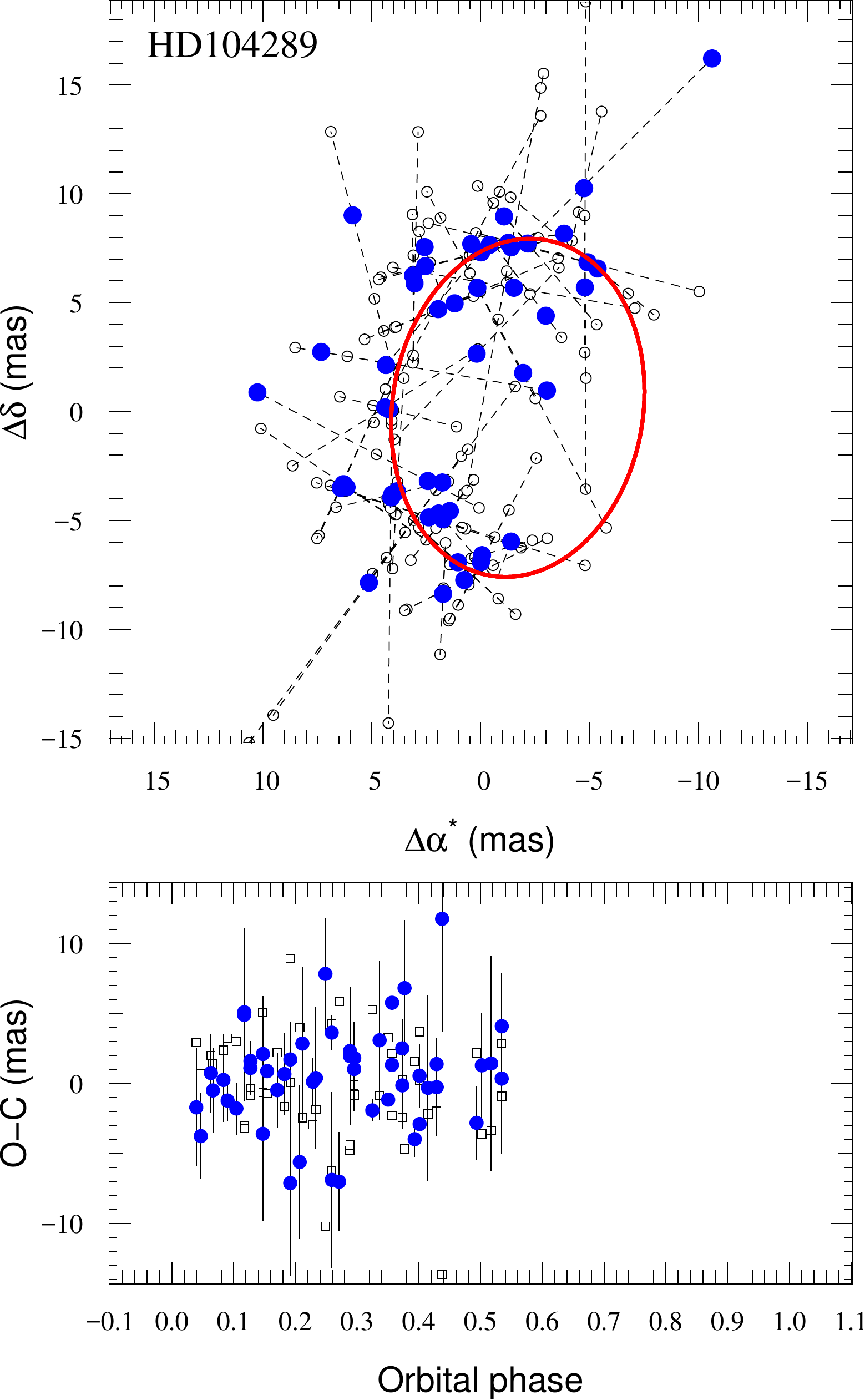}\vspace{0.5cm}
\includegraphics[width= 0.3\linewidth]{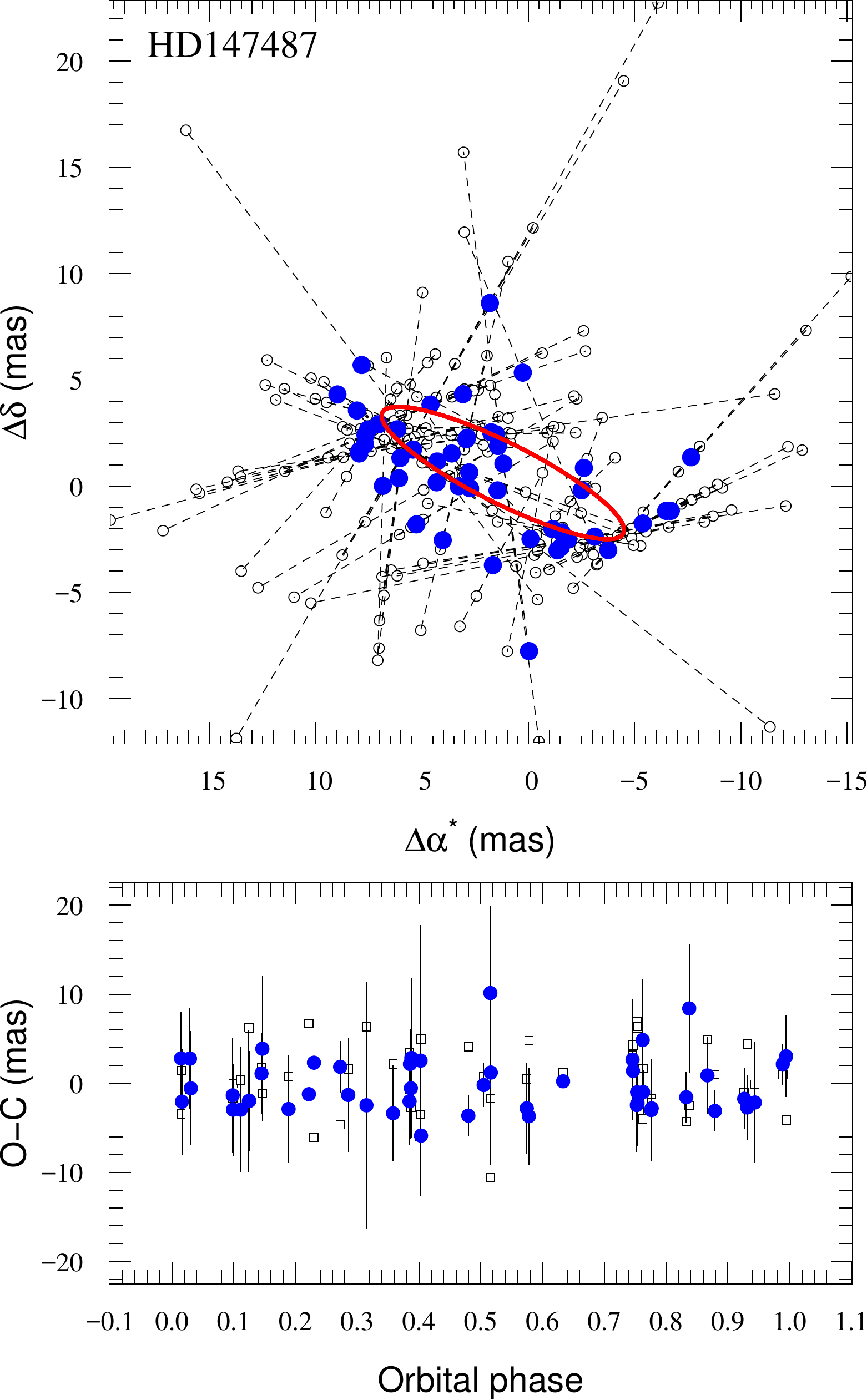} 
\includegraphics[width= 0.3\linewidth]{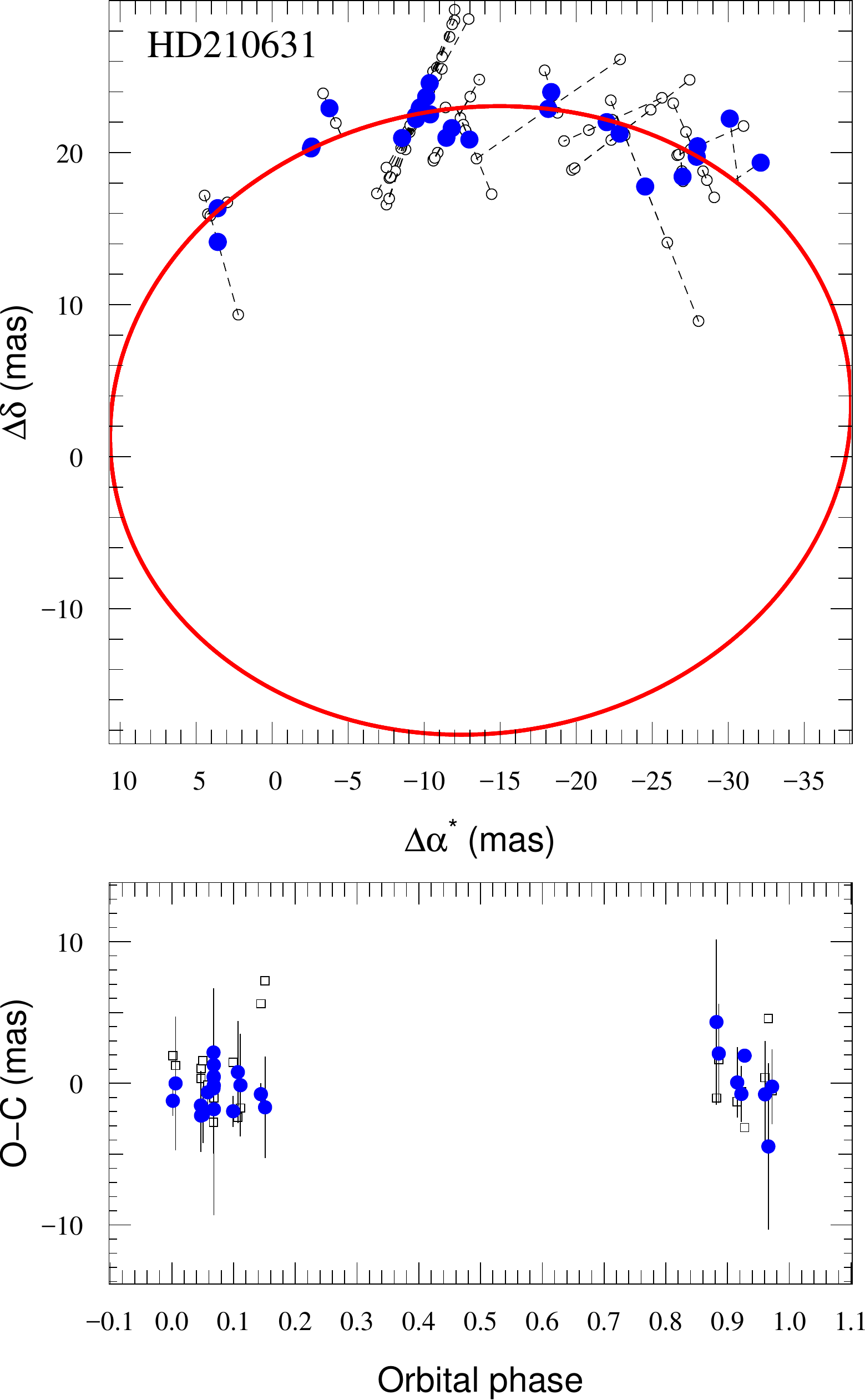} 
 \caption{Hipparcos astrometry 2-$\sigma$ detections in 5 systems. \emph{Top panels:} Modelled astrometric orbits projected on the sky, cf. Figure \ref{fig:orbits_3sigma}} 
\label{fig:orbits_2sigma}
 \end{center}
 \end{figure}
 
\end{appendix}

\end{document}